\documentclass[usenatbib]{mn2e}

\usepackage{amsmath}
\usepackage{natbib}
\usepackage{epsfig}
\usepackage{txfonts}

\newcommand{\xe}{x_{\rm e}}

\newcommand{\id}{{\,\rm d}}

\newcommand{\beq}{\begin{equation}}   %

\newcommand{\eeq}{\end{equation}}   %

\newcommand{\beqa}{\begin{eqnarray}}   %

\newcommand{\eeqa}{\end{eqnarray}}   %

\newcommand{\beal}{\begin{align}}

\newcommand{\enal}{\end{align}}

\newcommand{\bspl}{\begin{split}}

\newcommand{\espl}{\end{split}}

\newcommand{\bsub}{\begin{subequations}}

\newcommand{\esub}{\end{subequations}}

\newcommand{\bmulti}{\begin{multline}}   %

\newcommand{\beqm}{\begin{mathletters}}   %

\newcommand{\eeqm}{\end{mathletters}}   %

\newcommand{\kB}{k_{\rm B}}

\newcommand{\me}{m_{\rm e}}

\newcommand{\Ne}{N_{\rm e}}

\newcommand{\Te}{T_{\rm e}}

\newcommand{\Tg}{T_{\gamma}}

\newcommand{\The}{\theta_{\rm e}}

\newcommand{\sigT}{\sigma_{\rm T}}

\newcommand{\nPl}{n_{\rm Pl}}

\newcommand{\vek} [1]{\mbox{\boldmath${#1}$\unboldmath}}

\newcommand{\pd}{\partial}

\newcommand{\pAb}[2]{\frac{\displaystyle\pd #1}{\displaystyle\pd #2}}

\newcommand{\Abl}[2]{\frac{{\rm d} #1}{{\rm d} #2}}

\newcommand{\pot}[2]{#1 \times 10^{#2}}

\newcommand{\Thz}{\theta_{z}}

\newcommand{\ion}[2]{{\text{{\sc #1}\,{\sc #2}}}}

\newcommand{\Tz}{{T_{z}}}
\newcommand{\Tav}{{T_{\rm av}}}

\newcommand{\TCMB}{{T_{\rm CMB}}}

\newcommand{\zmu}{{z_{\mu}}}
\newcommand{\zmuy}{{z_{\mu,y}}}
\newcommand{\ye}{{y_{\rm e}}}
\newcommand{\yff}{{y_{\rm ff}}}

\newcommand{\zs}{{z_{\rm s}}}
\newcommand{\ze}{{z_{\rm e}}}

\newcommand{\nS}{n_{\rm S}}

\newcommand{\nrun}{n_{\rm run}}

\newcommand{\Ngpl}{N^{\rm pl}_{\gamma}}

\newcommand{\nut}{{\tilde{\nu}}}

\newcommand{\vbetac}{{{\boldsymbol\beta}_{\rm p}}}
\newcommand{\vbetach}{{\hat{{\boldsymbol\beta}}_{\rm p}}}
\newcommand{\betac}{\beta_{\rm p}}

\newcommand{\betacsq}{{\beta^2_{\rm p}}}

\newcommand{\muc}{\mu_{\rm p}}

\newcommand{\vecJ}[1]{{\bf #1}}
\newcommand{\vecx}{{\vecJ{x}}}
\newcommand{\vecxc}{{\vecJ{x}_{\rm c}}}

\newcommand{\gh}{{\hat{\gamma}}}

\newcommand{\vgh}{{\hat{\boldsymbol\gamma}}}
\newcommand{\vghp}{{\hat{\boldsymbol\gamma}'}}

\newcommand{\vbh}{{\boldsymbol{\hat{\beta}}}}

\newcommand{\vbc}{{\boldsymbol{\beta}_{\rm p}}}

\newcommand{\gammac}{{\gamma_{\rm p}}}

\newcommand{\vp}{{\vecJ{p}}}
\newcommand{\vpp}{{\vecJ{p'}}}

\newcommand{\mus}{{\mu_{\rm s}}}

\newcommand{\Deltaxe}{\Delta \xe}
\newcommand{\Deltaye}{\Delta y_{\rm e}}

\newcommand{\xg}{x_{\gamma}}
\newcommand{\xav}{x_{\rm av}}

\newcommand{\Theav}{\theta_{\rm e, av}}
\newcommand{\Teav}{T_{\rm e, av}}
\newcommand{\Thez}{\theta_{z}}
\newcommand{\taudot}{\dot{\tau}}
\newcommand{\taudotc}{\dot{\tau}_{\rm c}}
\newcommand{\kD}{k_{\rm D}}
\newcommand{\rs}{r_{\rm s}}
\newcommand{\cs}{c_{\rm s}}

\newcommand{\ygl}[1]{y_{\geq #1}^{\rm g}}

\usepackage{color}

\usepackage{hyperref}
\usepackage{grffile}
\usepackage{graphics}
\usepackage{booktabs}

\title[Acoustic Damping]
{CMB at 2$\times$2 order: the dissipation of primordial 
acoustic waves and the observable part of the associated energy release}

\author[Chluba, Khatri and Sunyaev]{J. Chluba$^{1}$\thanks{E-mail:
  jchluba@cita.utoronto.ca},  R. Khatri$^{2}$\thanks{E-mail:
 rkhatri@mpa-garching.mpg.de} and R.~A. Sunyaev$^{2,3,4}$ 
  \\
  \\
$^{1}$ Canadian Institute for Theoretical Astrophysics, 60 St. George Street,
Toronto, ON M5S 3H8, Canada
\\
$^{2}$ Max-Planck Institut f\"ur Astrophysik, Karl-Schwarzschild-Str. 1,
D-85740 Garching, Germany
\\
$^{3}$ Space Research Institute, Russian Academy of Sciences, Profsoyuznaya 84/32,
117997 Moscow, Russia
\\
$^{4}$ Institute for Advanced Study, Einstein Drive, Princeton, New Jersey 08540, USA
}

\begin{document}

\date{Submitted 2012 January 31; Accepted 2012 June 7}

\maketitle

\begin{abstract}
Silk damping of primordial small-scale perturbations in the photon-baryon fluid due to 
diffusion of photons inevitably creates spectral distortions in the CMB. With 
the proposed CMB experiment PIXIE it might become possible to measure these distortions 
and thereby constrain the primordial power spectrum at comoving wavenumbers 
$50\,{\rm Mpc}^{-1}\lesssim k \lesssim 10^4\,\rm{Mpc}^{-1}$. 
Since primordial fluctuations in the CMB on these scales are completely erased by Silk damping, these distortions may provide the only way to shed light on otherwise unobservable aspects of inflationary physics. 
A consistent treatment of the primordial dissipation problem requires going to
second order in perturbation theory, while thermalization of these distortions 
necessitates consideration of second order in Compton scattering energy
transfer. 
Here we give a full 2$\times$2 treatment for the creation and evolution of spectral distortions due to the acoustic dissipation process, consistently including the effect of polarization and photon mixing in the free-streaming regime.
We show that 1/3 of the total energy (9/4 larger than previous estimates) stored in small-scale temperature
perturbations imprints observable spectral distortions, while the remaining 2/3 only 
raises the average CMB temperature, an effect that is unobservable.
At high redshift dissipation is mainly mediated through the quadrupole anisotropies, 
while after recombination peculiar motions are most important. During recombination
 the damping of the higher multipoles is also significant.
We compute the average distortion for several examples using {\sc CosmoTherm}, 
analyzing their dependence on parameters of the primordial power spectrum.
For one of the best fit WMAP7 cosmologies, with $\nS=1.027$ and $\nrun=-0.034$, the cooling 
of baryonic matter practically compensates the heating from acoustic dissipation in 
the $\mu$-era.
We also derive the evolution equations for anisotropic spectral
  distortions in first order perturbation theory. We furthermore argue
  that the first order anisotropies of spectral distortions may dominate over the corresponding second order contributions from recombination if an average fractional distortion $\simeq 10^{-5}$ is already present before recombination.
\end{abstract}



\section{Introduction}
\label{sec:Intro}
The energy spectrum of the cosmic microwave background (CMB), as measured with with 
COBE/FIRAS \citep{Mather1994, Fixsen1996}, is Planckian with extremely high precision.
This is direct evidence that the Universe was dense and hot at early times, a fact that constitutes one of the most important foundations of big bang cosmology. 
Possible deviations from a blackbody spectrum were constrained to be $\mu\lesssim\pot{9}{-5}$ and $y \lesssim\pot{1.5}{-5}$ \citep{Fixsen1996} for chemical potential and Compton-$y$ distortions \citep{Zeldovich1969, Sunyaev1970mu, Sunyaev1970SPEC, Illarionov1975b,Danese1977,Burigana1991, Hu1993}.
At lower frequencies a similar limit on $\mu$ was recently obtained by ARCADE \citep{arcade2}, and $\mu\lesssim\pot{6}{-5}$ at $\nu \simeq 1\,{\rm GHz}$ is derived from TRIS \citep{tris1, tris2}.

Two orders of magnitude improvement in sensitivity 
over COBE/FIRAS have in principle been possible for 
several years \citep{Fixsen2002}, and the recently proposed CMB experiment
PIXIE \citep{Kogut2011PIXIE} would be able to detect distortions at the level of $\Delta I_\nu/I_\nu \simeq 10^{-8}-10^{-7}$ in the frequency range $30\,{\rm GHz}\lesssim \nu \lesssim 6\,{\rm THz}$. 
Technologically even higher sensitivity might be possible, but PIXIE is already close to what is required to detect the spectral distortions arising from Silk damping \citep{Silk1968} of primordial perturbation due to photon diffusion in the primordial plasma \citep{Sunyaev1970diss, Daly1991, Barrow1991, Hu1994}. 
Although acoustic dissipation creates a mixture of $\mu$ and $y$-type distortions \citep{Chluba2011therm}, the $\mu$-type contribution is especially important since it can only be introduced at redshifts $\pot{5}{4}\lesssim z\lesssim \pot{2}{6}$ and is distinguishable from other types of distortions due to its unique spectral dependence. 
However, there are many other more strong sources of $y$-type distortions arising during reionization of the Universe and in the later epoch (for example, shock heated intracluster gas or hot gas in the galaxies and clusters of galaxies; see Sect.~\ref{sec:Pixie}). This is the reason why below we concentrate on early energy injection.

The modes which produce the $\mu$-type distortion due to Silk damping dissipate their energy close to the photon diffusion scale, corresponding to comoving wavenumbers of $50\,{\rm Mpc}^{-1} \lesssim  k \lesssim 10^4\,{\rm Mpc}^{-1}$.
Since photon diffusion completely erases perturbations on these scales, the
resulting spectral distortions may provide the only way to constrain the
primordial power spectrum at these wavenumbers. 
This could help shed light on the high energy physics responsible for generating the initial conditions of our Universe, an observational possibility that cannot be ignored. 
The erasure of perturbations on small scales also has important consequences for structure formation and this was one of the earlier motivations to study this process \citep{Silk1968,Peebles1970b,Chibisov1972,DZS1978}.

Motivated by the above considerations we carefully formulate and solve 
the problem describing the dissipation of primordial acoustic modes and the
 subsequent evolution of the resulting average spectral distortion in the early Universe. 
This requires calculation of the photon Boltzmann
equation at 2$\times$2 order, i.e. second order in perturbations and
second order in the energy transfer by Compton scattering.
Previous estimates \citep{Sunyaev1970diss, Daly1991, Barrow1991, Hu1994} for the effective energy release rate were based on simplifying assumptions ignoring microscopic aspects of the heating process.
Also, it is clear that previous estimates for the effective heating rate
are only valid in certain limits. For example, the tight coupling  expressions given by \citet{Hu1994} are restricted to the radiation dominated era and  overestimate the heating rate during and after recombination \citep{Chluba2011therm}.
Recently, \citet{Khatri2011} generalized these estimates to also include the possibility for a running spectral index, $\nrun$, and an approximate treatment of the heating rate during and after hydrogen recombination.
This analysis indicates that the energy released by acoustic dissipation could balance the cooling of photons by adiabatically expanding baryonic matter \citep{Chluba2005, Chluba2011therm}.
The latter leads to Bose-Einstein (BE) condensation \citep{Khatri2011}, creating a $\mu$-type spectral distortion with {\it negative} chemical potential.

Since the net amplitude of the distortion caused by both processes strongly depends on the time-dependence of the energy release, it is important to consider the cosmological dissipation problem rigorously, starting from first principles.
This removes the uncertainties connected with previous estimates, allowing accurate predictions for a given primordial power spectrum.
We show here that only 1/3 of all the energy stored in small-scale perturbations of the photon-baryon-electron fluid leads to spectral distortions, while the remaining 2/3 just increases the average photon temperature.
Microscopically one should think about the problem as \emph{redistribution} of energy inside the radiation field, where energy release is caused by the damping of small-scale temperature perturbations, which affects the average radiation field.
At low redshifts, after recombination, the peculiar
    motions of baryons also become important, as we show here.
Consistently taking into account all microscopic aspects, we find that in comparison to previous estimates the effective heating rate at high redshifts is 9/4 higher, but the observable part that causes a spectral distortion is 3/4 smaller.

To write down the photon Boltzmann equation in second order we can rely on several previous works \citep{Hu1994pert, Bartolo2007, Pitrou2009}.
In \citep{Hu1994pert} a detailed derivation of the second order collision
term for Compton scattering can be found. However, there the second order energy transfer is  considered for the average monopole by means of the usual Kompaneets equation \citep{Kompa56, Weymann1965}, while the scattering of anisotropies is only treated in the Thomson limit, assuming no energy exchange.
To consistently formulate the thermalization process at high redshifts ($z\simeq 10^6$), well before recombination, it is however necessary to also include scattering of anisotropies with energy transfer. 
Thus before integrating over electron momenta for the Compton collision integral, 4th order terms in the total electron momentum must be included.
Here we obtain the photon Boltzmann equation in 2$\times$2 order perturbation theory. We provide a detailed derivation in Appendix~\ref{sec:AnisoCS} using two independent methods. One is based on the Lorentz-boosted Boltzmann equation, to account for the effect of peculiar motions on the scattering process, while the other uses explicit Lorentz transformation of the Thomson scattering and generalized Kompaneets terms to derive the required expressions.
The latter allows us to obtain all the second order terms that have been discussed in the literature so far just from the expression for Thomson scattering in a very simple and compact manner. Lorentz transformation of the generalized Kompaneets term allows one to derive all 2$\times$2 terms.
The final expressions for the temperature-dependent corrections to the Compton scattering collision integral are given in Appendix~\ref{app:fin_exp_coll}.
In addition we include an approximate treatment of anisotropic Bremsstrahlung (BR) and double Compton (DC) emission (see Appendix~\ref{sec:AnisoBRDC}).

Armed with these results, we start our discussion of the primordial dissipation problem by explaining
some of the basic aspects leading to distortions (Sect.~\ref{sec:simple_pic} and \ref{sec:formulation}).
In particular, Sect.~\ref{sec:simple_pic} gives a distilled physical picture for the processes leading to the spectral distortion, which is based on the main results of our detailed second order Boltzmann treatment.
We then discuss the cosmological thermalization problem in zeroth, first and second order perturbation theory, highlighting important aspects that are necessary to understand the primordial dissipation process (Sect.~\ref{sec:lhs_photon}).
We point out that spatial variations of spectral distortions of the CMB, which
we call {\it spatial-spectral distortions}, also have the potential of constraining new physics responsible for uniform energy release  in the early Universe. 
An average distortion of the order of $10^{-5}-10^{-6}$ during
recombination would result in spatial-spectral perturbations of order $\simeq
10^{-10}$, which dominate over the second order spatial-spectral
perturbations, parts of which were calculated by \cite{Pitrou2010}.
We also explain that the time-dependence of the energy release process provides a way to {\it slice} the Thomson visibility function, if energy release occurs during recombination.

For the cosmological dissipation problem only the average distortion really matters, as spatial-spectral distortions are at least $100$ times smaller \citep[cf.][]{Pitrou2010}.
Since at second order in perturbation theory the frequency-dependence of
the photon Boltzmann equation cannot be factored out, all calculations have
so far relied on defining an effective temperature by integrating the
Boltzmann equation over frequency \citep{Khatri2009, Khatri2009b, Senatore2009, Nitta2009, Pitrou2010b}, ignoring energy exchange with the matter.
However, CMB experiments measure flux as a function of frequency and
consequently these calculations are not directly comparable with observations. We point
out a unique separation of the Boltzmann equation into uncoupled  spectral
distortion and blackbody temperature perturbation evolution equations in which the spectral
dependence does factor out. Thus it is straightforward to apply
second order perturbation theory without resorting to ad-hoc temperature variables.
These considerations might have implications for the second order
calculations of non-Gaussian signatures \citep[e.g.,][]{Nitta2009, Pitrou2010b, Bartolo2012}, which to some extent mix temperature and distortion parts.
Our separation shows the mentioned 2:1 division of the energy release.
We also find that perturbations in the potentials only matter for the
average photon temperature (see Eq.~\ref{eq:monopole_potential}), while second order Doppler terms only enter the evolution equation of the spectral distortions.

We close our analysis of the cosmological dissipation problem by explicitly computing the effective heating rates for different primordial power spectra (Sect.~\ref{sec:sources_dist} and \ref{sec:dist_results}), analyzing their dependence on the spectral index of scalar perturbations, $\nS$, and its possible running, $\nrun$.
For pure power-law primordial power spectra we present simple expressions for the effective heating rate during the $\mu$-era in Sect.~\ref{sect:approx_SAC}, e.g., Eq.~\eqref{eq:heat_SZ_appr_final_nS}.
The results are then included into the cosmological thermalization code, {\sc CosmoTherm} \citep{Chluba2011therm} to compute the final spectral distortion (Sect.~\ref{sec:dist_results}).
Our computations show that for one of the best fit WMAP7 cosmologies with running spectral index, $\nS=1.027$ and $\nrun=-0.034$ \citep{Larson2011, Komatsu2010}, we indeed find that the cooling of baryonic matter practically compensates the heating from acoustic dissipation in the $\mu$-era, and the net distortion is mainly characterized by a $y$-type distortion, as anticipated earlier \citep{Khatri2011}. 
However, the sign of the $\mu$-parameter is very sensitive to the values of $\nS$ and $\nrun$. 
For example, for a Harrison-Zeldovich spectrum ($\nS=1$) and even for $\nS=0.96$ (another best fit WMAP7 value) the observable part of the energy dissipated by sound waves exceeds the adiabatic cooling of baryonic matter, so that the net distortion is positive.
Therefore,
even a  null result  provides a valuable and sensitive measure of the primordial small-scale power spectrum.
We also show here that for a given value of $\nS$ one can find a critical value of $\nrun^{\rm bal}$ at which the net $\mu$-parameter changes sign, models that we call balanced injection scenarios (Sect.~\ref{sec:balanced}).

\subsection{Metric, gauge and some notation}
\label{sec:gauge_metric}
Throughout, we use the conformal Newtonian gauge to carry out the computations.
We use the metric $\id s^2 =  a^2 e^{2\phi} \id x^2 - e^{2\psi}c^2 \id t^2$, where $a(t)$ is the usual scale factor, $\psi$ describes Newtonian potential perturbations, and $\phi$ perturbations to the spatial curvature.
We neglect vector and tensor perturbations modes.
We refer to the reference frame defined by this metric as Newtonian frame.

In the text we denote temperature perturbations in different orders by $\Theta^{(i)}\equiv (\Delta T/T)^{(i)}$. Peculiar velocities are given by $\vbetac = \vecJ{v}_{\rm p}/ c$.
In the following bold font denotes 3-dimensional vectors and an additional hat means that the vector is normalized to unity, e.g., $\vgh$ for the direction of a photon.
The spherical harmonic coefficient for $Y_{lm}(\vgh)$ of some function $X(\vgh)$ are denoted by the projection integral $[X]_{lm}\equiv X_{lm}$. 
We also use the abbreviation, $X_l=\sum_{m=-l}^{l}X_{lm}Y_{lm}$, whenever applicable.
The Legendre transforms of the temperature field are defined by
$\hat{\Theta}_{l} = \frac{1}{2}\,\int \Theta(\mu)P_l(\mu)\id \mu\equiv
\Theta_{l0}/\sqrt{4\pi(2l+1)}$, where $P_l(\mu)$ is a Legendre
polynomial and $\mu=\vgh\cdot \vbetach$.

\section{Entropy production due to Silk damping}
\label{sec:simple_pic}
In this section we provide a distilled version of the
    main physics responsible for the  spectral distortions of the CMB
    as a result of  the dissipation of  small-scale acoustic waves in the early Universe. 
Part of this section simply provides some background about the physical conditions during the periods of interest here, but a large part of this section follows directly from our more detailed Boltzmann treatment presented in subsequent sections.

\subsection{The baryon-photon-electron fluid at high redshifts}
At high redshifts ($z\gg 925$) we encounter an extremely interesting period when the primordial plasma  is radiation dominated, i.e. not only the pressure $p\simeq p_\gamma=\rho_{\gamma}/3$, but also the energy density of the baryon-photon-electron fluid, $\rho\simeq \rho_\gamma\approx
0.26 [1 + z]^4 {\rm eV cm^{-3}}$, are determined by the CMB photons, with the contribution of baryons to the energy density of this primordial fluid being negligible.
Obviously, some baryonic matter (here the definition includes
  electrons) is also present, setting the relatively small mean free path for the photons because of Compton scattering. 
This permits us to consider photons strongly interacting with free
electrons along with protons and ions as a single fluid with pressure $p=\rho_{\gamma}/3$ and adiabatic index $4/3$. 
We  also encounter adiabatic perturbations which, after crossing the sound horizon, $\rs$,
are converted into standing sound waves. 
During the radiation dominated era the velocity of sound, $\cs=1/\sqrt{3(1+R)}$, is close to
$c/\sqrt{3}$, where $c$ is the speed of light, and
$R=3\rho_{\rm b}/4\rho_{\gamma}$ is the weighted ratio of baryon to photon energy
density, also known as baryon loading. 
This picture is  very similar to normal Jeans instability with
the sound horizon corresponding to the Jeans wavelength, which is only
a factor of $\sqrt{3}$ smaller than the cosmological horizon \citep{lifshitz, bonnor}.

It has been known for a long time that perturbations on small scales, $\lambda\ll \rs$,
are damped due to  thermal conductivity and viscosity
\citep{lifshitz,LK1963}. 
\citet{Silk1968} was the first to compute the dissipation scale due to
thermal conductivity  in the  cosmological picture, when the CMB was discovered, and predicted what came to be known as Silk damping. Shear viscosity, which dominates over thermal conductivity in a radiation dominated plasma, was included by \citet{Peebles1970}, and \citet{Kaiser1983} included the effects of photon polarization, (see \citet{Weinberg1971,WeinbergBook} for a detailed discussion).
The effect of Silk damping today is observed by CMB anisotropy experiments \citep{Larson2011,acbar, Keisler2011,Dunkley2010}. This damping  of small-scale perturbations liberates energy stored as sound waves, transferring it to the isotropic part of the radiation field, a process that does not proceed silently, but imprints spectral distortions of $\mu$ and $y$-type onto the CMB.
Observing these characteristic distortions is probably the
only way to constrain the initial power spectrum of perturbations in
a very broad range of scales which were completely erased due to Silk
damping in the radiation dominated era.

For the evolution of primordial perturbations two moments are of great
importance: the first is encountered when the energy density of radiation in the strongly interacting baryon-electron-photon fluid  (due to Thomson scattering) dropped below the energy density of baryonic matter. 
This occurs at $z_{\rm b\gamma}\simeq 925$, but the effects of baryon loading on the plasma become important as early as $z\simeq 5000$.
The second moment is marked by the recombination of hydrogen at $z_{\rm rec}\simeq 1100$ \citep{Zeldovich68, Peebles68, SeagerRecfast1999},
when the mean free path of photons becomes comparable to the horizon size. 
Before recombination, the
pressure is fully defined by the pressure of radiation,
$p_{\gamma}=0.9N_{\gamma}kT_{\gamma} \gg 2 \Ne k \Te \simeq p_{\rm gas}$, 
because the specific entropy of our Universe (or the ratio of photon to electron number
density) is $N_{\gamma}/\Ne \simeq \pot{1.6}{9}$ with $\Te\simeq T_{\gamma}$ \citep[e.g.,][]{WMAP_params, Larson2011}.
Recombination causes a huge drop in Jeans wavelength, at that point solely defined
by the pressure of baryonic gas, but much more importantly the gravity of dark matter
begins to play a major role.

\subsection{Thermalization of spectral distortions and definition of the blackbody surface}
\label{sec:thermalization}
It is well known that the energy release in the early Universe causes spectral distortions in the CMB spectrum \citep{Zeldovich1969, Sunyaev1970mu, Sunyaev1970SPEC}.
Depending on the epoch at which the energy release occurs and the total amount of energy, the distortions partially thermalize under the combined action of Compton scattering, double Compton and Bremsstrahlung emission.
The thermalization process has been studied in great detail by several independent groups, both analytically and numerically \citep{Illarionov1975, Illarionov1975b, Danese1982, Burigana1991b, Burigana1991, Hu1993, Hu1993b, Burigana1995, Burigana2003, Lamon2006, Procopio2009, Chluba2011therm}.

As mentioned above, hydrogen recombination at $z_{\rm rec}\simeq 1100$ defines the last scattering surface of photons in our Universe. However, there are additional, very important {\it surfaces} in the history of our Universe:
practically any significant perturbation of the photon spectrum 
due to energy release before $z\simeq  2\times 10^6$ is wiped out
by low frequency photon production because of double Compton emission and
bremsstrahlung, supported by the comptonization mechanism, which redistributes
photons over frequency and shapes the final blackbody spectrum.
We refer to this surface as the {\it blackbody surface}.
The second boundary around $z\simeq \pot{5}{4}$ defines the point at which, within a Hubble time, comptonization is unable to re-establish kinetic equilibrium between photons and electrons, so that the photon distribution can no longer relax towards a Bose-Einstein spectrum, with the $\mu$-parameter defining the lack of photons relative to blackbody spectrum of same temperature.
This redshift indicates the transition between the $\mu$ and $y$-era.

\subsection{Mixing of blackbodies: energy and entropy arguments}
\label{mix_bb}
A large part of this paper is devoted to considering
thermal conductivity and viscosity in the radiation dominated fluid. 
Sound waves in such a fluid decay as diffusion of photons leads to a mixture of
blackbodies with tiny but finite differences in temperature. It is well
known \citep{Zeldovich1972} that a mixture of blackbodies should lead to
a $y$-type spectral distortion, which is rapidly converted to $\mu$-type
distortions at $z\gtrsim \pot{5}{4}$.
Let us consider the sum of two
blackbodies resulting from an anisotropy of the radiation field, with
temperatures $T_1=T+\delta T$ and $T_2=T-\delta T$, assuming $\delta T \ll T$ and keeping terms of order $(\delta T/T)^2$. The average temperature
is clearly equal to $T$. 
If we mix these two blackbodies, the resulting
spectrum has an average energy density equal to 
$\rho_{\rm av}=a_{\rm R} (T_1^4+T_2^4)/2 \approx a_{\rm R} T^4[1+6(\delta T/T)^2]>a_{\rm R} T^4$, where $a_{\rm R}$
is the radiation constant, and $a_{\rm R} T^4$ is the energy density of a blackbody
at temperature $T$.
Similarly, the average photon number density is 
$N_{\rm av}=b_{\rm R} (T_1^3+T_2^3)/2 \approx b_{\rm R} T^3[1+3(\delta
 T/T)^2]>b_{\rm R} T^3$, where $b_{\rm R} T^3$ 
 gives the number density of photons for a blackbody of temperature $T$.
 Thus we can readily see that the resultant spectrum has a larger energy
 and number density than a blackbody at the average temperature of the original blackbodies. 
 
On the other hand, we can determine the effective
temperature of a blackbody with average number density $N_{\rm av}$. It is given by 
$T_{\rm BB}=(N_{\rm av}/b_{\rm R})^{1/3}\approx T [1+(\delta T/T)^2]$.
The energy density of this
blackbody is $\rho_{\rm BB}=a_{\rm R} T_{\rm BB}^4 \approx a_{\rm R} T^4[1+4 (\delta T/T)^2]$, which  is smaller than $\rho_{\rm av}$.
The excess energy density, $\Delta \rho_{y|\mu}=\rho_{\rm av}-\rho_{\rm BB}=a_{\rm R} T^4[2 (\delta T/T)^2]$ thus has to cause a spectral distortion. 
Initially this distortion is of $y$-type with
$y\simeq \frac{1}{4}(\Delta \rho/\rho)=\frac{1}{2} (\delta T/T)^2$ \citep{Zeldovich1969}, however, at high redshifts ($z\gtrsim \pot{5}{4}$) this spectrum rapidly relaxes to a Bose-Einstein equilibrium
spectrum with chemical potential $\mu\simeq 1.4\,\Delta \rho/\rho=2.8(\delta T/T)^2$ \citep{Sunyaev1970mu, Illarionov1975b}.
Thus the resultant spectrum is not a blackbody, but has to be distorted.

The statements made here are proven below with the second order
  Boltzmann equation. In particular, it turns out that at second order, the first order temperature perturbations give rise to a source function for the average photon field which has two contributions, one related to a simple temperature shift, and a second which is a $y$-distortion. 
The $y$-distortion does not change the number of photons, but does carry extra energy. 
The momentary energy density of the source function is $4 a_{\rm R} T^4 (\delta T/T)^2$ for the temperature term, and $2 a_{\rm R} T^4 (\delta T/T)^2$ for the $y$-distortion part.
The ratio of these two source terms implies that only $1/3$ of the energy dissipated by the mixing of blackbodies gives rise to a distortion, while the remaining $2/3$ only raises the average CMB temperature.

\subsubsection{Generalizing to the CMB}
Along the same lines, we can easily write down the correct expressions for the average energy density of the CMB 
\begin{align}
\rho^{\rm temp}_{\gamma}&=\left<a_{\rm R} T^4\right>
=\left<a_{\rm R} \bar{T}^4(1+\Theta)^4\right>\nonumber\\
&=\rho_{\rm pl}(\bar{T})\left<1+4\Theta+6\Theta^2+\rm{h.o.}\right>\nonumber\\
&\approx \rho_{\rm pl}(\bar{T}) \left(1+6\sum_{l=0}^{\infty}(2l+1)\left<\hat{\Theta}_{l}^2\right> \right),\label{correct_relation}
\end{align}
with\footnote{In the following subscript `pl' refers to a Planckian photon distribution.} $\rho_{\rm pl}(\bar{T})=a_{\rm R} \bar{T}^4$ denoting the energy
density of a blackbody at average temperature $\bar{T}$,
which we have defined such that the average of the temperature perturbations, $\left<\Theta\right>=\left<\Delta T/T\right>$ vanishes. The last line was obtained using the Legendre-transform of the temperature field (see Sect.~\ref{sec:gauge_metric} for definitions).
We will see below from the solution of the Boltzmann equation
that at high redshifts we dissipate exactly the above amount of {\it total} energy:
\begin{align}
\label{eq:total_energy_release_estimate}
\frac{1}{a^4\rho_{\gamma}}\Abl{a^4\rho_{\gamma}}{t}
\approx
-\frac{\id}{\id t}\frac{\rho^{\rm temp}_{\gamma}}{\rho_{\rm pl}(\bar{T})}
&=
-6\frac{\id }{\id t} \sum_{l=0}^{\infty}(2l+1)\left<\hat{\Theta}_{l}^2\right>.
\end{align}
The factors of the scale factor $a=1/(1+z)$ arise since we only need to take the comoving energy release into account.
Similarly we can write down the average number density in terms of
blackbody number density at average temperature $N_{\rm pl}(\bar{T})$
\begin{align}
N^{\rm temp}_{\gamma}
&\approx N_{\rm pl}(\bar{T}) \left(1+3\sum_{l=0}^{\infty}(2l+1)\left<\hat{\Theta}_{l}^2\right> \right),
\end{align}
Again comparing with the energy density of a blackbody at temperature $T_{\rm BB}=(N_{\rm av}/b_{\rm R})^{1/3}$, we thus have a total energy release, available for creation of distortions, given by$-2\frac{\id }{\id t}
\sum_{l=0}^{\infty}(2l+1)\left<\hat{\Theta}_{l}^2\right>$ implying $\frac{\id }{\id t}\mu=-2.8\frac{\id }{\id t}
\sum_{l=0}^{\infty}(2l+1)\left<\hat{\Theta}_{l}^2\right>$.

The outlined procedure also naturally divides a  spectrum into  a blackbody part
corresponding to the number density of photons and a spectral
distortion part  defined as pure {\it redistribution} of
photons.  Thus the total number density of photons, $\int \id x\, x^2 n(x)$,
where $n(x)$ is the occupation number and $x=h\nu/k T$ the dimensionless frequency, is zero for
the spectral distortion part of the spectrum\footnote{We note that this
  condition is true for $y$-type distortions or the SZ effect and that
  $\mu$-type  distortions (or any general distortion) can also be defined to
  satisfy this condition (see Sect.~\ref{sec:approx_mu_y}).}.  This definition also
 naturally separates out the evolution equation for the spectral
 distortions from the full Boltzmann equation, as will become clearer below. 

\subsubsection{Entropy production}
In a similar manner we can understand how the specific entropy of our Universe is affected 
by the dissipation of acoustic modes.
The average entropy density of the anisotropic blackbody radiation is given
by an expression similar to the average number density above, since both
have the same $T^3$-dependence on temperature for a blackbody:
\begin{align}
s_{\rm pl}^{\rm initial}&=\frac{4}{3}a_{\rm R}\left<T^3\right>
=\frac{4}{3}a_{\rm R}\bar{T}^3\left< (1+\Theta)^3\right>
\nonumber\\
&\approx s_{\rm pl}(\bar{T}) \left(1+3\sum_{l=0}^{\infty}(2l+1)\left<\hat{\Theta}_{l}^2\right> \right),
\end{align}
with $\bar{s}_{\rm pl}=4/3a_{\rm R} \bar{T}^3$ denoting the entropy density of a blackbody at average temperature $\bar{T}$.
As shown in this work, 2/3 of the energy in perturbations lead to an increase in the temperature of the average blackbody by $\Delta T/T\simeq \sum_{l=0}^{\infty}(2l+1)\left<\hat{\Theta}_{l}^2\right>$, resulting in an entropy of the average blackbody $s_{\rm pl}^{\rm final} \equiv s_{\rm pl}^{\rm initial}$. 
Thus, the final entropy in the average blackbody part of the spectrum after decay of perturbations is equal to the initial entropy in the temperature perturbations.
The remaining 1/3 of energy creates an average $y$-type distortion and consequently results in net entropy production of 
\begin{align}
\frac{\id s}{\id t}=\frac{1}{\bar{T}}\frac{\id Q}{\id t}
\approx -2\frac{\rho_{\gamma}}{\bar{T}}\frac{\id}{\id t} \sum_{l=0}^{\infty}(2l+1)\left<\hat{\Theta}_{l}^2\right>.
\end{align}
Thus only the pure  $y$-type distortion part of the mixing of blackbodies represents true
entropy production and is the 1/3 observable part of the energy released by
dissipation of acoustic waves.  
The change in the average temperature, accounting for the remaining 2/3 of the energy in sound waves, just transfers entropy from small-scale perturbations to the average blackbody and is unobservable. We should emphasize that the total number and energy of photons is conserved in the mixing of blackbodies/dissipation of acoustic waves but the destruction of regularity in the sound waves/photon field configurations creates  entropy.

\subsubsection{Why is the  estimate based on the classical formula for
  energy in sound waves inapplicable here?}
\label{sec:estimate_doesnotwork}
In  fluid mechanics theory  the energy density stored in the sound waves is
given by the simple formula 
\begin{align}
\label{classicalformula}
&\frac{1}{2}\rho {\rm v}^2+\frac{1}{2}\rho \cs^2 \left(\frac{\delta \rho}{\rho}\right)^2
= \rho {\rm v}^2=\rho \cs^2 \left(\frac{\delta \rho}{\rho}\right)^2,
\end{align}
where $\rho$ is the average \emph{mass} density in the
fluid, $\rm{v}$ the fluid velocity, and the equalities are valid
for the waves satisfying the relation $\rm{v}=c_s\delta \rho/\rho$.
This formula \citep[e.g., see][]{ll1987} is  valid only for massive (non-relativistic)  particles. 
However, in earlier estimates this expression was directly used for the relativistic (massless) photon fluid (neglecting baryon loading) with $\cs^2\simeq c^2 / 3$ and $\delta \rho / \rho\simeq 4 (\delta T_0/T_0)$; this turns out to be too simplistic (see Sect.~\ref{sect:approx_SAC} for explicit comparison and simple expressions).

In the  radiation dominated plasma in the early Universe  we can consider
sound waves on scales smaller than the Jeans length (sound horizon) and
larger than the mean free path of the photons. The baryon diffusion length
is much smaller than the diffusion length of photons and dissipation on
small scales is therefore dominated by photon
diffusion\footnote{The baryon-electron component of the plasma, however, does in addition transfer energy to the photon field through the second order Doppler effect which we will find in the Compton collision term.}. 
We calculate the energy in sound waves by averaging the total energy density in radiation at every point in space. For this we use the fact that when the dissipative length becomes comparable with the wavelength of the sound wave, the photons from
different part of the sound wave form a mixture of photons belonging to
blackbodies with different temperatures due to different degrees of
compression in different phases of the sound waves. 
As shown below, the computation of the {\it total energy} in this perturbed photon field gives a value of energy density in the dissipated sound waves, which is 9/4 times higher than the value given by Eq. \eqref{classicalformula} after replacing 
$\delta\rho/\rho$ (and $\rho$) by the perturbation in radiation energy density
$\delta \rho_{\gamma}/\rho_{\gamma}$ (and $\rho_{\gamma}$) as used
by \citet{Sunyaev1970diss} and \citet{Hu1994}.  
The considerations of \citet{Sunyaev1970diss} and \citet{Hu1994} also did not allow to obtain the 2:1 ratio of the temperature to $y$-type source terms mentioned above.
Taking it into account we find that the final observable amount of energy going to $\mu$-distortions coincidentally is equal to 3/4 of the value used by \citet{Sunyaev1970diss} and \citet{Hu1994}.

\citet{Sunyaev1970diss} and \citet{Hu1994} used the
  non-relativistic formula in the form $Q_{\rm ac}\simeq \rho c_{\rm s}^2 (\delta \rho / \rho)^2$.
\citet{zn1983}, on the other hand, based their estimate on the
non-relativistic expression, $Q_{\rm ac}\simeq \rho v^2$. These two formulae, of course, are
identical for the non-relativistic fluid, but when generalizing to the
relativistic plasma, the latter gives an additional coefficient of 9/16 and is, thus, 4 times smaller than our result. 
This factor is connected with the relation between the adiabatic baryon
and photon density perturbations, $\delta \rho_{\rm b}/\rho_{\rm
  b}=\frac{3}{4} \delta \rho_{\gamma}/\rho_{\gamma}$.   
These formulae can be compared with the dipole term in Eq. \eqref{correct_relation} 
which is twice the \citet{zn1983} result.

Another problem of estimates based on Eq. \eqref{classicalformula} is also immediately apparent: the dipole is a gauge-dependent quantity even on small
scales \citep[for related discussion see][]{Zibin2008}. 
But the microphysics responsible for photon diffusion and
damping of small-scale perturbation is local and cannot depend on the
choice of reference frame. Furthermore, there is more to
the CMB than just the monopole and the dipole. 
In fact during the $\mu$-era ($z\gtrsim \pot{5}{4}$), the main role in Silk damping and the resulting energy release and spectral distortions is not played by the dipole but the
quadrupole. The dipole is affected only through coupling to the quadrupole
via the Boltzmann hierarchy, which before recombination in the electron rest frame is almost zero in the tight coupling limit.
In addition, the simple estimate according to Eq.~\eqref{classicalformula}
(as well as the estimate, Eq.~\eqref{eq:total_energy_release_estimate}, put forward here) does not include the second order Doppler effect which becomes important after recombination.
Addition of these terms makes the source function for the spectral distortion gauge-independent, as we see below.

\section{Basics of the problem}
\label{sec:formulation}
In this section we discuss some of the basic ingredients that are required to formulate the cosmological thermalization problem in different orders of perturbation theory.
We reformulate part of the statements made in the previous section using common language related to spherical harmonics and perturbation theory.
In particular we generalize to photon occupation number, rather than integrated energy or number density.
We also explain how electrons up-scatter in the distorted radiation field, leading to local Compton equilibrium between electrons and photons.

\subsection{The local superposition of blackbodies and the up-scattering of electrons}
\label{sec:sup_bb_dist}
Before thinking in detail about spectral distortions and their evolution in the expanding Universe, let us assume that the photon distribution in every direction $\vgh$ and at spatial position $\vecx$
is given by a pure blackbody with temperature
%
\beal
\nonumber
T(t, \vecx, \vgh)&=T_{\rm av}(t)\left[1+\Theta(t, \vecx, \vgh) \right].
\end{align}
Here $T_{\rm av}(t)$ is the spatially averaged global temperature of the photon field, and by definition one has
$\left<T\right>=T_{\rm av}$ and $\left<\Theta\right>=0$, where $\left<...\right>$ denotes the spatial average\footnote{Below we define $T_{\rm av}$ with respect to the temperature fluctuations in first order perturbation theory, i.e., $\left<\Theta^{(1)}\right>=0$ but generally $\left<\Theta^{(i)}\right>\neq 0$ for $i>1$. This is a normal consequence of the perturbative treatment of the problem which also reflects the fact that to higher order in $\Theta$ the definition of temperature becomes more convoluted.}.
Introducing the dimensionless frequency $\xav=h\nu/k \Tav$ for $\Theta\ll 1$ we can then express the photon occupation number, $n(t, \xav,\vecx, \vgh)$, at each point by
\beal
\label{eq:def_n_av}
n(t, \xav, \vecx, \vgh)&=\frac{1}{e^{\xav/[1+ \Theta(t, \vecx, \vgh)]}-1}
\nonumber\\[1mm]
&\approx \nPl(\xav)+\mathcal{G}(\xav)\left(\Theta +\Theta^2\right)
+ \frac{1}{2}\mathcal{Y}_{\rm SZ}(\xav)\,\Theta^2.
\end{align}
Here $\nPl(x)=1/[e^x-1]$ is the Planckian occupation number, and the functions $\mathcal{G}$ and $\mathcal{Y}_{\rm SZ}$ are defined in Appendix~\ref{sec:def_GY}.
This expression shows that to second order in $\Theta$ two contributions appear, one that is equivalent to a temperature perturbation $(\propto \mathcal{G})$ and the other has a non-thermal, 
$y$-type spectral dependence ($\propto \mathcal{Y}_{\rm SZ}$).

Still, so far not much has happened. 
The second line of Eq.~\eqref{eq:def_n_av} is an alternative way of approximating a blackbody with varying temperature in different directions, but nothing more.
If we now compute multipoles of the photon distribution we have
\beal
\nonumber
n_{lm}(t, \xav, \vecx)&\approx \sqrt{4\pi} \delta_{l0}\delta_{m0} \, \nPl
+\mathcal{G}\left(\Theta_{lm}+\left[\Theta^2\right]_{lm}\right)
+ \frac{1}{2}\mathcal{Y}_{\rm SZ}\, [\Theta^2]_{lm}.
\end{align}
%
The projection integral $[...]_{lm}$ essentially leads to a superposition of blackbodies with different temperatures.
This superposition {\it no longer} is a pure blackbody \citep[e.g., see][]{Zeldovich1972, Chluba2004}:
if we define the local mean temperature, $\bar{T}(\vecx)=\Tav[1+\Theta_{0}(\vecx)]$, then naively one could have expected $n_{00}(t, \xav, \vecx)\equiv \bar{n}_{00}(t, \xav, \vecx)$ with
\beal
\nonumber
\bar{n}_{00}(t, \xav, \vecx)&=\sqrt{4\pi} \,\nPl(\xav/[1+\Theta_{0}])
\\
&\approx \sqrt{4\pi} \nPl
+\mathcal{G}\left(\Theta_{00}+\frac{\Theta^2_{00}}{\sqrt{4\pi}}\right)
+ \frac{1}{2}\mathcal{Y}_{\rm SZ}\, \frac{\Theta^2_{00}}{\sqrt{4\pi}}
\nonumber
\end{align}
for the monopole coefficient of the local spectrum.
The first line is just a blackbody of temperature $\bar{T}$.
As before, the extra terms $\propto \mathcal{G}$ and $\propto \mathcal{Y}_{\rm SZ}$ only appear when we express the spectrum of this blackbody with respect to a reference blackbody at different temperature, $\nPl(\xav)$. 
From Eq.~\eqref{eq:def_n_av}, on the other hand we find
\bsub
\label{eq:def_nlm_comp_exp}
\beal
n_{00}(t, \xav, \vecx)&= 
\bar{n}_{00}(t, \xav, \vecx) + \mathcal{G}(\xav)\,y_{00} 
+ \frac{1}{2}\mathcal{Y}_{\rm SZ}(\xav) \,y_{00} 
\\
\label{eq:def_nlm_comp_exp_b}
y_{0}&=y_{00}Y_{00} = \sum_{l=1}^{\infty}\sum_{m=-l}^{l}\frac{\left|\Theta_{lm}\right|^2}{4\pi}
=\sum_{l=1}^{\infty}(2l+1)\hat{\Theta}^2_l,
\end{align}
\esub
which has an extra temperature and distortion term. 
In the last step for Eq.~\eqref{eq:def_nlm_comp_exp_b} we also assumed that the temperature field is azimuthally symmetric about the $z$-axis.
The extra $\mathcal{G}$ term in Eq.\eqref{eq:def_nlm_comp_exp} can be absorbed by additionally shifting the reference temperature (see Sect.~\ref{sec:disc_energy_number_general}), however, it is impossible to absorb the last term, which corresponds to a Compton $y$-distortion with $y$-parameter $y_{0}/2$ \citep{Chluba2004}.
Electrons inside this averaged radiation field are up-scattered, since $y_{0}>0$ (see Sect.~\ref{sec:up_scatt_sup}). This leads to heating of the medium and evolution of the local spectrum because of energy exchange between electrons and photons.

\subsubsection{Simple estimate for the distortion of the CMB sky from the superposition of blackbodies}
The CMB temperature anisotropies today have a typical amplitude $\Theta_{lm}\simeq 10^{-5}-10^{-4}$ for $l>1$ \citep{WMAP_params}, implying a tiny average $y$-parameter $y_{0}\simeq 10^{-10}-10^{-9}$ \citep{Chluba2004}. For an ensemble of universes one can write
\beal
\label{eq:y_0_CMB}
\left<y_{0}\right>&=\sum_{l=1}^{\infty} \frac{[2l+1] C_l}{4\pi},
\end{align}
using the CMB temperature power spectrum, $C_l$, which in general is a
function of time. For a snapshot of the CMB today this implies (excluding dipole) a $y$-type distortion with effective $y$-parameter $y= \left<y_{0}\right>/2\approx \pot{8}{-10}$, and an increase in monopole temperature of $\left<y_{0}\right> \times T_{\rm{CMB}}\approx \pot{1.6}{-9}\times T_{\rm{CMB}}=4.4\,\rm{nK}$, with the aforementioned $1:2$ division of total energy due to superposition of blackbodies.
We used the output of {\sc Cmbfast} \citep{CMBFAST} for the best fit cosmology of WMAP7 without running \citep{Larson2011} for this simple estimate. 70\% of the contribution to $\left<y_{0}\right>$ is from $l \gtrsim 70$, where cosmic variance is already less important. However, this uncertainty can be avoided using the observed CMB sky, as explained below.

This simple estimate does not include the effect of energy release that occurred at high redshifts before recombination, e.g., because of acoustic damping. 
This implies that the spectrum in every direction of the sky is already distorted by some small amount. It turns out that this additional distortion is a few times larger, having the characteristics of both $\mu$ and $y$-type distortion (see Sect.~\ref{sec:dist_results}).
Also, the dipole anisotropy was neglected for the estimate, being a
frame-dependent quantity related to the local peculiar velocity with
respect to the CMB. For our own rest frame with $\betac\approx \pot{1.24}{-3}$ \citep{Fixsen1996} we find a $y$-type distortion of $y\approx \betac^2/6 \approx 2.56\times 10^{-7}$ \citep{Chluba2004}. For a local electron this motion-induced anisotropy implies a distortion which leads to heating of electrons.

It is also important to mention that the distortion from the superposition of blackbodies depends on the averaging procedure. Indeed the average spectrum of the CMB is distorted by the superposition of blackbodies, but by observing the CMB spectrum in small patches and subtracting the main temperature anisotropy in every direction one can in principle avoid the distortion caused by this local superposition \citep{Chluba2004}.
Alternatively, one can directly use the high resolution maps of the CMB temperature anisotropies obtained, for example, with WMAP and knowledge of the CMB dipole to predict the associated $y$-distortion with very high accuracy. 
However, the dipole needs to be measured with slightly higher precision to achieve the required  accuracy  at the level $y\simeq 10^{-10}-10^{-9}$.
Although for us, as observers, this separation is possible, electrons inside the anisotropic radiation field, at every time, are directly affected by the distortion from the superposition of blackbodies, as explained above. 

\subsubsection{Mean photon number and energy density}
\label{sec:disc_energy_number_general}
The terms $\propto \mathcal{G}$ in Eq.~\eqref{eq:def_nlm_comp_exp} imply a larger mean number of photons for the radiation field than expected just from using the mean temperature $\bar{T}=\Tav[1+\Theta_{0}]$.
If we multiply Eq.~\eqref{eq:def_nlm_comp_exp} by $x_{\rm av}^2$ and integrate over all frequencies and directions, we obtain 
\beal
\label{eq:N_bar_gamma}
N_\gamma(\vecx)=\Ngpl(\bar{T})[1+3\,y_{0}],
\end{align}
since $\int \xav^2 \mathcal{Y}_{\rm SZ}(\xav)\id\xav =0$. Here $\Ngpl(T)$ is the number density of photons for a blackbody of temperature $T$.
As we see, at second order in $\Theta$ the number density of photons is increased.

The extra term $\propto \mathcal{G}$ can in principle be absorbed by defining the average effective temperature $\bar{T}^\ast=\Tav[1+\Theta_{0} + y_{0}]$. This practically does not alter the distortion part, which is a consequence of the superposition of blackbodies:
with this redefinition one finds $N_\gamma(\vecx)\equiv \Ngpl(\bar{T}^\ast)$ and 
\bsub
\label{eq:def_nlm_comp_exp_redef}
\beal
n_{00}(t, \xav, \vecx)&= \sqrt{4\pi}\,\nPl(\xav/[1+\Theta_{0}+y_{0}]) 
+ \frac{1}{2}\mathcal{Y}_{\rm SZ}(\xav) \,y^\ast_{00} 
\\
y^\ast_{0}&=y^\ast_{00}Y_{00} = y_{0}-(2\Theta_{0}y_{0}+y^2_{0}) \approx y_{0},
\end{align}
\esub
showing that the $y$-type part is affected at the level $y_{0}^2\approx 0$.
Also by introducing $\bar{x}^\ast=\xav/[1+\Theta_{0}+y_{0}]$ we can alternatively write
\beal
\label{eq:def_nlm_redef_xbar}
n_{00}(t, \bar{x}^\ast, \vecx)&\approx \sqrt{4\pi}\,\nPl(\bar{x}^\ast) 
+ \frac{1}{2}\mathcal{Y}_{\rm SZ} (\bar{x}^\ast)\,y_{00},
\end{align}
since once the linear terms are gone every other shift in the reference temperature  
enters at higher order \citep[see][for more details on higher order terms]{Chluba2004}.

With this definition we can also compute the energy density of the photon field. Multiplying Eq.~\eqref{eq:def_nlm_redef_xbar} by $(\bar{x}^\ast)^3$ and integrating over all $\id \bar{x}^\ast$ yields
\beal
\label{eq:rho_bar_gamma}
\rho_\gamma(\vecx)
&=\rho_{\rm pl}(\bar{T}^\ast)[1+2\,y_{0}] \approx \rho_{\rm pl}(\bar{T})[1+6\,y_{0}],
\end{align}
where $\rho_{\rm pl}(T)$ is the energy density of a blackbody at temperature $T$.
Again this formula shows that the average energy density of the photon field is not just defined by $\bar{T}$, but is larger by about $\Delta \rho_\gamma / \rho_{\rm pl}(\bar{T})\approx 6\,y_{0}$, with $\Delta \rho_\gamma / \rho_{\rm pl}(\bar{T})\approx 2\,y_{0}$ of this additional energy density contributed by the $y$-distortion, while $\Delta \rho_\gamma / \rho_{\rm pl}(\bar{T})\approx 4\,y_{0}$ is caused by the average increase of the temperature term, $\propto\mathcal{G}$.
This already indicates the 1/3 to 2/3 separation of source terms in the second order perturbation problem, however, we give a more rigorous argument below (Sect.~\ref{sec:change_Ng_second}).

It is important to emphasize that the local monopole of the {\it temperature} field does not play any role for the distortion caused by a superposition of blackbodies.  If no anisotropies in the photon temperature are present then naturally no distortion arises.
At high redshifts, the temperature field only has spatially varying monopole, dipole and quadrupole, while the other multipoles can normally be neglected, being strongly suppressed by Thomson scattering \citep{Sunyaev1970}.
This means that {\it locally} the main sources of distortions are the dipolar and quadrupolar temperature anisotropies on different scales.
Also it is clear that in the presence of peculiar motions, velocity-dependent terms appear from the Lorentz transformation to the local rest frame of the baryon-electron fluid.

We furthermore mention that for the (ensemble) averaged total energy density of the CMB the variation of the monopole does matter: with Eq.~\eqref{eq:rho_bar_gamma} we can write 
$\left<\rho_\gamma\right>=\rho_{\rm pl}(\Tav)[1+6\left<\Theta^2_0+y_{0}\right>]$.
The monopole therefore does play a role for the spatial average of blackbodies with different temperatures, however, taking the time derivative of this expression the monopole terms cancel, being unaffected by Thomson scattering events (see Sect.~\ref{sec:second_order}).

\subsubsection{Compton equilibrium temperature from the superposition}
\label{sec:up_scatt_sup}
One can also directly compute the Compton equilibrium temperature, $\Te^{\rm C, eq}$, for electrons in the distorted radiation field given by Eq.~\eqref{eq:def_nlm_redef_xbar}.
With \citep{Levich1969}
\beal
\nonumber
\Te^{\rm C, eq}&=\frac{h\int \nu^4 n_0 (1+n_0) \id \nu}{4k\int \nu^3 n_0 \id \nu}
\end{align}
we obtain
\beal
\label{eq:Te_Compton_sup}
\Te^{\rm C, eq}& \approx\bar{T}^\ast\left[1+\left(\frac{10\pi^2}{21}-2\right) y_0 \right] 
\approx \Tav\left[1+\Theta_{0}+3.70\, y_0 \right].
\end{align}
This expression shows that electrons attain a temperature that is slightly higher than $\bar{T}^\ast=\Tav[1+\Theta_{0} + y_{0}]$.
Consequently, also the baryons heat up because of Coulomb scattering with electrons.
Comptonization causes CMB photons to down-scatter, thermalizing
  $y$-type distortions into $\mu$-type distortions, while at low frequencies Bremsstrahlung and the double Compton process lead to photon production. 
The interaction of electrons with the anisotropic photon field is one of the real sources of distortions which is active throughout the whole history of the Universe and cannot be avoided by changing the observing strategy.

\subsection{Multipole expansion of the distorted occupation number}
\label{sec:n_lm_expand}
In addition to the variations of the temperature, we now allow for a small spectral distortion, $\Delta n(t, \xav,\vecx, \vgh)$, to the photon occupation number in different directions.
Here it is assumed that $\Delta n$ excludes distortions caused by the instantaneous
superposition of different blackbodies but may have contribution from the
superposition of blackbodies at earlier times.
Its spherical harmonic coefficients can be expressed as 
\beal
\label{eq:def_Dnlm_spec}
\Delta n_{lm}&\equiv\sqrt{4\pi} \delta_{l0}\delta_{m0} \, \Delta n_{\rm av} + \Delta n_{{\rm s}, lm},
\end{align}
where $\Delta n_{\rm av}$ defines the possible spectral distortion of the globally averaged spectrum, $n_{\rm av}=\nPl+\Delta n_{\rm av}$, which has {\it no} spatial dependence, while $\Delta n_{{\rm s}, lm}$ is the spherical harmonic coefficient of the spatially varying distortion. 
Adding this term to Eq.~\eqref{eq:def_n_av} and using\footnote{There is no term $\Theta^{(0)}$ as per definition in zeroth order perturbation theory $\Theta^{(0)}\equiv 0$, since there are no spatial variations.} 
$\Theta \approx \Theta^{(1)} +\Theta^{(2)}$, the spherical harmonic coefficients for the photon occupation number in different perturbation orders are
\bsub
\label{eq:def_nlm_perturbed}
\beal
\label{eq:def_nlm_perturbed_a}
n^{(0)}_{lm}(t, \xav, \vecx)&\approx \sqrt{4\pi} \delta_{l0}\delta_{m0} \, 
\left[ \nPl(\xav)+\Delta n^{(0)}_{\rm av}(\xav)\right]
\\
\label{eq:def_nlm_perturbed_b}
n^{(1)}_{lm}(t, \xav, \vecx)&\approx \sqrt{4\pi} \delta_{l0}\delta_{m0} \, \Delta n^{(1)}_{\rm av}+ \Delta n^{(1)}_{{\rm s}, lm}+\mathcal{G}\, \Theta^{(1)}_{lm}
\\
\label{eq:def_nlm_perturbed_c}
n^{(2)}_{lm}(t, \xav, \vecx)
&\approx 
\sqrt{4\pi} \delta_{l0}\delta_{m0} \, \Delta n^{(2)}_{\rm av}+\Delta n^{(2)}_{{\rm s}, lm}
\nonumber\\
&\qquad+\mathcal{G}\, \left[ \Theta^{(2)}+(\Theta^{(1)})^2\right]_{lm} + \frac{1}{2}\mathcal{Y}_{\rm SZ}\, [(\Theta^{(1)})^2]_{lm}.
\end{align}
\esub
In zeroth order perturbation theory one can describe the evolution of the globally averaged spectrum if some energy release is present. This problem is briefly discussed in Sect.~\ref{sec:zeroth_order} and was also recently studied in detail by \citet{Chluba2011therm} for several thermal histories of the Universe.
In first order perturbation theory it is possible to describe the co-evolution of temperature and spectral perturbations. If in zeroth order perturbation no spectral distortions are produced then mainly temperature perturbations appear, however if $\Delta n^{(0)}_{\rm av}\neq 0$ then also small spatial-spectral distortions arise, as we explain in Sect.~\ref{sec:first_order}.
Finally, second order perturbation theory allows us to describe additional modifications of temperature perturbations as well as corrections to spectral distortions. 
In the present work we are not interested in corrections to the lower order solutions, but rather want to work out how on average the superposition of blackbodies and the motion of the baryonic matter affect the CMB spectrum. 
This problem is discussed in detail in Sect.~\ref{sec:second_order}.

\subsubsection{Choice of the reference temperature}
\label{sec:reference_temp}
To write down the expansions Eq.~\eqref{eq:def_nlm_perturbed} we have to define a global reference blackbody, which we chose to be $\nPl(\xav)$.
In general (e.g., with uniform energy release) $\Tav$ does not scale linearly with redshift. However, when writing down the photon Boltzmann equation it is useful to absorb the effect of the Hubble expansion on the photon occupation number by introducing the dimensionless frequency $\xg=h\nu/k\Tz$, where $\Tz=T_{\rm ref}[1+z]$ with constant temperature $T_{\rm ref}$, which defines a uniform reference blackbody.
Naturally, one might say $T_{\rm ref}=T_{\rm av}(z=0)\approx \TCMB=2.726\,$K according to the measurement of COBE/FIRAS  \citep{Fixsen2009}, however, because of cosmic variance our local value of the monopole is generally not equal to $T_{\rm av}(z=0)$. 
In addition, in the presence of energy release the effective temperature of the CMB increases with time, since part of the released energy fully thermalizes, just leading to entropy production, without causing distortions.
The increase in the total energy density of the CMB has to be taken into account when defining $T_{\rm ref}$, and also matters when interpreting the results of the computation (Sect.~\ref{sec:approx_mu_y}).

Here we define $T_{\rm ref}$ at the initial redshift, where the energy release starts, such that after the energy release is completed the effective temperature of the spectrum is equal to $\TCMB$.
As explained in \citet{Chluba2011therm} this means that $T_{\rm ref}<\TCMB$ and the difference depends on the total energy that is released (see Sect.~\ref{sec:dist_results}).
Although in general $\TCMB\neq T_{\rm av}(z=0)$ this does not introduce any significant difference for the final distortion.

\section{Perturbing the photon Boltzmann equation} 
\label{sec:lhs_photon}
In this section we obtain the evolution equations describing the cosmological thermalization problem in different perturbations orders.
Basic introduction on the photon Boltzmann equation in the expanding medium can be found in \citet{deGroot1980}, \citet{Bernstein1988}, and \citet{DodelsonBook}.
Since we account for deviations from the blackbody spectrum we start with the zeroth order equation and then subsequently generalize keeping all terms that are needed in the end.
The zeroth order equations describe the standard thermalization problem (see Sect.~\ref{sec:thermalization}) when spatially uniform energy release is present. 
In first order perturbation theory one mainly obtains temperature perturbations if no zeroth order distortions were present, however, small spatial-spectral distortions with zero mean are created even in this case, although they are suppressed by a factor $\simeq k\Tz/\me c^2$ compared to first order temperature perturbations.
Second order perturbation theory is needed to describe the dissipation of acoustic modes in the early Universe.
For notation and different definitions of functions we refer the reader to the Appendices. However, as far as possible we define quantities once they appear in the text.

\subsection{Zeroth order equation} 
\label{sec:zeroth_order}
We can write the Boltzmann equation in zeroth order, neglecting any perturbations of the stress-energy tensor or metric. Since the spectrum of the global reference blackbody is not affected by the expansion of the Universe (in any order), and since we ignore spatial variations in zeroth order, using Eq.~\eqref{eq:def_nlm_perturbed_a} we only have
\bsub
\beal
\label{eq:BoltzEq_perturbed_zero}
\partial_t\Delta n^{(0)}_{\rm av}&=\mathcal{C}^{(0)}_{\rm th}[n^{(0)}_{\rm av}].
\end{align}
with $n^{(0)}_{\rm av}=\nPl+\Delta n^{(0)}_{\rm av}$.
The collision term, $\mathcal{C}^{(0)}_{\rm th}[n^{(0)}_{\rm av}]$,  just has the normal contributions arising in the thermalization problem of global CMB spectral distortions, which vanish identically if thermal equilibrium between radiation and matter is never perturbed\footnote{Even in the case of no global energy release the adiabatic cooling of baryonic matter continuously extracts some small amount of energy from the CMB, which inevitably results in some small spectral distortions \citep{Chluba2005, Chluba2011therm, Khatri2011}.}.
For the homogeneous Universe without anisotropies in the radiation field we have \citep{Sunyaev1970mu, Illarionov1975b, Burigana1991, Hu1993b, Chluba2011therm}
\beal
\label{eq:BoltzEq_perturbed_zero_b}
\dot{\tau}^{-1}\,\mathcal{C}_{\rm th}[n_{\rm av}^{(0)}]
&=\frac{\Theav}{x^2_\gamma}\partial_{\xg} x^4_\gamma \left\{\partial_{\xg} n^{(0)}_{\rm av}+\frac{\Tz}{\Teav}n^{(0)}_{\rm av} [1+n^{(0)}_{\rm av}]\right\}
\nonumber
\\&\qquad\qquad + \frac{K(\xg, \Theav)}{x^3_{\gamma}}\left\{ 1- n^{(0)}_{\rm av}[e^{\xe}-1]\right\}
\end{align}
\esub
where $\dot{\tau}=\Ne^{(0)}\sigma_{\rm T}\,c$ is the time derivative of the Thomson optical depth, $\The=k\Te/\me c^2$, and $K(\xg, \Theav)$ is the emission coefficient of thermal Bremsstrahlung (BR) and double Compton emission (DC).
For the BR coefficients accurate fitting formulae are given by \citet{Nozawa1998} and \citet{Itoh2000}, while for DC we follow the approach of \citet{Chluba2011d}.
In addition to Eq.~\eqref{eq:BoltzEq_perturbed_zero} one has to solve for the temperature, $\Te$, of the electrons, which because of heating or cooling processes could be pushed away from the radiation temperature.
As in principle non-uniform heating could be possible here we defined the average, spatially uniform electron temperature $\Theav=\The^{(0)}$.

For Eq.~\eqref{eq:BoltzEq_perturbed_zero_b}, we assumed that there is no additional source of photons other than BR and DC.
Detailed calculations of the particle cascade caused by energy release in the form of decaying or annihilating relic particles \citep[e.g.,][]{Slatyer2009} could allow the inclusion of additional processes that lead to photon production. However, usually the primary photons are emitted at energies $h\nu \gg k\Tg$ and in addition the total number of photons produced directly is expected to be much smaller than those created by BR and DC.
We leave a more detailed discussion for future work.

The first term in curly bracket describes the repeated scattering of photons by thermal electrons, while the second accounts for photon emission and absorption by Bremsstrahlung and double Compton scattering.
Assuming that no energy release occurs, it is clear that the cooling of baryonic matter leads to a distortion $\Delta n^{(0)}_{\rm s}/\nPl \lesssim 10^{-9}-10^{-8}$ \citep{Chluba2011therm, Khatri2011}. 
Furthermore, the cosmological recombination process introduces an average distortion $\Delta n_{\rm s}/\nPl \simeq 10^{-9}$, which reaches $\Delta n_{\rm s}/\nPl \simeq 10^{-7}$ at low frequencies \citep[e.g.,][]{Chluba2006}.
In both cases, spatial-spectral distortions are $\lesssim 10^{-14}-10^{-12}$, and we shall neglect those below.
However, current constraints from COBE/FIRAS \citep{Fixsen2011} still allow $\Delta n_{\rm s}/\nPl \simeq 10^{-5}$, which could be produced by global energy release in the early Universe \citep{Sunyaev1970mu}. In this situation spatial-spectral variations could be expected at a level $\simeq 10^{-5}-10^{-4}$ relative to the uniform spectral distortion. 
This appears to be of the same order of magnitude as the uniform distortion caused by the dissipation of acoustic modes in the early Universe, and hence is considered more carefully below.
Also for a uniform initial distortion corresponding to the COBE upper bound
the first order spatial-spectral distortions created during recombination are about two orders of magnitude larger than the spatial-spectral variations recently discussed by \citep{Pitrou2010}.

\subsubsection{Source of distortion} 
\label{sec:zeroth_sources}
To understand slightly better what is the source of the distortions let us explicitly insert $n^{(0)}_{\rm av}=\nPl+\Delta n^{(0)}_{\rm av}$ into the zeroth order equation, Eq.~\eqref{eq:BoltzEq_perturbed_zero_b}, and carry out the perturbative expansion:
\beal
\label{eq:BoltzEq_perturbed_zero_inserted}
\dot{\tau}^{-1}\,\mathcal{C}^{(0)}_{\rm th}[n_{\rm av}]
&\approx 
%
[\Theav-\Thz] \mathcal{Y}_{\rm SZ}-\frac{\Theav-\Thz}{\Thz}\,\frac{K(\xg, \Thz)}{x^3_{\gamma}}(1-e^{\xg}) \mathcal{G}
\nonumber
\\&\quad
+\frac{\Thz}{x^2_\gamma}\partial_{\xg} x^4_\gamma \left\{\partial_{\xg} \Delta n^{(0)}_{\rm av}+\Delta n^{(0)}_{\rm av} [1+2\nPl]\right\}
\nonumber
\\&\qquad
+ \Delta n^{(0)}_{\rm av} \frac{K(\xg, \Thz)}{x^3_{\gamma}}[1-e^{\xg}],
\end{align}
where we used $x^{-2}_{\gamma}\partial_{\xg} x^4_{\gamma}\partial_{\xg}\nPl \equiv \mathcal{Y}_{\rm SZ}$.
The first two terms only appear when $\Teav\neq\Tz$. These are the real source of distortions, while the other two just describe additional redistribution of photons and absorption/emission. 
We linearized the induced scattering process, as terms $(\Delta n^{(0)}_{\rm av})^2$ are of higher order.
For small $\tau$ only the first two terms in Eq.~\eqref{eq:BoltzEq_perturbed_zero_inserted} matter, however, at sufficiently small $\xg$ even for $\tau\ll 1$ the last term has to also be taken into account, because the efficiency of photon absorption/emission increases as $1/x^3_\gamma$.

\subsubsection{Increase of specific entropy and photon production} 
Integrating the Boltzmann equation, Eq.~\eqref{eq:BoltzEq_perturbed_zero}, over $\xg^2\id\xg\id^2\vgh$ one can compute the net production of photons.
For convenience we define the integrals
\bsub
\beal
\label{eq:G_I_H_Intes}
\mathcal{G}^{(i)}_k&= \int x^k_\gamma n^{(i)}_0 \id\xg 
\\
\Delta \mathcal{G}^{(i)}_k&= \int x^k_\gamma \Delta n^{(i)}_{0} \id\xg 
\\
\Delta \mathcal{I}^{(i)}_k&= \int x^k_\gamma \Delta n^{(i)}_{0} (1+2\nPl)\id\xg.  
\end{align}
\esub
Here $\Delta n^{(i)}_0=\Delta n^{(i)}_{\rm av}+\Delta n^{(i)}_{\rm s, 0}$, and $n^{(i)}_0$ is simply the total monopole of the occupation number.
We also sometimes use $\mathcal{G}^{\rm pl}_k$, which just means that $n^{(i)}_0\equiv \nPl$ is assumed.

Since Compton scattering conserves photon number, by integrating Eq.~\eqref{eq:BoltzEq_perturbed_zero} over $x^2 \id x$, we find
\beal
\label{eq:Ngamma_zero}
\pAb{\Delta N^{(0)}_\gamma}{\tau}
&\approx \frac{\Ngpl}{\mathcal{G}^{\rm pl}_2} 
\int K(\xg, \Thz)\left[ 1- n_{\rm av}(e^{\xe}-1)\right]\id\ln\xg,
\end{align}
where $\Ngpl $ is the number density of photon for a blackbody of temperature $\Tz$ and $\mathcal{G}^{\rm pl}_2=\int \xg^2 \nPl(\xg)\id\xg \approx 2.404$.
Emission and absorption can increase or decrease the number of photons. If the photon distribution has a Planckian spectrum then the number and energy density are both described by one temperature, however, if spectral distortions are present the {\it effective} temperatures for the number and energy density differ from each other.
In this situation the radiation temperature can no longer be uniquely defined.

\subsubsection{Change in the photon energy density and energy exchange with baryonic matter} 
The electron temperature depends critically on the energy exchange with the CMB photon field. Also BR and DC emission contribute to the total energy balance, but these terms can usually be neglected \citep{Chluba2011therm}. 
Multiplying Eq.~\eqref{eq:BoltzEq_perturbed_zero_inserted} by $x^3_\gamma$ and integrating over $\id\xg$ and the directions of the scattered photons we obtain the net change in the energy density of the photon field
\bsub
\label{eq:Energy_zero}
\beal
\label{eq:Energy_zero_a}
\pAb{\rho^{(0)}_{\gamma}}{\tau}
&\approx4\rho^{\rm pl}_\gamma [\rho^{(0)}_{\rm e}-\rho^{\rm eq,(0)}_{\rm e}]
\\
\label{eq:Energy_zero_b}
\rho^{\rm eq,(0)}_{\rm e}&= 1 
+ \frac{\Delta \mathcal{I}^{(0)}_4}{4\mathcal{G}^{\rm pl}_3}
-  \frac{\Delta \mathcal{G}^{(0)}_3}{\mathcal{G}^{\rm pl}_3},
\end{align}
\esub
where $\rho^{\rm pl}_\gamma\propto \mathcal{G}^{\rm pl}_3$ is the energy density of the distorted CMB blackbody photon field and $\rho^{(i)}_{\rm e}=\Te^{(i)}/\Tz$ defines the ratio of electron temperature to $\Tz$.
We neglected the additional term caused by photon emission and absorption.

This equation shows that if the temperature of the electrons (plus baryons) differs from $\Tz$ then the number and energy density of the photon field can change. For global heating one usually expects $\Te>\Tz$ so that the energy density of the photon field increases by a small amount and the matter is cooled by the Compton interaction.
Equation~\eqref{eq:Energy_zero} therefore implies the zeroth order Compton cooling term for the matter ($\Te\approx T_{\rm m}$ because of fast Coulomb interactions)
\bsub
\beal
\label{eq:Compton_cooling_zero}
\left.\pAb{\rho^{(0)}_{\rm e}}{\tau}\right|_{\The}&\approx
\frac{4\tilde{\rho}^{\rm pl}_\gamma}{\alpha_{\rm h}}[\rho^{\rm eq,(0)}_{\rm e}-\rho^{(0)}_{\rm e}],
\end{align}
\esub
where $\tilde{\rho}^{\rm pl}_\gamma= \rho^{\rm pl}_\gamma/\me c^2$ is the energy density of the distorted CMB blackbody photon field in units of the electron rest mass energy and $k\alpha_{\rm h}$ is the heat capacity of  baryonic matter (electrons plus baryons).

As we show below, higher order corrections to the Compton cooling process can all be brought into the same form, however, the expressions for the corrections to the equilibrium temperature ratio, $\rho^{{\rm eq},(i)}_{\rm e}$, differ.
Also, in the case of no energy release it is clear that $\rho^{\rm eq,(0)}_{\rm e}\approx 1$, although the adiabatic cooling of normal matter does lead to a very small distortion itself \citep{Chluba2011therm}, which we can safely neglect in first order perturbation theory.
Note that with global energy release $\rho^{\rm eq,(0)}_{\rm e}$ differs from unity at the level of $\simeq \Delta n^{(0)}_{\rm s}$.

\subsection{First order equation with spectral distortions} 
\label{sec:first_order}
%
In first order perturbation theory two types of equations are obtained. One describes changes in the photon field that can be interpreted as temperature perturbations, for which the spectral dependence is $\mathcal{G}(\xg)$. 
The other equation describes the evolution of spatial-spectral distortions which initially start as $y$-type distortions.
If no spectral distortion is produced at zeroth order, usually only the temperature perturbations have to be evolved.
However, allowing the possibility of global energy release, one already in first order perturbation theory introduces spatial-spectral contributions to the CMB spectrum, as we discuss below. 

We shall limit our discussion of the first order equations to the basic aspects, as a general solution of the problem is more demanding.
However, one of the main results of this section is that in first order perturbation theory no average distortion is created. This greatly simplifies the consideration for the average distortion created by the dissipation of acoustic modes in second order perturbation theory (see Sect.~\ref{sec:second_order}).

Using the dimensionless frequency, $\xg=h\nu/k\Tz$, it is clear that the total derivative of the photon occupation number with respect to time up to first order perturbation theory has the form \citep[e.g., see][]{Hu1996, DodelsonBook}
\beal
\label{eq:BoltzEq_general}
\partial_t n^{(1)} + \frac{\gh_i}{a}\,\partial_{x_i} n^{(1)} 
- \xg\partial_{\xg} n^{(0)}\left[\partial_t\phi^{(1)}+ \frac{\gh_i}{a}\,\partial_{x_i} \psi^{(1)}\right]
&=\mathcal{C}^{(1)}[n],
\end{align}
where $a$ is the usual scale factor,  $\phi$ and $\psi$ are potential perturbations, and $\mathcal{C}[n]$ denotes the collision term.
Inserting Eq.~\eqref{eq:def_nlm_perturbed_b} for the photon occupation number and grouping terms corresponding to the usual thermal perturbations ($\propto \mathcal{G}$) and those of the non-thermal parts we find (for detailed derivation see Appendix~\ref{sec:AnisoCS}):
\bsub
\label{eq:BoltzEq_perturbed_first}
\beal
\label{eq:BoltzEq_perturbed_first_a}
\partial_t \Theta^{(1)}+\frac{\gh_i}{a}\,\partial_{x_i} \Theta^{(1)}
&=-\left[\partial_t\phi^{(1)}+ \frac{\gh_i}{a}\,\partial_{x_i} \psi^{(1)}\right] 
+ \mathcal{\tilde{C}}^{(1)}_{\rm T}[n_{\rm bb}]
\\
\label{eq:BoltzEq_perturbed_first_a2}
\partial_t\Delta n^{(1)}_{\rm s}+\frac{\gh_i}{a}\,\partial_{x_i} \Delta n^{(1)}_{\rm s}
&= -\left[\partial_t\phi^{(1)}+ \frac{\gh_i}{a}\,\partial_{x_i} \psi^{(1)}\right]\mathcal{T}
\nonumber
\\
&\qquad\qquad
+  \mathcal{C}^{(1)}_{\rm T}[\Delta n_{\rm s}] 
+ \mathcal{C}^{(1)}_{\rm th}[n] 
\\[1.5mm]
\label{eq:BoltzEq_perturbed_first_b}
\dot{\tau}^{-1}\mathcal{\tilde{C}}^{(1)}_{\rm T}[n_{\rm bb}]
&= \Theta^{(1)}_{0} -\Theta^{(1)}+\frac{1}{10} \Theta^{(1)}_{2}+\betac\muc 
\\[1.5mm]
\label{eq:BoltzEq_perturbed_first_d}
\dot{\tau}^{-1}\mathcal{C}^{(1)}_{\rm T}[\Delta n_{\rm s}] 
&= \! \Delta n^{(1)}_{\rm s, 0} -\Delta n^{(1)}_{\rm s}+\frac{1}{10} \Delta n^{(1)}_{\rm s, 2}
+\betac\muc \mathcal{T},
\end{align}
\esub
with $\muc=\vbetach\cdot\vgh$ and $\mathcal{T}(\xg)=-\xg \partial_{\xg} \Delta n^{(0)}_{\rm av}$. 
Equation~\eqref{eq:BoltzEq_perturbed_first_a} with Eq.~\eqref{eq:BoltzEq_perturbed_first_b} is just the familiar evolution equation for the brightness temperature of the radiation field \citep{Hu1994pert}.
The direction of the bulk velocity has been aligned with the $z$-axis, since in first order perturbation theory the $k$ vector of the different modes in Fourier space is parallel to $\vbetac$. 
This choice of coordinates therefore simplifies the analysis significantly.

Here one comment should be made immediately. When writing Eq.~\eqref{eq:BoltzEq_perturbed_first} we directly omitted $\Delta n_{\rm av}^{(1)}$. It turns out that on average no additional distortion is created, if we ignore the small correction $\propto (\Delta n^{(0)}_{\rm av})^{2}$. This is because all the source terms that we discuss below have zero mean.
As we see in second order perturbation theory this is no longer the case.

In addition, we mention that when writing down the equations~\eqref{eq:BoltzEq_perturbed_first} we separated spectral from temperature perturbations knowing that first order spectral distortions are several orders of magnitude smaller than the temperature perturbations themselves. This means that one can first solve for the first order temperature perturbations and then discuss the spectral distortion part.
A more rigorous argument can be given by integrating the Boltzmann equation over frequency.
The Thomson scattering terms all have spectral dependence $\propto \mathcal{G}$ resulting in $\int \mathcal{G} \xg^2 \id \xg = 3 \mathcal{G}^{\rm pl}_{2}$, while the frequency integral over all scattering terms with energy exchange vanishes (see Appendix~\ref{sec:AnisoCS} for detailed explanations).
Then only the BR and DC emission terms are left in addition to the Thomson scattering terms. However the former two lead to very small corrections, so that Eq.~\eqref{eq:BoltzEq_perturbed_first_a} follows.

If we ignore the term $\mathcal{C}^{(1)}_{\rm th}[n] $ for a moment, then we can see that the equation for the temperature perturbations looks rather similar to the one for the spectral distortions. 
The only difference is that the sources of spatial-spectral distortions have the frequency-dependence $\mathcal{T}(t, \xg)$, where this source function is also an explicit function of time. 
This introduces another modulation of the anisotropies in the distortions in addition to the normal Thomson visibility function:
as long as $\mathcal{T}\simeq 0$ no spatial-spectral distortions arise, and only temperature perturbations appear in first order perturbation theory.
If  energy release occurs well before last scattering then the spatial-spectral distortions should have very similar statistics to those of the temperature perturbations. If energy release occurs well after last scattering, only those parts of the evolution of perturbations happening afterward matter.
This implies that the time-dependence of the energy release introduces another possibility to {\it slice} the Thomson visibility function at different epochs.
This is particularly interesting when energy is released during the hydrogen recombination epoch, where most of the CMB temperature anisotropies are formed.
A more detailed discussion of this aspect of the problem is beyond the scope of this work.

\subsubsection{Local change of the specific entropy} 
Integrating the Boltzmann equation, Eq.~\eqref{eq:BoltzEq_general}, over $\xg^2\id\xg\id^2\vgh$ one can compute the net change in the local number of photons.
Since scattering conserves photon number one has
\beal
\label{eq:Ngamma_first}
\pAb{N^{(1)}_\gamma}{\tau}
&\approx -a^{-1}\taudot^{-1}\partial_z N^{(1)}_{\gamma, 1} -3 N^{(0)}_\gamma \pAb{\phi^{(1)}}{\tau} 
+ \left.\pAb{N^{(1)}_\gamma}{\tau}\right|_{\rm e/a},
\end{align}
where $N^{(i)}_\gamma= \Ngpl \mathcal{G}^{(i)}_2/\mathcal{G}^{\rm pl}_2$ denotes the photon number density in different orders, and $N^{(i)}_{\gamma, 1}=(\Ngpl/\mathcal{G}^{\rm pl}_2) \,\int \xg^2 \,\hat{n}^{(i)}_{1}(\xg) \id \xg$ with the Legendre transforms $\hat{n}^{(i)}_l(\xg)=\frac{1}{2} \int  n^{(i)}(\mu, \xg)\,P_l(\mu)\id\mu$. Here we used the fact that in first order perturbation theory the photon field is azimuthally symmetric around the $z$-axis $\parallel \vbetac$. 
The effect of photon emission and absorption is captured by the last term in Eq.~\eqref{eq:Ngamma_first}, which can be calculated with Eq.~\eqref{eq:BR_DC_first}.
However, it is a small correction in comparison to the other two contributions and can usually be neglected.
This implies that local changes in the photon number are dominated by the flows of photons and potential perturbations. On average there is no photon production in first order perturbation theory. This statement is also correct when additional emission and absorption terms  are included, as everything only depends linearly on perturbations in the medium.

\subsubsection{First order thermalization term}
Let us now consider the perturbed thermalization term, $\mathcal{C}^{(1)}_{\rm th}[n]$. 
The first obvious contribution is related to the spatial variation in the electron density and temperature
\beal
\label{eq:BoltzEq_perturbed_first_c}
\delta \mathcal{C}^{(0)}_{\rm th}[n_{\rm av}]
&\approx N_{\rm e}^{(1)} \partial_{\Ne} \mathcal{C}^{(0)}_{\rm th}[n_{\rm av}]
+  \The^{(1)}\partial_{\The} \mathcal{C}^{(0)}_{\rm th}[n_{\rm av}].
\end{align}
This term {\it only} matters if already in zeroth order some significant (in our metric $\Delta n/n \simeq 10^{-5}$) distortion is created.
Perturbations in the electron density are directly related to perturbations in the baryon density \citep[$N_{\rm e}^{(1)}\propto 5 \rho_{\rm b}^{(1)}$ during recombination][]{Khatri2009}, and because of the Compton process electron temperature perturbations are identical to photon temperature perturbations, until low redshifts, where the energy exchange between electrons and photons becomes inefficient.
The electron density variations lead to spatially varying thermalization efficiency, while one expects the electron temperature perturbations to be practically canceled by corresponding terms with photon temperature perturbations of the monopole (see below).
%

The remaining contributions to $\mathcal{C}^{(1)}_{\rm th}[n]$ are more complicated. For Compton scattering of an anisotropic radiation field we found the first order perturbed terms in Appendix~\ref{sec:AnisoCS}, while 
an approximate treatment for BR and the DC process in first order perturbation theory is given in Appendix~\ref{sec:AnisoBRDC}.

Inserting $n^{(1)}\approx \Delta n^{(1)}_{\rm s}+\mathcal{G}\,\Theta^{(1)}$ into the first order collision integral for Compton scattering, Eq.~\eqref{eq:BoltzEqlm_The_pert_first_final}, and collecting terms that are independent of distortions we find
\bsub
\label{eq:sss}
\beal
\label{eq:sss_a}
\mathcal{C}^{(1)}_{{\rm th}, \Theta}[n]
&\approx
\Delta \Te^{(0)} \left[\bar{\Theta}^{\rm g}+\betac\muc\right] \, \mathcal{D}_{\xg} \mathcal{G}
+\left[\The^{(1)}-\Thez \bar{\Theta}^{\rm g} \right] \mathcal{Y}_{\rm SZ}
\nonumber\\
&\!\!\!
+\Delta \Te^{(0)} \left[\frac{N^{(1)}_{\rm e}}{N^{(0)}_{\rm e}}-\betac\muc\right] \mathcal{Y}_{\rm SZ}
+2 \Thez\, \mathcal{G} \,\mathcal{B} \left[\Theta^{\rm g} - \bar{\Theta}^{\rm g}   \right],  
\end{align}
where $\Delta \Te^{(0)}=\Theav-\Thez$ and 
\beal
\label{eq:Theta_bar_first}
\bar{\Theta}^{\rm g}=\Theta^{(1)}_0-\frac{2}{5}\left[\Theta^{(1)}_1-\betac\muc\right]+\frac{1}{10}\Theta^{(1)}_2-\frac{3}{70}\Theta^{(1)}_3.
\end{align}
Also, in $\Theta ^{\rm g}$ one has to replace the dipolar temperature term by $\Theta^{(1)}_1-\betac\muc$.
Furthermore, we used $\partial_{\xg} \mathcal{G}  + \mathcal{G} [1+2\nPl] =-\partial_{\xg} \nPl$ and  defined $\mathcal{D}_{\xg} = x_{\gamma}^{-2} \,\partial_{\xg} x_{\gamma}^4 \,\partial_{\xg}$, $\mathcal{B}= x_\gamma[ \mathcal{A} -\mathcal{G} ]$ and $\mathcal{A}=1+2 \nPl$.
The first and third term in Eq.~\eqref{eq:sss_a} are only present if in zeroth order perturbation theory spectral distortions are present, since otherwise $\Delta \Te^{(0)}\approx 0$.
We also see that temperature dipole terms only appear as $\Theta^{\rm g}_1=\Theta^{(1)}_1-\betac\muc$, which is gauge-independent.

Those contributions involving spectral distortions are
\beal
\label{eq:sss_b}
\mathcal{C}^{(1)}_{{\rm th}, \Delta n}[n]
&\approx
\frac{\Thz}{x_{\gamma}^{2}} \,\partial_{\xg} x_{\gamma}^4 
\left\{ \partial_{\xg} \Delta \bar{n}^{\rm g}_{\rm s} +  \Delta \bar{n}^{\rm g}_{\rm s} \mathcal{A}\right\}
\nonumber\\
&\!\!\!\!\!\!\!\!\!\!\!
+\left[\frac{N^{(1)}_{\rm e}}{N^{(0)}_{\rm e}} - 2\betac\muc \right]
\frac{\Thz}{x_{\gamma}^{2}} \,\partial_{\xg} x_{\gamma}^4 
\left\{ \partial_{\xg} \Delta n^{(0)}_{\rm av} +   \Delta n^{(0)}_{\rm av} \mathcal{A}\right\}
\nonumber\\[1mm]
&\!\!\!\!\!
+ [\The^{(1)}+ \betac\muc ] \mathcal{D}_{\xg} \Delta n^{(0)}_{\rm av}
+2 \Thez \mathcal{B} \left[\Delta n^{\rm g}_{\rm s} - \Delta \bar{n}^{\rm g}_{\rm s}   \right]
\nonumber\\[1mm]
&\,
+\betac\muc\frac{\Thz}{x_{\gamma}^{2}} \,\partial_{\xg} x_{\gamma}^4 
\left\{ \partial_{\xg} \mathcal{T} +  \mathcal{T}\mathcal{A}\right\},
\end{align}
\esub
with $\mathcal{T}=-\xg \partial_{\xg} \Delta n^{(0)}_{\rm av}$ and where we introduced 
\beal
\label{eq:n_bar_first_gauge}
\Delta \bar{n}^{\rm g}_{\rm s}= \Delta n^{(1)}_{\rm s, 0}-\frac{2}{5}\left[\Delta n^{(1)}_{\rm s, 1}-\betac\muc\mathcal{T}\right]+\frac{1}{10}\Delta n^{(1)}_{\rm s, 2}-\frac{3}{70}\Delta n^{(1)}_{\rm s, 3}.
\end{align}
In addition, the dipolar spectral distortion term in $\Delta n^{\rm g}_{\rm s}$ has to be replaced by $\Delta n^{(1)}_{\rm s, 1}-\betac\muc\mathcal{T}$.

An approximate first order expression for BR and DC emission and absorption is given by Eq.~\eqref{eq:BR_DC_first}, resulting in:
\beal
\label{eq:BR_DC_first_insert}
\mathcal{C}^{(1)}_{{\rm th}, \rm e/a}[n]
&\approx
\left[\left(2-\xg\partial_{\xg} \ln K \right) \betac \muc + \The^{(1)}\partial_{\The} \ln K\right]
\mathcal{C}^{(0)}_{{\rm th}, \rm e/a}[n]
\nonumber\\
&\qquad
+\frac{K}{\xg^3} (1-e^{\xg}) \left[ \Delta n^{(1)}_{\rm s, 0}+ \Theta^{(1)}_0\mathcal{G} 
+\betac\muc \!\left(\mathcal{G}+\mathcal{T}\right) \right]
\nonumber\\
&\qquad\quad
+\frac{K}{\xg^2} n^{(0)}_0 \varphi ^{(0)}\left(\frac{\The^{(1)}}{\The^{(0)}} +\betac\muc\right) e^{\xg},
\end{align}
where $\mathcal{C}^{(0)}_{{\rm th}, \rm e/a}[n]$ is the zeroth order emission and absorption term.

For Eq.~\eqref{eq:sss_b} we neglected the $(\Delta n^{(0)}_{\rm av})^2$ term because of induced scattering, as this only provides a small additional spatially uniform correction to the zeroth order distortion.
We furthermore omitted terms $\propto \Delta \bar{n}^{(0)}_{\rm av}\Delta \bar{n}^{(1)}_{\rm s}$, which are of order $\Delta n/n\simeq 10^{-15}$.
We also approximated $1+2n^{(0)}_0+\xg\partial_{\xg} n^{(0)}_0 \approx \mathcal{A} -\mathcal{G}$.
As explained in Appendix~\ref{sec:AnisoCS}, we dropped terms 
$\propto \Theav \Theta^{(1)}_{lm} \mathcal{G}$ and $\propto \Theav \Delta \bar{n}^{(1)}_{\rm s}$, 
which appear as small temperature corrections to the scattering cross section of the dipole, quadrupole and octupole in Eq.~\eqref{eq:BoltzEqlm_The}.

Looking at Eq.~\eqref{eq:sss_a} we can see that the first two terms are related to the effect of the Kompaneets operator on the background spectrum. In zeroth order we had $\mathcal{D}_{\xg} \nPl = \mathcal{Y}_{\rm SZ}$, however, $\nPl$ is now replaced by $\Theta^{(1)} \mathcal{G}$, but since the diffusion operator only couples to monopole, dipole, quadrupole and octupole (see Appendix~\ref{eq:BoltzEqlm_The_pert_b}), one is left with $\simeq \bar{\Theta}^{(1)}$.
Gauge-dependent corrections are given by $\propto\betac\muc$.
The last term in Eq.~\eqref{eq:sss_a} is caused by the recoil effect (including stimulated recoil) and first order Klein-Nishina corrections to the Compton cross section.
We can identify similar terms in Eq.~\eqref{eq:sss_b}.

\begin{figure}
\centering
\includegraphics[width=\columnwidth]{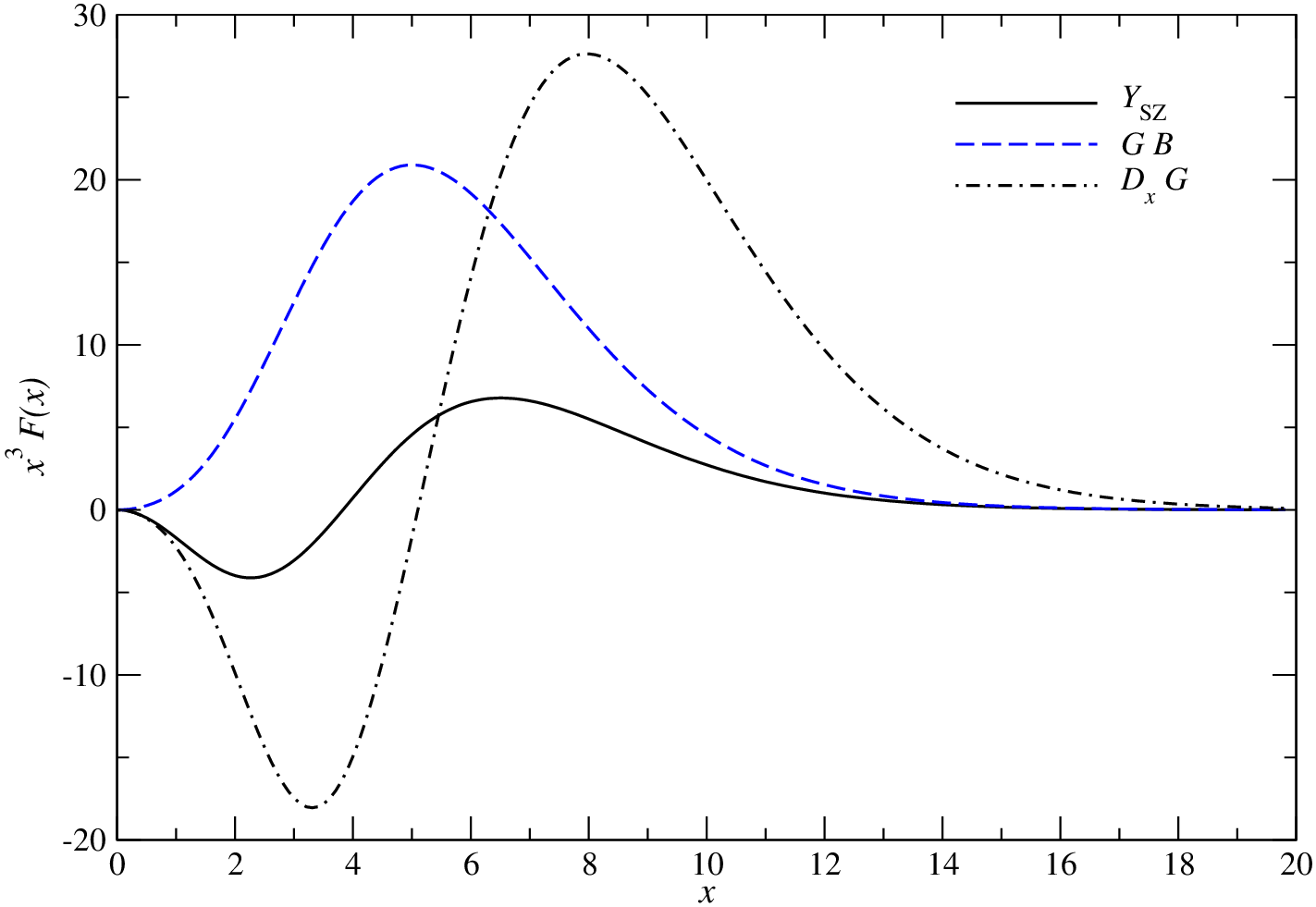}
\caption{Comparison of different source functions for spectral distortions according to Eq.~\eqref{eq:sss_a}.
$\mathcal{G} \,\mathcal{B}$ arises because of the recoil effect and the first order Klein-Nishina correction, while $\mathcal{D}_x \mathcal{G}$ describes the effect of Doppler broadening and boosting on temperature perturbations.}
\label{fig:functions}
\end{figure}
At this point it is worth mentioning that if no distortions are present at zeroth order, then Eq.~\eqref{eq:sss_a} still leads to distortions because of recoil once temperature perturbations in the photon fluid start appearing.
However, these distortions are suppressed by $\Thz$ and only appear for $l>0$.
Also the gauge-dependent terms start causing the dipolar part of the spectrum to deviate from full equilibrium once bulk flows are important. 
The spectral dependence of the corresponding source terms is not just of $y$-type.
To illustrate the differences in Fig.~\ref{fig:functions} we present a comparison of the source functions appearing in Eq.~\eqref{eq:sss_a}.
%
%
Both $\mathcal{G} \,\mathcal{B}$ and $\mathcal{D}_{\xg} \mathcal{G}$ tend to be slightly larger than $\mathcal{Y}_{\rm SZ}$. 
%
%
We find that rescaling $\mathcal{G} \,\mathcal{B}$ by a factor $\simeq 1/3$ and $\mathcal{D}_{\xg} \mathcal{G}$ by $\simeq 1/4$ makes the height of the maxima comparable to $\mathcal{Y}_{\rm SZ}$.

One of the interesting aspects of the thermalization problem with spatial-spectral anisotropies is that the monopole, dipole, quadrupole and octupole have different effective $y$-parameters.
If we define $y_{\gamma}=c \int \Thez \sigT \Ne \id t$, then because $\alpha_1=-2/5$ the effective $y$-parameter for the dipole is $0.4$ smaller than for the monopole and has opposite sign. This means that if photons of the monopole are up-scattering, those in the dipole distortion are down-scattering. 
Similarly, for the quadrupole we have $\alpha_2=1/10$, which implies that when $y_\gamma \simeq 10$ in the monopole, it is only about $y_\gamma\simeq 1$ for the quadrupole. 
This suggests that the redistribution of photons in the quadrupole distortion is significantly slower than in the monopole. 
For the octupole we have $\alpha_3= -3/70$, so that the amplitude of the effective $y$-parameter is only about $4\%$ of the one in the monopole, again with opposite sign like for the dipole.
These peculiar aspects of the scattering problem could lead to interesting differences in the types of distortions in the monopole, dipole, quadrupole and octupole, which we plan to investigate in future work.
However, for the main problem of interest here this is not relevant.

\subsubsection{Energy exchange term with baryonic matter} 
Multiplying Eq.~\eqref{eq:sss} by $x^3_\gamma$ and integrating over $\xg$ and the directions of the scattered photons the velocity-dependent terms drop out and we are left with
\bsub
\beal
\label{eq:Compton_cooling_zero}
\left.\pAb{\rho^{(1)}_{\rm e}}{\tau}\right|_{\The}&\approx
\frac{4\tilde{\rho}^{\rm pl}_\gamma}{\alpha_{\rm h}}[\rho^{\rm eq,(1)}_{\rm e}-\rho^{(1)}_{\rm e}]
\\
\rho^{\rm eq,(1)}_{\rm e}&\approx\Theta^{(1)}_{0}+\frac{\Delta \mathcal{I}^{(1)}_4}{4\mathcal{G}^{\rm pl}_3}
-  \frac{\Delta \mathcal{G}^{(1)}_3}{\mathcal{G}^{\rm pl}_3} 
+ [\rho^{\rm eq,(0)}_{\rm e}-\rho^{(0)}_{\rm e}] \,\frac{N^{(1)}_{\rm e}}{N^{(0)}_{\rm e}}
%
%
%
\end{align}
\esub
for the energy exchange term between electrons and photons. 
Assuming quasi-stationarity from this, one can directly deduce the solution $\The^{(1)}\approx \Thz\, \rho^{\rm eq,(1)}_{\rm e}$ for the electron temperature in first order perturbation theory (see Sect.~\ref{sec:Te_evolution_sec} for explanation).

However, in general one should allow for a spatially varying energy release term in the electron temperature equation, which could directly source spatial variations in the CMB spectrum. The statistics of these spatial-spectral variations are determined by the statistics of the energy release process, which, if associated with non-linear perturbations, could be non-Gaussian.
For example, the efficiency of energy release by annihilating particles scales as $N^2_{\chi}$ with the annihilating particle number density, $N_{\chi}$. 
If these particles are non-baryonic the growth of density perturbation could lead to small-scale fluctuations in the energy release even at early times.

\subsubsection{Evolution with vanishing zeroth order distortions} 
If we assume that in zeroth order no significant distortion is produced then from Eq.~\eqref{eq:sss} and \eqref{eq:BR_DC_first_insert} with $\Theav\approx \Thez$ and $\Delta n^{(0)}_{\rm av}=0$ we find the first order thermalization term
\bsub
\label{eq:sss_no_zero_dist}
\beal
\mathcal{C}^{(1)}_{\rm th}[n]
&\approx
\frac{\Thez }{x_{\gamma}^{2}} \,\partial_{\xg} x_{\gamma}^4 
\left\{ \partial_{\xg} \Delta \bar{n}^{\rm g}_{\rm s} + \Delta \bar{n}^{\rm g}_{\rm s} \mathcal{A} \right\}
+  \left[\theta ^{(1)}_{\rm e} - \Thez \bar{\Theta}^{\rm g} \right] \mathcal{Y}_{\rm SZ} 
\nonumber\\
&\quad
+2 \Thez \mathcal{B} \left\{\mathcal{G} \left[\Theta ^{\rm g} - \bar{\Theta}^{\rm g}   \right]  
+   \Delta n^{\rm g}_{\rm s} - \Delta \bar{n}^{\rm g}_{\rm s}    \right\}
\nonumber\\
&\qquad
-\frac{K}{\xg^3} (e^{\xg}-1) \left[ \Delta n^{(1)}_{\rm s, 0} 
+ \mathcal{G}\left(\Theta^{(1)}_{0}-\frac{\theta ^{(1)}_{\rm e}}{\Thz}\right)\right],
\\[1mm]
\theta ^{(1)}_{\rm e}&\approx
\Thz\left[\Theta^{(1)}_{0} + \frac{\Delta \mathcal{I}^{(1)}_4}{4\mathcal{G}^{(0)}_3}
-  \frac{\Delta \mathcal{G}^{(1)}_3}{\mathcal{G}^{(0)}_3}\right].
\end{align}
\esub
Here we have still included first order spectral distortions, however, it turns out that at least for the monopole there is no direct source of non-thermal distortions arising from the collision term. This can be seen by projecting the above expression onto $Y_{00}$ and realizing that $\theta ^{(1)}_{\rm e} \approx \Thez \Theta^{(1)}_{0}$ if initially $\Delta n^{(1)}_{\rm s, 0}=0$.
Then only the monopole part of the diffusion term remains, but no source of distortions.
However, dipole, quadrupole and octupole do have non-vanishing source terms.
These can lead to distortions also for the monopole part once photons start mixing  (see below).

If no distortions are present in zeroth order, the overall evolution equation for the spectral distortion part reads 
\beal
\label{eq:BoltzEq_perturbed_first_final_spec}
\partial_\tau\Delta n^{(1)}_{\rm s}+\frac{\gh_i}{a \dot{\tau}}\,\partial_{x_i} \Delta n^{(1)}_{\rm s}
&\approx  
\Delta n^{(1)}_{\rm s, 0} +\frac{1}{10} \Delta n^{(1)}_{\rm s, 2} -\Delta n^{(1)}_{\rm s}
+ \mathcal{C}^{(1)}_{\rm th}[n].
\end{align}
This equation is linear in the variables that have to be solved for. 
The temperature fluctuations source distortions at $l>0$, however due to the tight coupling of $\betac$ and the local dipole anisotropy at high redshifts, distortions only start appearing at the level of the quadrupole. 
Redistribution by diffusive scattering is only working at $l<4$.
No potential perturbation terms are present, since $\mathcal{T}\approx 0$. However, different modes couple because of the gradient on the left hand side of the Boltzmann equation. This allows the redistribution of power among the multipoles. 
Although the monopole itself has no source terms, this implies that also the monopole will start developing a distortion once the higher multipoles start deviating from a Planckian spectrum. This also means that $\theta^{(1)}_{\rm e}$ will depart from $\Theta^{(1)}_{0}\Thez$, with the difference being of order of $\Delta n^{(1)}_0$.

It is clear that at high redshifts all higher multipole temperature perturbations are exponentially suppressed by Thomson scattering. Since the monopole has no source for distortions, deviations from a blackbody can only arise at rather low redshifts, when the visibility becomes larger.
At those times $\Thez \simeq \pot{\text{few}}{-7}$, so that the distortions in the considered situation are expected to have a typical amplitude $\simeq \pot{\text{few}}{-12}$ and hence are small.
In addition, the distortions in first order perturbation theory vanish on average, since all sources appear in linear order.
This means that for the average distortion in second order perturbation theory we can omit 
the first order average distortion in the calculation, unless a significant distortion was 
produced in zeroth order. 
Furthermore, we generally expect that it is possible to neglect terms $\propto \Thz \mathcal{B}$ or $\propto \Thz \mathcal{B}\mathcal{G}$, as their effect will only be important at high redshifts.
However, we leave a more detailed discussion for future work.

\subsection{Second order equation for spectral distortions} 
\label{sec:second_order}
The aim of this section is to provide the second order equation for the evolution of the CMB spectral distortion caused by the dissipation of acoustic modes in the early Universe. 
We shall assume that in zeroth and first order no significant distortions are created.
Furthermore, we are only interested in the {\it average distortion} caused by the dissipation of acoustic modes in the early Universe. This simplifies the problem significantly.
Two main aspects are important in this context: (i) the dissipation process increases the average energy density of the CMB by a small amount without causing a distortion; 
(ii) it results in an average spectral distortion, $\Delta n_{\rm
  av}^{(2)}$, while the second order spatial-spectral departures from a blackbody spectrum are at least $\simeq 100$ times smaller \citep[cf.][]{Pitrou2010}.

In Appendices~\ref{sec:AnisoCS} and \ref{sec:AnisoBRDC} we provide a detailed derivation for all the required parts of the collision term.
In particular, Eq.~\eqref{eq:Final_EQ} and \eqref{eq:BR_DC_second} are important for the dissipation problem.
We start our discussion with the left hand side of the Boltzmann equation in second order and then write down the various sources of distortions.

\subsubsection{Boltzmann equation for the average occupation number}
Denoting the collision term in second order perturbation theory by $\mathcal{C}^{(2)}[n]$ the Boltzmann equation for the photon occupation number reads \citep[cf.][]{Bartolo2007, Pitrou2009, Nitta2009}
\beal
\label{eq:Boltzmann_second_order_general}
\partial_t n^{(2)} + \frac{\gh_i}{a}\,\partial_{x_i} n^{(2)} 
- \xg\partial_{\xg} n^{(0)}\left[\partial_t\phi^{(2)}+ \frac{\gh_i}{a}\,\partial_{x_i} \psi^{(2)}  \right]
&
\nonumber\\
&\!\!\!\!\!\!\!\!\!\!\!\!\!\!\!\!\!\!\!\!\!\!\!\!\!\!\!\!\!\!\!\!\!\!\!\!\!\!\!\!\!\!\!\!\!\!\!\!\!\!\!\!\!\!\!\!\!\!\!\!\!\!\!\!\!\!\!\!\!\!\!\!\!\!\!\!\!\!\!\!
\!\!\!\!\!\!\!\!\!\!\!\!\!\!\!\!\!\!\!\!\!\!\!\!\!\!\!
- \xg\partial_{\xg} n^{(1)}\left[\partial_t\phi^{(1)}+ \frac{\gh_i}{a}\,\partial_{x_i} \psi^{(1)}\right]
-\left(\phi^{(1)}-\psi^{(1)}\right) \, \frac{\gh_i}{a}\,\partial_{x_i} n^{(1)} 
\nonumber\\
&\!\!\!\!\!\!\!\!\!\!\!\!\!\!\!\!\!\!\!\!\!\!\!\!\!\!\!\!\!\!\!\!\!\!\!\!\!\!\!\!\!\!\!\!\!\!\!\!\!\!\!\!\!\!\!\!\!\!\!\!\!\!\!\!\!\!\!\!\!\!\!\!\!\!\!\!\!\!\!\!
\!\!\!\!\!\!\!\!\!\!\!\!\!
+
\frac{1}{a}\,\left[\partial_{x_i} \left(\phi^{(1)}-\psi^{(1)}\right)
-\gh_i\gh_j \,\partial_{x_j} \left(\phi^{(1)}-\psi^{(1)}\right) 
 \right] \partial_{\gh_i} n^{(1)}
\nonumber\\
+ \xg\partial_{\xg} n^{(0)}
\left(\phi^{(1)}-\psi^{(1)}\right)\frac{\gh_i}{a}\,\partial_{x_i}\psi^{(1)}
&= \mathcal{C}^{(2)}[n].
\end{align}
In the above expression the Hubble term was absorbed using the frequency variable $\xg\propto \nu/(1+z)$. Also, we directly made use of $\partial_t \nPl(\xg)=0$, $\partial_{x_i}\nPl=0$, and neglected vector and tensor perturbations.
Averaging over $\id^2 \gh$ the last term on the left hand side vanishes. The term $\propto \partial_{\gh_i} n^{(1)}$ gives only one contribution for $i=j=3$, since $n^{(1)}$ is azimuthally symmetric around the $z$-axis.
With \citep{Stegun1972}
\beal
(1-\mu^2)\,\partial_{\mu} n^{(1)}&=\sum_{l=1} l(l+1)\,\hat{n}^{(1)}_l 
\left[P_{l-1}(\mu) - P_{l+1}(\mu)\right]
\nonumber
\end{align}
it is straightforward to show that upon averaging over the photon directions the remaining term is $\frac{2}{a}\,\hat{n}^{(1)}_1 
 \partial_{z} \left(\phi^{(1)}-\psi^{(1)}\right)$.
%

Introducing the abbreviation $D_t = \partial_t+\frac{\gh_i}{a}\,\partial_{x_i}$ and inserting $n^{(1)}=\mathcal{G} \Theta^{(1)}$, $n^{(2)}=\Delta n^{(2)}+ \mathcal{G} [\Theta^{(2)}+(\Theta^{(1)})^2]+\frac{1}{2}\mathcal{Y}_{\rm SZ}(\Theta^{(1)})^2$, after averaging over $\id^2 \gh$ we can rewrite Eq.~\eqref{eq:Boltzmann_second_order_general} as 
\bsub
\label{eq:Boltzmann_second_order_general_II}
\beal
\label{eq:Boltzmann_second_order_general_II_a}
\left[D_t \Delta n^{(2)}  \right]_0
+ \mathcal{G} \,\mathcal{S}_{\rm temp} 
-  \mathcal{Y}_{\rm SZ} \,\mathcal{S}_{\rm dist} 
&= \left[\mathcal{C}^{(2)}[n] \right]_0,
\end{align}
with the temperature and distortion source functions $\mathcal{S}_{\rm temp}$ and $\mathcal{S}_{\rm dist}$:
\beal
\label{eq:Boltzmann_second_order_general_II_b}
\mathcal{S}_{\rm temp}
&=
\left[D_t \Theta^{(2)}\right]_0  + \partial_t\phi^{(2)}
+ 2 \left[\Theta^{(1)} D_t \Theta^{(1)} \right]_0 
\nonumber\\
&
\qquad
+3\left[\Theta^{(1)} \left(\partial_t\phi^{(1)}+ \frac{\gh_i}{a}\,\partial_{x_i} \psi^{(1)}\right)  \right]_0
\nonumber\\
&
\qquad\quad
+\frac{2}{a}\,\hat{\Theta}^{(1)}_1 \partial_z  \left(\phi^{(1)}-\psi^{(1)}\right)
-\frac{1}{a}\,\left(\phi^{(1)}-\psi^{(1)}\right)\,\partial_z\hat{\Theta}^{(1)}_1
\nonumber\\
&=
%
\left[D_t \Theta^{(2)}\right]_0
+ \partial_t\phi^{(2)}
- 3 \mathcal{S}_{\rm dist} - \frac{1}{2}\left[D_t (\Theta^{(1)})^2 \right]_0
\\
&
\qquad\quad
+\frac{2}{a}\,\hat{\Theta}^{(1)}_1 \partial_z  \left(\phi^{(1)}-\psi^{(1)}\right)
-\frac{1}{a}\,\left(\phi^{(1)}-\psi^{(1)}\right)\,\partial_z\hat{\Theta}^{(1)}_1,
\nonumber\\[1.5mm]
\mathcal{S}_{\rm dist}
&=- \left[\Theta^{(1)} D_t \Theta^{(1)} \right]_0 
- \left[\Theta^{(1)} \left(\partial_t\phi^{(1)}+ \frac{\gh_i}{a}\,\partial_{x_i} \psi^{(1)}\right)  \right]_0
\nonumber\\
%
&=\taudot\left[\sum_{l=1}(2l+1)(\hat{\Theta}^{(1)}_l)^2  - \frac{1}{2}\,(\hat{\Theta}^{(1)}_{2})^2 
-\betac \hat{\Theta}^{(1)}_1\right],
\label{eq:Boltzmann_second_order_general_II_c}
\end{align}
\esub
where we used our notation $[...]_0$ to denote projection onto the monopole.
In the last step for $\mathcal{S}_{\rm dist}$ we have used the first order collision term for temperature perturbation to simplify the expression.
Without scattering of radiation in first order perturbation theory this term would vanish identically.
Since $\mathcal{S}_{\rm dist}$ appears on the left hand side of the Boltzmann equation this means that physically it is connected with the mixing of scattered photons from different regions in the medium.
We can see that for $\mathcal{S}_{\rm dist}$ only {\it local} photon mixing is important, while $\mathcal{S}_{\rm temp}$ also has contributions caused by local density gradients and potential perturbations.
Furthermore, as expected the local monopole cancels from the distortion source term, being unaffected by Thomson scattering.

\subsubsection{Local and global change of the specific entropy} 
\label{sec:change_Ng_second}
Integrating the Boltzmann equation, Eq.~\eqref{eq:Boltzmann_second_order_general_II_a}, over $\xg^2\id\xg$ one can compute the net change in the local number of photons.
Since scattering conserves the photon number and $\int \xg^2 \mathcal{Y}_{\rm SZ} \id \xg =0$ one has
\beal
\label{eq:Ngamma_second}
\pAb{N^{(2)}_\gamma}{\tau}
&= -a^{-1}\taudot^{-1}\partial_z N^{(2)}_{\gamma, 1} 
- 3 N^{\rm pl}_\gamma \taudot^{-1} \mathcal{S}_{\rm temp} + \left.\pAb{N^{(2)}_\gamma}{\tau}\right|_{\rm e/a}.
\end{align}
The subscript `e/a' denotes emission/absorption.
This results is very similar to Eq.~\eqref{eq:Ngamma_first}. However, $\mathcal{S}_{\rm temp}$ not only depends on second order potential perturbations, but also mixing terms from first order perturbation theory.
The photon emission and absorption term can be computed using Eq.~\eqref{eq:BR_DC_second} but the precise form is not important here.
Unlike in first order perturbation theory locally the last term is comparable to the first two terms, and globally it even dominates.
This can be seen if we take the spatial average of Eq.~\eqref{eq:Ngamma_second}. Since $\mathcal{S}_{\rm temp}$ arises directly from the Liouville operator it is clear that globally $\left<\mathcal{S}_{\rm temp}\right>= 0$, implying that
\beal
\label{eq:Ngamma_second_av}
\left<\pAb{N^{(2)}_\gamma}{\tau}\right>
&=  \left<\left.\pAb{N^{(2)}_\gamma}{\tau}\right|_{\rm e/a}\right> > 0.
\end{align}
This expression shows that in second order perturbation theory the specific entropy of the Universe changes because of dissipative processes, leading to net photon production.
From this argument with Eq.~\eqref{eq:Boltzmann_second_order_general_II_b} it also follows that 
\beal
\label{eq:monopole_potential}
\partial_t\left<\Theta^{(2)}_0+\phi^{(2)} - \frac{1}{2} y^{\rm s}_0 \right>
&=  
\frac{3}{a}\,\left<\left(\phi^{(1)}-\psi^{(1)}\right)\,\partial_z\hat{\Theta}^{(1)}_1\right>
+ 3 \left<\mathcal{S}_{\rm dist}\right>,
\end{align}
where $y^{\rm s}_0=\sum_{l=0}(2l+1)(\hat{\Theta}^{(1)}_l)^2$. This implies
\beal
\partial_t\left<\Theta^{(2)}_0+\phi^{(2)}\right>
&=  
\frac{3}{a}\,\left<\left(\phi^{(1)}-\psi^{(1)}\right)\,\partial_z\hat{\Theta}^{(1)}_1\right>
+ 3 \left<\mathcal{S}_{\rm dist}\right>+\frac{1}{2}\partial_t\left<y^{\rm s}_0\right>
\nonumber\\
&=
\frac{3}{a}\,\left<\left(\phi^{(1)}-\frac{2}{3}\psi^{(1)}\right)\,\partial_z\hat{\Theta}^{(1)}_1
-\hat{\Theta}^{(1)}_0\partial_t\phi^{(1)}\right>
+ 2 \left<\mathcal{S}_{\rm dist}\right>,
\nonumber
\end{align}
which is equivalent to an energy increase
\beal
\nonumber
\frac{\id a^4 Q_{\rm temp}}{ a^4 \rho_\gamma \id t}
&
\approx 
\frac{12}{a}\,\left<\left(\phi^{(1)}-\frac{2}{3}\psi^{(1)}\right)\,\partial_z\hat{\Theta}^{(1)}_1
-\hat{\Theta}^{(1)}_0\partial_t\phi^{(1)}\right>
+ 8 \left<\mathcal{S}_{\rm dist}\right>
\nonumber\\
&\approx
\frac{12}{a}\,\left<\left(\phi^{(1)}-\frac{2}{3}\psi^{(1)}\right)\,\partial_z\hat{\Theta}^{(1)}_1
-\hat{\Theta}^{(1)}_0\partial_t\phi^{(1)}\right>
- 4 \frac{\id}{\id t}\left<y^{\rm s}_0\right>.
\nonumber
\end{align}
The energy release does not lead to any distortion, as it is injected `adiabatically', in form of a temperature perturbation, $\propto\mathcal{G}$, increasing the average contribution $\left<\Theta^{(2)}_0\right>\mathcal{G}$ to the photon occupation number.
What matters is not only the change in the mixture of blackbodies (last term), but also energy extraction from the potential perturbation. The second order Doppler term is absent here but does appear as a source of distortions (see Eq.~\ref{eq:Final_EQ_elim_Te}). 
On the other hand, we will see that the potentials only lead to an adiabatic increase of the photon energy density but no distortion.
In fact already here we can see that only $2/3$ of the energy stored in the superposition of blackbodies, $\Delta \rho_{\rm sup}=6\left<y^{\rm s}_0\right>\,\rho_\gamma$, goes into the increase of the average CMB temperature. 
Consequently, $1/3$ should lead to the production of distortions, as we confirm in Sect.~\eqref{sec:comp_Hu_appr}.

\subsubsection{Correction to the local electron temperature in second order perturbation theory}
\label{sec:energy_ex_2_order}
As explained in Appendix~\ref{sec:Te_evolution_sec}, a good approximation for the second order correction to the local electron temperature can be given by computing the balance between Compton heating and cooling in the {\it local} rest frame of the moving volume element.
Integrating Eq.~\eqref{eq:integrated_Kompaneets} over $\xg^3 \id \xg$ with the integrals of Eq.~\eqref{eq:energy_int_GYHE} it is straightforward to show that
\beal
\label{eq:The_sol_2}
\The^{(2)}&\approx 
\Thz\left(\frac{\Delta \mathcal{I}^{(2)}_4}{4\mathcal{G}^{\rm pl}_3}
-  \frac{\Delta \mathcal{G}^{(2)}_3}{\mathcal{G}^{\rm pl}_3}\right)
+\Thz\left(\Theta^{(2)}_0+\ygl{1}
+\betac\hat{\Theta}^{\rm g}_{1} +\frac{\betacsq}{6}\right)
\nonumber\\
&\quad
+ 2.70 \, \Thz  \, \ygl{1}  
+ 0.30\,  \Thz\left[ - \frac{6}{5}(\hat{\Theta}^{\rm g}_{1})^2 + \frac{1}{2}\hat{\Theta}^2_{2} - \frac{3}{10}\hat{\Theta}^2_{3} \right],
\end{align}
where $\ygl{1}$ is defined by Eq.~\eqref{eq:y_Theta_10_gauge_b}.
Here we neglected small corrections due to the adiabatic cooling by the expansion of the Universe, and also assumed that no extra energy release occurred.
We also ignored the cooling of matter by BR and DC emission and assumed that Compton cooling and heating are always in balance.

We mention that the first three terms can be obtained just from the standard Kompaneets equation. The last term arises as correction from the effect of dipole through octupole scattering with velocity-dependent corrections.
If we furthermore compare with Eq.~\eqref{eq:Te_Compton_sup} and assume $\betac=0$ we see that the terms $\propto \ygl{1}$ just arise because of electrons up-scattering in the anisotropic blackbody field. The first term is the correction because of deviations from the simple superposition of blackbodies.
Contributions related to $\betac$ give the correct gauge-dependence.

\subsubsection{Second order collision term for the locally averaged photon distribution}
\label{sec:C_second_average}
With expression \eqref{eq:The_sol_2} it is possible to eliminate the dependence of the collision term on $\The^{(2)}$.
Inserting into Eq.~\eqref{eq:Final_EQ} and defining
\bsub
\label{eq:Final_EQ_elim_Te_all}
\beal
\label{eq:Compton_source_terms}
\mathcal{S}^{\The}_{\Delta n}
&=
\frac{\Delta \mathcal{I}^{(2)}_4}{4\mathcal{G}^{\rm pl}_3}
-  \frac{\Delta \mathcal{G}^{(2)}_3}{\mathcal{G}^{\rm pl}_3}
\\[1.5mm]
\mathcal{S}^{\The}_{\mathcal{Y}_{\rm SZ}}
&=
12.16\, \hat{\Theta}^{\rm g}_{1}\betac
+2.70\,\mathcal{S}^{\The}_{\mathcal{H}}
+0.30\,\mathcal{S}^{\The}_{\mathcal{E}}
\\[1.5mm]
\label{eq:Compton_source_terms_H}
\mathcal{S}^{\The}_{\mathcal{H}}
&=
\ygl{1} - \frac{16}{5} \hat{\Theta}^{\rm g}_{1}\betac 
\\[1.5mm]
\label{eq:Compton_source_terms_E}
\mathcal{S}^{\The}_{\mathcal{E}}
&=
 - \frac{6}{5}(\hat{\Theta}^{\rm g}_{1})^2 + \frac{1}{2}\hat{\Theta}^2_{2} - \frac{3}{10}\hat{\Theta}^2_{3}
- \frac{2}{5}\hat{\Theta}^{\rm g}_{1}\betac 
\end{align}
\esub
we finally obtain
\beal
\label{eq:Final_EQ_elim_Te}
\left.\mathcal{C}^{(2)}[n]\right|_{\rm scatt}&
\approx
\frac{\Thz}{x_{\gamma}^{2}} \,\partial_{\xg} x_{\gamma}^4 
\left\{ \partial_{\xg} \Delta n^{(2)}_0 + \Delta  n^{(2)}_0 \mathcal{A}\right\}
-\betac\hat{\Theta}^{\rm g}_{1}\mathcal{Y}_{\rm SZ}
\nonumber
\\[1mm]
&\qquad
\Thz\left[\mathcal{S}^{\The}_{\Delta n}+\mathcal{S}^{\The}_{\mathcal{Y}_{\rm SZ}}\right]\mathcal{Y}_{\rm SZ}
-\Thz\mathcal{S}^{\The}_{\mathcal{H}} \mathcal{H}  
-\Thz\mathcal{S}^{\The}_{\mathcal{E}}  \mathcal{E} 
%
\end{align}
for the Boltzmann collision term of the average photon distribution.
The first term just describes the changes of the spectrum caused by photon scattering, leading to redistribution of photons over energy.
The energy exchange with electrons leads to additional source terms that have different contributions, each with their own spectral behaviour.
%
\begin{figure}
\centering
\includegraphics[width=\columnwidth]{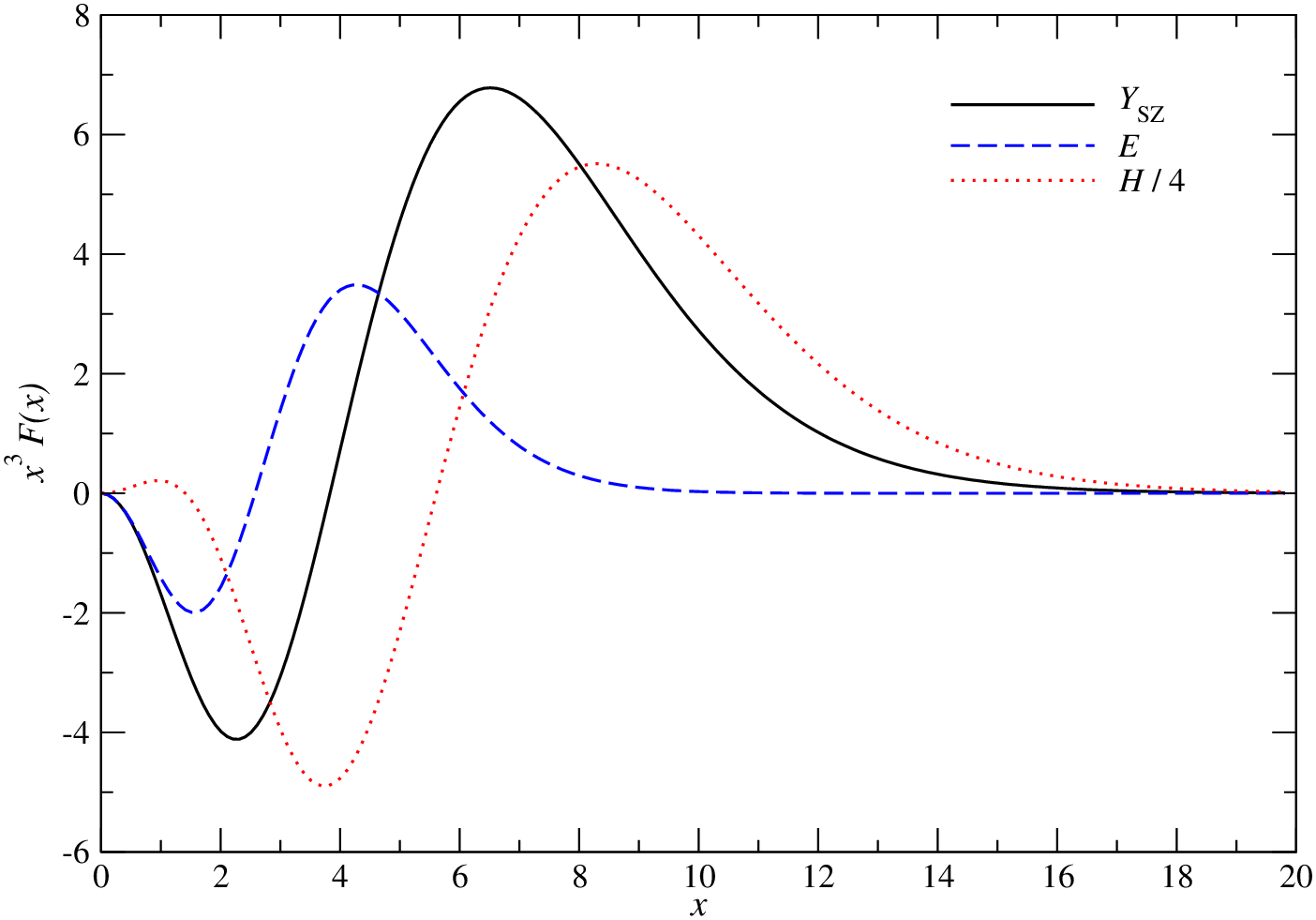}
\caption{Comparison of different source functions for spectral distortions according to Eq.~\eqref{eq:Final_EQ_elim_Te}. To make things more comparable we rescaled $\mathcal{H}$ by a factor of $4$.}
\label{fig:functions.III}
\end{figure}
We show a comparison in Fig.~\ref{fig:functions.III}.
The typical amplitude of $\mathcal{H}$ is $\simeq 4$ times larger than $\mathcal{Y}_{\rm SZ}$, while $\mathcal{E}$ is a factor of $\simeq 2$ smaller than $\mathcal{Y}_{\rm SZ}$.
For the BR and DC emission term with Eq.~\eqref{eq:BR_DC_second} we find 
\beal
\label{eq:BR_DC_second_insert}
\left.\pAb{n^{(2)}_0(\xg)}{\tau}\right|_{\rm e/a}
&\approx
-\frac{K}{\xg^3} (e^{\xg}-1) 
\left[ \Delta n^{(2)}_0 + \mathcal{G} \left(\mathcal{S}^{\The}_{\Delta n}+\mathcal{S}_{\rm e/a}\right)
+ \frac{1}{2}\mathcal{Y}_{\rm SZ}\,\ygl{1} 
\right]
\nonumber\\
\mathcal{S}_{\rm e/a}&= 
0.12\, \hat{\Theta}^{\rm g}_{1}\betac
+ 2.70 \, \ygl{1}  
+ 0.30\, \mathcal{S}^{\The}_{\mathcal{E}}
%
%
\end{align}
%
It is important to mention that the final collision term has no dependence on $\Theta^{(2)}_0$. This fact allows us to further simplify the calculation of the average distortion.

\subsubsection{Evolution equation for the average spectral distortion}
We now have all the pieces to write down the final evolution equation for the average spectral distortions.
For the whole Universe we can only give meaningful answers for a statistical ensemble.
This means that we have to perform an ensemble average of the equations, which is equivalent to a spatial average.
This all boils down to computing the average of the source functions $\mathcal{S}$ we defined above.

We already know from Eq.~\eqref{eq:Boltzmann_second_order_general_II_a} that the resulting spectrum has a distortion but also a contribution $\propto \mathcal{S}_{\rm temp}\,\mathcal{G}$. 
The latter implies a small increase of the average photon energy density, but does not lead to any net source term for $\Delta n^{(2)}_{\rm av}$, since $\left<\mathcal{S}_{\rm temp}\right>\approx 0$ (cf. Sect.~\ref{sec:change_Ng_second}).
With Eq.~\eqref{eq:Boltzmann_second_order_general_II_a},  \eqref{eq:Final_EQ_elim_Te_all}, and \eqref{eq:BR_DC_second_insert} it therefore follows that
\beal
\label{eq:final_EQUATION}
\pAb{\Delta n^{(2)}_{\rm av}}{\tau}&
\approx
\frac{\Thz}{x_{\gamma}^{2}} \,\partial_{\xg} x_{\gamma}^4 
\left\{ \partial_{\xg} \Delta n^{(2)}_{\rm av} + \Delta  n^{(2)}_{\rm av} \mathcal{A}\right\}
+\left<\mathcal{S}_{\rm ac}\right> \mathcal{Y}_{\rm SZ}
\\[1mm]
&\qquad
+\frac{K}{\xg^3} (1-e^{\xg}) 
\left[ \Delta n^{(2)}_{\rm av} + \mathcal{G} \left< \mathcal{S}^{\The}_{\Delta n}+ \mathcal{S}_{\rm e/a}\right>  
+ \frac{1}{2}\left<\ygl{1}\right> \mathcal{Y}_{\rm SZ} 
\right]
\nonumber
\\[1mm]
&\qquad\quad
+\Thz\left<\mathcal{S}^{\The}_{\Delta n}+\mathcal{S}^{\The}_{\mathcal{Y}_{\rm SZ}}\right>\mathcal{Y}_{\rm SZ}
-\Thz\left<\mathcal{S}^{\The}_{\mathcal{H}}\right> \mathcal{H}  
-\Thz\left<\mathcal{S}^{\The}_{\mathcal{E}}\right>  \mathcal{E}
\nonumber
\\[3mm]
\mathcal{S}_{\rm ac}& \equiv
\frac{\mathcal{S}_{\rm dist}}{\taudot} - \betac\hat{\Theta}^{\rm g}_{1} 
=\ygl{1} - \frac{1}{2}\,(\hat{\Theta}^{(1)}_{2})^2 
= 3(\hat{\Theta}^{\rm g}_{1})^2+\frac{9}{2}(\hat{\Theta}^{(1)}_{2})^2 + \ygl{3}.
\nonumber
\end{align}
The main source of distortions is determined by $\mathcal{S}_{\rm ac}$, with $\ygl{1}$ defined by Eq.~\eqref{eq:y_Theta_10_gauge_b}. The first term in Eq.~\eqref{eq:final_EQUATION} describes how the distortion evolves under Compton scattering and the third term accounts for BR and DC emission.
The last three terms include modifications of the source function that are of similar order as $\Thz \Delta n^{(2)}_{\rm av}$. 
All these could potentially be important as long as $y\simeq \int  \Thz \id \tau \gtrsim 1$, but certainly can be neglected at low redshifts.
However, our analysis shows that even at high redshifts only $\left<\mathcal{S}_{\rm ac}\right>$ and $\left<\mathcal{S}_{\Delta n}\right>$ really matter for the distortions (Sect.~\ref{sec:corr_source}).

It is also interesting to realize that the term $\left<\ygl{1}\right> \mathcal{Y}_{\rm SZ}$ is mainly caused by photon mixing (from the left hand side of the Boltzmann equation), while only $-\left<\betac \hat{\Theta}^{\rm g}\right> \mathcal{Y}_{\rm SZ}= \left<\betac (\betac/3- \hat{\Theta}^{(1)})\right> \mathcal{Y}_{\rm SZ}$ arises from scattering in second order perturbation theory.
All other terms on the right hand side of Eq.~\eqref{eq:final_EQUATION} are from collisions of photons with the baryonic matter (electrons and baryons).
As we show below, at high redshifts photon mixing terms dominate, while at low redshifts the second order Doppler terms dominate (Sect.~\ref{sec:source_S_av}).

\section{Source terms for the cosmological dissipation problem}
\label{sec:sources_dist}
In the previous section we derived the second order Boltzmann equation that allows describing the evolution of the average spectral distortion of the CMB caused by the dissipation of acoustic modes in the Universe.
The main problem is now to compute the average source terms in Eq.~\eqref{eq:final_EQUATION}. 
All sources are caused by first order perturbations, for which precise
transfer functions can be computed numerically \citep[e.g.,][]{Peebles1970,BE1984,Ma1995}. Therefore, the ensemble average of $X\,Y$ is given by
\beal
\label{eq:def_ens_av_XY}
\left< X\,Y\right>&= \int \frac{k^2\id k }{2\pi^2} P(k) \, T_{X}\,T_{Y},
\end{align}
where $P(k)$ is the primordial power spectrum and $T_i$ the transfer function for the variable $i\in \{X,Y\}$.
Below we use both numerical as well as analytical results to perform this task.

\subsection{Source of distortion in the tight coupling limit}
\label{sect:approx_SAC}
At high redshifts the tight coupling approximation can be used to estimate the sources of distortions.
In this limit only monopole, dipole and quadrupole temperature perturbations are present. Higher order anisotropies can be neglected. 
In addition the local dipole and peculiar motion are directly related and practically cancel each other until close to the cosmological recombination epoch.

\subsubsection{Main source of $y$-distortions}
First looking at the main source term in Eq.~\eqref{eq:final_EQUATION}, which is $\mathcal{Y}_{\rm SZ} \mathcal{S}_{\rm ac}\simeq\mathcal{Y}_{\rm SZ}\left[3\hat{\Theta}^2_{1}-2 \betac\hat{\Theta}_{1}+\betacsq/3+\frac{9}{2}\hat{\Theta}^2_{2}\right]$, we have to compute the ensemble average of 
\beal
\nonumber
&\frac{\left(3\hat{\Theta}_{1}-\betac\right)^2}{3}+\frac{9}{2}\hat{\Theta}^2_{2}.
\end{align}
In Fourier space squared quantities are replaced by convolutions. However, because of isotropy and homogeneity we can directly replace these convolutions with the product of the transformed variables, after taking the ensemble average. 
In the tight coupling limit one has the simple relations \citep[e.g., see][]{DodelsonBook}
\beal
\nonumber
\betac&\approx \frac{3\hat{\Theta}_{1}}{1-i\frac{kc_{\rm s} R }{\taudotc}},
&\hat{\Theta}_{2}&\approx i\frac{4k}{9\taudotc}\,\hat{\Theta}_{1},
\end{align}
where $\cs\approx 1/\sqrt{3(1+R)}$ is the effective sound speed of the photon-baryon fluid and $R=3\rho_{\rm b}/4\rho_\gamma$ is the baryon loading. 
Also here $\taudotc$ is the derivative of $\tau$ with respect to conformal time, and $\hat{\Theta}_{l}$ and $\betac $ are now Fourier variables, but we omit writing $\tilde{X}$.
With these expressions we can compute the source function in Fourier space:
\beal
\label{eq:S_ac_k_dependence}
\mathcal{S}_{\rm ac}
& \rightarrow 
\frac{k^2}{\taudotc^2}\left[\frac{R^2}{1+R}+\frac{8}{9}\right]\,|\hat{\Theta}_{1}|^2 \mathcal{Y}_{\rm SZ}.
\end{align}
The first contribution in brackets can be identified as  heat
  conduction, while the second term is due to shear viscosity \citep{WeinbergBook}.
Additional corrections because of polarization have been neglected, but would lead to $8/9\rightarrow 16/15$ \citep[cf.][]{Kaiser1983, Zaldarriaga1995}.
Expression \eqref{eq:S_ac_k_dependence} shows that at high redshifts the main source of distortions arises solely from the quadrupole anisotropy, because $R\rightarrow 0$.
The approximation by \citet{Hu1994} was also obtained in this limit, however, as we see below an additional factor $3/4$ arises here.

Neglecting the variation of the potentials close to the diffusion scale, $\kD$, we have 
\citep[cf.][]{DodelsonBook}
\beal
\nonumber
|\hat{\Theta}_{1}|^2
& \approx\frac{\sin^2\left(k\rs\right)}{3(1+R)} \, [3\hat{\Theta}_{0}(0)]^2 \, e^{-2k^2/\kD^2}.
\end{align}
According to \citet{Ma1995} we can furthermore write
\beal
\nonumber
[3\hat{\Theta}_{0}(0)]^2
& = P_\zeta(k)/[1+4R_\nu/15]^2 \equiv \alpha_\nu \, P_\zeta(k),
\end{align}
where $R_\nu=\rho_\nu/(\rho_\gamma+\rho_\nu)\approx 0.41$ 
and $P_\zeta(k)=2\pi^2 A_\zeta k^{-3} (k/k_0)^{\nS-1}$, with $A_\zeta=\pot{2.4}{-9}$, and $k_0=0.002\,{\rm Mpc}^{-1}$ \citep{Komatsu2010}.
We therefore have to calculate the integral
\beal
\label{eq:S_ac_int}
\left<\mathcal{S}_{\rm ac}\right>
& \approx 
\frac{\alpha_{\nu}\cs^2}{2 \taudotc^2}\left[\frac{R^2}{1+R}+\frac{8}{9}\right]\,
\int \frac{\id^3 k}{(2\pi)^3}\,k^2 P_\zeta(k)\, 2\sin^2\left(k\rs\right)\, e^{-2k^2/\kD^2}
\nonumber
\\
&=\frac{A_\zeta}{k_0^{\nS-1}}\,\frac{\alpha_{\nu}}{\taudotc} \left(\partial_\eta \kD^{-2}\right)
\,\int \id k \,k^{\nS} \, 2\sin^2\left(k\rs\right)\, e^{-2k^2/\kD^2}
\nonumber
\\
&\approx 
\frac{A_\zeta}{k_0^{\nS-1}}\,\frac{\alpha_{\nu}}{\taudotc} \left(\partial_\eta \kD^{-2}\right)
2^{-(\nS+3)/2}\,	\kD^{\nS+1}\,\Gamma\left(\frac{1+\nS}{2}\right),
\end{align}
where we used $\partial_\eta \kD^{-2}=\cs^2[R^2/(1+R)+8/9]/(2\taudotc)$ \citep[cf.][]{DodelsonBook}.
To obtain the effective heating rate from this we have to multiply by $\taudotc \, Y_{\rm SZ}$ and integrate over $\xg^3 \id \xg\id^2 \vgh$. This gives $\partial_\eta Q_{\rm ac}$, which upon conversion to redshift reads
\beal
\label{eq:heat_SZ_appr}
\frac{1}{a^4 \rho_\gamma}\frac{\id a^4 Q_{\rm ac}}{\id z}
\approx 
\frac{4\alpha_{\nu}\,A_\zeta}{k_0^{\nS-1}} \left(\partial_z \kD^{-2}\right) 
2^{-(\nS+3)/2}\,	\kD^{\nS+1}\,\Gamma\left(\frac{1+\nS}{2}\right).
\end{align}
The factor of $4$ arises from the integral $\int \xg^3 \mathcal{Y}_{\rm SZ} \id\xg = 4 \rho_\gamma$, so that $(a^4 \rho_\gamma)^{-1}\id a^4 Q_{\rm ac} / \id z\equiv -4\taudot \left<\mathcal{S}_{\rm ac}\right>/H(1+z)$, where $H$ is the Hubble expansion factor.
Furthermore, we have $4\alpha_{\nu}\simeq 3.25$. 
We will see below that Eq.~\eqref{eq:heat_SZ_appr} provides a very useful approximation for the effect of dissipation from acoustic modes, however, small modifications of the transfer function lead to an even better agreement with the full numerical result (see Sect.~\ref{sec:improve}).

At high redshifts we can give an explicit expression for the dependence of the effective heating rate on $\nS$. During the $\mu$-era we find $\kD \approx A^{-1/2}_{\rm D} (1+z)^{3/2}$ with \citep[compare also][]{Khatri2011}
\beal
\nonumber
A_{\rm D}\approx \frac{ (8/9) \, c } {18 H_0 \Omega^{1/2}_{\rm r} N_{\rm e, 0} \sigT} \approx 
\pot{5.92}{10}\,{\rm Mpc^{2}}.
\end{align}
Here $H_0$ is the Hubble parameter, $\Omega_{\rm r}$ is the density of relativistic species, and $N_{\rm e}=N_{\rm e, 0}(1+z)^3$ is the number density of electrons (bound and free).
In this limit $\partial_z \kD^{-2}=-3A_{\rm D}/(1+z)^{4}$, such that 
\beal
\label{eq:heat_SZ_appr_final_nS}
\frac{1}{a^4 \rho_\gamma}\frac{\id a^4 Q_{\rm ac}}{\id z}
\approx 
-\frac{3\alpha_{\nu}\,A_\zeta}{1+z} 
\;\Gamma\left(\frac{1+\nS}{2}\right)
\left[ \frac{(1+z)^3}{2k_0^2 A_{\rm D}}\right]^{\frac{\nS-1}{2}}.
\end{align}
This expression shows that for $\nS \simeq 1$ the effective heating rate scales as $1/(1+z)$, and just depends on $A_\zeta$ and $\alpha_\nu=(1+4 R_\nu/15)^{-2}$.
The CMB spectrum therefore also provides an interesting probe for the effective number of neutrino species.
On the other hand the additional cosmology dependence entering via $A_{\rm D}$ is much weaker.

Expression~\eqref{eq:heat_SZ_appr_final_nS} also shows, that for nearly scale-invariant power spectra the dependence on the exact coefficient for the shear viscosity is small.
In the definition of $A_{\rm D}$ we used $8/9$. Replacing this with $16/15$ \citep{Kaiser1983} leads to a $\simeq 9\%$ change of $\kD$; however, for the effective heating rate it implies a modification by a factor of $(5/6)^{(\nS-1)/2} \approx 1+\frac{1}{2}\ln(5/6) \, (\nS-1)\approx 1 - 0.09 (\nS-1)$.
For $\nS=0.96$ this means that the energy release is overestimated by $\simeq 0.4\%$ when using $8/9$ instead of $16/15$.
Numerically we also found this difference.

\subsubsection*{Comparison with earlier estimates}
\label{sec:comp_Hu_appr}
The result Eq.~\eqref{eq:heat_SZ_appr_final_nS} is $3/4$ of the amplitude obtained by \citet{Khatri2011} and \citet{Hu1994}.
Their estimates are based on the formula for average energy
density in sound waves in massive particles applied to photons, 
\beal
\nonumber
Q
\approx \frac{1}{3} \rho_\gamma \left< \delta^2_{0,\gamma}\right>.
\end{align}
The factor of $1/3$ arises from the sound speed of the photon fluid, $c_{\rm s}=1/\sqrt{3}$, and $\delta_{0,\gamma}=\Delta \rho_0/\rho_\gamma$ is the perturbation in the energy density of the monopole part of the photon field.
The effective heating rate in this picture is
\beal
\nonumber
\frac{\id }{\id t}\frac{Q_{\rm Classical}}{\rho_\gamma}
&
\approx -\frac{1}{3}\frac{\id}{\id t} \left<\left(\frac{\Delta\rho_0}{\rho_\gamma}\right)^2\right>
=-\frac{16}{3}\frac{\id}{\id t} \left<\hat{\Theta}_0^2\right>
\equiv\frac{4}{3}\,\frac{\id }{\id t}\frac{Q_{\rm ac}}{\rho_\gamma}.
\end{align}
where we compared with Eq.~\eqref{eq:heat_SZ_appr}.
As we see this is 4/3 times the result obtained with the full Boltzmann equation.

The estimate above does not full reflect the microphysics of the problem, in
which the dissipation/heating is mainly due to the scattering of the quadrupole and
resulting superposition of blackbodies. In particular we should be
using the total energy in the photon field  with varying temperature by means of the
Stephan-Boltzmann law, $\rho_{\gamma}\propto T^4$, as already explained in Sect.~\ref{sec:estimate_doesnotwork}.
Then we can write
\beal
\nonumber
\left<\rho^{\rm temp}_\gamma\right>&
=\rho_{\rm pl}(T_{\rm av})\left[1+6\left<\Theta^2_0+y_{0}\right>\right].
\end{align}
We already know that $2/3$ of $6\left<\Theta^2_0+y_{0}\right>=6\left<y^{\rm s}_0\right>$ directly appears as a temperature perturbation without creating any distortion (cf. Sect.~\ref{sec:change_Ng_second}). This statement is even true when considering the subsequent evolution, since injection of perturbations with spectral shape $\mathcal{G}$ allows movement along a sequence of quasi-equilibrium states with blackbody spectra. 
This only leads to a slow increase of the average CMB temperature, but no distortion.

On the other hand $2\left<y^{\rm s}_0\right>$ appears directly as $y$-type distortion.
We therefore have the effective energy release
\beal
\nonumber
\frac{\id}{\id t}\frac{Q}{\rho_\gamma}
&
\approx -2\frac{\id}{\id t}\left<y^{\rm s}_0\right>= -4 \left<[\Theta\id_t \Theta]_0\right> = 4\left<\mathcal{S}_{\rm dist}\right> \approx 4\taudot \left<\mathcal{S}_{\rm ac}\right>,
\end{align}
where $\mathcal{S}_{\rm dist}$ is defined by Eq.~\eqref{eq:Boltzmann_second_order_general_II_b}.
The last step is possible since at high redshifts the difference $\left<\taudot^{-1}\mathcal{S}_{\rm dist}-\mathcal{S}_{\rm ac}\right>=\left<\betac \hat{\Theta}^{\rm g}_1\right>\approx 0$.
This shows that there only changes in the mixture of blackbodies are important. However at low redshifts the term $\left<\betac \hat{\Theta}^{\rm g}_1\right>$ starts dominating. This contribution to the heating arises from second order scattering terms (dipole and velocity cross-term $\simeq \betac \Theta_1$ and second order Doppler effect, $\propto \betacsq$) which are not captured by the simple estimates given above.
Furthermore, {\it all} multipoles contribute to the energy release.
This naturally defines the real dissipation (without counting power interchange among the multipoles) from the system and not just the total change of the monopole amplitude.
Finally, the energy release caused by second order scattering terms can be added by simply replacing $\hat{\Theta}_1\rightarrow\hat{\Theta}^{\rm g}_1$. This then recovers the full source term $\left<\mathcal{S}_{\rm ac}\right>$ as obtained with the Boltzmann equation.

\color{black}
\subsubsection{Other sources of distortions in the tight coupling limit}
\label{sec:other_S_appr}
So far we only considered the main source term in Eq.~\eqref{eq:final_EQUATION}.
However, we discovered additional source terms, which we discuss now.
The ensemble average of $\mathcal{S}_{\Delta n}$ is trivial, simply meaning the replacement $\Delta n^{(2)}_0 \rightarrow \Delta n^{(2)}_{\rm av}$ inside the corresponding integrals.
With $\left<\mathcal{S}_{\rm ac}\right>$ and Eq.~\eqref{eq:final_EQUATION} we also have 
\beal
\nonumber
\left<\ygl{1}\right>
& \approx 
\frac{\frac{10}{9}+\frac{27}{8} \cs^2\,R^2}{1+\frac{27}{8} \cs^2\,R^2 }
\left<\mathcal{S}_{\rm ac}\right>
\rightarrow  \frac{10}{9}\left<\mathcal{S}_{\rm ac}\right>,
\end{align}
where we also indicate the limiting behaviour at very high redshift.
For the source term associated with $\mathcal{H}$ according to \eqref{eq:Compton_source_terms_H} we only require $\left<\betac \hat{\Theta}^{\rm g}_1\right>$ in addition.
In Fourier space one has
\beal
\nonumber
\betac \hat{\Theta}^{\rm g}_1
& \rightarrow 
\frac{k^2}{\taudot^2} \frac{\sqrt{3}R}{\sqrt{1+R}}\,|\hat{\Theta}_{1}|^2 .
\end{align}
Comparing with the above expression we find
\bsub
\beal
\left<\betac \hat{\Theta}^{\rm g}_1\right>
& \approx 
\frac{27}{8} \frac{\cs\,R}{1+\frac{27}{8} \cs^2\,R^2 } \left<\mathcal{S}_{\rm ac}\right> \rightarrow  0
\\
\left<\mathcal{S}^{\The}_{\mathcal{H}}\right>
& \approx 
\left[
\frac{\frac{10}{9}-\frac{54}{5}\cs\,R+\frac{27}{8} \cs^2\,R^2}{1+\frac{27}{8} \cs^2\,R^2 }
\right] \left<\mathcal{S}_{\rm ac}\right> 
\rightarrow  \frac{10}{9}\left<\mathcal{S}_{\rm ac}\right>.
\end{align}
\esub
%
For the source function related to $\mathcal{E}$ using the same arguments and neglecting the octupole we find
\beal
\left<\mathcal{S}^{\The}_{\mathcal{E}}\right>
& \approx 
\left[
\frac{\frac{1}{9}-\frac{27}{20}\cs\,R+\frac{54}{25} \cs^2\,R^2}{1+\frac{27}{8} \cs^2\,R^2 }
\right] \left<\mathcal{S}_{\rm ac}\right> 
\rightarrow  \frac{1}{9}\,\left<\mathcal{S}_{\rm ac}\right>
\end{align}
This result shows that the source term related to $\mathcal{E}$ is suppressed by $1/10$ in comparison to the one for $\mathcal{H}$.

With these expression we can also compute the average of $\mathcal{S}^{\The}_{\mathcal{Y}_{\rm SZ}}$ and $\mathcal{S}_{\rm e/a}$.
At high redshifts we find the limiting behaviour
\beal
\left<\mathcal{S}^{\The}_{\mathcal{Y}_{\rm SZ}}\right> 
\approx \left<\mathcal{S}_{\rm e/a}\right>
\rightarrow 3.03 \left<\mathcal{S}_{\rm ac}\right>.
\end{align}
These expressions determine all the source terms appearing in Eq.~\eqref{eq:final_EQUATION}. At high redshifts these approximations work very well, while around hydrogen recombination they start breaking down. However, we mention here that source terms other than $\left<\mathcal{S}_{\rm ac}\right>$ appear with additional (small) coefficients $\simeq \Thz$ and frequency-dependent functions. This renders all these terms negligible as corrections to the main source functions, as we discuss in Sect.~\ref{sec:corr_source}.

\begin{figure}
\centering
\includegraphics[width=\columnwidth]{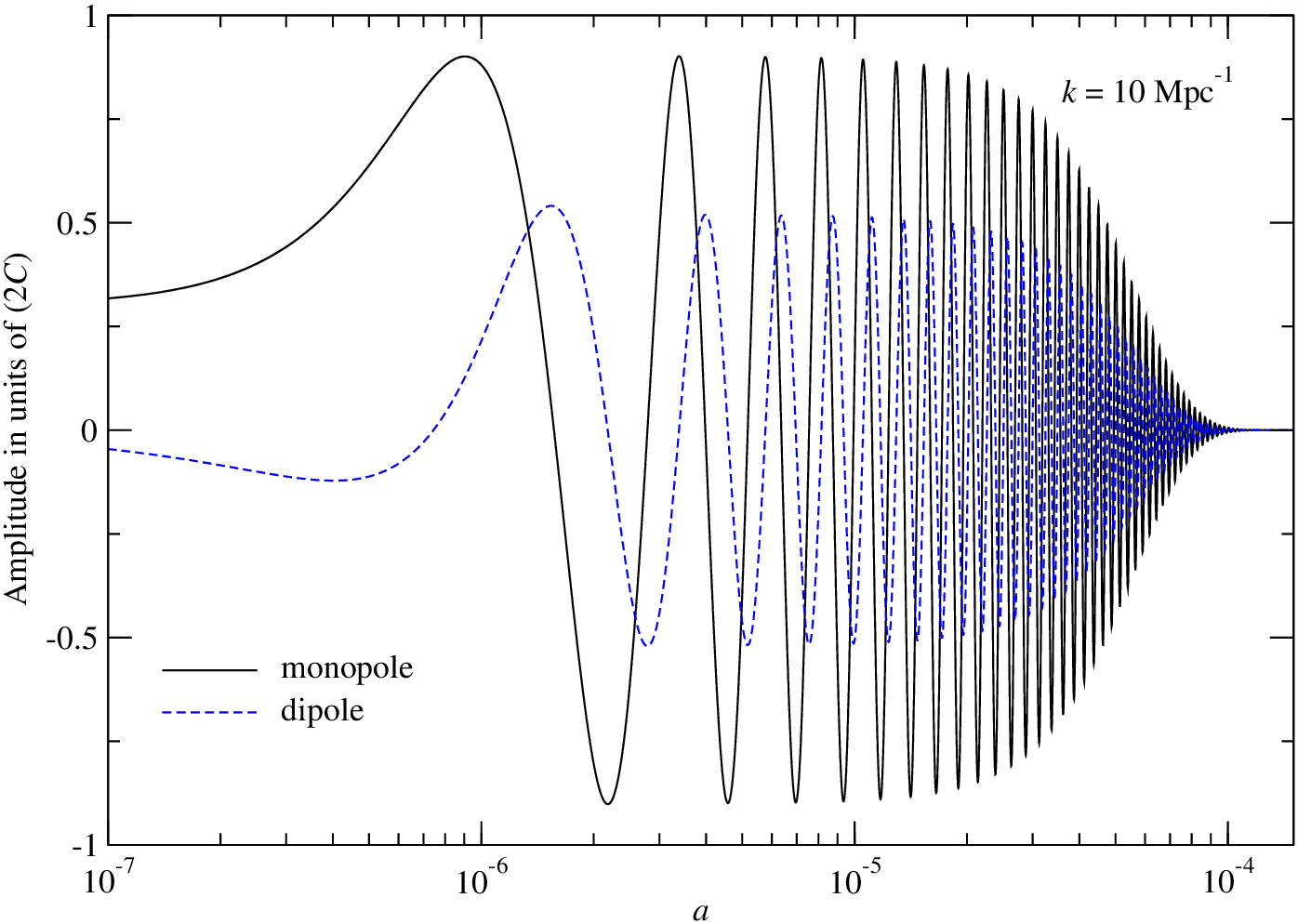}
\\[1mm]
\includegraphics[width=\columnwidth]{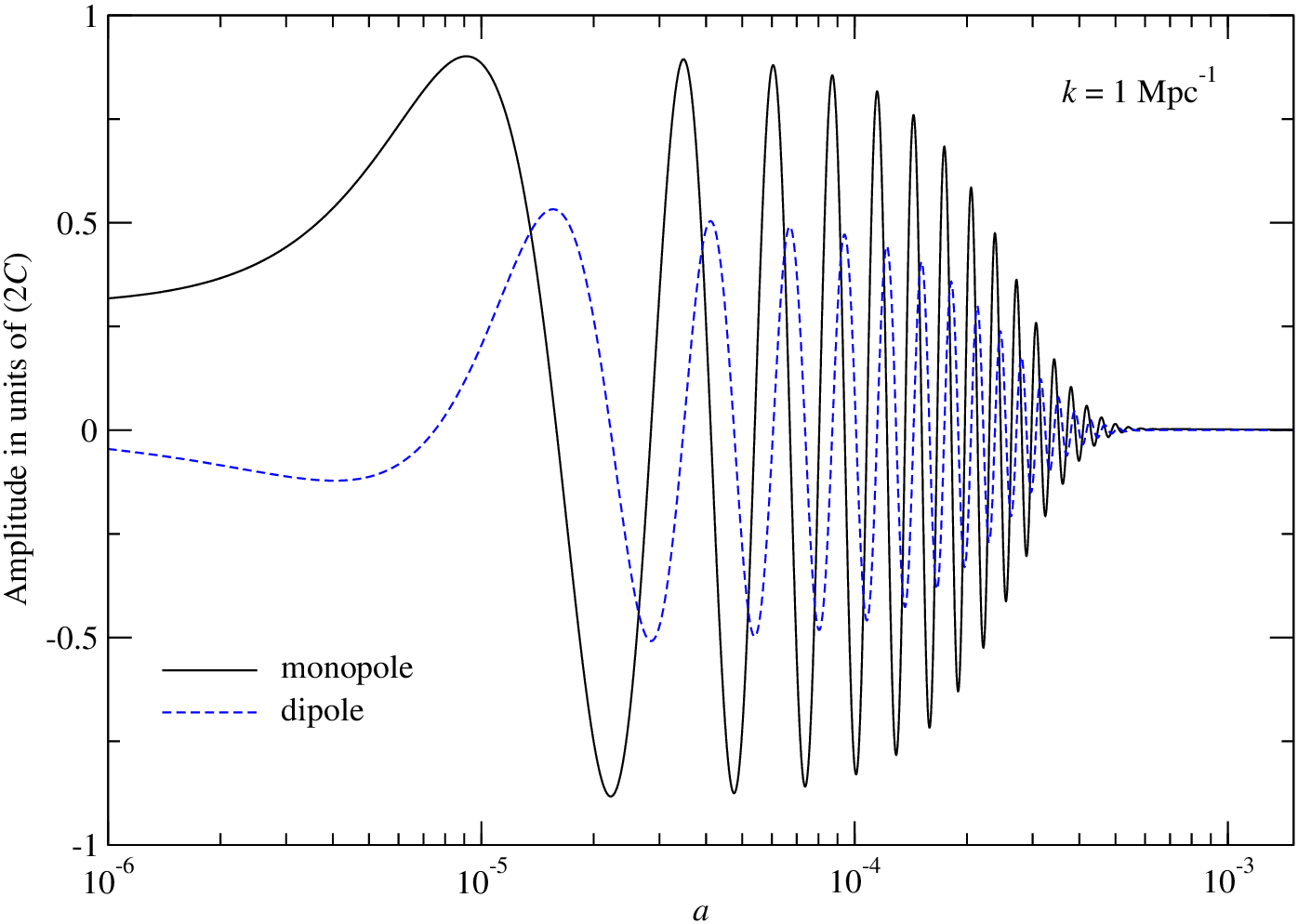}
\\[1mm]
\includegraphics[width=\columnwidth]{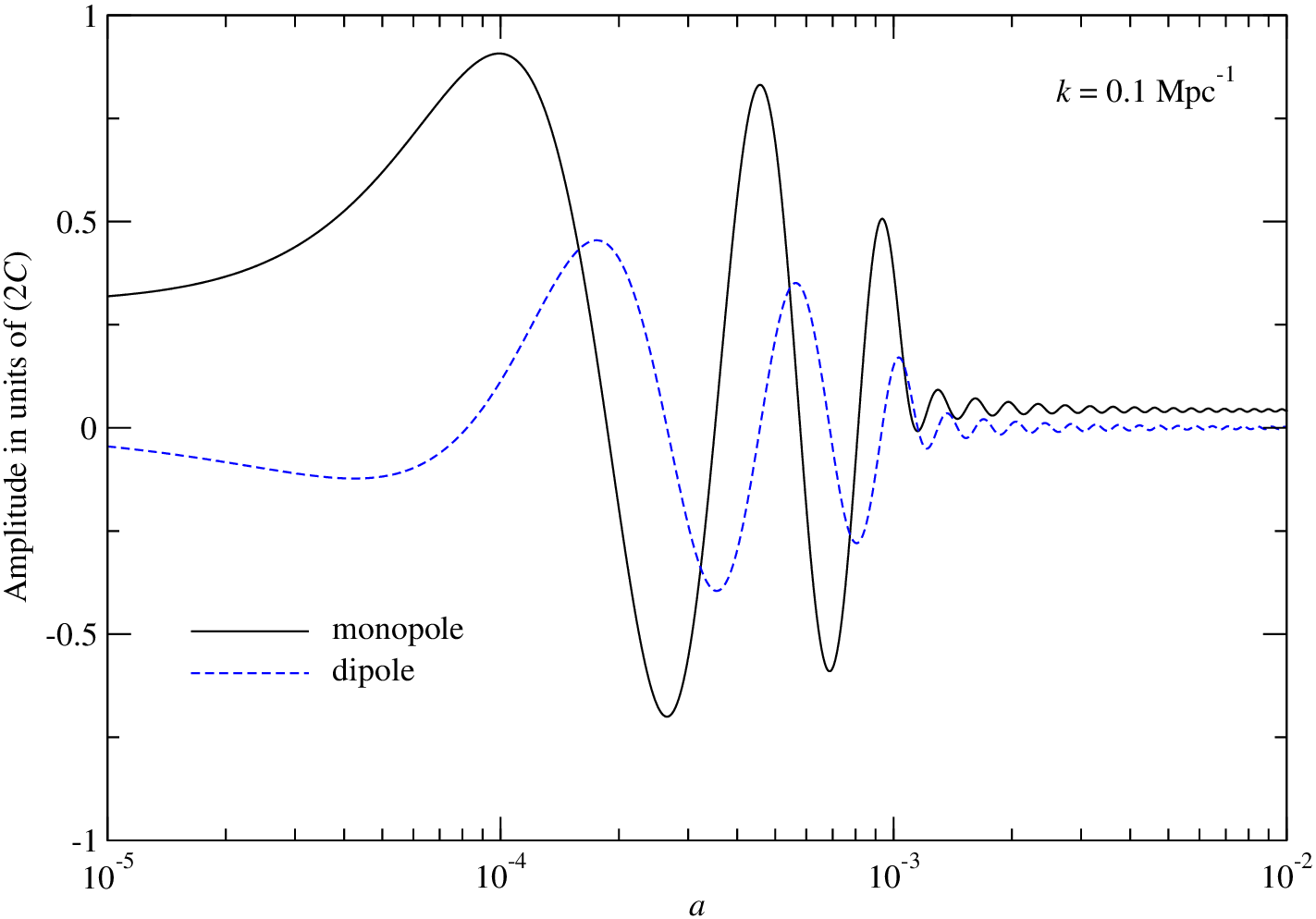}
\caption{Monopole and dipole transfer functions for different values of
  $k$, normalized to initial curvature perturbation in comoving gauge
    $\zeta\equiv 2C=1$. We compared all of the cases with {\sc CMBfast} and found agreement at the level of $\lesssim 1\%$. 
In the free-streaming regime many multipoles needed to be included to avoid errors caused by truncation of the Boltzmann hierarchy.
}
\label{fig:modes_MD}
\end{figure}
\begin{figure}
\centering
\includegraphics[width=\columnwidth]{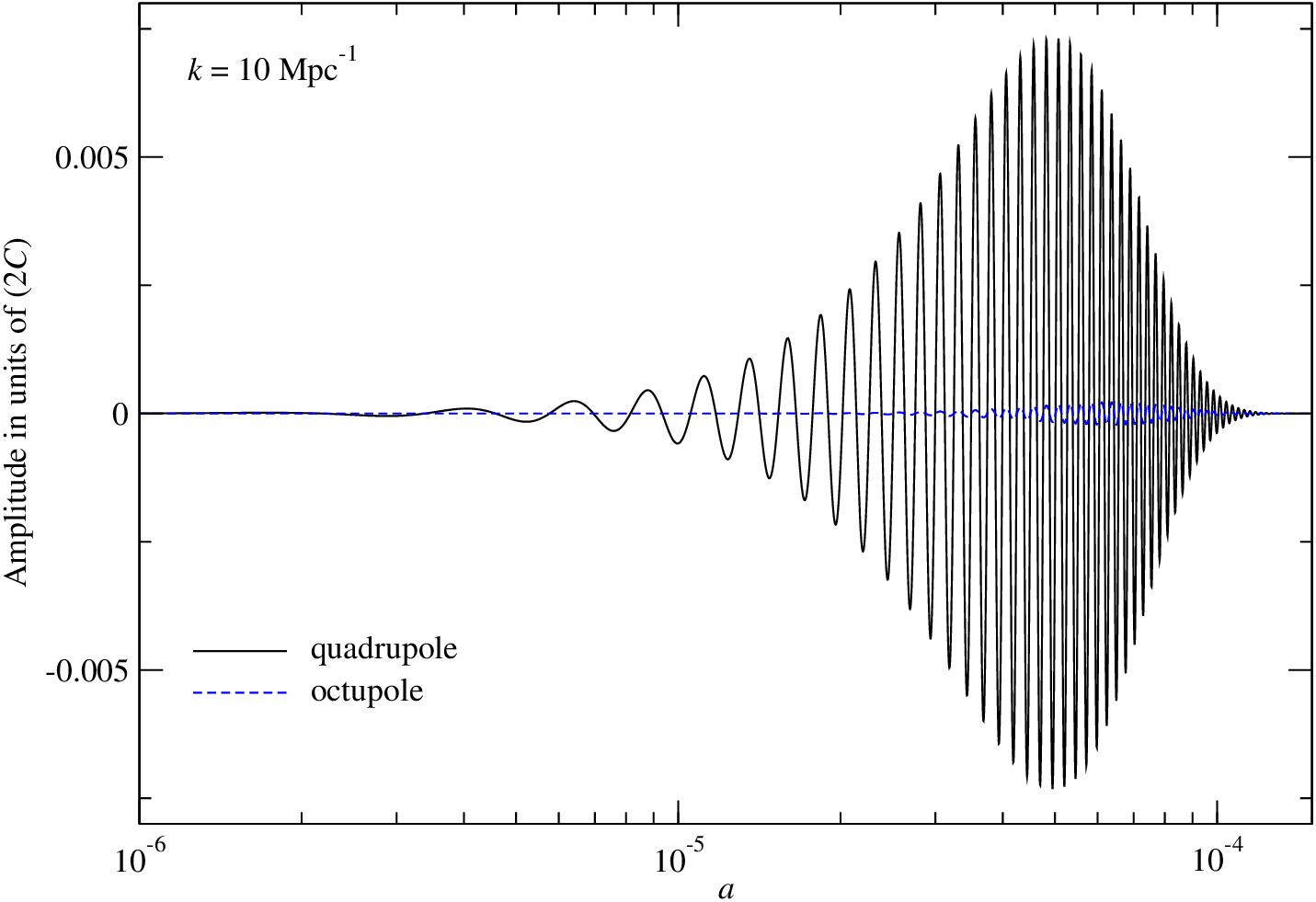}
\\[1mm]
\includegraphics[width=\columnwidth]{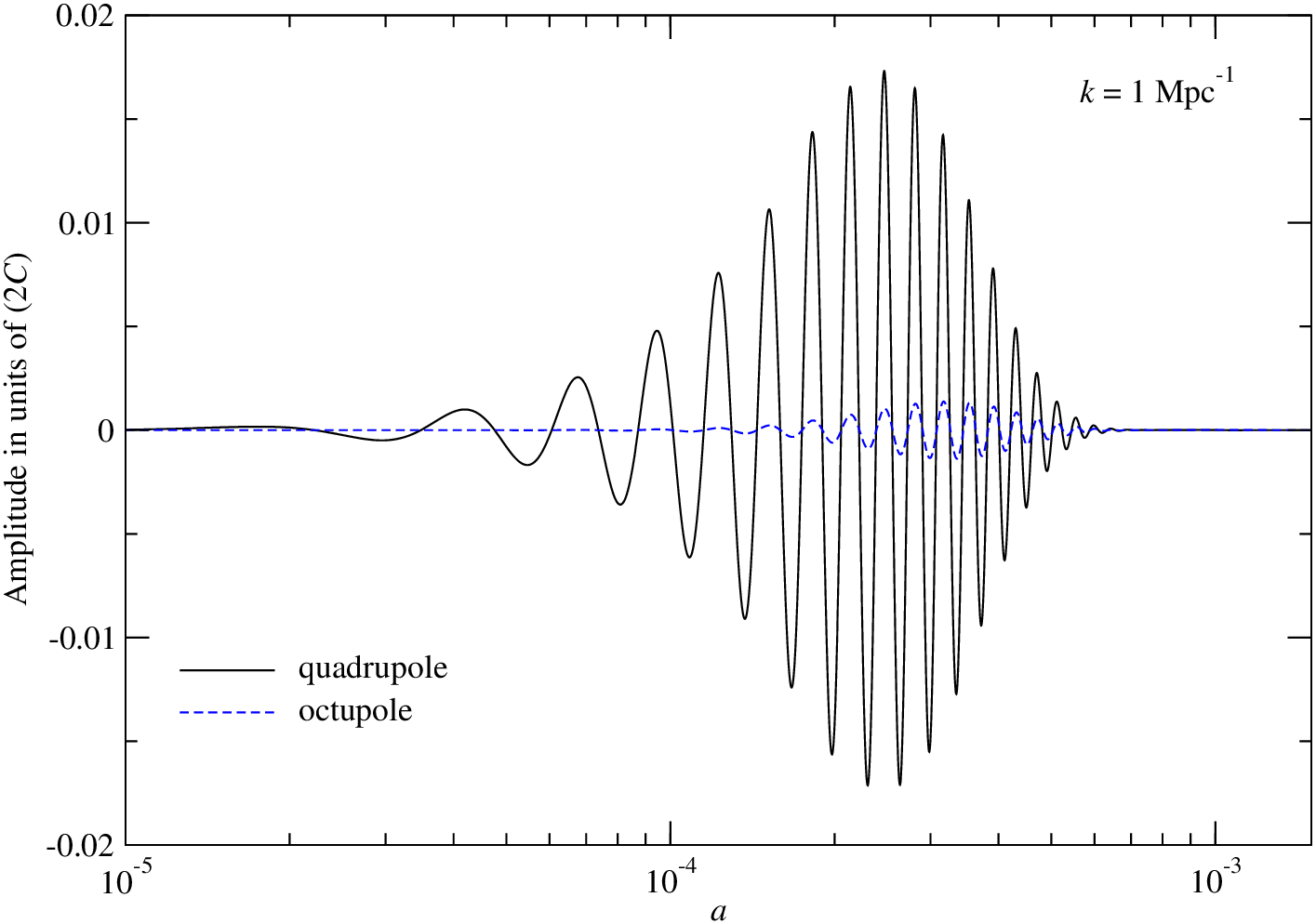}
\\[1mm]
\includegraphics[width=\columnwidth]{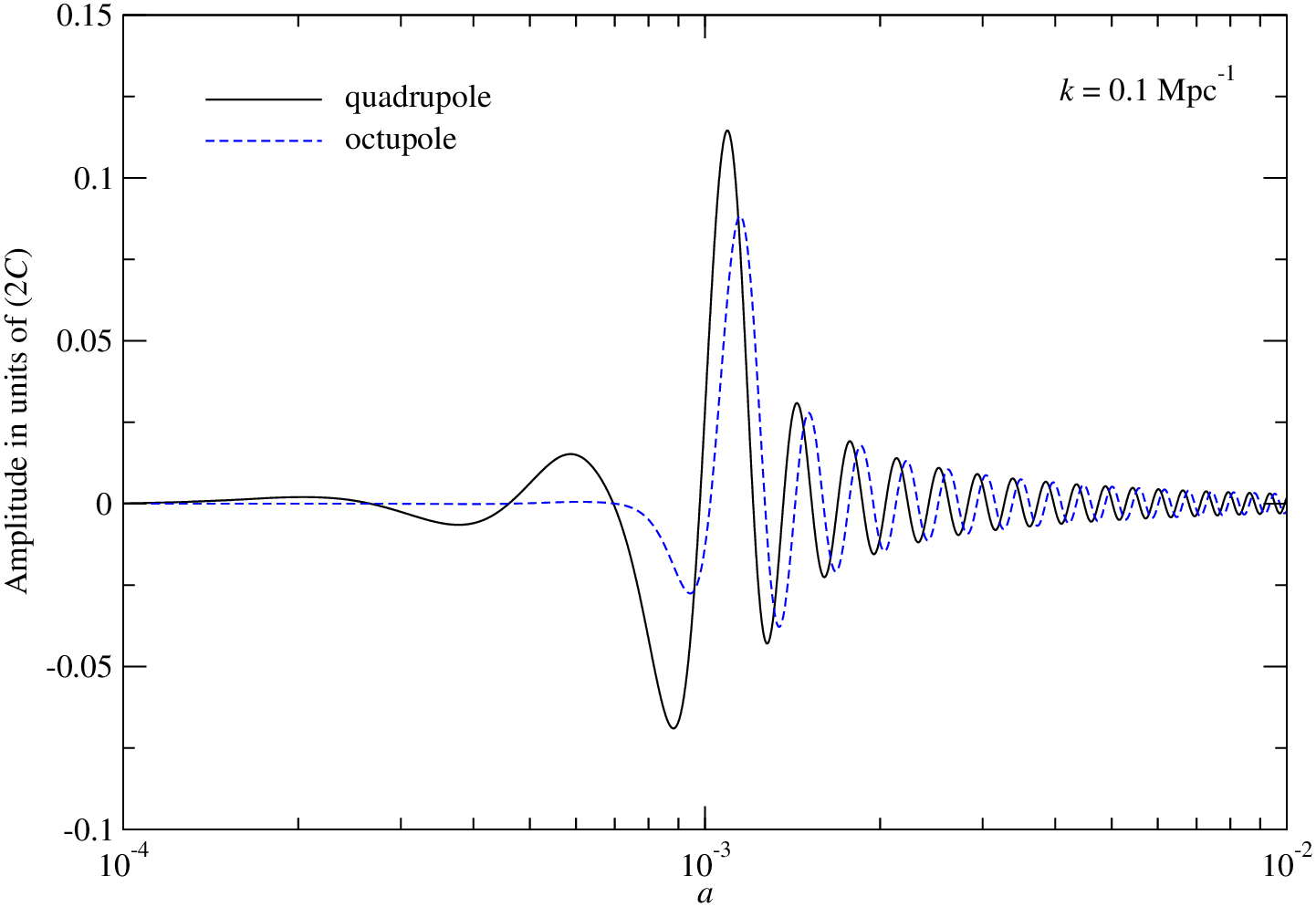}
\caption{Same as Fig.~\ref{fig:modes_MD} but for the quadrupole and octupole.}
\label{fig:modes_QO}
\end{figure}
\subsection{Results for the transfer functions}
\label{sec:transfer_funcs}
At high redshifts well before the recombination epoch one expects the tight coupling approximation to work very well in describing the source terms.
However, during and after recombination it is clear that this approximation breaks down and a numerical computation of the perturbation equations becomes necessary.
The main reason is that once the Thomson optical depth decreases higher multipoles start to contribute, and when scattering ceases the photon distribution evolves by free-streaming.

To understand the amplitudes of the different source terms as a function of
time we numerically solved the perturbation equations following
\citet{Ma1995}. For the calculations we use the conformal Newtonian gauge. We utilized the stiff ordinary equation (ODE) solver implemented for {\sc CosmoRec} \citep{Chluba2010b}.
To obtain accurate results we split the range in $k$ up into several parts. Within these regions we densely sampled uniformly in $\log k$.
For the evolution of modes that enter the horizon at very early times, only a few multipoles had to be followed, while during and after recombination we included up to $l=80$ multipoles for both photons and massless neutrinos.

In Figs.~\ref{fig:modes_MD} and \ref{fig:modes_QO} we show some examples for the transfer functions obtained up to $l=3$. 
We observe the well known behaviour: once modes enter the horizon they
oscillate; monopole and dipole anisotropies dominate at early times; 
once dissipation sets in quadrupole, octupole and  higher multipoles start
appearing; and for high values of $k$ usually only the monopole through quadrupole are important at all times, while for smaller $k$ higher multipoles also matter.
For the acoustic dissipation at high redshifts the quadrupole terms dominate, while at low redshift the net dipole/Doppler is largest (see next section).

To confirm our numerical treatment, we compared our transfer functions with outputs from {\sc Cmbfast} \citep{CMBFAST}, finding excellent agreement (better than $1\%$ before free-streaming starts). 
In the free-streaming regime we included up to 80 multipoles to avoid errors caused by truncation of the Boltzmann hierarchy \citep[see][for related discussion]{Ma1995}.
For the curves given in Fig.~\ref{fig:modes_MD} and \ref{fig:modes_QO} we accounted for polarization, however, for the computation of the source functions presented in this section we neglected it. This only affects the results at the level of $\simeq 5 \%$ (see Sect.~\ref{sec:polarization}).

\begin{figure}
\centering
\includegraphics[width=\columnwidth]{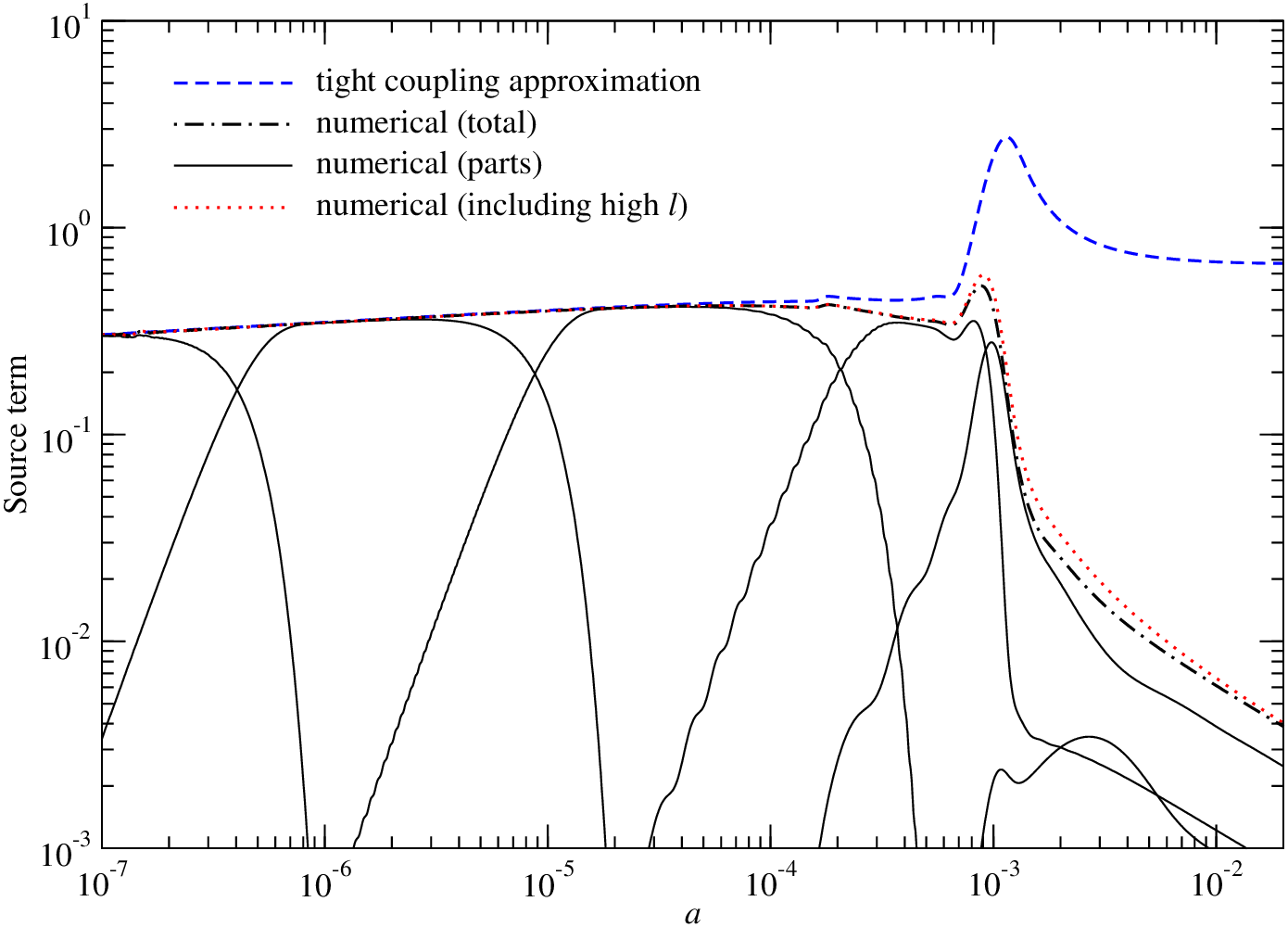}
\caption{Main source function, $\left<\mathcal{S}_{\rm ac}\right>$, in units of $A_\zeta H(z) / \taudot$ for $\nS=0.96$. 
For the dotted/red line multipoles up to $l=80$ were included, while for all the other curves only the moments up to the quadrupole were accounted for. 
The result obtained with Eq.~\eqref{eq:heat_SZ_appr} (dashed/blue) is compared with the total result from the perturbation code (dash-dotted/black). The solid lines give partial contributions to the source function for different ranges in $k$ (in units of ${\rm Mpc}^{-1}$), starting from the right with $10^{-4}\leq k\leq 10^{-2}$, $10^{-2}\leq k\leq 10^{-1}$, $10^{-1}\leq k\leq 1$, $1\leq k\leq 10^{2}$, $10^{2}\leq k\leq 10^{4}$, and $10^{4}\leq k\leq 10^{6}$.}
\label{fig:heating_SZ}
\end{figure}
%
\begin{figure}
\centering
\includegraphics[width=\columnwidth]{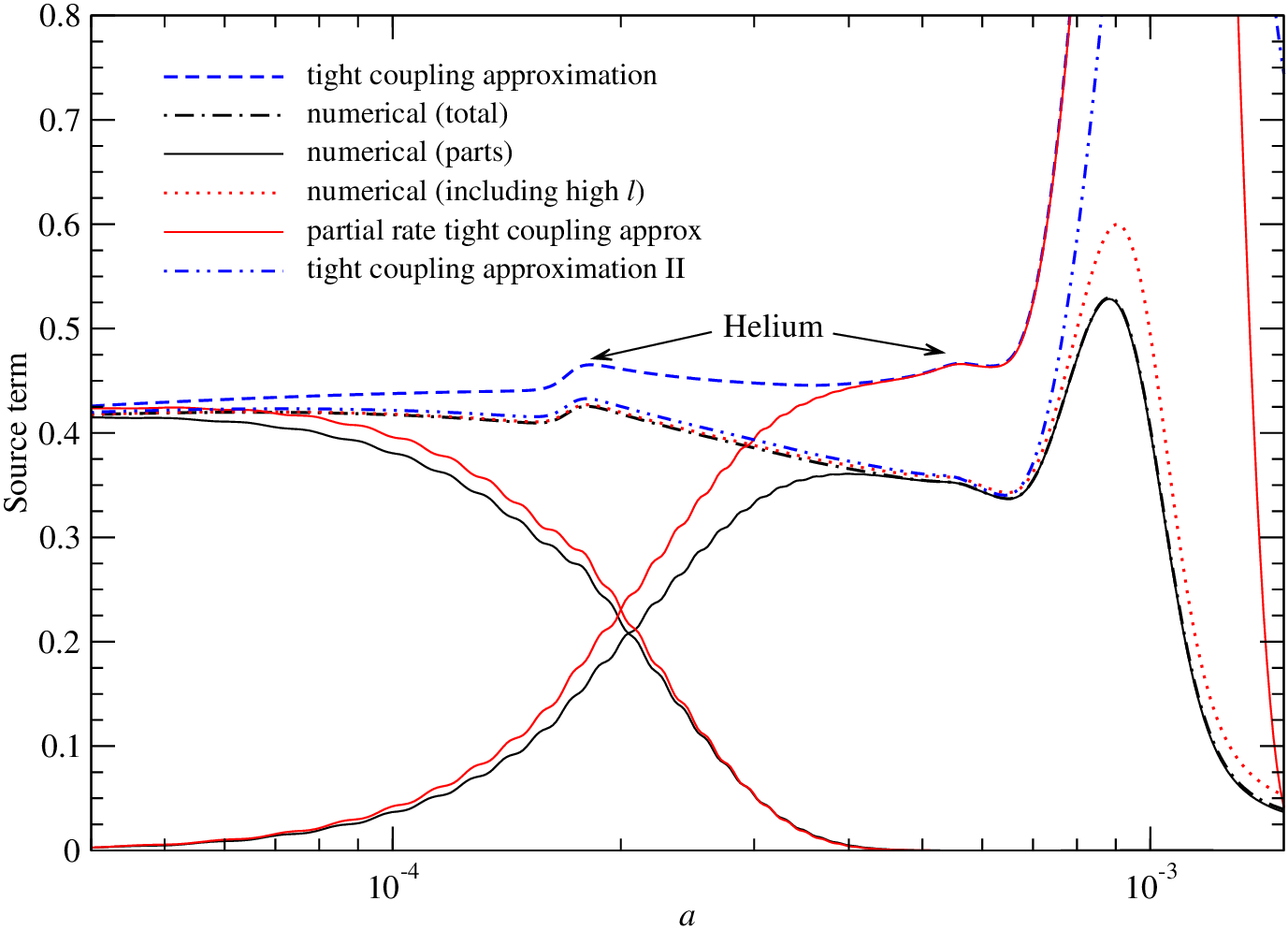}
\caption{Same as Fig.~\ref{fig:heating_SZ} but zoomed in on the region $a\simeq \pot{4}{-5} - 10^{-3}$.
The solid lines show the two partial contributions from $k$ modes in the
range $10^{-2}\leq k\leq 1$ and $1\leq k\leq 10^2$ (right to left). One can
see slight oscillations of the partial rates around $a\simeq \pot{2}{-4}$,
which average out in the sum. We also present the analytic approximation for the partial rates based on the simple tight-coupling solution. Furthermore, we have marked the two helium recombination features, which are visible in both the simple analytic estimate as well as the full numerical result. The blue/dashed-double-dotted line shows an improved analytical approximation which includes some higher order corrections (see Sect.~\ref{sec:improve} for details).
}
\label{fig:heating_SZ_zoom}
\end{figure}
\subsection{Source term $\left<\mathcal{S}_{\rm ac}\right>$}
\label{sec:source_S_av}
To begin let us consider the source term $\left<\mathcal{S}_{\rm ac}\right>$. In Fig.~\ref{fig:heating_SZ} we present the rate as a function of scale factor $a$. 
We scaled $\left<\mathcal{S}_{\rm ac}\right>$ by $\taudot/H$ to make it comparable to $(1+z)\, \id (Q_{\rm ac}/\rho_\gamma)/\id z$ defined by Eq.~\eqref{eq:heat_SZ_appr}.
We compare the results obtained using the simple tight coupling formulae to those computed with our Boltzmann code.
At high redshifts we see excellent agreement between the two, while close to hydrogen recombination ($z\simeq 10^3$) the estimate using the tight-coupling approximation yields a larger result.
After recombination ends ($z\lesssim 800$) free-streaming starts and the source term drops very fast, a point that is unsurprisingly not captured with the tight coupling solutions.
At these late times the contribution from higher multipoles and bulk motions of the baryons also become important, as our numerical results indicate.

In Fig.~\ref{fig:heating_SZ} we also present the partial rates. At redshifts $z\simeq 10^5-10^6$ modes with $k\simeq 10^2-10^4\,\rm Mpc^{-1}$ dissipate their energy, while at redshifts $z\simeq 10^4-10^5$ those with $k\simeq 1-10^2\rm \,Mpc^{-1}$ contribute most. During and after recombination modes in the range $k\lesssim 1 \,\rm Mpc^{-1}$ are responsible for the heating.
We can also see that the partial rates have a small time-dependent modulation caused by the oscillatory behaviour of the transfer function. This is even more visible in Fig.~\ref{fig:heating_SZ_zoom}, where we have zoomed in on the region $a\simeq \pot{4}{-5} - 10^{-3}$.
The modulation is also found when using the tight-coupling solution. 
However, when adding the contributions of neighbouring modes the oscillations disappear. This shows that the heating rate for individual modes is actually varying in time, but the superposition of modes that are separated by about half a sound horizon removes this variation.

\begin{figure}
\centering
\includegraphics[width=\columnwidth]{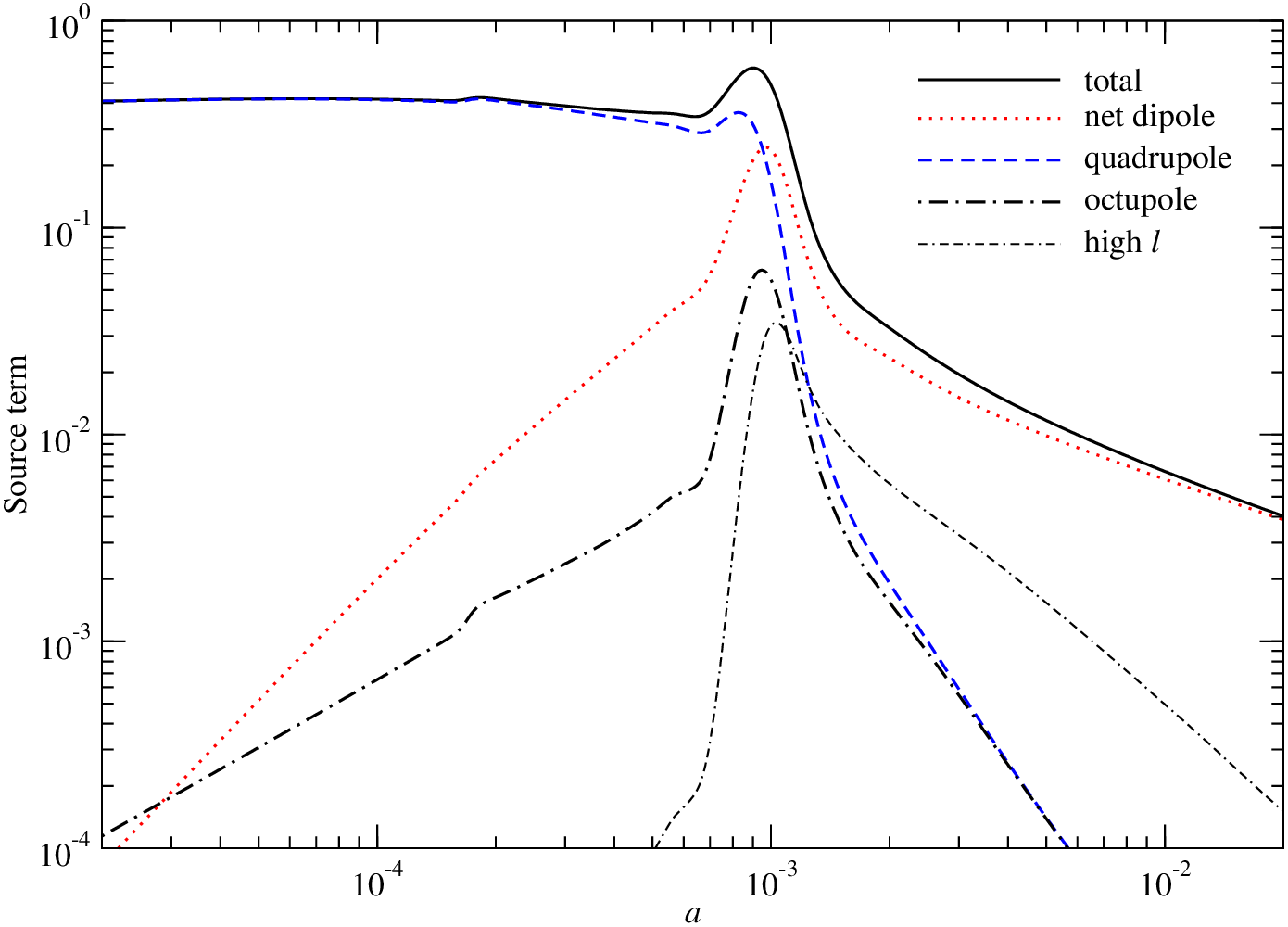}
\caption{Different contributions to the main source function, $\left<\mathcal{S}_{\rm ac}\right>$, in units of $A_\zeta H(z) / \taudot$ for spectral index, $\nS=0.96$. 
At high redshifts the quadrupole term clearly dominates, while at low redshifts the net dipole/Doppler term, i.e., $(\hat{\Theta}^{\rm g}_1)^2$, is most important. During recombination the higher multipoles contribute $\simeq 10\%$, but are negligible otherwise.
}
\label{fig:heating_terms}
\end{figure}
In Fig.~\ref{fig:heating_terms} we illustrate the importance of the different terms for $\left<\mathcal{S}_{\rm ac}\right>$. 
At high redshifts the quadrupole term clearly dominates, while at low redshifts after hydrogen recombination the net dipole/Doppler term is most important. During recombination the higher multipoles contribute $\simeq 10\%$, but are negligible otherwise.

\subsubsection{Improving the analytic estimate}
\label{sec:improve}
As Fig.~\ref{fig:heating_SZ} indicates, even around \ion{He}{ii} and \ion{He}{i} recombination, differences are visible between approximation~\eqref{eq:heat_SZ_appr} for the source term and the numerical result.
The tight coupling approximation is expected to work at redshifts $z \gtrsim 10^3$, so that this differences is slightly surprising. 
However, in the derivation of Eq.~\eqref{eq:heat_SZ_appr} additional approximation were used.
Firstly, for the $k$ integral Eq.~\eqref{eq:S_ac_int} $\sin^2\left(k\rs\right)$ was replaced by the average value $1/2$. This approximation is not too bad and indeed only breaks down at redshifts below $\simeq 10^3$.
The other simplification is more severe. For the expression~\eqref{eq:S_ac_k_dependence} we approximated 
\beal
\nonumber
3(\hat{\Theta}^{\rm g}_1)^2 \approx 3\frac{k^2 c^2_{\rm s} R^2}{\taudot^2}\left[1+\frac{k^2 c^2_{\rm s} R^2}{\taudot^2}\right]^{-1} \approx
\frac{k^2}{\taudot^2}\,\frac{R^2}{1+R}.
\end{align}
The second step turns out to be too imprecise at $10^3 \lesssim z\lesssim 10^4$. If we include the full expression and numerically carry out the $k$ integral, we find an improved agreement, with the difference decreasing roughly by 2.
This indicates that already at $z\lesssim 10^4$ higher order powers of $k/\taudot$ become important for the dispersion relation of the modes, which are consistently captured by the numerical treatment, but can only be partially accounted for with the above expressions.

Finally, in the more accurate approximations for the transfer functions of $\hat{\Theta}_1$ given by \citet{Hu1995CMBanalytic}, again neglecting corrections because of the potentials far inside the horizon (where the dissipation is happening),
there are additional factors of $(1+R)$ accounting for the finite energy density
of baryons. 
We note that even though the baryon energy density becomes comparable to the photon energy density before recombination, the pressure is dominated by photons until recombination occurs and the photons enter the free-streaming regime. The transfer function for $\hat{\Theta}_1$ has to be multiplied by $(1+R)^{-1/4}$, implying another factor $(1+R)^{-1/2}$ 
in $\left<\mathcal{S}_{\rm ac}\right>$. 
If we include these modifications we obtain the dash-dot-dotted line in Fig.~\ref{fig:heating_SZ_zoom}, clearly indicating an improvement over the more simple estimate, Eq.~\eqref{eq:heat_SZ_appr}.

\begin{figure}
\centering
\includegraphics[width=\columnwidth]{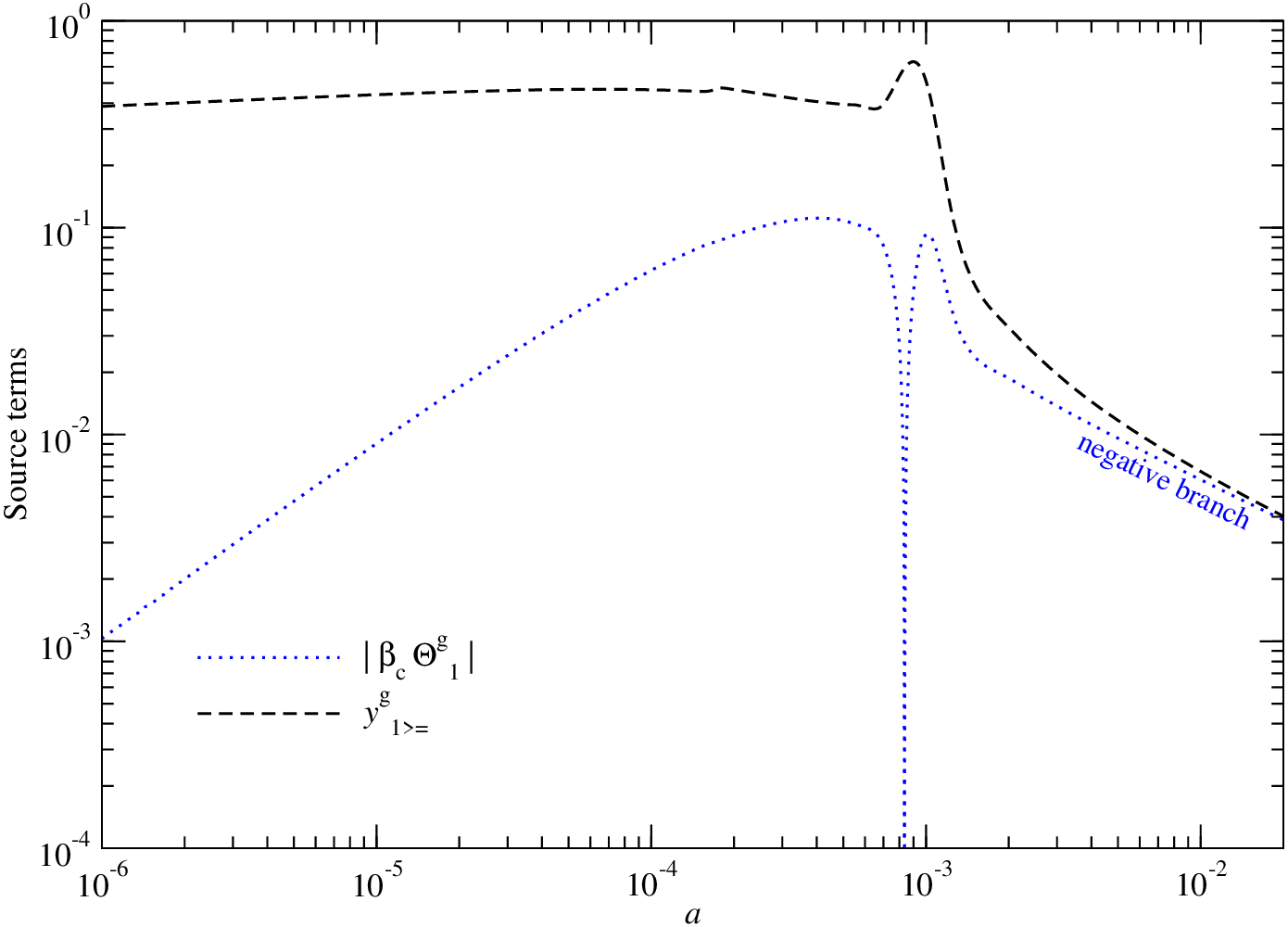}
\\[1mm]
\includegraphics[width=\columnwidth]{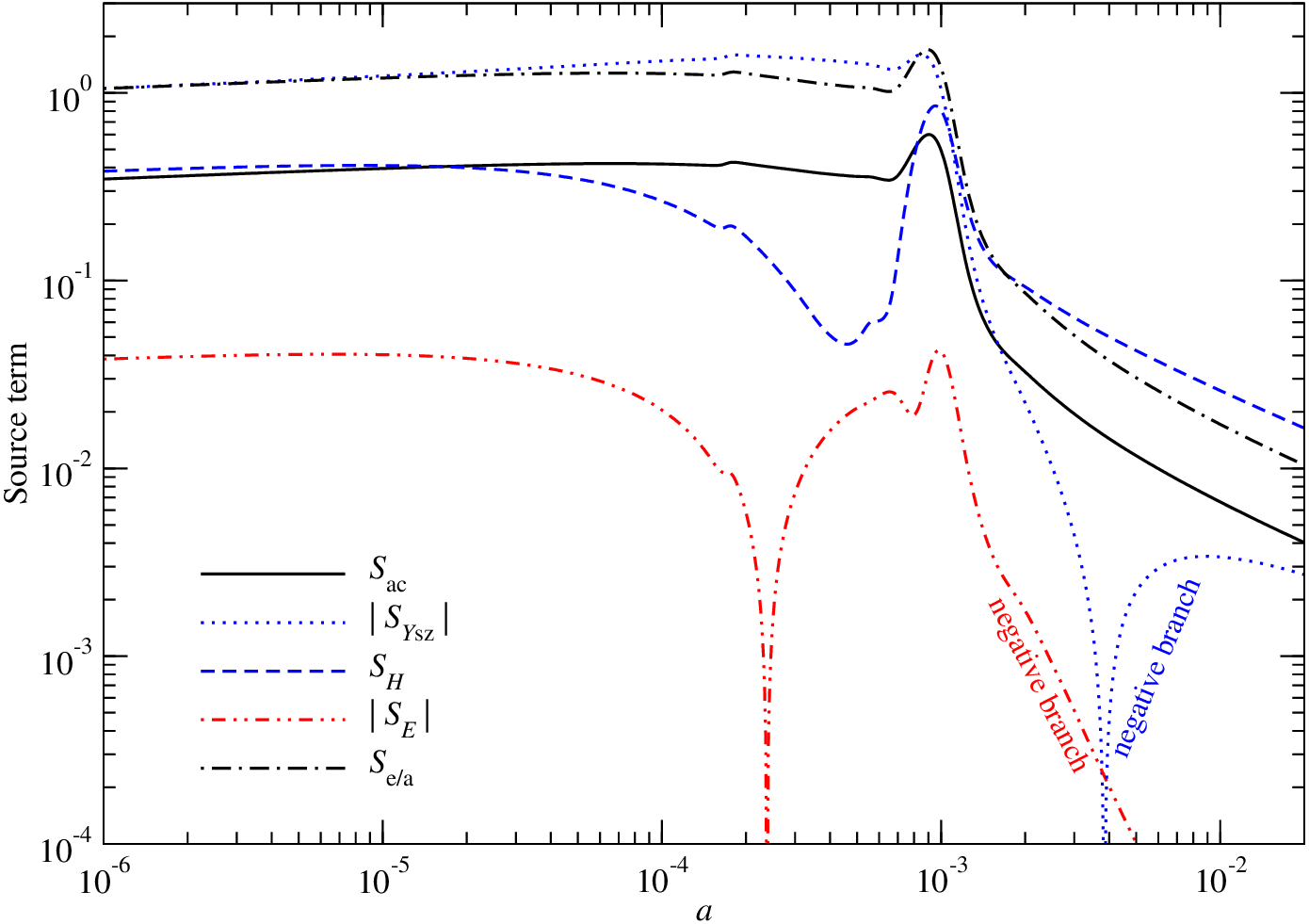}
\caption{Other source terms as defined by Eq.~\eqref{eq:Final_EQ_elim_Te_all} and \eqref{eq:BR_DC_second_insert} in units of $A_\zeta H(z)/\taudot$ for spectral index, $\nS=0.96$. For comparison we also show $\left<\mathcal{S}_{\rm ac}\right>$. Notice that the relative importance of the source functions for the spectral distortions is not reflected in this figure, since the source terms are modulated by additional (small) factors and frequency-dependent functions.
}
\label{fig:heating_other}
\end{figure}
\subsection{Other source functions}
\label{sec:corr_source}
We can similarly compute the amplitude of the other source terms appearing in Eq.~\eqref{eq:final_EQUATION}. In Fig.~\ref{fig:heating_other} we show the comparison.
The functions $\mathcal{S}$ all look rather comparable, with noticeable differences only around recombination and after. One can also clearly see the limiting behaviour for high redshifts (cf. Sect.~\ref{sec:other_S_appr}). 

The relative importance of the source functions for the spectral distortions is not reflected in Fig.~\ref{fig:heating_other}, since the source terms are modulated by additional (small) factors and frequency-dependent functions.
The additional source terms related to scattering have frequency-dependence $\mathcal{Y}_{\rm SZ}$, $\mathcal{H}$, and $\mathcal{E}$.
The function $\mathcal{H}$ is typically 4 times larger than $\mathcal{Y}_{\rm SZ}$, while $\mathcal{E}$ is smaller by $\simeq 2$ (cf. Fig.~\ref{fig:functions.III}).
In addition the related source functions for the spectral distortions are suppressed by a factor of $\Thz\simeq \pot{4.6}{-10}(1+z)$.
This renders the extra source terms $\Thz\left<\mathcal{S}^{\The}_{\mathcal{Y}_{\rm SZ}}\right>\mathcal{Y}_{\rm SZ}\simeq \pot{1.4}{-9}(1+z) \left<\mathcal{S}_{\rm ac}\right>\mathcal{Y}_{\rm SZ}$, $\Thz\left<\mathcal{S}^{\The}_{\mathcal{H}}\right>\mathcal{H}\simeq \pot{2.0}{-9}(1+z) \left<\mathcal{S}_{\rm ac}\right>\mathcal{Y}_{\rm SZ}$ and $\Thz\left<\mathcal{S}^{\The}_{\mathcal{E}}\right>  \mathcal{E}\simeq \pot{2.6}{-11}(1+z) \left<\mathcal{S}_{\rm ac}\right>\mathcal{Y}_{\rm SZ}$, a correction that is smaller than $\simeq 2\%$ at $z\lesssim 10^7$ in all cases.

The terms $\frac{1}{2}\left<\ygl{1}\right>\mathcal{Y}_{\rm SZ}$ and $\left<\mathcal{S}^{\The}_{\rm e/a}\right>\mathcal{G}$ related to BR and DC emission enter the problem without an extra suppression $\Thz$.
They arise as corrections to the electron temperature (see Eq.~\eqref{eq:The_sol_2}) and are actually related to parts of the functions $\left<\mathcal{S}^{\The}_{\mathcal{H}}\right>$ and $\left<\mathcal{S}^{\The}_{\mathcal{E}}\right>$.
Without the extra terms $\propto \betac \The$ and $\propto \betacsq \The$, these would not have the correct velocity-dependence and hence yield erroneous results for the net energy exchange and equilibrium between photons and baryonic matter.
However, in comparison to $\left<\mathcal{S}_{\Delta n}\right>$ they all turn out to have a small effect on the final spectrum. 
We checked this statement explicitly and found that they can be safely neglected.

We also mention here that at redshifts $z\gtrsim 10^7$ the extra source terms can become important. However, energy release in such early epochs will only cause a net increase of the specific entropy of the Universe, but lead to no distortions. Furthermore, at those redshifts other processes become important, but a more detailed analysis of this problem is beyond the scope of the present work.

\subsection{Correction because of polarization}
\label{sec:polarization}
The effect of polarization in the Thomson limit is readily included by adding $\frac{1}{2} \taudot[ \hat{\Theta}^{(1), \rm P}_0 + \hat{\Theta}^{(1), \rm P}_2]P_2(\mu)\,\mathcal{G}$ to the first order collision term, Eq.~\eqref{eq:BoltzEq_lim_0_t_a}.
Here the superscript `P' denotes polarization multipoles.
For the simple analytic estimate of $\mathcal{S}_{\rm ac}$ according to Eq.~\eqref{eq:S_ac_k_dependence} this means a replacement $8/9\rightarrow 16/15$ \citep[cf.][]{Kaiser1983, Zaldarriaga1995}.
Furthermore, the addition of polarization in the Thomson limit only affects the definition of $\mathcal{S}_{\rm temp}$, Eq.~\eqref{eq:Boltzmann_second_order_general_II_b}, and consequently $\mathcal{S}_{\rm ac}$.
For the numerical computations one therefore has to calculate the ensemble average $\left<\hat{\Theta}^{(1)}_2\left[\hat{\Theta}^{(1), \rm P}_0+\hat{\Theta}^{(1), \rm P}_2\right]\right>$ in addition, once polarization is included in the first order computation.
This means
\beal
\mathcal{S}_{\rm ac}\rightarrow \mathcal{S}_{\rm ac} - \frac{1}{2}
\left[\left<\hat{\Theta}^{(1)}_2\hat{\Theta}^{(1), \rm P}_0\right> + \left<\hat{\Theta}^{(1)}_2\hat{\Theta}^{(1), \rm P}_2\right>\right].
\end{align}
Because of Thomson scattering some power leaks from the temperature multipoles into the polarization multipoles. This increases the apparent dissipation rate of the temperature multipoles, however, the power appearing as polarization has to be subtracted again. The net change therefore is expected to be small. 
A detailed numerical computation shows that around the hydrogen recombination epoch the effective heating rate is smaller by about $\simeq 5\%-10\%$. At both lower and higher redshifts the effect of polarization is negligible.
Since polarization only becomes important during hydrogen recombination, where $\Thz\lesssim 10^{-7}$, the other source terms can be computed without including the effect of polarization.
%

\subsection{Fast computation of the heating rates}
\label{sec:fast_heating}
To compute the results for the effective heating rates that are presented above we pushed the accuracy of our Boltzmann code to extreme settings. 
The corresponding calculations were rather expensive ($\simeq 2$ hours on 80 cores), however, now that we have obtained good analytic approximations for the heating rates it is possible to simplify the numerical computation significantly.
First of all, the analytic approximation can be used for modes with $k\geq 1\, {\rm Mpc^{-1}}$, since these dissipate their energy at $a\lesssim \pot{5}{-4}$ where according to Fig.~\ref{fig:heating_SZ_zoom} the analytic approximation works well.
We therefore only have to treat more carefully modes with $k\leq 1\, {\rm Mpc^{-1}}$.

Furthermore we can reduce the $k$-sampling by a large amount. We used about $10^5$ points in the previous calculations, however, for the smaller $k$-range a sampling of a few hundred is sufficient. 
Using Clenshaw-Curtis rules \citep{Clenshaw1960} for integration in log-$k$ we were able to obtain converged results for $n=256$.
We also increased the maximal step-size in $a$ several times, without affecting the results significantly. 
In addition, it was sufficient to include only about $\simeq 10$ multipoles in the computation of the heating rates. Although details of the transfer functions are different in the free-streaming period, this had negligible effect on the final heating rate.
With all these modifications the computation of the source functions reduces to about 1 minute on a standard laptop.

We included this simplified treatment of the perturbation calculation into {\sc CosmoTherm}. Prior to the thermalization calculation, for given parameters of the power spectrum and cosmology one can solve for the transfer functions and then load the corresponding effective heating rates for the thermalization problem.
%

\section{Spectral distortions from the dissipation of acoustic modes}
\label{sec:dist_results}
We are now in a position to directly compute the resulting spectral distortion caused by the dissipation of acoustic modes in the early Universe.
From recent estimates \citep{Chluba2011therm, Khatri2011} it is clear that
the distortion is a mixture of $\mu$ and $y$-type distortion with typical
amplitude $\Delta n/n \simeq 10^{-8}-10^{-9}$, depending sensitively
  on the spectral index.
The negative $\mu$-type distortion caused by the down-scattering
  of CMB photons because of the adiabatically cooling baryonic matter
  contributes at the level of $\Delta n/n \simeq 10^{-9}$  \citep{Chluba2005,
    Chluba2011therm, Khatri2011}, diminishing or even dominating the distortion
  from the dissipation process according to the total acoustic power on small scales.
Also, if in zeroth order perturbation theory some small distortion is created by uniform energy release this should be computed simultaneously for accurate predictions of the average signal.

All the mentioned effects can be simultaneously included by just adding the corresponding source terms from the zeroth order perturbation problem to Eq.~\eqref{eq:final_EQUATION}. 
Since in first order perturbation theory no average distortion is created, this should have no significant effect on the consistency of the second order perturbation problem.
To formulate the problem one also has to follow the zeroth order temperature equation for the electrons and baryons to determine the average temperature of electrons. 
The source term $\left< \mathcal{S}^{\The}_{\Delta n} \right>$ from the dissipation problem can be absorbed by adding it to the temperature equation, since a similar term appears in zeroth order perturbation theory, cf. Eq.~\eqref{eq:Energy_zero_b}. 
This makes the computation simpler, but does not affect the result at an important level.

With all these comments in mind, the evolution equation for the average spectral distortion, $\Delta n_{\rm av}$, can be written including: (i) energy release from the dissipation of acoustic modes; (ii) the extraction of energy from the photon field by the adiabatic cooling of baryonic matter; and (iii) a possible uniform energy release directly leading to heating of matter. It reads
\bsub
\label{eq:average_distortion_evol_eq}
\beal
\label{eq:final_EQUATION_cooling}
\Abl{\Delta n_{\rm av}}{\tau}&
\approx
\frac{\Thz}{x_{\gamma}^{2}} \,\partial_{\xg} x_{\gamma}^4 
\left\{ \partial_{\xg} \Delta n_{\rm av} + \Delta  n_{\rm av} \mathcal{A}\right\}
+\frac{K(\xg, \Thz)}{\xg^3} (1-e^{\xg}) \Delta n_{\rm av}
\nonumber
\\[1mm]
&\!\!\!\!\!\!\!\!\!\!\!\!\!
%
+[\The-\Thz] \mathcal{Y}_{\rm SZ}-\frac{\The-\Thz}{\Thz}\,\frac{K(\xg, \Thz)}{x^3_{\gamma}}(1-e^{\xg}) \mathcal{G}
%
%
+\left<\mathcal{S}_{\rm ac}\right>\mathcal{Y}_{\rm SZ}
\\[2mm]
\label{eq:Te_equation}
\Abl{\rho_{\rm e}}{\tau}&\approx
\frac{\taudot^{-1}\dot{Q}}{\alpha_{\rm h}\theta_{\gamma}}
+\frac{4\tilde{\rho}^{\rm pl}_\gamma}{\alpha_{\rm h}}[\rho^{\rm eq}_{\rm e}-\rho_{\rm e}] 
%
-\taudot^{-1} H\,\rho_{\rm e}.
\end{align}
\esub
Here $\rho_{\rm e}=\Te/\Tz$, $\dot{Q}$ parametrizes the uniform heating rate of the matter for which different examples are given in \citet{Chluba2011therm}, and $\rho^{\rm eq}_{\rm e}=1+\left< \mathcal{S}^{\The}_{\Delta n} \right>$ with $\Delta n^{(2)}_{\rm av}$ in Eq.~\eqref{eq:Compton_source_terms} replaced by $\Delta n_{\rm av}$.
We neglected additional cooling by BR and DC emission, which can be easily added, but does not lead to any significant change \citep{Chluba2011therm}.
Furthermore, we omitted additional corrections to the source function for the acoustic dissipation problem (see Sect.~\ref{sec:corr_source}). However, it is also straightforward to add these terms, if needed.

This set of equations can be readily solved using the thermalization code {\sc CosmoTherm} \citep{Chluba2011therm}, once the source function for the cosmological dissipation problem is known.
We define the initial condition of the problem such that the effective temperature, $T^\ast_{\gamma}$, of the photon field at the end is close to $\TCMB=2.726\,{\rm K}$. For this value we calculate the evolution of the total energy budget of the thermalization problem being considered, prior to the computation of the spectral distortions.
Neglecting the distortions themselves, as well as BR and DC terms and higher
order corrections to the source function for the acoustic dissipation
problem (without loss of  precision in final results), this requires solving
\beal
\label{eq:final_rho_cooling}
\frac{1}{a^4 }\Abl{a^4 \rho_{\gamma}}{\tau}&
\approx
4 \rho_\gamma \left<\mathcal{S}_{\rm ac}\right> 
-4 \rho_\gamma \Thz[1-\rho_{\rm e}]
\nonumber
\\[2mm]
\Abl{\rho_{\rm e}}{\tau}&\approx
\frac{\dot{Q}\,\taudot^{-1}}{\alpha_{\rm h}\theta_{\gamma}}
+\frac{4\tilde{\rho}_\gamma}{\alpha_{\rm h}}[1-\rho_{\rm e}] 
-\taudot^{-1} H\,\rho_{\rm e}.
\end{align}
Once $\Delta \rho_{\gamma}(\ze)=\rho_{\gamma}(\ze)-\rho^{\rm pl}_\gamma(z_{\rm e})$ at the final redshift, $\ze$, is obtained we directly have $\Tz(\zs)=\TCMB(\zs)\left[1-\Delta \rho_{\gamma}(\ze)/\rho^{\rm pl}_\gamma(z_{\rm e})\right]^{1/4}\equiv \Te(\zs)$ as the initial temperature of the CMB blackbody at $\zs$.
This procedure was explained in more detail by \citet{Chluba2011therm} and implies $T_{\rm ref}\equiv \TCMB \left[1-\Delta
    \rho_{\gamma}(\ze)/\rho^{\rm pl}_\gamma(z_{\rm e})\right]^{1/4}$ for
  our computation. It simply means that we start with a blackbody which has
  an effective temperature lowered by $\simeq \frac{1}{4}\Delta
  \rho_{\gamma}/\rho^{\rm pl}_\gamma$, so that after adding  the
    total amount of released energy that ends up in the photon field we get
    the final CMB temperature equal to $T_{\rm{CMB}}$.

Using {\sc CosmoTherm} we ensure that total energy is conserved to very high precision. One simple way to confirm the precision of the computation is to check if after the full evolution the final effective temperature of the photon field reaches $\TCMB$. Notice that although the heat capacity of baryonic matter is rather small, at the end a fraction of the released energy leads to heating of matter, but no net entropy production.
This is especially noticeable in the case of uniform heating, i.e., $\dot{Q}\neq 1$ in Eq.~\eqref{eq:final_rho_cooling}.

\begin{figure}
\centering
\includegraphics[width=\columnwidth]{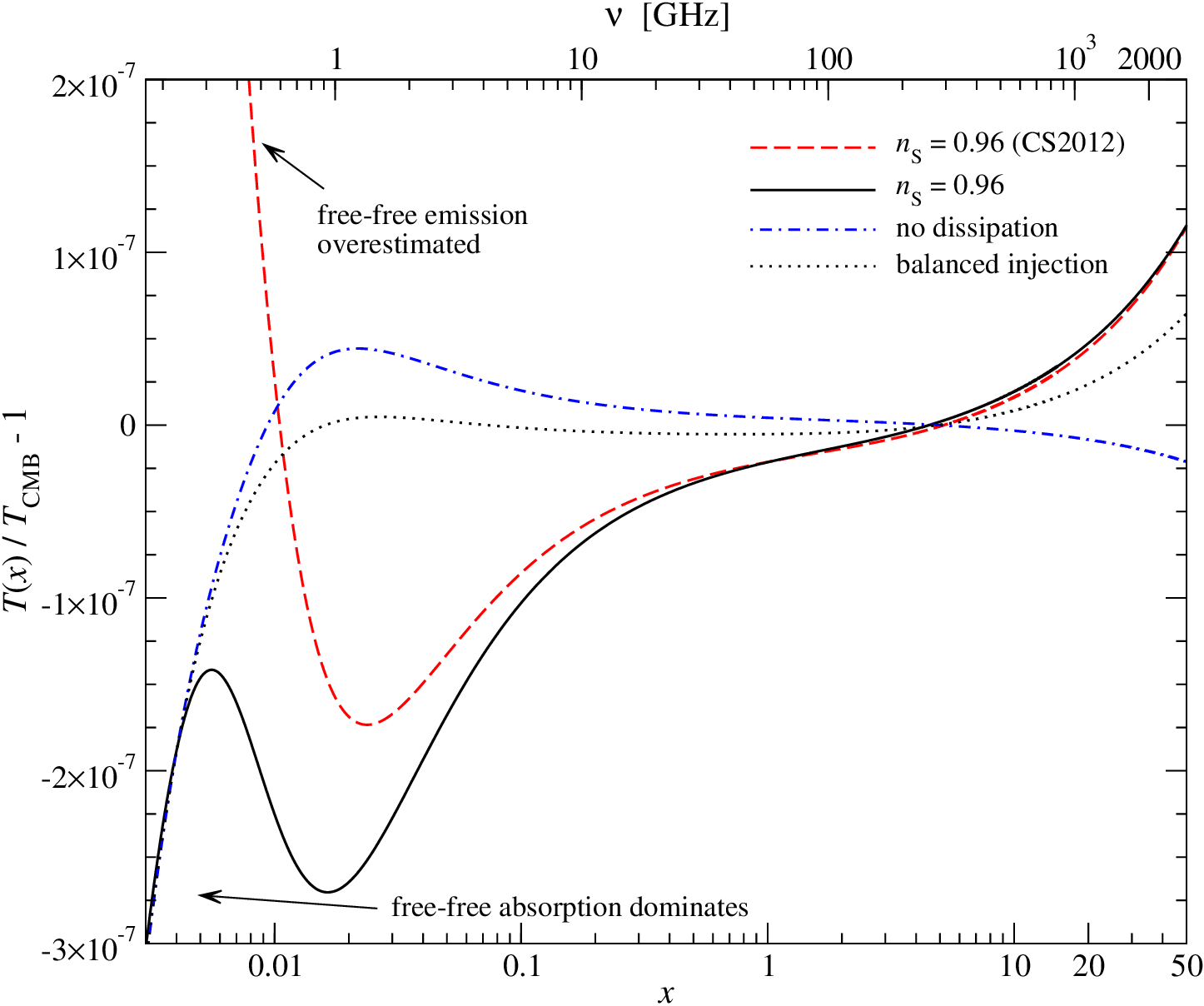}
\caption{Comparison of total distortion from acoustic damping and BE condensation with the previous computation of \citet{Chluba2011therm}. The amplitudes of the $\mu$ and $y$-type contributions are rather similar, however, in the previous computation the heating of baryonic matter after recombination was overestimated, leading to free-free emission instead of absorption at very low frequencies.
We also show the case for balanced injection (Sect.~\ref{sec:balanced}), $\nS=0.96$ and $n^{\rm bal}_{\rm run}=-0.0265$, for which practically only the free-free and $y$-type distortions are visible.}
\label{fig:Tg_old_comp}
\end{figure}

\begin{figure}
\centering
\includegraphics[width=\columnwidth]{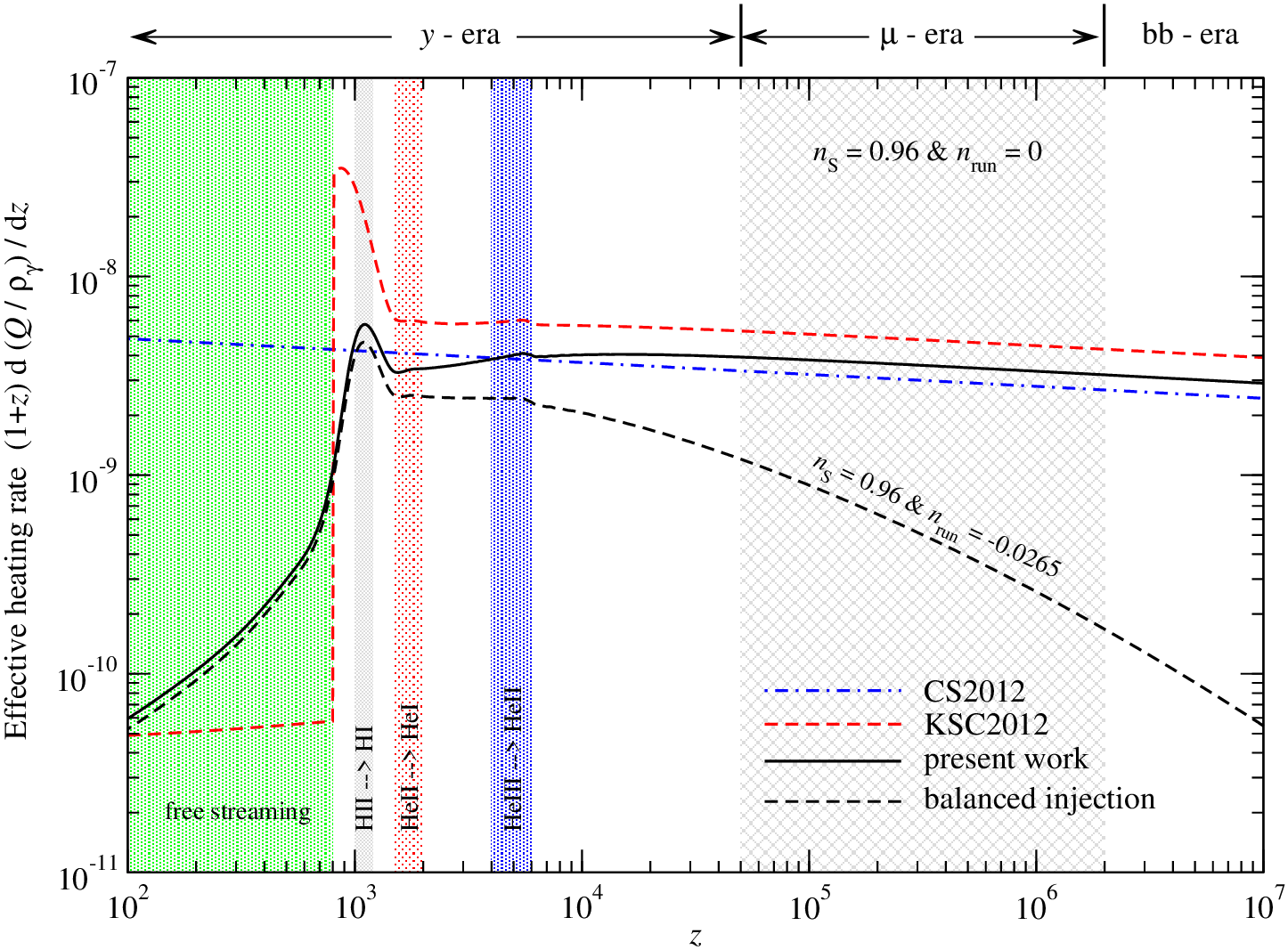}
\caption{Comparison of different estimates for the heating rates during the different cosmological epochs for $\nS=0.96$. The blue/dash-dotted line shows the estimate from \citet{Chluba2011therm}, while the red/dashed line gives the approximation of \citet{Khatri2011}. The solid black line is the result obtained here.
The dashed/black line shows the heating rate for the balanced injection scenario (Sect.~\ref{sec:balanced}).
}
\label{fig:comp_heating}
\end{figure}

\subsection{Comparison with previous computations}
\label{sec:comp_old}
Recently, \citet{Chluba2011therm} used the estimates derived from the
  classical formula to compute the CMB spectral distortion caused by acoustic damping and BE condensation of photons.
Here we compare the results of our more detailed computation with those obtained by \citet{Chluba2011therm}.
In Fig.~\ref{fig:Tg_old_comp} we consider the case $\nS=0.96$ and $\nrun=0$, as example.
We present the brightness temperature, $T(\xg)=\Tz \xg / \ln[1+n(\xg) ]$, relative to $\TCMB$ as a function of frequency.
The thermalization problem is solved starting at $\zs=\pot{4}{7}$ and ending at $\zs=200$ using {\sc CosmoTherm}.
As the figure indicates, the amplitudes of the $\mu$ and $y$-type contributions are rather similar in both approaches, however, at low frequencies the distortion is very different.
In \citet{Chluba2011therm} the effective heating rate caused by the dissipation of acoustic modes was parametrized by
\beal
\label{eq:DE_Dt_waves}
\frac{\id}{\id z}\frac{ |Q_{\rm ac}|}{\rho_\gamma}
& \approx 
\mathcal{F}(n_{\rm S})\, [1+z]^{(3n_{\rm S}-5)/2},
\end{align}
with $\mathcal{F}(x)\approx \pot{5.2}{-9}\,x^{-2.1}\,[0.045]^{x-1}$ for
$x\simeq 1$ (see Fig.~\ref{fig:comp_heating}). 
This expression is based on the formulae given by \citet{Hu1994}, where the power spectrum normalization was obtained by calculating the rms fluctuation on 10 degree scales. This method gives a slightly lower normalization compared to the detailed fit of the power spectrum used by WMAP \citep{Bunn1997}.
Clearly, Eq.~\eqref{eq:DE_Dt_waves} is rather rough, but it does capture the basic scaling of the effective heating rate with $\nS$ and redshift.

In the work of \citet{Chluba2011therm}, this expression for $\dot{Q}$ was used as heating for the matter. 
It was pointed out that this approach does not reflect the physics of the dissipation problem. However, especially at high redshifts, one expects no real difference in the effect on the spectrum, because energy exchange between photons and electrons is very rapid.
At low redshifts on the other hand this approach does make an important difference. The above procedure actually overestimates the heating of baryonic matter, so that the temperature of electrons was found to rise above the temperature of the photon field \citep{Chluba2011therm}. 
One reason for this is that the dissipated energy was assumed to directly heat the matter, but the other is that Eq.~\eqref{eq:DE_Dt_waves} breaks down once free-streaming becomes important.
Within the approach of \citet{Chluba2011therm} this implies strong
free-free emission at low frequencies, which is absent in the correct, more detailed treatment presented here.

In Fig.~\ref{fig:comp_heating} we also show the effective heating rate as given by \citet{Khatri2011}. At high redshifts one can see the factor of $4/3$ difference with respect to our detailed calculation.
The post-recombination evolution is also only captured approximately in this
  approach using the free-streaming solution of the Boltzmann equation,
  but the second order Doppler effect is ignored. Using this rate we obtain a slightly larger $\mu$-type distortion as well as an overestimated $y$-type contribution.

\subsection{Balanced energy release in $\mu$-era}
\label{sec:balanced}
The cooling of baryonic matter in the expanding Universe diminishes the net distortion of the CMB caused by the dissipation of acoustic modes \citep{Chluba2011therm}.
Depending on the initial power spectrum, $P_\zeta(k)$, this can even lead to a complete cancellation of the overall distortion from the $\mu$-era \citep{Khatri2011}.
Adding the possibility for a running spectral index, the primordial power spectrum can be parametrized as \citep{Kosowsky1995}
\beal
P_\zeta(k)&=2\pi^2 A_\zeta k^{-3} (k/k_0)^{\nS-1+\frac{1}{2} n_{\rm run} \ln(k/k_0)},
\end{align}
with $n_{\rm run}=\id \nS / \id \ln k$. 
As we will see, the total power that is dissipated strongly depends on $\nrun$ and to a smaller extend on $\nS$.
The dependence on $A_\zeta$ is only linear, and also current constraints indicate $A_\zeta\simeq \pot{2.4}{-9}$ \citep{Komatsu2010, Dunkley2010, Keisler2011}, which we use as the fiducial value.
Without running, the value for $\nS$ from WMAP7 alone is $\nS=0.963 \pm 0.014$, while with running the currently favoured values are $\nS=1.027\pm 0.051$ and $\nrun=-0.034 \pm 0.026$ \citep{Larson2011, Komatsu2010}.
More recent measurements of the damping tail of the CMB power spectrum by {ACT} \citep{Dunkley2010} and {SPT} \citep{Keisler2011} yield\footnote{For both experiments we quote the constraint derived in combination with $\rm WMAP7+BAO+H_0$.} $\nS=1.017\pm 0.036$ and $\nrun=-0.024 \pm 0.015$ and $\nS=0.9758\pm 0.0111$ and $\nrun=-0.020 \pm 0.012$, respectively.

Using Eq.~\eqref{eq:S_ac_int} it is possible to compute the total energy release, $\Delta \rho_\gamma/\rho_\gamma=\int (a^4 \rho_\gamma)^{-1} \id_z (a^4 Q) \id z $, caused by acoustic damping during the $\mu$-era ($\pot{5}{4}\lesssim z \lesssim \pot{2}{6}$). Equating this with the energy extracted by the adiabatically cooling matter one can determine a relation between $\nS$ and $\nrun$ for which the net $\mu$-type distortion is expected to vanish.
The total energy extracted from the CMB by the cooling of baryonic matter in the redshift range $\zs = \pot{2}{6}$ and $\ze=\pot{5}{4}$ is $\Delta \rho_\gamma/\rho_\gamma \simeq -\pot{5.6}{-10} \ln(\zs/\ze) \simeq -\pot{2.1}{-9}$ \citep{Chluba2011therm}.
For given value of $\nS$ this is balanced by the energy release due to acoustic damping for
\beal
\label{eq:n_run_bal}
n^{\rm bal}_{\rm run}&\approx-0.0336[1+5.22 (\nS-1) +0.58(\nS-1)^2].
\end{align}
This estimate is valid in the range $0.5\leq \nS \leq 1.5$. 
For $\nrun \geq n^{\rm bal}_{\rm run}$ the net $\mu$-type distortion has $\mu\geq 0$, while for $\nrun \leq n^{\rm bal}_{\rm run}$ the cooling process dominates and the net distortion has $\mu\leq 0$.
As an example, the balanced injection scenario, $\nS=0.96$ and $n^{\rm bal}_{\rm run}=-0.0265$, is presented in Fig.~\ref{fig:Tg_old_comp} and \ref{fig:comp_heating}.
Indeed one can see that the distortion has only a very small (negative)
$\mu$-type contribution, and is dominated by free-free absorption at
low frequencies and $y$-type distortion at high frequencies. Reducing $\nrun$ even further one expects the $\mu$-type contribution to approach the value without extra dissipation. 
This value is set by the BE condensation of photons caused by the adiabatic cooling process, which is close to $\mu_{\rm BE} \simeq -\pot{2.2}{-9}$ and depends on the total heat capacity of baryonic matter in the Universe \citep{Chluba2011therm}.
This part of the distortion can in principle be predicted with very high precision, which allows subtraction to isolate the $\mu$-distortion caused by the dissipation of acoustic modes.

\begin{figure}
\centering
\includegraphics[width=\columnwidth]{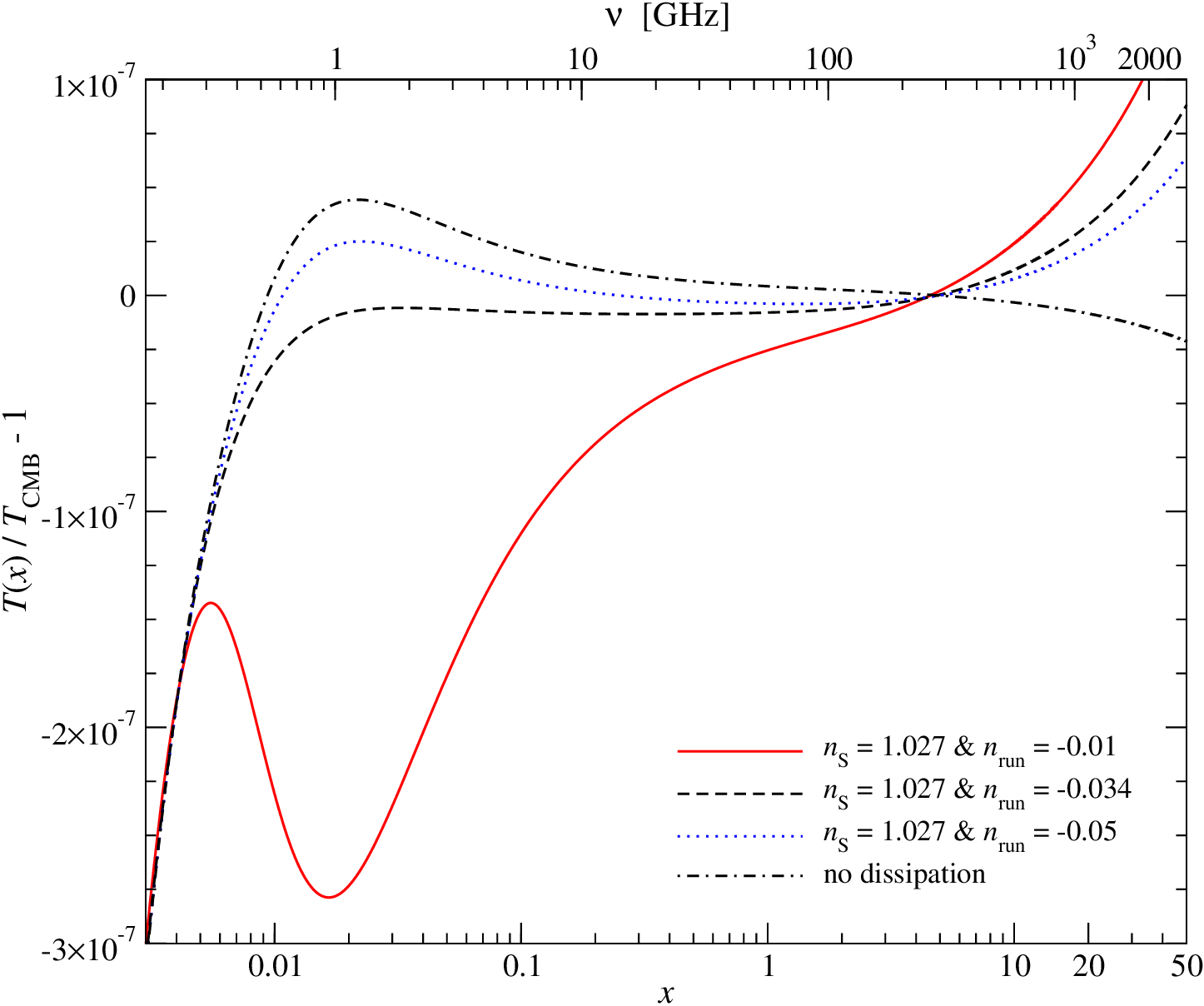}
\caption{Spectral distortion from acoustic damping and BE condensation for $\nS=1.027$ and $\nrun=-0.01, -0.034, -0.05$. The balanced scenario is very close to $\nrun=-0.034$.
For comparison we also show the case without any dissipation.}
\label{fig:Tg_examp_WMAP7}
\end{figure}
\begin{figure}
\centering
\includegraphics[width=\columnwidth]{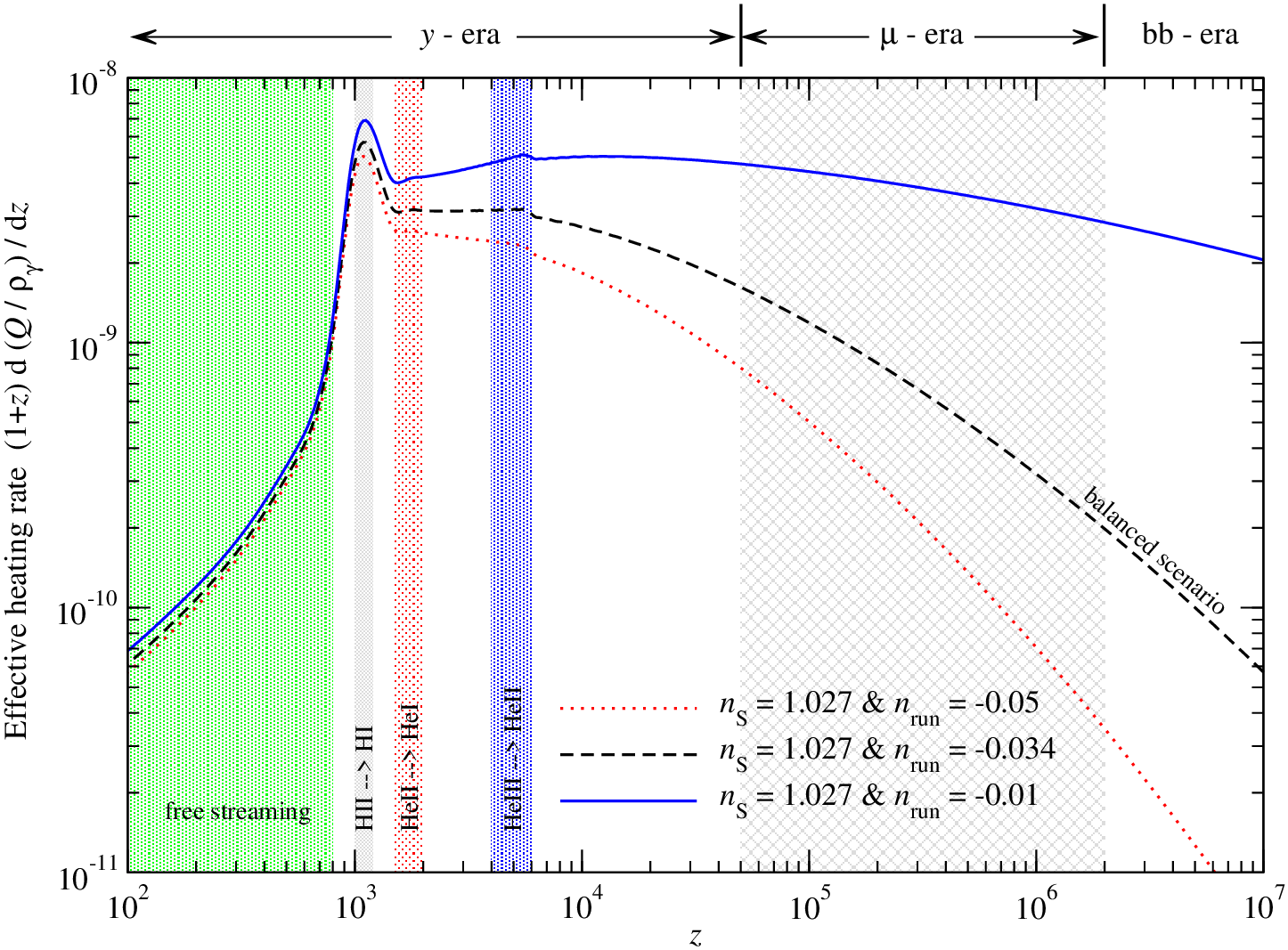}
\caption{Effective heating rates caused by the dissipation of acoustic modes for $\nS=1.027$ and $\nrun=-0.01, -0.034, -0.05$. The balanced scenario is very close to $\nrun=-0.034$.
}
\label{fig:heating_examp_WMAP7}
\end{figure}
\subsubsection{WMAP7 case: an example for balanced energy release}
\label{sec:balanced_WMAP7}
From Eq.~\eqref{eq:n_run_bal} it also follows that for $\nS=1.027$ and $\nrun=-0.034$ from WMAP7 the net $\mu$-type distortion should be rather close to zero.
For the {SPT} and {ACT} central values on the other hand a positive $\mu$-type distortion is expected.
In Figs.~\ref{fig:Tg_examp_WMAP7} and \ref{fig:heating_examp_WMAP7} we show the distortion and heating rate for the WMAP7 case, as well as examples with larger and smaller values of $\nrun$.
For the best fit values for a running spectrum from WMAP7 the $\mu$-distortion indeed practically vanishes.
For $\nrun=-0.05$ the heating rate in the $\mu$-era greatly decreases, so that a negative $\mu$ distortion caused by BE condensation appears. 
On the other hand for $\nrun=-0.01$ a significant positive $\mu$ distortion is found.
Also the $y$-type contribution changes significantly, since the effective heating rate at $z\lesssim \pot{5}{4}$ does vary noticeably with $\nrun$.
The free-free absorption at low frequencies is mainly caused by the cooling of matter during and after recombination, when $\Te<\TCMB$, and hence is practically unaffected by changes in the acoustic dissipation process.
%

\begin{figure}
\centering
\includegraphics[width=\columnwidth]{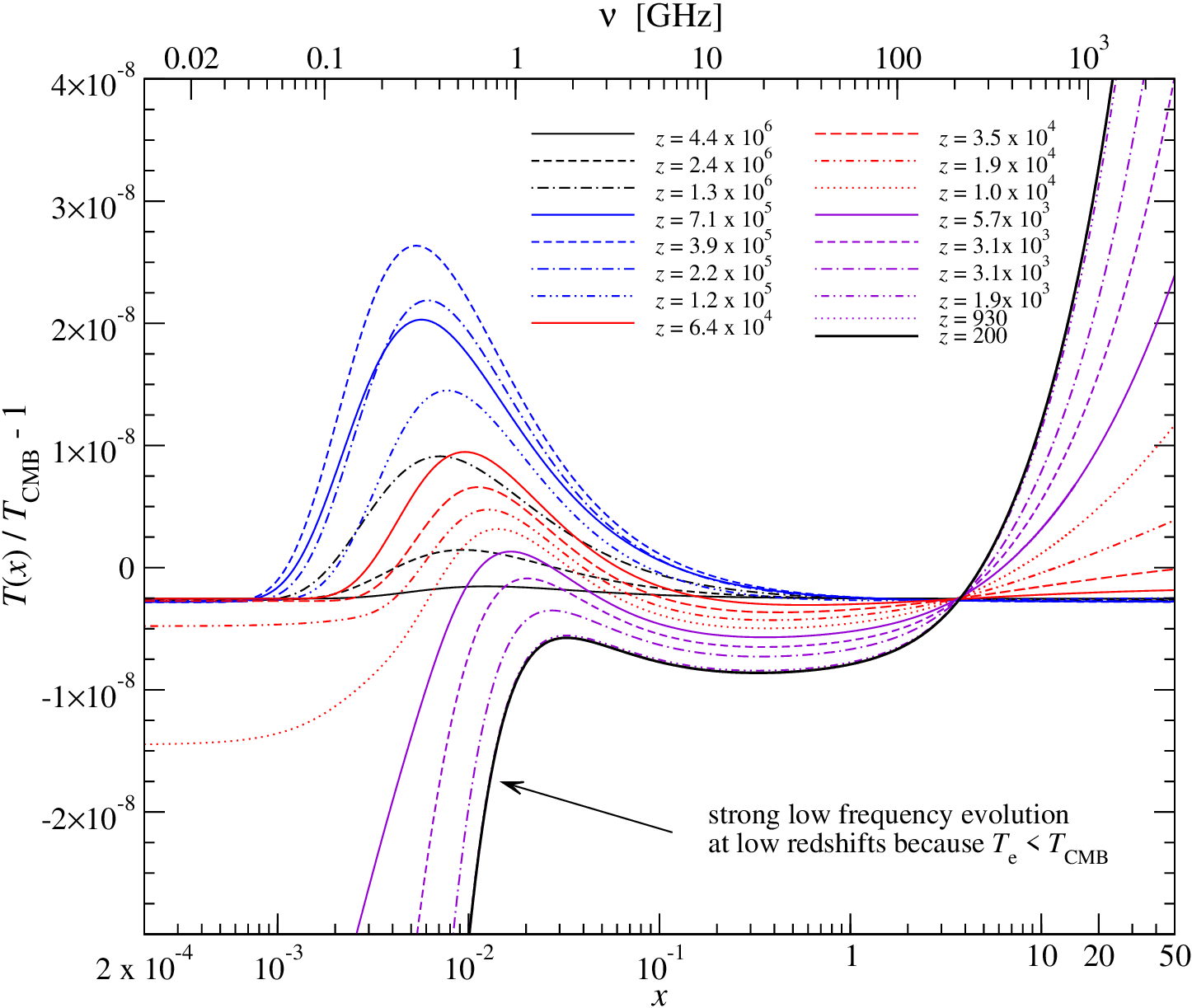}
\caption{Snapshots of the spectral distortion at different redshift. We show the balanced injection scenario for $\nS=1.027$ and $\nrun=-0.034$.}
\label{fig:Tg_evolution}
\end{figure}
In Fig.~\ref{fig:Tg_evolution} we show a sequence of the spectral distortion at different stages, illustrating the overall evolution.
At high redshifts, BE condensation of photons dominates and the distortions starts out with $\mu<0$.  At redshift $z\simeq \pot{4}{5} - \pot{5}{5}$ the heating caused by dissipation of acoustic modes starts dominating and $\mu$ slowly approaches zero.
Around $z\simeq \pot{6}{4}-\pot{7}{4}$ one can see the development of a $y$-type distortion at high frequencies, while at low frequencies the spectrum reaches equilibrium with the electrons at $T\simeq \Te\lesssim \TCMB$ by the free-free process.
The subsequent evolution further increases the $y$-type distortion because of superposition of blackbodies with different temperatures, while the $\mu$-distortion practically disappears. The low frequency part is strongly dominated by free-free absorption since $\Te<\TCMB$ during and after hydrogen recombination. Notice that we zoomed in very significantly to show the small distortion in detail (compare Fig.~\ref{fig:Tg_examp_WMAP7}).

We also checked some additional balanced injection scenarios and found the estimate given by Eq.~\eqref{eq:n_run_bal} to work well even for more extreme cases.
Notice also that the negative $y$-type distortion from BE condensation is always overcompensated by the positive $y$-distortion from the dissipation process. 
The effective heating from acoustic waves for $\nS\simeq 1$ is simply much larger than the cooling effect, unless $A_\zeta$ is decreased significantly.

\subsection{Representation of the final spectrum}
\label{sec:approx_mu_y}
The net spectrum discussed in the previous sections has four main contributions with respect to the initial blackbody, $\nPl(\xg)$ at temperature $T_{\rm ref}$ \citep[see also][]{Chluba2011therm}: (i) a free-free distortion, $\Delta n_{\rm ff}= \yff \,e^{-\xg}/\xg^{3}$, which dominates at very low frequencies; (ii) a temperature term, $\Delta n_{\phi}=\mathcal{G}(\xg)\Delta\phi$; (iii) a $\mu$-type distortion, $\Delta n_\mu=-\mathcal{G}(\xg)\,\mu(\xg)/\xg$, with a characteristic dip in the brightness temperature at low to intermediate frequencies; and (iv) a $y$-type contribution, $\Delta n_{y}=\mathcal{Y}_{\rm SZ}(\xg)\,\ye$, which is most important at high frequencies.
We now discuss different approximations and their interpretation regarding the amount of energy release.

\subsubsection{Low frequency free-free distortion}
The free-free part is always very similar in all considered cases. It only arises at rather low redshifts, where the dissipation of acoustic modes already has little effect on the matter temperature and everything is determined by the adiabatic cooling process (e.g., see Fig.~\ref{fig:Tg_examp_WMAP7}).
We find $\yff \approx -\pot{2.5}{-12}$ to work rather well, however, below we now focus on the high frequency distortion, for which the free-free part does not contribute significantly.

\subsubsection{Temperature shift and increased specific entropy}
The temperature shift $\Delta \phi=\Delta T/ T_{\rm ref}$ caused by energy release can be determined by considering the total energetics of the problem. Given the total energy release, $\Delta\rho_\gamma/\rho_\gamma$, and integrating $\Delta n(\xg) \approx \Delta n_\phi+\Delta n_\mu+\Delta n_y$ over $\xg^3 \id \xg\id^2\vgh$ it is evident that at the end of the evolution one should find 
\beal
\label{eq:phi_equ}
\Delta \phi &\approx \frac{1}{4}\frac{\Delta\rho_\gamma}{\rho_\gamma} + \frac{\zeta(3)}{4 \zeta(4)}\mu_\infty - \ye.
\end{align}
We neglected the free-free distortion, since it does not contribute much to the total energy density. We also set $\mu(\xg)\approx \mu_\infty\equiv {\rm const}$ for a similar reason.
This expression shows that $\Delta \phi$ depends on how much of the injected energy was fully thermalized between the initial and final stages. 
However, the $\Delta \rho_\gamma/\rho_\gamma$ part of $\Delta \phi $ is actually not separable, as it corresponds to a pure temperature shift with respect to the initial blackbody spectrum for which one does not have a direct measurement.
For the $\mu$ and $y$ terms, the distortion parts allow them to be discerned.
If we compare to a blackbody with effective temperature $\TCMB\neq T_{\rm ref}$ we have the remaining temperature term
\beal
\label{eq:phi_equ_prime}
\Delta \tilde{\phi} &\approx \frac{\zeta(3)}{4 \zeta(4)}\mu_\infty - \ye.
\end{align}
This part accounts for contributions to the total energy density of the CMB spectrum from  the $\mu$- and $y$-distortion, which depend on the specific definition for the spectral shape of these distortions:
using $x=h\nu/k\TCMB$ and adding all terms we have
\beal
\label{eq:n_x_final}
\Delta n(x) &= n(x)- \nPl(x)
\approx\Delta\tilde{\phi} \,\mathcal{G}(x) 
-\mathcal{G}(x) \, \frac{\mu_\infty}{x} 
+ \mathcal{Y}_{\rm SZ}(x)\,\ye
\nonumber\\
&\equiv \mathcal{M}(x)\,\mu_\infty +\mathcal{Y}_{\rm alt}(x)\,\ye,
\end{align}
with $\mathcal{M}(x)=\mathcal{G}(x)[\zeta(3)/4 \zeta(4) - 1/x]$ and $\mathcal{Y}_{\rm alt}(x)=\mathcal{Y}_{\rm SZ}(x)-\mathcal{G}(x)$.
When taking the {\it total} energy density as a proxy for the effective temperature of the CMB (our choice here), $\mathcal{M}(x)$ and $\mathcal{Y}_{\rm alt}(x)$ are the $\mu$ and $y$-type spectral templates that can be directly compared with observations. 
Notice that $\int \mathcal{M}(x) x^3 \id x = \int \mathcal{Y}_{\rm alt}(x) x^3 \id x \equiv 0$, i.e. these types of distortion do not affect the total energy density of the CMB.
On the other hand, $\mathcal{Y}_{\rm SZ}(x)$, is constructed such that the total number density of photons is not affected ($\int \mathcal{Y}_{\rm SZ}(x) x^2 \id x \equiv 0$).
One can similarly find $\mathcal{M}_{N}(x)=\mathcal{G}(x)[\zeta(2)/3 \zeta(3) - 1/x]$ for which $\int \mathcal{M}_{N}(x) x^2 \id x \equiv 0$, however, we do not use this definition here.

Notice that the above statements are only a matter of presenting the spectrum obtained from the thermalization calculation. The difference is just connected with the way $\TCMB$ is defined.
In both our calculation and the interpretation above, $\TCMB$ is equal to the effective temperature of the distorted CMB spectrum.  
Alternatively, when using $\mathcal{Y}_{\rm SZ}(x)$ and $\mathcal{M}_{N}(x)$ as templates for the distortion parts, the CMB temperature determined by simultaneously fitting a blackbody and the distortions would approximately be 
\beal
\label{eq:TCMB_obser}
\tilde{T}_{\rm CMB} 
&= \TCMB
+\left[\frac{\zeta(3)}{4 \zeta(4)}-\frac{\zeta(2)}{3 \zeta(3)}\right]\mu_\infty 
- \ye,
\end{align}
while when using the templates $\mathcal{Y}_{\rm SZ}(x)$ and $\mathcal{M}(x)=-\mathcal{G}(x)/x$ one finds $\tilde{T}_{\rm CMB} = \TCMB+\Delta \tilde{\phi}$.
To obtain a measurement of $\TCMB$ one can use these expressions to
interpret the result. 

\subsubsection{Effective $\mu$ and $y$-parameters}
The $\mu$ and $y$-type contributions depend on the total energy release in the related epochs.
It is well known \citep{Sunyaev1970mu, Danese1982, Hu1993} that energy released at redshifts $z\gtrsim \pot{1.98}{6}\equiv \zmu$ is suppressed by the {\it visibility function for spectral distortions} 
\beal
\label{eq:J_vis}
\mathcal{J}_{\rm bb}(z)
& \approx 
\exp\left(-\left[z/\zmu\right]^{5/2}\right).
\end{align}
At $z\gg\zmu$ all the released energy just increases the specific entropy
of the Universe, and hence only increases  the average
  temperature of the CMB without causing significant spectral distortions.
The weighted total energy release in the $\mu$- and $y$-era is therefore
\bsub
\label{eq:def_mu_y_Dr_r}
\beal
\label{eq:def_mu_y_Dr_r_a}
\left.\frac{\Delta\rho_\gamma}{\rho_\gamma}\right|_{\mu}
&\approx 
\int_\zmuy^\infty 
\frac{\mathcal{J}_{\rm bb}(z)}{a^4 \rho_\gamma}
\frac{\id a^4 Q}{\id z} 
\id z 
\\[1mm]
\left.\frac{\Delta\rho_\gamma}{\rho_\gamma}\right|_{y}
&\approx 
\int^\zmuy_0    
\frac{1}{a^4 \rho_\gamma}
\frac{\id a^4 Q}{\id z}
\id z,
\end{align}
\esub
where $\zmuy\approx \pot{5}{4}$ \citep[cf.][]{Hu1993}. With the simple expressions from 
\citet{Sunyaev1970mu}, 
$\mu_\infty \approx 1.4 \,\Delta \rho_\gamma/\rho_\gamma|_\mu$ and $\ye \approx \frac{1}{4} \Delta \rho_\gamma/\rho_\gamma|_y$, 
this can be directly used to estimate the expected distortion at high frequencies.
We emphasize again that $\Delta \rho_\gamma/\rho_\gamma|_\mu+\Delta \rho_\gamma/\rho_\gamma|_y\neq \Delta \rho_\gamma/\rho_\gamma$, since part of the released energy does not appear as distortion but thermalizes, leading to $\Delta T/T_{\rm ref}\neq 0$.
The shape of the distortion is given by Eq.~\eqref{eq:n_x_final} when comparing the final spectrum to a blackbody with effective temperature $\TCMB$.

\begin{figure}
\centering
\includegraphics[width=\columnwidth]{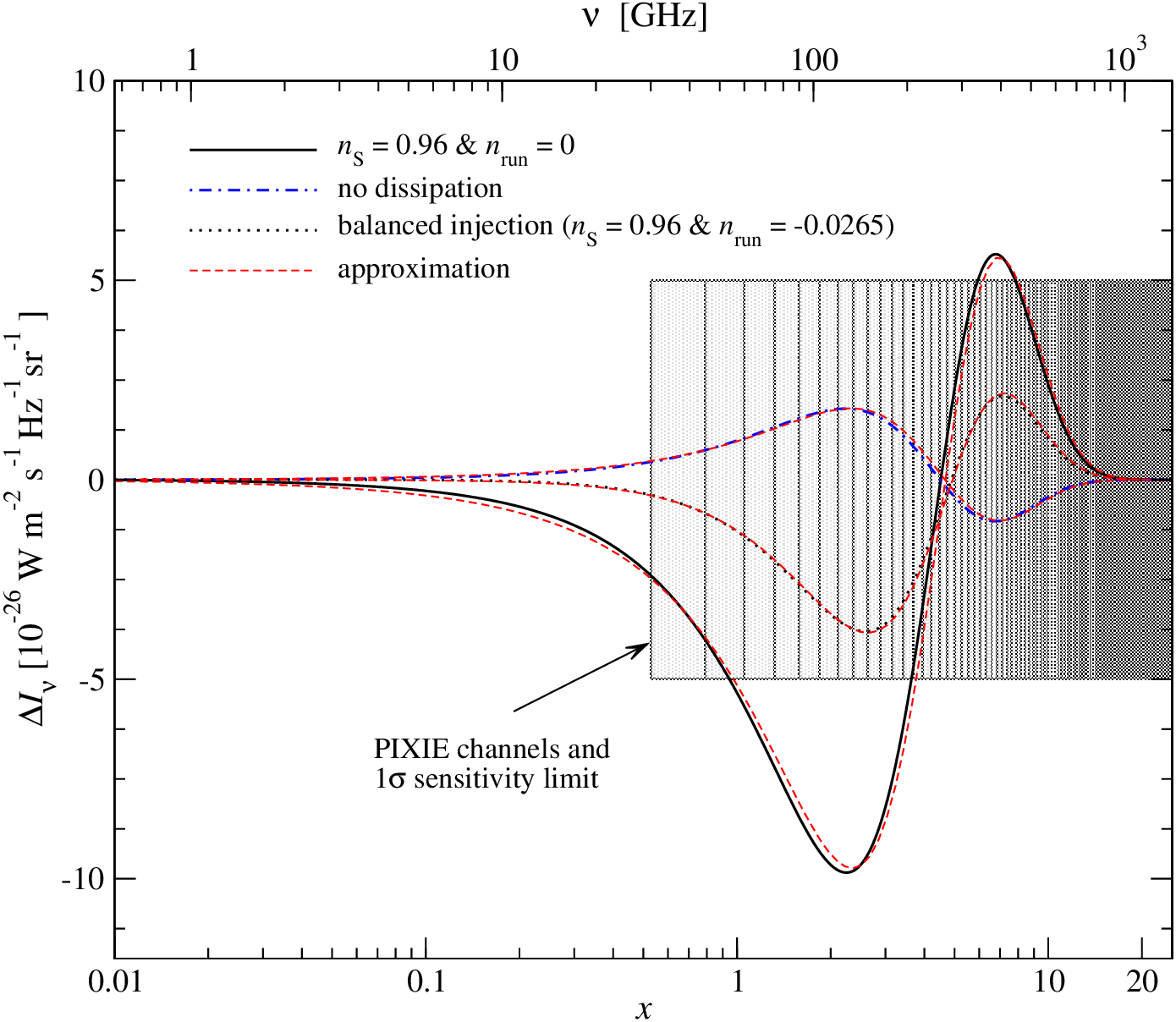}
\caption{CMB distortion at high frequencies. The approximations are explained in Sect.~\ref{sec:approx_mu_y}. For the case without dissipation $\mu_\infty \simeq -\pot{2.7}{-9}$ and $\ye \simeq-\pot{6.8}{-10}$. The case $\nS=0.96$ and $\nrun=0$ is represented by $\mu_\infty \simeq \pot{1.4}{-8}$ and $\ye \simeq\pot{3.8}{-9}$.
The balanced injection scenario is well approximated by a pure $y$-distortion, $\mu_ \infty \simeq 0$ and $\ye \simeq\pot{1.8}{-9}$.
}
\label{fig:DI_0.96}
\end{figure}
\subsubsection{Distortions from Bose-Einstein condensation of photons}
For the $\mu$-distortion introduced by BE condensation of photons the above estimate implies $\mu_{\rm BE}\simeq -\pot{2.7}{-9}$, which is a bit larger than the value $\mu_{\rm BE}\simeq -\pot{2.2}{-9}$ given by \citet{Chluba2011therm}.
However, in \citet{Chluba2011therm} $\mu_{\rm BE}$ was determined to best
approximate the low frequency part of the distortion. Indeed for the
high frequency part, around the crossover frequency of the $y$-distortion,
we find $\mu_{\rm BE} \simeq -\pot{2.7}{-9}$ and  $\ye=-\pot{6.8}{-10}$ to
work well (cf. Fig.~\ref{fig:DI_0.96}). The total energy extraction caused
by the cooling process is $\Delta\rho_\gamma/\rho_\gamma\simeq
-\pot{6.4}{-9}$, implying a change in the specific entropy density of the CMB by
$\Delta s_\gamma/s_\gamma\simeq \frac{3}{4} \Delta\rho_\gamma/ \rho_\gamma\simeq -\pot{4.8}{-9}$.
In  Fig.~\ref{fig:DI_0.96} we also show the different frequency channels of PIXIE and indicate the $1\sigma$ sensitivity band. The channels are separated by $\Delta \nu \simeq15\,{\rm GHz}$ in the frequency range $30\,{\rm GHz}<\nu<6\,{\rm THz}$  \citep{Kogut2011PIXIE}.
It is clear that the distortion caused by BE condensation alone is below the currently proposed sensitivity of PIXIE, however, we will give a more detailed discussion below. 
%

\begin{figure}
\centering
\includegraphics[width=0.97\columnwidth]{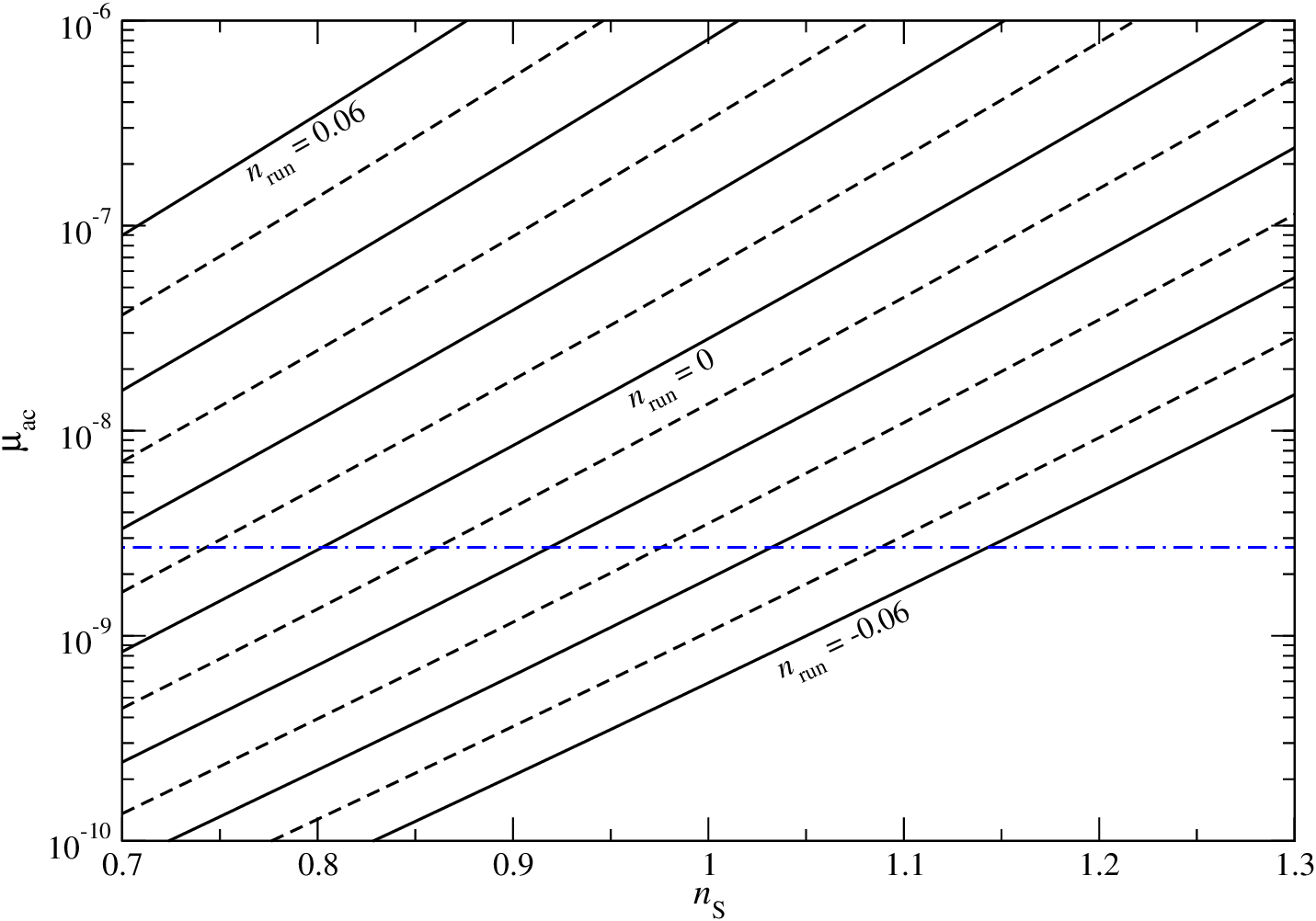}
\\
\includegraphics[width=0.97\columnwidth]{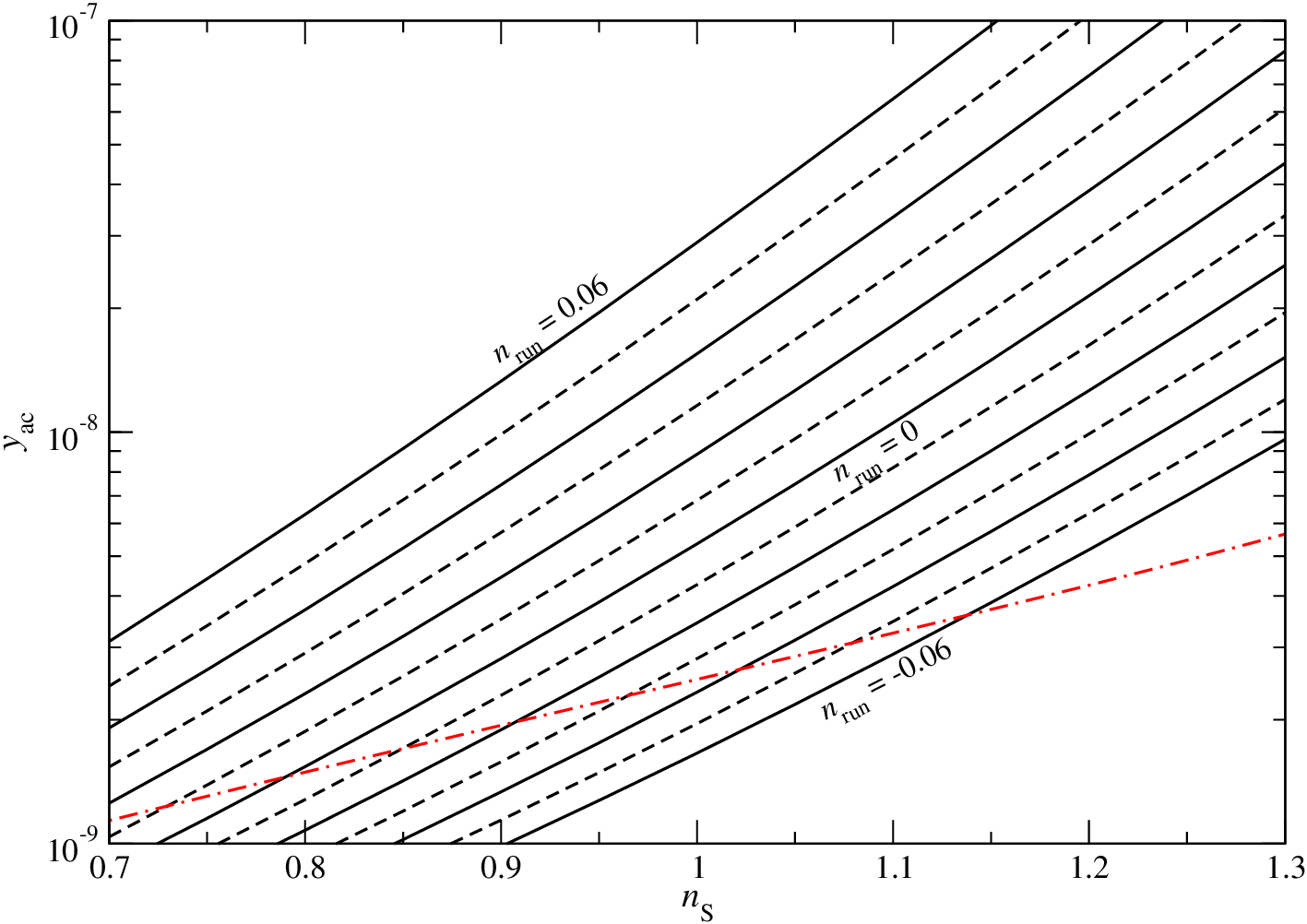}
\\
\includegraphics[width=0.97\columnwidth]{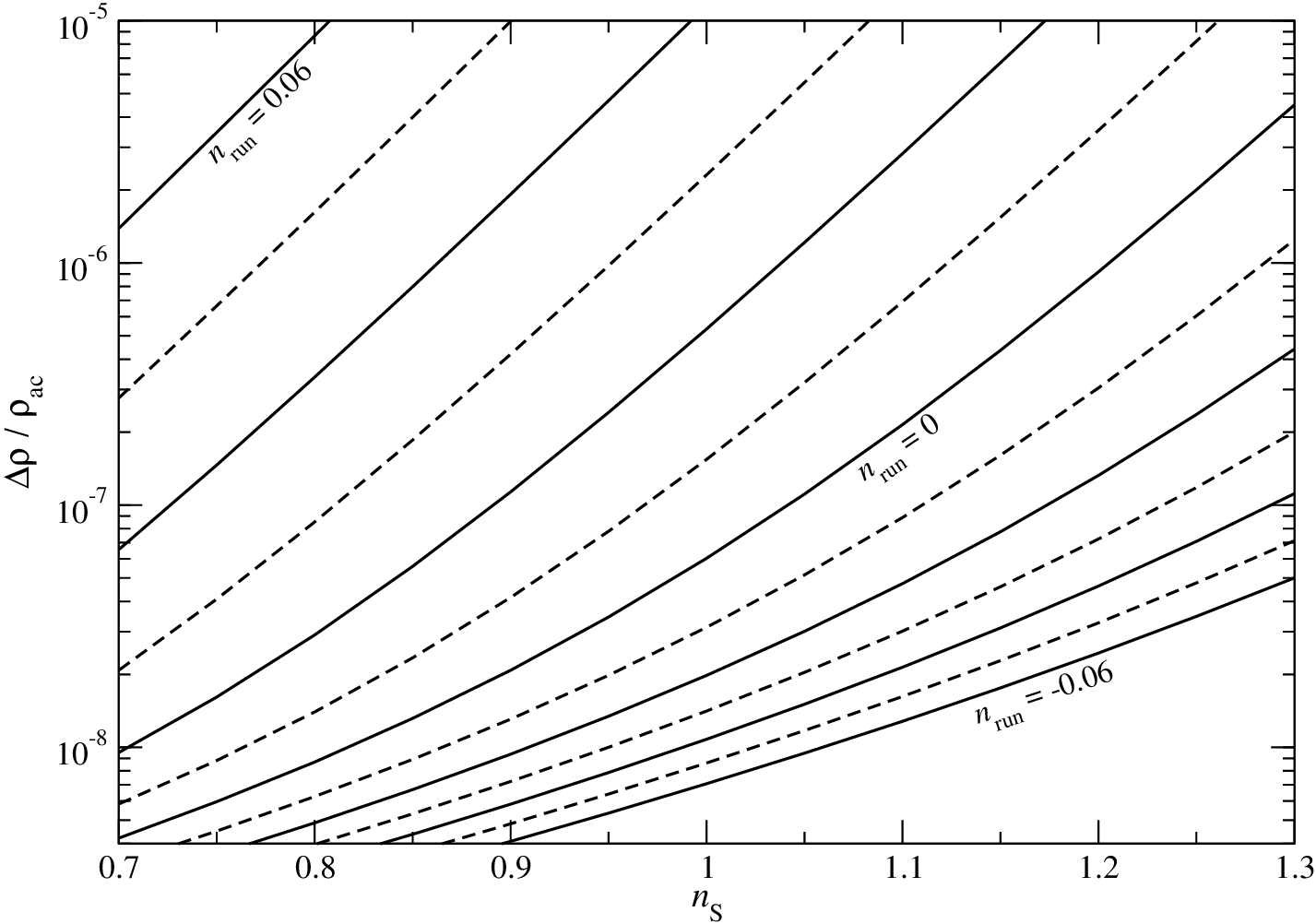}
\caption{Approximate values for the chemical potential, $\mu_{\rm ac}$, the Compton $y$-parameter, $y_{\rm ac}$, and total energy release, $\left. \Delta\rho_\gamma/\rho_\gamma\right|_{\rm ac}$, caused by the dissipation of acoustic modes in the redshift range $200\lesssim z\lesssim \pot{4}{7}$ for different values of $\nS$ and  $\nrun$. 
All curves are for $A_\zeta=\pot{2.4}{-9}$. The lines are separated by $\Delta \nrun = 0.01$ (alternating solid/dashed). 
The dash-dotted/blue line shows the negative chemical potential caused by BE condensation of photons $-\mu_{\rm BE}=\pot{2.7}{-9}$. Models along that line correspond to balanced injection scenarios (Sect.~\ref{sec:balanced}).
In the central panel the dash-dotted/red line indicates the approximate $y$-parameter $y_{\rm ac}$ for those cases.}
\label{fig:estimate_mu}
\end{figure}
\subsection{Numerical estimates for $\mu$, $y$, and $\Delta\rho_\gamma/\rho_\gamma$}
\label{sec:estimates_mu_y_rho}
Armed with the above formulae we can compute the expected values for $\mu_{\rm ac}$, $y_{\rm ac}$, and $\left. \Delta\rho_\gamma/\rho_\gamma\right|_{\rm ac}$ caused by the dissipation of acoustic modes.
For energy release well before recombination we can use the simple approximations for the effective heating rate discussed in Sect.~\ref{sect:approx_SAC}. However, using the same approximations in the $y$-era the resulting $y$-parameter is overestimated by a factor of $\simeq 2-3$ in comparison with the perturbation code.
This is because the breakdown of the tight-coupling approximation and the beginning of free-streaming are not captured correctly by these simple expressions, which leads to a significantly higher effective heating rate (see Fig.~\ref{fig:comp_heating}).

In Fig.~\ref{fig:estimate_mu} we show the numerical estimates obtained with our perturbation code.
One can see that the values depend strongly on both $\nS$ and $\nrun$. 
For all curves we used $A_\zeta=\pot{2.4}{-9}$, however, the obtained values are directly proportional to this and can easily be rescaled.
The curves for $\mu_{\rm ac}$ shown in Fig.~\ref{fig:estimate_mu} can be represented using the simple fitting formula
\bsub
\label{eq:mu_y_ac_approx}
\beal
\label{eq:mu_ac_approx}
%
\mu_{\rm ac}
&\approx \pot{5.54}{-4} A_\zeta
\,\exp\left(9.92\,\gamma_0\,\nS^{1.23 \gamma_1}+47.2\,\nrun\right),
\nonumber\\
\gamma_0&=1+2.91\nrun+23.6\nrun^2,
\nonumber\\
\gamma_1&=1-0.2\nrun-17.8\nrun^2,
\end{align}
which clearly indicates the strong dependence on $\nrun$.
The results for the effective $y$-parameter are approximately given by 
\beal
\label{eq:y_ac_approx}
y_{\rm ac}
&\approx \pot{2.85}{-2} A_\zeta
\,\exp\left(4.32\,\gamma_0\,\nS^{1.53 \gamma_1}+3.51\,\gamma_2\,\nrun\right),
\nonumber\\
\gamma_0&=1+4.68\nrun+22.8\nrun^2,
\nonumber\\
\gamma_1&=1-1.24\nrun-14.2\nrun^2,
\nonumber\\
\gamma_2&=1-7.7\nrun,
\end{align}
and the total energy release can be represented with 
\beal
\label{eq:Drho_ac_approx}
\left.\frac{\Delta\rho_\gamma}{\rho_\gamma}\right|_{\rm ac}
&\approx \pot{2.78}{-1} A_\zeta
\,\exp\left(4.54\,\gamma_0\,\nS^{2.5 \gamma_1}+3.14\,\gamma_2\,\nrun\right),
\nonumber\\
\gamma_0&=1+32.9\nrun+792\nrun^2-1677\nrun^3,
\nonumber\\
\gamma_1&=1-6.5\nrun-294\nrun^2+1173\nrun^3,
\nonumber\\
\gamma_2&=1+35.8\nrun-147\nrun^2-7025\nrun^3.
\end{align}
\esub
%
The formula for $\mu_{\rm ac}$ works best, while the one for $\Delta\rho_\gamma/\rho_\gamma |_{\rm ac}$ has the poorest performance.
The typical agreement with the full numerical computation is $5\% - 10\%$.
Although one execution of the perturbation code does not take very long, these simple formulae should be very useful for estimates.

\subsubsection{Comparison with {\sc CosmoTherm}}
To confirm the precision of the simple approximation for $\mu_{\rm ac}$ and $y_{\rm ac}$, we directly compare with the output from the thermalization code, {\sc CosmoTherm}.
In Fig.~\ref{fig:DI_0.96} we show the cases $(\nS, \nrun)=(0.96, 0)$,  and $(0.96, -0.0265)$.
The latter corresponds to a balanced injection scenario.
Notice that one expects $\mu_\infty=\mu_{\rm ac}+\mu_{\rm BE}$ and $\ye=y_{\rm ac}+y_{\rm BE}$.
For $(\nS, \nrun)=(0.96, 0)$ using the simple fitting formulae we find $\mu_\infty\simeq \pot{1.4}{-8}$ and $\ye \simeq\pot{3.3}{-9}$. 
For $(\nS, \nrun)=(0.96, -0.0265)$ we have $\mu_\infty\simeq 0$ and $\ye \simeq\pot{1.6}{-9}$. 
In both cases, the value for $\mu_\infty$ is very close to the result obtained with {\sc CosmoTherm}, while the best fitting $y$-parameter is slightly underestimated.
Nevertheless the agreement is very good, and close to the expected precision of the approximations.
\begin{figure}
\centering
\includegraphics[width=\columnwidth]{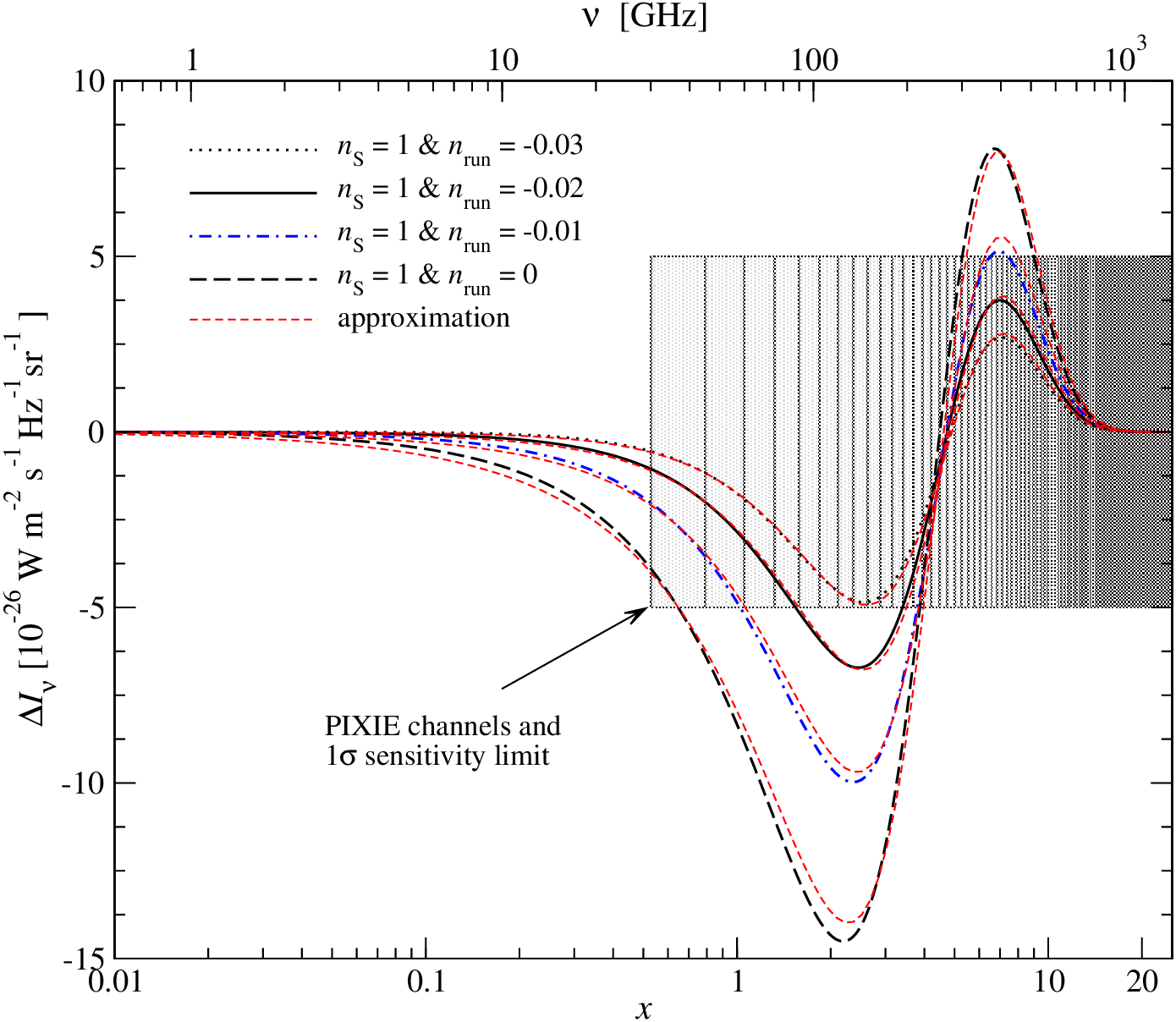}
\caption{CMB distortion at high frequencies for $\nS=1$ and different values of $\nrun$.
The approximations are explained in Sect.~\ref{sec:approx_mu_y}, with parameters given in Table~\ref{tab:mu_y_est}.
}
\label{fig:DI_1}
\end{figure}
\begin{table}
\centering
\caption{Values for $\mu$ and $y$ for different cases. $\mu_{\rm est}$ and $y_{\rm est}$ were obtained with Eq.~\ref{eq:mu_y_ac_approx}. For the approximations in the figures we used $\mu_{\rm est}$ and $\ye$, where $\ye$ was determined to improve the match with the numerical result. The typical difference between $\ye$ and $y_{\rm est}$ is $10\%-15\%$.}
\begin{tabular}{lccccc}
\hline
$(\nS,\nrun)$ & $\mu_{\rm est}$ & $y_{\rm est}$ &$\ye$ 
\\
\hline
   $(1, 0)$ & $\pot{2.4}{-8}$ &$\pot{4.5}{-9}$ & $\pot{5.1}{-9}$
\\
   $(1, -0.01)$ & $\pot{1.0}{-8}$ &$\pot{3.4}{-9}$ & $\pot{4.0}{-9}$
\\
   $(1, -0.02)$ & $\pot{3.8}{-9}$ &$\pot{2.6}{-9}$ & $\pot{3.0}{-9}$
\\
   $(1, -0.03)$ & $\pot{7.1}{-10}$ &$\pot{2.0}{-9}$ & $\pot{2.3}{-9}$
\\
\hline
\end{tabular}
\label{tab:mu_y_est}
\end{table}
In Fig.~\ref{fig:DI_1} we illustrate the performance of the approximation for additional cases with $\nS=1$.
Clearly a good representation of the numerical result is found, however, the matching value for $\ye$ is typically underestimated by $10\%-15\%$.

\subsection{Observability with PIXIE}
\label{sec:Pixie}
As our discussion indicates, the amplitude of the spectral distortion caused by the dissipation of acoustic modes in the early Universe strongly depends on the parameters describing the primordial power spectrum.
The total power at scales $50\,{\rm Mpc}^{-1} \lesssim  k \lesssim 10^4\,{\rm Mpc}^{-1}$ is crucial for the amplitude of the $\mu$-distortion, while power at $k \lesssim  50\,{\rm Mpc}^{-1}$ affects the $y$-distortion part.

In terms of future observability it is also very important to distinguish between the $\mu$ and $y$-parts: $\mu$-type distortions can only be introduced at early times, well before the recombination epoch, and, because of  its peculiar spectral shape, should be more easily separable from other distortions.
One the other hand, $y$-type distortions are also caused by several astrophysical processes at low redshifts.
For example energy release because of supernovae \citep{Oh2003}, or shocks during large scale structure formation \citep{Sunyaev1972b, Cen1999, Miniati2000}, have to be considered.
Also, the effect of unresolved SZ clusters \citep{Markevitch1991}, the
thermal SZ effect and the second order Doppler effect from reionization \citep[e.g., see][and references therein]{McQuinn2005}, and the integrated signals from dusty galaxies {\citep[e.g., see][]{Righi2008, Viero2009, Vieira2010, Dunkley2010, Lagache2011}} will contribute at a significant level.
These signals are orders of magnitudes larger, and hence render a measurement of the $y$-type distortion from the cosmological dissipation process more difficult.
%

\begin{figure}
\centering
\includegraphics[width=\columnwidth]{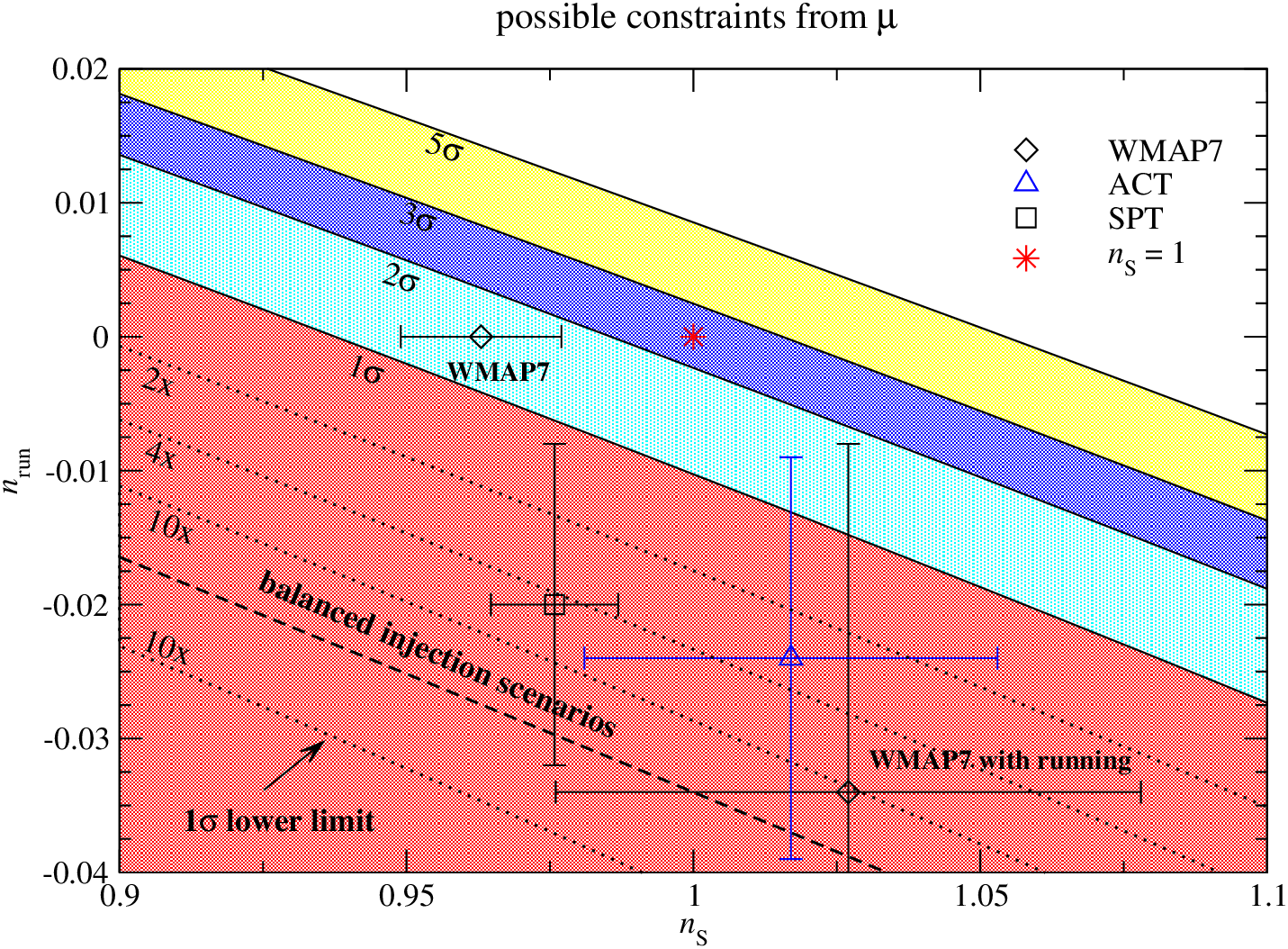}
\\[2mm]
\includegraphics[width=\columnwidth]{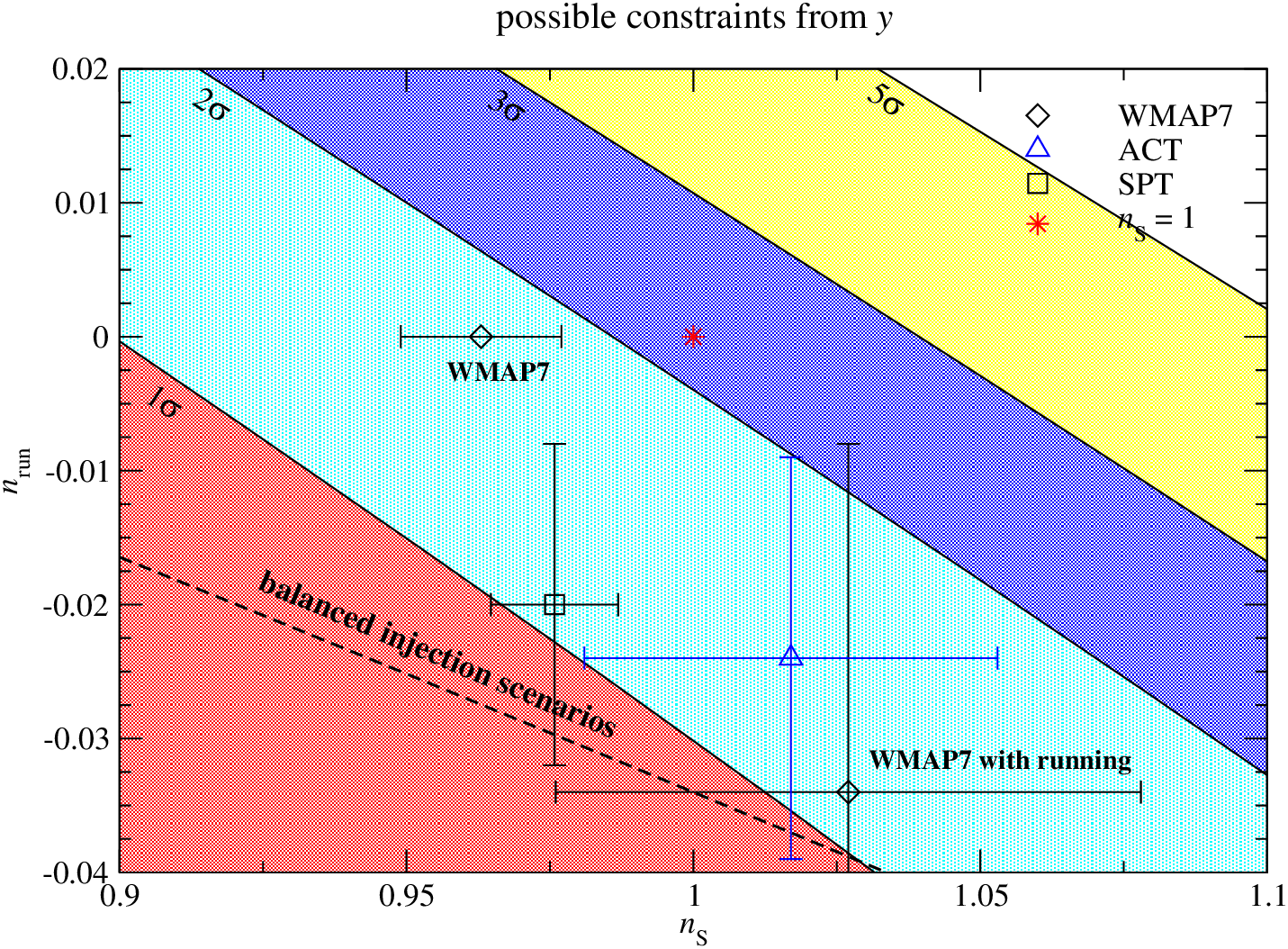}
\caption{Possible limits on $\nS$ and $\nrun$ derived from measurements of $\mu$ and $y$-type distortions with PIXIE. The coloured regions show $1\sigma$ through $5\sigma$ detection limits. Models in agreement with WMAP7 \citep{Komatsu2010}, ACT \citep{Dunkley2010} and SPT \citep{Keisler2011} are indicated. The dashed/black line shows models with balanced energy release during the $\mu$-era.
Above this line the net $\mu$-parameter caused by the dissipation and BE condensation process is expected to be positive, while below the line it will be in the range $-\pot{2.7}{-9}\lesssim \mu\lesssim 0$.
Also in the upper panel the dotted lines indicate how much the $1\sigma$ limit from $\mu$ would improve with sensitivities increased by the annotated factors. 
Notice that for 10 times PIXIE sensitivity a {\it lower} bound appears, since the negative $\mu$-distortion from BE condensation of CMB photons becomes observable at the $\simeq 2.7\sigma$ level.
%
}
\label{fig:mu_limits}
\end{figure}
Let us take an optimistic point of view and just ask by what factor the amplitude of the distortion exceeds the sensitivity of PIXIE.
According to \citet{Kogut2011PIXIE}, $y\simeq 10^{-8}$ and $\mu\simeq \pot{5}{-8}$ should be detectable at $5\sigma$-level.
In Fig.~\ref{fig:DI_0.96} and \ref{fig:DI_1} we already indicated the $1\sigma$ sensitivity level for some cases, however, using the approximations of the previous section we can turn this into a prediction for the possible limits on $\nS$ and $\nrun$.
In Fig.~\ref{fig:mu_limits} we show the results of this exercise.
For the balanced energy release scenarios the currently proposed sensitivities \citep{Kogut2011PIXIE} imply a null detection for both $\mu$ and $y$-distortion. 
As pointed out by \citet{Khatri2011}, for fixed value of $\nS$ this still is a rather sensitive measurement of $\nrun$ at very small scales, providing a strong upper limit.
Our analysis also shows that for $\nrun\lesssim \nrun^{\rm bal}$ the distortion asymptotes towards the case without any dissipation, so that the effective chemical potential is bounded by $\mu\gtrsim -\pot{2.7}{-9}$.
This implies that the sensitivity of PIXIE would need to be increased by about one order of magnitude to discern balanced injection scenarios from cases with $\nrun\ll \nrun^{\rm bal}$. 
At this level of sensitivity the cosmological recombination spectrum
\citep{Dubrovich1975, Chluba2006b, Sunyaev2009} also starts to contribute significantly to the signal in the PIXIE bands.
For detailed forecasts these contributions should all be considered simultaneously.

From Fig.~\ref{fig:mu_limits} we can also conclude that for the currently favoured WMAP7 model without running, the $\mu$-distortion might be observable at the $1.5\sigma$ level, however, for the other models shown only upper limits will be obtained, unless $\nrun$ lies in the upper part of the allowed $1\sigma$ range for these cases.
When increasing the sensitivity relative to PIXIE by one order of magnitude these models should be distinguishable at the $1\sigma-2\sigma$ level.
Also, once about ten times the sensitivity of PIXIE can be achieved, a lower limit on $\nrun$ can be derived, since the distortion from BE condensation of CMB photons becomes observable at the $\simeq 2.7\sigma$ level.

The lower panel of Fig.~\ref{fig:mu_limits} illustrates possible constraints derived from PIXIE using information from the $y$-type distortion.
For the $y$-part itself no balanced injection scenario is expected in reasonable ranges of parameters, because the heating by Silk-damping is normally several times larger than the cooling from BE condensation in the $y$-era.
For the currently proposed sensitivity PIXIE is slightly more sensitive to the $y$-distortion from the dissipation process than the $\mu$-part.
However, as mentioned above, the $y$-distortion part has several competing astrophysical processes at low redshifts that make a separation more difficult.
We mention that increasing the sensitivity by a factor of ten relative to PIXIE could in principle allow $\simeq 5\sigma$ detections of the $y$-distortion caused by the damping of acoustic waves for all models shown in Fig.~\ref{fig:mu_limits}.

In this work we have only considered the simplest parametrization of the primordial power spectrum, allowing running of the spectral index, $\nS$. However, given the large variety of inflationary models \citep[some examples of possible interest are][]{mukhanov,Silk1987, Salopek1989, Lehners2007, Ido2010, Barnaby2011}, the energy release in the $\mu$- and $y$-era can differ by a large amount.
Models with extra power on scales $50\,{\rm Mpc}^{-1}\lesssim k \lesssim 10^4\,{\rm Mpc}^{-1}$, corresponding to the photon diffusion scale at $z\simeq \pot{5}{4}$ to $z\simeq \pot{2}{6}$, will enhance the chemical potential distortion, while suppression of power on these scales leads to less energy release.
Measurements of the CMB spectrum may therefore allow constraints on inflationary models at scales well below those of the CMB today.
The amplitude of the net $\mu$-distortion is therefore very uncertain and could even be several times larger.
However, a detailed analysis for more exotic inflationary models is beyond the scope of this paper.

We also mention that to give the source terms for the average distortion in second
  order perturbation theory one has to compute ensemble averages. This
  implies that predictions for the associated spectral distortions are in principle limited by
  cosmic variance. However, the dissipation of perturbations happens close
  to the diffusion length, which at high redshifts is about two orders of
  magnitude smaller than the horizon, such that cosmic variance on these scales
  is negligible. 
On the other hand, during and after hydrogen recombination the
  $y$-type distortions get contributions from scales up to the horizon. 
  These contributions will have cosmic variance similar to the CMB
  power spectrum on these scales. However $y$-type distortions generated
  after recombination, in particular during reionization due to thermal SZ
  and second order Doppler effects, are expected to be more than  an order of magnitude larger, and separating the contribution of Silk damping would be challenging in any case. Thus for $y$-type distortions uncertainties in the low redshift contributions by far outweigh cosmic variance.
%

\section{Summary}
\label{sec:conc}
We have presented, for the first time, a consistent and accurate
  calculation for the spectral distortions of the CMB arising from the
  dissipation of acoustic waves (Silk damping) in the early Universe. 
 All previous calculations \citep[except for][]{Daly1991} used the classical expression
  for energy density  stored by acoustic waves in the radiation part of the
  primordial plasma, which turns out to omit some details.
Here we derive the evolution equation for the average CMB spectral distortions including uniform heating of baryonic matter and accounting for the effect of perturbations in the cosmological medium.
We consistently treat the problem in second order perturbation theory considering second order Compton energy exchange for the Boltzmann collision term.
The final evolution equation has order 2$\times$2 and is discussed in Sect.~\ref{sec:dist_results}. 

Unlike in first order perturbation theory, the frequency-dependence of the second order
Boltzmann equation cannot be directly factored out. We show, however, that a separation can be found using a definition of spectral distortions based on photon number conservation.  
With this definition only $y$-type distortion terms are present in the second order
Boltzmann equation. Thus multiplying the Boltzmann equation by $x^2$ and
integrating over $x$ makes the distortion vanish. 
The frequency-dependence of the surviving, pure temperature perturbation factors out (similar to to the case for the first order Boltzmann equation). 
After making use of the equation obtained above to isolate the pure temperature part, we also obtain an evolution equation for the $y$-part, in which again the frequency-dependence factors out. 
We utilize this procedure to separate the full 2$\times$2 Boltzmann equation into evolution equations for pure spectral distortions and second order average temperature.

Our detailed derivation (Appendix~\ref{sec:AnisoCS}) of the Boltzmann collision term starts with the Lorentz-boosted Boltzmann equation to include the effect of bulk motions. 
However, eventually we use a simpler, more compact approach based on direct Lorentz transformations of the generalized Kompaneets equation, Eq.~\eqref{eq:BoltzEqlm_The}, which includes the effect of energy exchange on the dipole through octupole anisotropy.
As we prove here, in a similar way one can derive all second order terms of the Boltzmann collision integral which have been previously discussed in the literature \citep[e.g.,][]{Hu1994pert, Bartolo2007} by just using Lorentz transformations of the simple Thomson scattering collision term, Eq.~\eqref{eq:BoltzEq_lim_0}.
We also provide an approximate treatment of second order corrections to the BR and DC process (Sect.~\ref{sec:AnisoBRDC}).

To arrive at the final evolution equation, Eq.~\eqref{eq:average_distortion_evol_eq}, we consider the zeroth, first and second order thermalization problems in a step-by-step manner (Sect.~\ref{sec:lhs_photon}).
We demonstrate that in first order perturbation theory spatial-spectral distortions appear if uniform distortions were created in zeroth order. In the future these distortion might allow constraints to be places on new physics responsible for uniform energy release and spectral distortions in the early Universe. 
An average distortion of the order of $10^{-5}-10^{-6}$ during
recombination could result in spatial-spectral perturbations of order $\simeq
10^{-10}$ that dominate over the second order spatial-spectral
perturbations, part of which was calculated by \cite{Pitrou2010}.

In second order perturbation theory we are only interested in the average
distortion, since spatial-spectral contributions have a much smaller  amplitude $\simeq 10^{-12}-10^{-11}$.
We provide a full microphysical description of how perturbations in the cosmological fluid affect the average CMB spectrum.
The creation of distortions by the dissipation of acoustic waves is explained in simple words in Sect.~\ref{sec:simple_pic}, but the most important aspects
can be understood as follows:
anisotropies of the photon temperature cause an increase of the CMB energy density by $\Delta \rho_\gamma /\rho_\gamma \simeq 6\sum_{l=0}^\infty (2l+1) \left<(\hat{\Theta}^{(1)}_l)^2\right>$ at every moment.
This energy density changes slowly with time as temperature
  anisotropies are erased on small scales by Silk-damping causing energy
  release. 
  Locally the dominant effect is that of mixing of blackbodies of different
temperatures (mostly in the quadrupole at early times) by Thomson
scattering. This mixing isotropizes the radiation in the
rest frame of the electrons and transfers energy which is stored in temperature
anisotropies or acoustic waves to the average/isotropic component of
the radiation. 
2/3 of this energy just raises the average temperature of
the CMB and is unobservable. The observable 1/3 of this acoustic energy gives
rise to an instantaneous $y$-type distortion which
evolves towards equilibrium by comptonization and photon production. 
The final spectrum is a mixture of $y$-type and $\mu$-type distortions.
The influence of gravity and gradients in the photon fluid lead to an additional  increase in the average temperature of the CMB, but no distortion (see
Eq.~\ref{eq:Boltzmann_second_order_general_II}).
This increase is again not directly measurable.


The final $\mu$-type spectral distortions depend sensitively on the amplitude of the primordial power
spectrum in the comoving wavenumber range of $50\,{\rm Mpc}^{-1}\lesssim k
\lesssim 10^4\,{\rm Mpc}^{-1}$. These primordial perturbations are
completely destroyed by Silk damping and the CMB 
spectral distortion is their only observable imprint. 
Measurement of spectral distortions thus provides
a new probe of cosmological initial conditions  which is complementary to
other cosmological probes. A measurement of the primordial power
spectrum on such small scales may allow us to shed light on aspects of high energy physics
(e.g. inflation or dark matter physics) responsible for creating the initial conditions which are otherwise hard to
constrain. 
We mention that here our analysis was focused on curvature perturbations, however, the expressions obtained here can be generalized to iso-curvature perturbations.

The adiabatic cooling of baryons creates a distortion which has an
opposite sign to those caused by Silk damping. The amplitude of this distortion is $\mu_{\rm BE}\simeq
-2.7\times 10^{-9}$, which is relatively stable to changes in cosmological
parameters within the current error bars. 
At an experimental sensitivity level of $\Delta I/I\simeq 10^{-9}$ one should be able to detect this signature of Bose-Einstein condensation of photons in the CMB spectrum. 
This distortion partially cancels the distortion caused by the dissipation of acoustic waves in the early Universe, however, the net effect strongly depends on the values for $\nS$ and $\nrun$ (see Fig.~\ref{fig:estimate_mu}).
PIXIE \citep{Kogut2011PIXIE} has a $5\sigma$ detection limit of $5\times
10^{-8}$ for $\mu$, so that for currently favoured models with running, null-detections are expected (see projected constraints on parameter space in Fig. \ref{fig:mu_limits}).
This in itself is a very sensitive measurement of the initial perturbation power on small scales, and for given $\nS$ it provides a strong upper limit on $\nrun$.
Improving the sensitivity of PIXIE by about one order of magnitude could, however, allow us to distinguish these models from cases with balanced energy release during the $\mu$-era (Sect.~\ref{sec:balanced}).

Achieving a sensitivity of $\Delta I/I\simeq 10^{-9}$ requires subtraction of
foregrounds due to synchrotron emission, free-free emission, dust emission,
and spinning dust emission at the same precision level. The simulations for PIXIE indicate
that an accuracy of 1\,nK may in principle be achievable in foreground subtraction \citep{Kogut2011PIXIE}.
There is however substantial uncertainty in our understanding of foregrounds and possible systematics, as indicated by an observed excess signal at $\simeq 3\,{\rm GHz}$ by ARCADE \citep{Fixsen2011}, which is not completely explained by the current galactic and extragalactic emission models. 
Nevertheless, given the important physics that can be learned from a measurement of primordial power on extremely small scales, a case can be made for attempting to increase the sensitivity of experiments like PIXIE.

\section*{Acknowledgments}
The authors cordially thank Douglas Scott for detailed comments and suggestions, which greatly helped improving the manuscript.
JC is very grateful for useful discussions with Richard Shaw, Eric Switzer, Adrienne Erickcek, Geoff Vasil, and Marcelo Alvarez.
Furthermore, the authors acknowledge the use of the GPC supercomputer at the SciNet HPC Consortium. 
SciNet is funded by: the Canada Foundation for Innovation under the auspices of Compute Canada; 
the Government of Ontario; Ontario Research Fund - Research Excellence; and the University of Toronto.

\begin{appendix}


\section{Functions $\mathcal{G}$, $\mathcal{Y}$, $\mathcal{Y}_{\rm SZ}$, and some of their properties} 
\label{sec:def_GY}
In the formulation of the thermalization problem one encounters the functions $\mathcal{G}$ and $\mathcal{Y}$ which  are defined as
\beal
\label{eq:def_GY_func}
\mathcal{G}(\xg)&\equiv \frac{\xg\,e^{\xg}}{[e^{\xg}-1]^2}= - \xg \, \partial_{\xg} \nPl(\xg) = \xg\nPl[1+\nPl],
\nonumber\\[1mm]
\mathcal{Y}(\xg)& \equiv\frac{1}{2}\,\mathcal{G}(\xg)\,\xg \frac{e^{\xg}+1}{e^{\xg}-1} = \frac{1}{2}\,\mathcal{G}(\xg)\,\xg [1+2\nPl],
\end{align}
where $\nPl=1/(e^{\xg}-1)$ is the occupation number of a blackbody.
The function $\mathcal{G}(\xg)$ has the shape of temperature perturbations to the blackbody spectrum, while $\mathcal{Y}(\xg)$ exhibits a $y$-type  behaviour.
In the limit of $\xg \rightarrow 0$ one has $\mathcal{Y}(\xg)\simeq \mathcal{G}(\xg)/ 2$ and $\mathcal{G}(\xg) \simeq 1/\xg$, while for
 $\xg \gg 1$ one has $\mathcal{Y}(\xg)\simeq \mathcal{G}(\xg) \, \xg / 2$ and $\mathcal{G}(\xg) \simeq \xg e^{-\xg}$.

To separate parts that are related to temperature perturbations ($\propto \mathcal{G}$) and real spectral distortions it is convenient  to define 
\beal
\label{eq:def_Yth_YSZ_func}
\mathcal{Y}(\xg)&=\frac{1}{2}\,\mathcal{Y}_{\rm SZ}(\xg)+2\mathcal{G}(\xg)
\nonumber\\
\mathcal{Y}_{\rm t}(\xg)&=\mathcal{Y}(\xg)-\mathcal{G}(\xg)
=\frac{1}{2}\,\mathcal{Y}_{\rm SZ}(\xg)+\mathcal{G}(\xg)
\nonumber\\[1mm]
\mathcal{Y}_{\rm SZ}(\xg)&=2[\mathcal{Y}(\xg)-2\mathcal{G}(\xg)].
\end{align}
In particular $\mathcal{Y}_{\rm SZ}$ is important since it characterizes redistribution of photons over energy, without affecting the total number of photons.
$\mathcal{Y}_{\rm SZ}$ is also known in connection with the thermal SZ effect of galaxy clusters \citep{Zeldovich1969}, and has the spectral shape of a $y$-distortion.
For the first derivative of $\mathcal{G}$
one finds
\beal
\label{eq:def_dGY_func}
\xg \partial_{\xg} \mathcal{G}(\xg)&=\mathcal{G}(\xg) - 2\mathcal{Y}(\xg) = - 3 \mathcal{G}(\xg) - \mathcal{Y}_{\rm SZ}(\xg).
\end{align}
%
Additional handy relations for derivatives of the $\nPl$ are
\bsub
\label{eq:relations_nPl}
\beal
\xg\partial_{\xg} \nPl &=-\mathcal{G}
\\
\xg^2\partial^2_{\xg} \nPl &=+\xg\mathcal{G} \mathcal{A} \equiv 2\mathcal{Y}
\\
\xg^3\partial^3_{\xg} \nPl &=-\xg\mathcal{G}\left[6 \mathcal{G} +\xg \right]
\\
\xg^4\partial^4_{\xg} \nPl &=+\xg^2\mathcal{G}\left[12 \mathcal{G} +\xg \right]\mathcal{A},
\end{align}
\esub
with $\mathcal{A}(x)=1+2\nPl(x) = (e^x+1)/(e^x-1)=\coth(x/2)$.

\subsection{Useful relations for the diffusion operator} 
\label{sec:diff_op}
In the derivation of the Boltzmann collision term one encounters terms with different orders of derivatives of the photon occupation number.
Since the scattering process conserves the number of photons the differential operators should have the generic form $x^{-2} \partial_x O[n]$, because then $\int x^2 ( x^{-2} \partial_x O[n] ) \id x= 0$.
It is therefore useful to write all derivatives in this way.

From the normal Kompaneets equation \citep{Kompa56, Weymann1965} we already know that the normal diffusion operator has the form $\mathcal{D}_x = x^{-2} \partial_x x^4 \partial_x$. We also expect that terms like $\mathcal{D}_{\xg}\mathcal{G}=-\mathcal{D}_{\xg}\xg \partial_{\xg} \nPl$ and $\mathcal{D}_{\xg}\mathcal{Y}_{\rm SZ}=\mathcal{D}_{\xg}\mathcal{D}_{\xg}\nPl$ should appear, the latter two being third and fourth order in terms of the derivatives of the photon distribution.
Therefore we give the useful operator relations
\bsub
\label{eq:diffusion_op}
\beal
x^2 \partial^2_{x} &= \mathcal{D}_{x} - 4 x \partial_{x}
\\
x^3 \partial^3_{x} &= \mathcal{D}_{x} x \partial_{x} - 6 \mathcal{D}_{x} + 20 x \partial_{x}
\\
x^4 \partial^4_{x} &= \mathcal{D}_{x}\mathcal{D}_{x}-12\mathcal{D}_{x} x \partial_{x} + 38 \mathcal{D}_{x} - 120 x \partial_{x}.
\end{align}
\esub
From this is also follows that
\bsub
\label{eq:diffusion_op_nPl}
\beal
\xg^2 \partial^2_{\xg} \nPl&= \mathcal{Y}_{\rm SZ} + 4 \mathcal{G}
\\
\xg^3 \partial^3_{\xg} \nPl&= - \mathcal{D}_{\xg} \mathcal{G} - 6 \mathcal{Y}_{\rm SZ} - 20 \mathcal{G}
\\
\xg^4 \partial^4_{\xg} \nPl&= \mathcal{D}_{\xg}\mathcal{Y}_{\rm SZ} +12\mathcal{D}_{\xg} \mathcal{G} + 38 \mathcal{Y}_{\rm SZ} + 120 \mathcal{G},
\end{align}
\esub
where we simply applied the operators on a Planck spectrum. Together with Eq.~\eqref{eq:relations_nPl} this also implies
\bsub
\label{eq:diffusion_op_nPl}
\beal
\mathcal{D}_{\xg}\nPl &= \xg \mathcal{G} \mathcal{A} - 4 \mathcal{G}\equiv \mathcal{Y}_{\rm SZ}
\\
\mathcal{D}_{\xg} \mathcal{G}&= 
- 6 \mathcal{Y}_{\rm SZ} - 20 \mathcal{G} + \xg \mathcal{G}\left(6\mathcal{G}+\xg\right)
\\
\mathcal{D}_{\xg}\mathcal{Y}_{\rm SZ}&= 34 \mathcal{Y}_{\rm SZ} + 120 \mathcal{G} 
\nonumber
\\
&\qquad
+12\xg \mathcal{G}^2\left[\xg\mathcal{A} - 6 \right] 
+ \xg^2\mathcal{G}\left[\xg\mathcal{A} -12 \right],
\end{align}
\esub
and
\bsub
\label{eq:def_dGY_func_rel_2}
\beal
\partial_{\xg} \mathcal{G}(\xg)+\mathcal{G}(\xg)\mathcal{A}&=-\partial_{\xg} \nPl
\\
\frac{1}{\xg^2}\partial_{\xg}\xg^4
\left\{\partial_{\xg} \mathcal{G}(\xg)+\mathcal{G}(\xg)\mathcal{A}\right\}&=-\mathcal{Y}_{\rm SZ}.
\end{align}
\esub
These expression are rather useful when rearranging terms of the Boltzmann equation.
In addition we encounter the functions
\bsub
\label{eq:def_HE_func}
\beal
\mathcal{H}&=-\frac{1}{2}\mathcal{D}^\ast_{\xg}
\left\{\partial_{\xg} \mathcal{Y}_{\rm SZ}(\xg)+\mathcal{Y}_{\rm SZ}\mathcal{A}\right\}
\\
\mathcal{E}&=2\xg \mathcal{G} [\mathcal{G}+\mathcal{Y}_{\rm SZ}] 
\equiv -\mathcal{D}^\ast_{\xg}\mathcal{G}^2
\\
\mathcal{B}&=\xg [\mathcal{A}-\mathcal{G}] 
\\
\mathcal{D}_{\xg}\mathcal{G}&=\mathcal{Y}_{\rm SZ}+\mathcal{H}+\mathcal{E},
\end{align}
\esub
with $\mathcal{D}^\ast_{\xg}=\xg^{-2}\partial_{\xg} \xg^4$. In particular the last relation is useful.

\subsection{Some integrals of $\mathcal{G}$, $\mathcal{Y}_{\rm SZ}$, $\mathcal{H}$ and $\mathcal{E}$} 
\label{sec:integrals_GYHE}
In our discussion we need the integrals of $\mathcal{G}$, $\mathcal{Y}_{\rm SZ}$, $\mathcal{H}$ and $\mathcal{E}$ over $\xg^2 \id \xg$ and $\xg^3 \id \xg$. 
For the integrals over $\xg^2 \id \xg$ one has
\beal
\int \mathcal{Y}_{\rm SZ}(\xg) \xg^2 \id \xg 
& 
= \int \mathcal{H}(\xg) \xg^2 \id \xg 
=\int \mathcal{E}(\xg) \xg^2 \id \xg =0
\nonumber\\
\int \mathcal{G}(\xg) \xg^2 \id \xg & = 6\zeta(3)= 3\mathcal{G}^{\rm pl}_2 \approx 7.2123, 
\label{eq:number_int_GYHE}
\end{align}
where $\zeta(n)$ is the Riemann $\zeta$-function.
This property of $\mathcal{G}$ and $\mathcal{Y}_{\rm SZ}$ allows us to define a separation of spectral distortions from temperature distortions (Sect.~\ref{sec:second_order}).

Defining $\mathcal{J} = \frac{4\pi^4}{15}\equiv 4\mathcal{G}^{\rm pl}_3$, it is also easy to show that 
\beal
\int \mathcal{G}(\xg) \xg^3 \id \xg & = \int \mathcal{Y}_{\rm SZ}(\xg) \xg^3 \id \xg  = \mathcal{J}
\approx 25.976
\nonumber\\
\int \mathcal{H}(\xg) \xg^3 \id \xg & = \left(\frac{10\pi^2}{21}-2\right) \mathcal{J} 
\approx 2.70 \,\mathcal{J}
\nonumber\\
\int \mathcal{E}(\xg) \xg^3 \id \xg & = \left(5-\frac{10\pi^2}{21}\right) \mathcal{J}
\approx 0.30 \,\mathcal{J}.
\label{eq:energy_int_GYHE}
\end{align}
These integrals become important when considering the energy exchange between electrons and photons in second order perturbation theory.
We also note that $\int \mathcal{H}(\xg) \xg^3 \id \xg+\int \mathcal{E}(\xg) \xg^3 \id \xg\equiv 3 \,\mathcal{J}$.

\section{Relations for spherical harmonic coefficients}
\label{sec:s_harm_relations}
\subsection{Projections of $\mu \, n(\nu, \vgh)$ and $\mu^2 \, n(\nu, \vgh)$}
\label{sec:proj_mu_n}
When decomposing the Boltzmann equation into spherical harmonics we have to compute the projections $[\mu \, n(\nu, \vgh)]_{lm}$.
Introducing
$c_{l,m}=\sqrt{(l^2-m^2)/(4l^2-1)}$
it is straightforward to show that
\beal
\label{eq:projection_mu}
[\mu \, n(\nu, \vgh)]_{lm}&= c_{l,m}\,n_{l-1,m}+ c_{l+1,m}\, n_{l+1,m}
\nonumber\\
[\mu \, Y_{l'm'}(\vgh)]_{lm}
&= \delta_{m m'}\left[ \delta_{l-1, l'} c_{l,m} + \delta_{l+1, l'}c_{l+1,m} \right].
\end{align}
Here $n_{lm}$ denotes the spherical harmonic coefficients of the photon occupation number.
A few examples are $[\mu \, n(\nu, \vgh)]_{00}=n_{10}/\sqrt{3}$, $[\mu \, n(\nu, \vgh)]_{10}=n_{00}/\sqrt{3}+2n_{20}/\sqrt{15}$, and $[\mu \, n(\nu, \vgh)]_{1\pm 1}=n_{2\pm1}/\sqrt{5}$.

In addition we need the projections $[\mu^2 \,  \, n(\nu, \vgh)]_{lm}$, which are easily shown to be
\beal
\label{eq:projection_mu2}
[\mu^2 \, n(\nu, \vgh)]_{lm}&= 
d_{l,m}\,n_{l-2,m}
+\left[c_{l,m}^2+c_{l+1,m}^2\right]n_{l,m}
+d_{l+2,m}\,n_{l+2,m}
\nonumber\\
[\mu^2 \, Y_{l'm'}(\vgh)]_{lm}&= 
\delta_{mm'}\left[ \delta_{l-2, l'} d_{l,m} 
+\delta_{l, l'}\left(c_{l,m}^2+c_{l+1,m}^2\right)\right] 
\nonumber\\
&
\qquad
+\delta_{mm'} \delta_{l+2, l'}d_{l+2,m}n_{l',m'}
\end{align}
using Eq.~\eqref{eq:projection_mu} and defining $d_{l, m}\equiv c_{l-1,m}\,c_{l,m}$. A few examples are, 
$[\mu^2 \, n(\nu, \vgh)]_{00}=n_{00}/3+2n_{20}/[3\sqrt{5}]$, $[\mu^2 \, n(\nu, \vgh)]_{10}=3n_{10}/5+2\sqrt{3}\,n_{30}/[5\sqrt{7}]$, and $[\mu^2 \, n(\nu, \vgh)]_{11}=n_{11}/5+2\sqrt{2}\,n_{31}/[5\sqrt{7}]$.

It turns out that most of the terms of the collision integral $\propto\betac$ and $\propto\betacsq$ can be neatly expressed using $c_{l,m}$ and $d_{l, m}$.
This is because physically bulk motion of the electrons leads to the relativistic aberration effect and Doppler boosting, which, when the $z$-axis is aligned with the bulk velocity vector, just results in mixing of spherical harmonic coefficients $(l, m)$ with $(l\pm1, m)$ to first order in $\betac$, and $(l\pm2, m)$ in $\mathcal{O}(\betac^2)$.
In essence this leads to coupling coefficients of the form $c_{l, m}$ and $d_{l, m}$ \citep{Challinor2002, Kosowsky2010, Amendola2010, Chluba2011ab}, which alternatively can be expressed in terms of Wigner-3$J$ symbols \citep{Stegun1972}.

\subsection{Lorentz transformation of the photon occupation number}
\label{sec:Lorentz_trafo}
The photon occupation number is Lorentz invariant. We therefore have $n'(\nu', \vghp)\equiv n(\nu(\nu', \vghp), \vgh(\vghp))$, where in this section a prime denotes the quantities in the moving frame. 
Aligning the peculiar velocity vector, $\vbetac$, with the $z$-axis means $\nu=\gammac \nu' (1+\betac\mu')$ and $\mu=(\mu'+\betac)/(1+\betac\mu')$.
Therefore one can write
\beal
\label{eq:n_Lorentz}
n'(\nu', \mu')&\approx n(\nu', \mu')+\betac \partial_{\betac} n(\nu, \mu) + \frac{1}{2}\betacsq \partial^2_{\betac} n(\nu, \mu)
\end{align}
up to second order in $\betac$. 
To first order in $\betac$ we have
\beal
\label{eq:n_dLorentz}
\betac \partial_{\betac} n(\nu, \mu)
&\approx \partial_{\nu'} n(\nu', \mu') \, \betac \partial_{\betac} \nu' 
+ \partial_{\mu'} n(\nu', \mu') \, \betac \partial_{\betac} \mu
\nonumber\\
&
\approx
\betac\mu'\, \nu' \partial_{\nu'} n(\nu', \mu') +  \betac (1-\mu'^2) \, \partial_{\mu'} n(\nu', \mu').
%
\end{align}
With the identity $(1-\mu^2)\partial_\mu Y_{lm}(\mu)=(2l+1)c_{l,m}Y_{l-1,m}(\mu)-l\,\mu Y_{lm}(\mu)$ this directly means
\beal
\label{eq:n_dLorentz_lm}
[\betac \partial_{\betac} n(\nu, \mu)]_{lm}
&
\approx
\betac
\left[
c_{l,m}\,\nu' \partial_{\nu'} n_{l-1,m}+c_{l+1,m}\,\nu' \partial_{\nu'} n_{l+1,m}
\right]
\\
&\quad
+\betac
\left[
(l+2) c_{l+1,m} n_{l+1,m}
-(l-1)\,c_{l,m} n_{l-1,m}
\right].
\nonumber
\end{align}
Terms $\propto \betacsq$ are second order in perturbation theory. We therefore only need those contributions that arise from the monopole of the photon distribution.
This implies that consistent up to second order perturbation theory we find
\beal
\label{eq:n_Lorentz_lm}
n'_{lm}(\nu')&\approx n_{lm}
+\delta_{l0}\frac{\betacsq}{2}\left[\nu' \partial_{\nu'}+\frac{1}{3}\nu'^2 \partial^2_{\nu'}\right] n_{00}
+\delta_{l2}\delta_{m0} \frac{\betacsq}{2}  \,d_{20}\,\nu'^2 \partial^2_{\nu'} n_{00}
\nonumber\\
&\qquad
-
\betac\,c_{l,m}
\left[(l-1) - \nu' \partial_{\nu'} \right] n_{l-1,m}
\nonumber\\
&\qquad\qquad
+\betac\,c_{l+1,m} 
\left[
(l+2) +\nu' \partial_{\nu'} 
\right] n_{l+1,m}.
\end{align}
A few examples for the transformed photon distribution are
\beal
\label{eq:n_Lorentz_examples}
n'_{00}(\nu')&\approx n_{00}
+\betac c_{10}\left[ 2 + \nu' \partial_{\nu'} \right] n_{10} 
+\frac{\betacsq}{6}\left[\mathcal{D}_{\nu'}- \nu' \partial_{\nu'}\right] n_{00},
\nonumber\\
n'_{1m}(\nu')
&\approx n_{1m}
+\betac c_{10} \nu' \partial_{\nu'} n_{00}
+\betac c_{2m}\left[ 3 + \nu' \partial_{\nu'} \right] n_{2m} 
\nonumber\\
n'_{2m}(\nu')
&\approx n_{2m}
-\betac c_{2m}\left[ 1 - \nu' \partial_{\nu'} \right] n_{1m} 
+\betac c_{3m}\left[ 4 + \nu' \partial_{\nu'} \right] n_{3m} 
\nonumber\\
&\qquad+\frac{\betacsq}{2} \,d_{20}\,\left[\mathcal{D}_{\nu'} - 4\nu' \partial_{\nu'}\right]n_{00}
\\
n'_{3m}(\nu')
&\approx n_{3m}
-\betac c_{3m}\left[ 2 - \nu' \partial_{\nu'} \right] n_{2m} 
+\betac c_{4m}\left[ 5 + \nu' \partial_{\nu'} \right] n_{4m}.
\nonumber
\end{align}
These expressions are useful for deriving parts of the Boltzmann collision term using explicit Lorentz transformations.
One simplifying aspect is that $\nu' \partial_{\nu'}$ and $\mathcal{D}_{\nu'}$ are Lorentz invariant.

\section{Anisotropic Compton scattering} 
\label{sec:AnisoCS}
In this section we present a derivation of the kinetic equation for Compton
scattering of initially anisotropic photons by thermal
electrons. We use the Lorentz-boosted Boltzmann equation to include bulk
motion. A similar formalism was applied to derive relativistic corrections to the SZ effect \citep{Challinor1998, Itoh98, Sazonov1998, Nozawa1998SZ, Challinor2000, Chluba2005b}.

The bulk flow effectively makes the electron distribution function anisotropic in the Newtonian frame. Alternatively one can also think of this problem in the rest frame of the moving volume element. 
In this case the CMB has to be transformed into the moving frame, so that because of aberration and Doppler boosting the multipoles $l=0, 1, 2, 3$ and $4$ are expected to couple directly to the scattering cross section in second order of the bulk velocity \citep{Chluba2011ab}.
This view allows one to obtain all kinematic terms using appropriate Lorentz transformations of the Boltzmann collision term and photon phase space distribution in and out of the moving frame.
We use this alternative approach to explicitly confirm our expressions.
In particular, we show that all velocity-dependent contributions to the collision term, which were discussed in second order perturbation theory thus far \citep[e.g.][]{Hu1994pert, Bartolo2007}, can be derived just from the simple result for Thomson scattering, without requiring any additional integrals or Taylor expansions of the scattering cross section and distribution functions.

To develop a coherent picture we first summarize the
basic equations used in this derivation although some of these can be found in
earlier publications \citep[e.g., see][]{Hu1994pert, Bartolo2007}.
However, here we obtain a generalization of the Kompaneets equation, finding that the monopole through octupole are affected by the thermal motions of electrons.
We also derive additional terms that reflect the correct gauge-dependence of the collision term in second order perturbation theory.

\subsection{Transforming of the electron distribution function}
One of the important ingredients for the computation of the Boltzmann collision term is the distribution function for the electron momenta.
In the rest frame of the moving volume element the electrons have a thermal phase space density that is
isotropic and may be described by a relativistic Maxwell-Boltzmann distribution,
\beq\label{eq:relMBD}
f_{\rm c}(t_{\rm c}, \vecxc, \vp_{\rm c})
=\frac{\Ne^{\rm c}\, e^{-\epsilon_{\rm c}(\vp_{\rm c})/\theta^{\rm c}_{\rm e}}}{4\pi (\me c)^3 K_2(1/\theta^{\rm c}_{\rm e})\,\theta^{\rm c}_{\rm e}},
\eeq
where $K_2(1/\theta^{\rm c}_{\rm e})$ is the modified Bessel function of second kind
with $\theta^{\rm c}_{\rm e}=\kB\Te^{\rm c}/\me c^2$, where $\Te^{\rm c}$ is the electron temperature defined in the rest frame of the moving element. Furthermore, $\Ne^{\rm c}$ is the electron number density  and
$\epsilon_{\rm c}(\vp_{\rm c})=\sqrt{1+\eta(\vp_{\rm c})^2}$ denotes the
dimensionless energy of the electrons with 
$\eta(\vp_{\rm c})=|\vp_{\rm c}|/\me c=\gamma \beta$. 
Notice that $\Te^{\rm c}$ and $\Ne^{\rm c}$ depend explicitly on the four-vector $x^\mu_{\rm c}=(t_{\rm c}, \vecxc)$ in the rest frame of the moving volume element, while the energy $\epsilon_{\rm c}$ only depends on the modulus of the momentum vector $\vp_{\rm c}$.

If we assume that the volume element is moving with velocity $\vbetac$ relative to the CMB, then because the phase space density is Lorentz invariant we have
$f(x^\mu, \vp)\equiv f_{\rm c}(x^\mu_{\rm c}(x^\mu), \vp_{\rm c}(\vp))$ in the Newtonian frame, where $x^\mu_{\rm c}(x^\mu)$ and $\vp_{\rm c}(\vp)$ are determined by Lorentz transformations.
In addition, one has to express $\Ne^{\rm c}$ and $\Te^{\rm c}$ by variables that are defined in the Newtonian frame. Here one can think of $\Ne^{\rm c}$ and $\Te^{\rm c}$ as parameters which describe the electron distribution function and only depend on $x^\mu_{\rm c}$ but not on $\vp_{\rm c}$.
The transformation of these parameters is non-trivial, in particular for the electron temperature (see Sect.~\ref{sec:Te_evolution_sec}). 
Since in the rest frame of the moving volume element the momentum-dependence of $f_{\rm
c}$ only enters through the energy of the electron, 
it follows that $f(x^\mu, \vp)\approx f_{\rm c}(x^\mu_{\rm c}(x^\mu), \epsilon(\vp))$, 
with energy $\epsilon(\vp)=\gammac\left[\epsilon_{\rm c}(\vp)-\vbc\cdot\vek{\eta}(\vp)
\right]$, where $\gammac=(1-\betacsq)^{-1/2}$ denotes the gamma factor related to the bulk velocity.
We note that $\betac$ is a function of position $\vecx$ and $t$ but we suppress this dependence here.

\subsubsection{Expressing $\Ne^{\rm c}$ and $\Te^{\rm c}$ in the Newtonian frame}
\label{sec:Te_evolution_sec}
Let us assume that in the Newtonian frame we can directly give $\Ne(t, \vecx)$, for example by $\Ne(t, \vecx)\propto N_{\rm b}$, where $N_{\rm b}$ is the baryon density which is calculated using cosmological perturbation theory. 
If we integrate $f(t, \vecx, \vp)$ over $\id^3 p$ it is straightforward to show that $\Ne^{\rm c}(t_{\rm c}, \vecxc)\equiv\Ne(t, \vecx)/\gamma_{\rm p}$.
As this relation indicates, for $\Ne$ bulk motion only matters at second order in $\betac$.

In the next step we express $\Te^{\rm c}$ using variables that we have at our disposal in the Newtonian frame.
Using the energy-momentum tensor of the electron gas it is easy to show that the energy density of the electrons in the Newtonian frame is \citep[cf. also][]{Juettner1911}
%
\beq
\label{eq:energy_dens}
\rho_{\rm e}=\gamma^2_{\rm p}\left[ \rho^{\rm c}_{\rm e} + \betacsq P^{\rm c}_{\rm e}\right]
\approx \gamma_{\rm p}\Ne \me c^2\left[ 1+\left(\frac{3}{2}+ \betacsq \right)\theta^{\rm c}_{\rm e}\right].
\eeq
This expression in principle can be used to determine $\theta^{\rm c}_{\rm e}$, however, in the standard perturbation calculations $\rho_{\rm e}$ is usually not computed explicitly, and since in the Newtonian frame
it is no longer possible to simply write $\rho_{\rm e}=\Ne \me c^2\left[ 1+\frac{3}{2}\theta_{\rm e}\right]$ in all perturbation orders, Eq.~\eqref{eq:energy_dens} does not help much.
However, we see that only to second order in $\betac$ do differences arise to the case $\betac=0$.
We therefore only need to include the effect of motion on $\theta^{\rm c}_{\rm e}$ in second order perturbation theory, but are allowed to set $\theta^{\rm c}_{\rm e}\approx \theta_{\rm e}$ in zeroth and first order perturbation theory.

In zeroth order perturbation theory the temperature of the electrons and baryons in the Newtonian frame is affected by (i) adiabatic cooling because of the universal expansion, (ii) possible heating of the matter by global sources of energy release, and (iii) the Compton interaction with the (distorted) background photon field.
The evolution of $\The^{(0)}\equiv\The^{\rm c, (0)}$ in this case can be explicitly computed, depending on the global thermal history of the Universe \citep[e.g., see][]{Chluba2011therm}.
At high redshifts it is clear that the matter temperature always is very close to the Compton equilibrium temperature, $\Te^{\rm C, eq}$, while at $z\lesssim 800$ photons and baryons start decoupling, allowing the electron temperature to differ from $\Te^{\rm C, eq}$.

In higher perturbation orders it is sufficient to assume that the matter temperature always reaches Compton equilibrium with the local (distorted) radiation field.
This is because once this assumption is no longer justified the dominant correction to the electron temperature is already captured by the zeroth order solution, $\The^{(0)}$. 
We therefore can use the quasi-stationary approximation to determine both $\The^{\rm c, (1)}$ and $\The^{\rm c, (2)}$ 

For situations without extra energy release one locally has $\theta^{\rm c, (1)}_{\rm e}\approx \Thez \Theta^{(1)}_{0}$ for electrons inside a moving volume element, while with spatially varying energy release distortions of the local radiation field also matter (see Sect.~\ref{sec:first_order}).
In second order perturbation theory it is also straightforward to give an expression for $\theta^{\rm c, (2)}_{\rm e}$ (see Sect.~\ref{sec:second_order}).
For convenience we shall just set $\theta^{\rm c}_{\rm e}\equiv \theta_{\rm e}$ in the discussion below, bearing in mind that the local electron temperature in the moving frame is required, which in second order perturbation theory does make a difference $\simeq \betacsq$.

With these comments in mind the final distribution function for the electrons in the Newtonian frame is given by
\beq\label{eq:relMBD_CMB}
f(x^\mu, \vp)
=\frac{\Ne\, e^{-\epsilon(\vp)/\theta_{\rm e}}}{4\pi\gammac (\me c)^3 K_2(1/\theta_{\rm e})\,\theta_{\rm e}},
\eeq
with $\epsilon(\vp)=\gammac\left[\epsilon_{\rm c}(\vp)-\vbc\cdot\vek{\eta}(\vp)\right]$. We emphasize again that to second order in $\betac$ the relation for the electron temperature $\theta_{\rm e}$ depends on the chosen reference frame, but its value can be readily determined in the different perturbation orders.
This expression is consistent with the one used in \citet{Sazonov1998}. In \citet{Nozawa1998SZ} the factor $1/\gammac$ is absent, since there the electron density measured in the cluster frame is used throughout.

\subsection{General form of the Boltzmann collision term for Compton scattering}
\label{sec:GenDef}
To describe the time evolution of the photon phase space density $n(t, \nu, \vecx, \vgh)$
at frequency $\nu$ in the direction $\vgh$ and spatial coordinate $\vecJ{x}$ under Compton scattering we need  the Boltzmann collision term
\beal
\label{eq:BoltzEq}
\left.\pAb{n(t, \nu, \vecx, \vgh)}{t}\right|_{\rm scatt}= c\int \frac{\id\sigma}{\id \Omega'}\,  \mathcal{F} \id^2 \hat{\gamma}' \id^3 p 
{,}
\end{align}
where $\id^2 \hat{\gamma}'$ is the solid angle of the incoming photon and
$\id\sigma/\id \Omega'$ denotes the differential Compton scattering cross section,
\beal
\label{eq:dsigdO}
\frac{\id\sigma}{\id \Omega'}=\frac{3\,\sigT}{16\pi}\,\left[\frac{\nu'}{\nu}\right]^2
\frac{X}{\gamma^2(1-\beta\mu)}.
\end{align}
Here $\gamma=1/\sqrt{1-\beta^2}$ is the usual Lorentz factor, $\beta$ is the
dimensionless velocity of the incoming electron, $\mu=\vbh\cdot\vgh$ is the
cosine of the angle between the incoming electron and photon, $\sigT$ is the
Thompson cross section and the squared matrix element reads \citep[e.g., see][]{Jauch1976}
\beal
X&=\frac{\kappa}{\kappa'}+\frac{\kappa'}{\kappa}
+2\left[\frac{1}{\kappa}-\frac{1}{\kappa'}\right]
+\left[\frac{1}{\kappa}-\frac{1}{\kappa'}\right]^2,
\end{align}
where for convenience the abbreviations 
\bsub
\beal
\kappa&= \tilde{\nu}\,\gamma (1-\beta\mu)
\\[1mm]
\kappa'&= \tilde{\nu}'\gamma (1-\beta\mu')
\\[1mm]
\frac{\nu'}{\nu}&=\frac{1-\beta\mu}{1-\beta\mu'+\frac{\tilde{\nu}}{\gamma}(1-\mus)}
\end{align}
\esub
were introduced. Furthermore, $\tilde{\nu}=\frac{h\nu}{\me c^2}$ and
$\tilde{\nu'}=\frac{h\nu'}{\me c^2}$ are the dimensionless energies of the
incoming and outgoing photon respectively, $\mu'=\vbh\cdot\vghp$ is the
cosine of the angle between the incoming electron and outgoing photon and
$\mus=\vgh\cdot\vghp$ is the cosine of the scattering angle between the
incoming and outgoing photon.

We assume that the temperature of the electron gas obeys
$\kB\Te\ll\me c^2$. Therefore Fermi-blocking is negligible and the statistical factor
$\mathcal{F}$ may be written as
\beq\label{eq:StatFact}
\mathcal{F}=f'\,n' (1+n)-f\,n (1+n')
{,}
\eeq
where the abbreviations $f=f(\vp, \vecx),\;f'=f(\vpp, \vecx)$ for the electron and
$n=n(\nu, \vecx,\vgh),\;n'=n(\nu', \vecx,\vghp)$ for the photon phase space
densities were introduced.

With Eq.~\eqref{eq:relMBD_CMB} the statistical factor can be rewritten in the form
\beq\label{eq:rewStatFact}
\mathcal{F}=f\left[e^{\Deltaye}\,n' (1+n)-n (1+n')\right]
{,}
\eeq
where we defined $\xe=\frac{h\nu}{\kB\Te}$, $\Deltaxe=\xe'-\xe$ and
\beal
\label{eq:dy}
\Deltaye&=\gammac\Deltaxe(1-\vbc\cdot\vghp)
+\gammac\,\xe\,\vbc\cdot(\vgh-\vghp).
\end{align}
The next step is to rewrite the collision term to include the effect of 
energy transfer in different orders.

\subsubsection{Rewriting the collision term for small energy transfer}
Physically we are interested in the scattering terms up to second order in energy transfer in the rest frame of the moving volume element ($\betac=0$). This means we are looking for terms $\mathcal{O}(1)$, $\mathcal{O}(\The)$, and $\mathcal{O}(\nut)$, describing Thomson scattering (no energy transfer), Doppler boosting ($\Delta \nu/\nu \propto \The$) and broadening ($\Delta \nu/\nu \propto \sqrt{\The}$), as well as first order recoil and stimulated scattering terms ($\Delta \nu/\nu \propto \nut$).
These are the physical effects that are required to obtain the Kompaneets equation for an isotropic incident radiation field.
In particular, these terms ensure that a blackbody distribution is conserved in full thermodynamic equilibrium.
We neglect higher order relativistic corrections which are $\propto \The^2$, $\propto \The\nut$, $\propto \nut^2$ and higher, in the rest frame of the moving volume element, as in terms of the dimensionless frequency $\xe$ they are all at least $\mathcal{O}(\The^2)$.

In the Newtonian frame we are seeking to be consistent up to second order in $\betac$. 
This implies that we need to consider the scattering process up to fourth order in the total electron momentum $\eta=|\vp|/\me c$, since the average of $\eta^4$ over the relativistic Maxwell Boltzmann distribution can give rise to terms $\propto \betac^2 \The$. These terms are required to obtain a consistent formulation for the gauge-dependence of the collision integral in second order perturbation theory with second order energy transfer.

To simplify the expression for the collision integral we perform a Taylor series expansion of $\mathcal{F}$ in terms of $\Deltaxe\ll \xe$ and $\Deltaye\ll \xe$. 
For the normal Kompaneets equation in the rest frame of the moving volume element this means that one has to go up to second order in $\Deltaxe$.
However, in the Newtonian frame even up to fourth order derivative terms contribute at second order in $\betac$.
We therefore have to keep all terms $\propto \Deltaxe^s\Deltaye^t$ with $s+t\leq4$. 
The statistical factor $\mathcal{F}$ then takes the form
\beal
\label{eq:Fappr}
\mathcal{F}/f(\vp)&\approx n(\xe,\vghp)-n(\xe,\vgh) 
+ \sum_{s=1}^4 \frac{\Deltaxe^s}{s!}\,\partial^s_{\xe} n(\xe,\vghp)
\nonumber\\
&\qquad
+[1+n(\xe,\vgh)]\,
\sum_{s=0}^4 \sum_{t=1}^{4-s} \frac{\Deltaxe^s}{s!}\,\frac{\Deltaye^t}{t!}\,\partial^s_{\xe} n(\xe,\vghp).
\end{align}
We now define the moments of the energy shifts
\beal
\label{eq:Ilmk}
I_{lm}^{st}= \frac{x^{-(s+t)}_{\rm e}}{\Ne\,\sigT}\int \frac{\id\sigma}{\id \Omega'}
\, f(\vp)\,\frac{\Delta x_{\rm e}^s}{s!}\,\frac{\Delta y_{\rm e}^t}{t!}\,Y_{lm}(\vghp) \,\id^2 \hat{\gamma}' \id^3 p 
\end{align}
over the scattering cross section. 
Since $Y_{l(-m)}(\vghp)=(-1)^m\,Y^\ast_{lm}(\vghp)$, it directly follows 
$I_{l(-m)}^{st}\equiv (-1)^m\,(I_{lm}^{st})^\ast$, so that only the integrals 
with $m\geq 0$ ever have to be explicitly computed.

If we write $n(\xe,\vghp)$ as a spherical harmonic expansion with spherical harmonic coefficients $n_{lm}(\xe)$, then from Eq.~\eqref{eq:Fappr} it is clear that we in principle need all moments $I_{lm}^{st}$ with $s+t\leq 4$. 
It is convenient to compute all those moment with $t=0$ separately, but introduce 
\beal
\label{eq:Klmk}
K_{lm}^{s}= \sum_{t=1}^{4-s} \xe^{t}\,I_{lm}^{st}
\end{align}
in addition, and evaluate the corresponding integrals in one go. 
Then the Boltzmann collision term takes the form
\beal
\label{eq:BoltzEqlm_lm}
\left.\pAb{n(\xe,\vgh)}{\tau}\right|_{\rm scatt}
&\approx
%
\sum^\infty_{l=0}\sum^{l}_{m=-l} I_{lm}^{00}\,n_{lm}(\xe)
-\sqrt{4\pi}\,I^{00}_{00}\,n(\xe,\vgh)
\nonumber\\
&\quad
+ [1+n(\xe,\vgh)] 
\sum_{s=0}^{4}\sum^\infty_{l=0}\sum^{l}_{m=-l} 
K_{lm}^{s} \, \xe^s \partial^s_{\xe} n_{lm}(\xe)
\nonumber\\
&\qquad
+ \sum_{s=1}^{4}\sum^\infty_{l=0}\sum^{l}_{m=-l} 
I_{lm}^{s0} \, \xe^s \partial^s_{\xe} n_{lm}(\xe),
\end{align}
with the total number of required moments reduced from 15 to 9 per multipole.
Notice that for simplicity we have suppressed the explicit dependence of $n$ on $\vecx$ and $t$.

To make further progress we now write down the Boltzmann collision term in different orders. 
To compute the moments $I_{lm}^{st}$ and $K_{lm}^{s}$ we turn the $z$-axis parallel to $\vbetac$. 
The moments have a slightly simpler form when the $z$-axis is parallel to $\vgh$, but for Fourier transforms this choice of coordinates is less convenient.
Furthermore, it is useful to first integrate over ${\rm d}^3 p$ and then take care of the $\id \Omega'$ part.
The required derivations were performed with the symbolic algebra package {\sc Mathematica}.

\subsection{Scattering by resting electrons ($\The=\betac = 0$) without recoil}
\label{sec:zero_The_betac_limit}
For the scattering of photons by resting electrons ($\The=\betac=0$) and neglecting recoil ($\nut=0$) the only non-vanishing moments are $I_{00}^{00}=Y_{00}$ and $I_{2m}^{00}=\frac{1}{10}Y_{2m}$, reflecting the symmetry (monopole and quadrupole scattering) of the Thomson cross section.
In this case the collision term reads
\beal
\label{eq:BoltzEq_lim_0}
\left.\pAb{n(\xe,\vgh)}{\tau}\right|_{\rm r}&\approx n_{0}(\xe) + \frac{1}{10} n_2(\xe,\vgh) - n(\xe,\vgh),
\end{align}
being the well known lowest order scattering term of the Boltzmann hierarchy, however, where no expansion of the photon distribution around the global monopole was performed so far.
Here and below we frequently use the abbreviation $f_{l}=\sum_{m=-l}^{m=l} f_{lm} Y_{lm}(\vgh)$, where $f$ can be $n$ or $\Theta$, but also $\Theta^2$ and $\Delta n_{\rm s}$. Furthermore, $f_{lm}\equiv [f]_{lm}$ denotes the projection of $f$ onto $Y_{lm}$.

\subsubsection{Perturbed collision term in $\mathcal{O}(1)$}
Inserting the expansions, Eq.~\eqref{eq:def_nlm_perturbed}, for $n$ and collecting terms in different perturbation orders, from Eq.~\eqref{eq:BoltzEq_lim_0} we have
\bsub
\label{eq:BoltzEq_lim_0_t}
\beal
\label{eq:BoltzEq_lim_0_t_a}
\left.\pAb{n^{(1)}(\xe,\vgh)}{\tau}\right|_{\rm r}
&\!\!\! \approx\mathcal{G}\left[ \Theta^{(1)}_{0} + \frac{1}{10}  \Theta^{(1)} _{2}- \Theta^{(1)} \right]
+\Delta n_{{\rm s}, 0}^{(1)} + \frac{1}{10} \Delta n_{{\rm s}, 2}^{(1)} - \Delta n_{\rm s}^{(1)}
\\
\label{eq:BoltzEq_lim_0_t_b}
\left.\pAb{n^{(2)}(\xe,\vgh)}{\tau}\right|_{\rm r}
&\!\!\!\approx \mathcal{G}\left[ \Theta^{(2)}_{0} + \frac{1}{10}  \Theta^{(2)} _{2} - \Theta^{(2)} \right]
+  \frac{N^{(1)}_{\rm e}}{N^{(0)}_{\rm e}} \! \left.\pAb{n^{(1)}(\xe,\vgh)}{\tau}\right|_{\rm r}
\nonumber\\
&\qquad+\mathcal{Y}_{\rm t}\left[ [(\Theta^{(1)})^2]_{0} + \frac{1}{10} [(\Theta^{(1)})^2]_{2} - (\Theta^{(1)})^2\right]
\nonumber\\
&\qquad\qquad+\Delta n_{{\rm s}, 0}^{(2)} + \frac{1}{10} \Delta n_{{\rm s}, 2}^{(2)} - \Delta n_{\rm s}^{(2)}.
\end{align}
\esub
In zeroth order perturbation theory $\partial_\tau n^{(0)}(\xe,\vgh) |_{\rm r}$ vanishes, so that no term $\propto N^{(2)}_{\rm e}$ appears. 
Furthermore, as the last equation shows, in second perturbation order a $y$-type spectral distortion from the first order temperature terms is generated for $l>0$.
In addition, possible spectral distortions that are present in first order perturbation theory because of global energy release are perturbed by first order variations in electron density.

\subsubsection{Angle averaged collision term in $\mathcal{O}(1)$ and energy exchange with baryonic matter}
Averaging the expressions Eq.~\eqref{eq:BoltzEq_lim_0_t} over all photon directions, it is easy to see that in the Thomson limit the collision term cancel identically, both for the temperature fluctuations as well as the spectral distortions, i.e., $\left.\partial_\tau n^{(1)}_0(\xe)\right|_{\rm r}=\left.\partial_\tau n^{(2)}_0(\xe)\right|_{\rm r}=0$.
This is expected from the fact that no energy is transferred by Thomson scattering events, but that photons are only redistributed in different directions. 
This also implies that for the global energetics of the scattering event Eq.~\eqref{eq:BoltzEq_lim_0_t} does not contribute.

\subsection{Generalized Kompaneets term}
\label{sec:zero_betac_limit}
The Kompaneets equation accounts for the effect of thermal electrons on the local monopole part of the photon distribution. 
However, here we obtain the generalized result, which also includes the effect of scattering on the dipole, quadrupole, and octupole part of the spectral distortion.
The relevant terms of the moments are
\beal
\label{eq:I_lm_The}
{I}^{00}_{lm}&= \left[\The\,\beta_l-2\tilde{\nu}\,\alpha_l\right]Y_{lm},
& K_{lm}^{0}&= 4\alpha_l \The Y_{lm},
\nonumber\\
I_{lm}^{10}&= \alpha_l\left[4\The -\nut \right] Y_{lm},
&K_{lm}^{1}&= 2\alpha_l \The Y_{lm},
\nonumber\\
I_{lm}^{20}&= \alpha_l \,\The Y_{lm}
\end{align}
with $\alpha_0=1$, $\alpha_1=-2/5$, $\alpha_2=1/10$, $\alpha_3=-3/70$, $\beta_0=0$, $\beta_1=\alpha_1$, $\beta_2=-6\alpha_2$, $\beta_3=-4\alpha_3$, and $\alpha_l=\beta_l=0$ for $l>3$.
We only kept terms $\propto \The$ and $\propto \nut$ for the above moments, while terms with $\betac\neq 0$ are considered below.

Defining $\bar{n}(\xe, \vgh)=\sum^3_{l=0} \alpha_l \, n_{l}(\xe, \vgh)$ we can introduce the flux
\beal
\label{eq:Flux_e}
\bar{F}(\xe, \vgh)&=\partial_{\xe} \bar{n}(\xe,\vgh)+\bar{n}(\xe,\vgh)[1+\bar{n}(\xe,\vgh)].
\end{align}
In the case of the normal Kompaneets equation for a fully isotropic medium and photon field one has $\partial_\tau n(\xe)=\The x^{-2}_{\rm e} \partial_{\xe} x^{4}_{\rm e}\,F(\xe)$, with $F=\partial_{\xe} n_0(\xe,\vgh)+n_0(\xe,\vgh)[1+n_0(\xe,\vgh)]$. 
One can bring the temperature-dependent terms of the Boltzmann equation for an anisotropic incident photon field into a similar form. Using $4x+x^2\,\partial_{x} \equiv x^{-2}\,\partial_{x} x^4$ and collecting terms we find
\beal
\label{eq:BoltzEqlm_The}
\left.\pAb{n(\xe,\vgh)}{\tau}\right|_{\The}&\approx
\The\sum^3_{l=1}\beta_{l}\,n_{l}(\xe)
+\frac{\The}{x_{\rm e}^{2}} \,\partial_{\xe} x_{\rm e}^4 \bar{F}(\xe, \vgh)
\\
&\;\;\;
+2\The \left[n(\xe,\vgh)- \bar{n}(\xe,\vgh) \right]\left[x_{\rm e}+\partial_{\xe} \xe^2 \bar{n}(\xe,\vgh) \right].
\nonumber
\end{align}
Here the first term accounts for temperature corrections to the total Compton scattering cross section of the dipole, quadrupole and octupole. For the monopole no such term exists, so that in zeroth order perturbation it does not appear.
In first order perturbation theory it leads to a correction to the evolution equation of temperature perturbations, but no additional spectral distortion.
Since at $z\lesssim \pot{2}{7}$ it is much smaller than one percent of the photon temperature perturbations\footnote{Close to the maximum of the Thomson visibility function it would introduce a correction $\simeq \pot{5}{-7}$ relative to the normal temperature perturbations in first order.}, in the end this term can be neglected for temperature perturbations. 
In second order perturbation theory it does lead to contributions $\propto \The [\Theta^2]_{lm} \mathcal{Y}_{\rm SZ}$, which we include for consistency, although for the average distortion this does not matter.

The second term in Eq.~\eqref{eq:BoltzEqlm_The} describes the diffusive scattering of photons by thermal electrons, however, with the inclusion of anisotropies in the photon distribution. If the photon distribution is isotropic one only has to worry about the monopole, resulting in the Kompaneets term for Compton scattering. 
If anisotropies exist, the monopole through octupole distortions matter in $\mathcal{O}(\The)$. 
Including higher order temperature correction one would also start to see the effect of photon diffusion for multipoles with $l>3$, however, the efficiency of photon diffusion drops strongly when compared to the scattering of the monopole.

The last term in Eq.~\eqref{eq:BoltzEqlm_The}  is related to the recoil effect with stimulated scattering and the first order Klein-Nishina correction ($\tilde{\nu}\simeq h\nu/\me c^2$) to the moments $I^{00}_{lm}$. 
At high frequencies ($\xe\gg 1$) stimulated scattering becomes negligible, since $\partial_{\xe} \xe^2 \bar{n}(\xe,\vgh)=\xe[2\bar{n} +\xe\partial_{\xe} \bar{n}]\approx \xe [2\nPl - \mathcal{G}]$ decays exponentially, while normal recoil remains significant.
It is also important to mention that the last term in Eq.~\eqref{eq:BoltzEqlm_The} vanishes for the monopole part of the distortion, however, even for the higher multipoles part of this term is present.
This means that at high redshifts the recoil effect in principle can alter the spectrum of anisotropic contributions of the radiation field. However, when no anisotropies with $l>0$ are present, the last term vanishes identically and only the  Kompaneets term for the monopole has to be considered.
At high redshifts one only expects contributions from the local monopole, dipole and quadrupole to matter, while, once higher multipoles appear in the Universe, $\The$ becomes so small that this term can be omitted.

\subsubsection{Perturbed collision term in $\mathcal{O}(\The)$}
We can now again write down the terms $\propto\The$ in different perturbation orders. 
With $\bar{n}^{(0)}\equiv n^{(0)}_{0}$ we have
\bsub
\label{eq:BoltzEqlm_The_pert}
\beal
\label{eq:BoltzEqlm_The_pert_a}
\left.\pAb{n^{(0)}(\xg,\vgh)}{\tau}\right|_{\The}&\approx
\frac{\The^{(0)}}{x_{\gamma}^{2}} \,\partial_{\xg} x_{\gamma}^4 
\left\{ \partial_{\xg} n^{(0)}_{0}+ \frac{\Tz}{\Te^{(0)}}\,n^{(0)}_{0} [1+n^{(0)}_{0} ]\right\}
\\[2mm]
\label{eq:BoltzEqlm_The_pert_b}
\left.\pAb{n^{(1)}(\xg,\vgh)}{\tau}\right|_{\The}&\approx
\frac{\The^{(0)}}{x_{\gamma}^{2}} \,\partial_{\xg} x_{\gamma}^4 
\left\{ \partial_{\xg} \bar{n}^{(1)} \!+\! \frac{\Tz}{\Te^{(0)}}  \bar{n}^{(1)} [ 1 + 2 n^{(0)}_{0} ]\right\}
\\
&\!\!\!\!\!\!\!\!\!\!\!\!\!\!\!\!\!\!\!\!\!\!\!\!
+\frac{N^{(1)}_{\rm e}}{N^{(0)}_{\rm e}} \left.\pAb{n^{(0)}(\xg,\vgh)}{\tau}\right|_{\The}
+ \frac{\The^{(1)}}{x_{\gamma}^{2}} \,\partial_{\xg} x_{\gamma}^4  \partial_{\xg} n^{(0)}_{0}
+\The^{(0)}\sum^3_{l=1}\beta_{l}\,n^{(1)}_{l}(\xe)
\nonumber\\
&\qquad
\nonumber
+2 \Thz x_\gamma\left[n^{(1)} - \bar{n}^{(1)}   \right]  \left[1+ 2 n^{(0)}_{0} +\xg \partial_{\xg} n^{(0)}_{0}\right]
\end{align}
\esub
for the zeroth and first order equations.
Here we transformed to the frequency $\xg=h\nu/k\Tz$.
In zeroth order one simply has the normal Kompaneets term for the local monopole, which in first order perturbation theory becomes a function of position.

To write down the terms in second order perturbation theory we resort to additional approximations.
In this work we have considered problems with global energy release, so that both $n^{(0)}$ and $n^{(1)}$ could be distorted, but $n^{(2)}$ is not of interest, being a small correction. In this situation Eq.~\eqref{eq:BoltzEqlm_The_pert} is enough to describe the problem.
On the other hand, when thinking about the dissipation of acoustic waves in the early Universe we can assume that $n^{(0)}$ and $n^{(1)}$ have no (significant) distortion (i.e., $n^{(0)}_{0}\approx \nPl$ and $n^{(1)}\approx \mathcal{G}\,\Theta^{(1)}$) and only $n^{(2)}$ has to be explicitly computed.
Therefore, the simplified second order collision term is given by 
\bsub
\label{eq:BoltzEqlm_The_pert_2}
\beal
\label{eq:BoltzEqlm_The_pert_2_a}
\left.\pAb{n^{(2)}(\xg,\vgh)}{\tau}\right|_{\The}&\approx
\frac{\The^{(0)}}{x_{\gamma}^{2}} \,\partial_{\xg} x_{\gamma}^4 
\left\{ \partial_{\xg} \bar{n}^{(2)} + \frac{\Tz}{\Te^{(0)}} \bar{n}^{(2)} \mathcal{A}\right\}
\\
&\quad
+2 \Thz \xg n^{(1)} [2\bar{n}^{(1)}+\xg\partial_{\xg}\bar{n}^{(1)}]
+2 \Thz \left[n^{(2)} - \bar{n}^{(2)}   \right] \mathcal{B} 
\nonumber\\
&
\qquad+ \The^{(1)} \mathcal{D}_{\xg} \bar{n}^{(1)}
+ \The^{(2)} \mathcal{Y}_{\rm SZ} 
+\frac{N^{(1)}_{\rm e}}{N^{(0)}_{\rm e}} \left.\pAb{n^{(1)}(\xg,\vgh)}{\tau}\right|^{\rm 0}_{\The}
\nonumber\\
&\qquad\quad
+ \The^{(1)} \sum^3_{l=1} \beta_{l}\,n^{(1)}_{l}
+ \The^{(0)} \sum^3_{l=1} \beta_{l}\,n^{(2)}_{l},
\nonumber
\end{align}
where $\mathcal{A}=1+2\nPl$ and $\mathcal{B}=x_\gamma[\mathcal{A} -\mathcal{G}]$ 
and we introduced the approximate first order term
\beal
\label{eq:BoltzEqlm_The_pert_2_b}
\left.\pAb{n^{(1)}(\xg,\vgh)}{\tau}\right|^{0}_{\The}
&\approx
\The^{(0)} \left[\Theta^{(1)}_0-\bar{\Theta}^{(1)}\right]\,\mathcal{Y}_{\rm SZ}+
\The^{(0)}\sum^3_{l=1}\beta_{l}\,\Theta^{(1)}_{l} \mathcal{G} 
\nonumber\\
&\quad
+2 \Thz \left[\Theta^{(1)} - \bar{\Theta}^{(1)}   \right] \mathcal{G} \mathcal{B}.
\end{align}
\esub
Here we used $\The^{(1)}\approx \The^{(0)}\Theta^{(1)}_0$
and $\partial_{\xg}\mathcal{G}+\mathcal{G}\mathcal{A}=-\partial_{\xg}\nPl$.
With the above assumptions, again no term $\propto \Ne^{(2)}$ appears.
The second and third terms in Eq.~\eqref{eq:BoltzEqlm_The_pert_2_a} are related to the recoil and stimulated recoil effect, which are mainly present at high redshifts.
The fifth term is only important if in second order perturbation theory temperature differences between the electrons and photon are present, sourcing $y$-type distortions of the local monopole.
The last two terms are caused by temperature corrections to the Compton cross section which only affect the dipole, quadrupole and octupole.

\subsubsection{Angle averaged collision term in $\mathcal{O}(\The)$}
If in addition we average Eq.~\eqref{eq:BoltzEqlm_The_pert_2} over all directions $\vgh$, by setting $\The^{(1)}\approx \The^{(0)}\Theta^{(1)}_0$, $\The^{(0)}\approx \Thz$, and $n^{(1)}\approx \Theta^{(1)} \, \mathcal{G}$ we obtain
\beal
\label{eq:BoltzEqlm_The_pert_2_ang_av}
\left.\pAb{n^{(2)}_0(\xg)}{\tau}\right|_{\The}&
\!\!\!\approx
\frac{\Thz}{x_{\gamma}^{2}} \,\partial_{\xg} x_{\gamma}^4 
\left\{ \partial_{\xg} n^{(2)}_0 + n^{(2)}_0 \mathcal{A}\right\}
+ \Thz |\Theta^{(1)}_0|^2 \mathcal{D}_{\xg} \mathcal{G}
+ \The^{(2)} \mathcal{Y}_{\rm SZ} 
\nonumber\\
&\qquad\quad
-2 \Thz \xg \mathcal{G} [\mathcal{G}+\mathcal{Y}_{\rm SZ}] 
\sum_{l=0}^3  \sum_{m=-l}^{l} \alpha_l\frac{|\Theta^{(1)}_{lm}|^2}{4\pi}.
\end{align}
%
This expression is important to describe the evolution of the average spectral distortion created from the dissipation of acoustic modes in the Universe.
It is good to mention that the integral of this expression over $\xg^2\id\xg$ vanishes, since the total number of electrons is conserved in each scattering event.
However, this part of the collision term does lead to some net energy exchange between electrons and photons, as we discuss in more detail in Sect.~\eqref{sec:energy_ex_2_order}.

If we in addition insert the second order expansion of the photon distribution function into Eq.~\eqref{eq:BoltzEqlm_The_pert_2_ang_av} we find
\beal
\label{eq:BoltzEqlm_The_pert_2_ang_av_exp}
\left.\pAb{n^{(2)}_0(\xg)}{\tau}\right|_{\The}&
\!\!\!\approx
\frac{\Thz}{x_{\gamma}^{2}} \,\partial_{\xg} x_{\gamma}^4 
\left\{ \partial_{\xg} \Delta n^{(2)}_0 + \Delta  n^{(2)}_0 \mathcal{A}\right\}
\\
&\quad+
\left[\The^{(2)}-\Thz\left(\Theta^{(2)}_0+[(\Theta^{(1)})^2]_0-|\Theta^{(1)}_0|^2\right) \right] \mathcal{Y}_{\rm SZ} 
\nonumber\\[1mm]
&\qquad
- \Thz\mathcal{H} 
\sum_{l=1}^\infty  \sum_{m=-l}^{l} \frac{|\Theta^{(1)}_{lm}|^2}{4\pi}
-\Thz \mathcal{E}
\sum_{l=1}^3  \sum_{m=-l}^{l} \alpha_l\frac{|\Theta^{(1)}_{lm}|^2}{4\pi},
\nonumber
\end{align}
where $\mathcal{H}$ and $\mathcal{E}$ are defined by Eq.~\eqref{eq:def_HE_func}, and we used the identity $\mathcal{Y}_{\rm SZ}\equiv \mathcal{D}_{\xg} \mathcal{G} - \mathcal{H} - \mathcal{E}$.
The first term describes the effect of scattering on the monopole of the spectral distortion, $\Delta n_0^{(2)}$, while the term $\propto \The^{(2)}$ is important for the energy exchange between photons and baryonic matter.
The other terms are all caused by the scattering of distortions that are sourced by temperature anisotropies in first order perturbation theory.
These terms all matter for the energy exchange between photons and electrons, leading to additional sources of distortions, however, those $\propto \mathcal{Y}_{\rm SZ}$ that are present in the above expression vanish once $\The^{(2)}$ is computed correctly.
We return to this point in more detail in Sec.~\ref{sec:energy_ex_2_order}.
%

\subsection{First order in $\betac$}
\label{sec:betac_1}
To first order in $\betac$ the only relevant terms in the moments are
\bsub
\label{eq:I_lm_betac}
\beal
{I}^{00}_{00}&= -\betac c_{10} Y_{10},
&
{I}^{00}_{1m}&= 2 \betac c_{1m} Y_{00} - \frac{\betac}{10} c_{2m} Y_{2m},
\nonumber\\[1mm]
{I}^{00}_{2m}&= -\frac{2}{5}\, \betac c_{2m} Y_{1m} + \frac{\betac}{10} c_{3m} Y_{3m},
&
{I}^{00}_{3m}&= \frac{2}{5}\,\betac c_{3m} Y_{2m},
\\[2mm]
{I}^{10}_{00}&= -\betac c_{10} Y_{10},
&
{I}^{10}_{1m}&= \betac c_{1m} Y_{00} + \frac{\betac}{10} c_{2m} Y_{2m},
\nonumber\\[1mm]
{I}^{10}_{2m}&= -\frac{\betac}{10} c_{2m} Y_{1m} - \frac{\betac}{10} c_{3m} Y_{3m},
&
{I}^{10}_{3m}&= \frac{\betac}{10} c_{3m} Y_{2m},
\end{align}
\esub
were we have defined $c_{l,m}=c_{l,-m}=\sqrt{(l^2-m^2)/(4l^2-1)}$ and only kept terms $\propto \betac$.
All other moments vanish. 
Notice also that $c_{l,l}\equiv 0$.

Collecting terms and projecting onto the different spherical harmonics with respect to $\vgh$ we can write
\beal
\label{eq:BoltzEqlm_beta_1}
\nonumber
\left.\pAb{n_{00}(\xg)}{\tau}\right|_{\betac}
&\!\!\approx
\betac c_{10}
\left[3 n_{10} + \xg \partial_{\xg} n_{10} \right]
\\[1mm]\nonumber
\left.\pAb{n_{1m}(\xg)}{\tau}\right|_{\betac}
&\!\!\approx
- \betac c_{1m} \xg \partial_{\xg} n_{00}
+\frac{\betac c_{20}}{10} 
\left[ 6 n_{2m} -\xg \partial_{\xg} n_{2m} \right]
\\[1mm]\nonumber
\left.\pAb{n_{2m}(\xg)}{\tau}\right|_{\betac}
&\!\!\approx
\frac{\betac c_{2m}}{10}
\left[ 9 n_{1m} + \xg \partial_{\xg} n_{1m} \right]
\!+\!\frac{\betac c_{3m}}{10} 
\!\left[ 14 n_{3m} + \xg \partial_{\xg} n_{3m} \right]
\\[1mm]\nonumber
\left.\pAb{n_{3m}(\xg)}{\tau}\right|_{\betac}
&\!\!\approx
\frac{\betac c_{3m}}{10}  
\left[ 11 n_{2m} - \xg \partial_{\xg} n_{2m} \right]
+ \betac c_{4m}  n_{4m}
\\[1mm]
\left.\pAb{n_{lm}(\xg)}{\tau}\right|_{\betac}
&\!\!\approx
\left[ \vbetac\!\cdot\vgh \,\, n(\vgh) \right]_{lm}
\!\!=
\betac \left[ c_{l, m}  n_{l-1 m}+c_{l+1, m}  n_{l+1 m}\right].
\;\;
\end{align}
The last equation is valid for all $l\geq 4$. 
Equations~\eqref{eq:BoltzEqlm_beta_1} describe the general scattering terms $\propto \betac$. In particular we have not yet made the transition to different perturbations.
However, in first order perturbation theory only temperature perturbations that are azimuthally symmetric around the direction $\vbetach$ are created. Since $\betac$ is already first order in perturbation theory, one can therefore neglect all $m\neq0$ terms. 
In this limit we can directly compare our result with the terms giving by \citet{Bartolo2007}.
There $f^{(1)}$ is used instead of $n^{(1)}$. Furthermore, in their notations one has $n^{(1)}_{l0}\equiv f^{(1)}_l/ \sqrt{4\pi(2l+1)}$.
If we take Eq. (5.34) from \citet{Bartolo2007} and perform the corresponding replacements, by projection onto $Y_{l0}(\vgh)$ we confirm that our expressions for $m=0$ and all $l$ are completely equivalent to theirs.

An alternative way of writing the expressions in Eq.~\eqref{eq:BoltzEqlm_beta_1} is
\bsub
\label{eq:BoltzEqlm_beta_1_alt}
\beal
\left.\pAb{n(\xg, \vgh)}{\tau}\right|_{\betac}
&\approx
\frac{\betac}{3} \left[\frac{5}{2}  + \xg \partial_{\xg}  \right] \sum_{m=-1}^{1}  Y_{1m} (\vbetach) \, n_{1m}
\nonumber\\
&\!\!\!\!\!\!\!\!\! +\vbetac\cdot\vgh \left\{
n-\left[1+\xg \partial_{\xg}\right] n_0-\frac{1}{2}n_1 + \frac{1}{10}\left[1-\xg \partial_{\xg}\right] n_2 
\right\}
\nonumber\\
&\!\!\!\!\!\!\!\!\!\!\!\!\!
-\sqrt{\frac{2\pi}{3}}\,\betac
Y^\ast_{1m'}(\vgh)Y^\ast_{1m''}(\vbetach)
\left(\!\!
\begin{array}{ccc}
2  & 1 & 1\\
m & m' & m''
\end{array}
\!\!\right)
n_{2m}
%
\nonumber\\
&\!\!\!\!\!\!\!\!\!\!\!\!\!\!\!\!\!
+\frac{1}{5}\sqrt{\frac{2\pi}{3}}\,\betac
Y^\ast_{2m'}(\vgh)Y^\ast_{1m''}(\vbetach)
\left(\!\!
\begin{array}{ccc}
1  & 2 & 1\\
m & m' & m''
\end{array}
\!\!\right)
\left[4+\xg\partial_{\xg}\right]
n_{1m}
%
\nonumber\\
&\!\!\!\!\!\!\!\!\!\!\!\!\!\!\!\!\!\!\!\!
-\frac{1}{5}\sqrt{\frac{3\pi}{7}}\,\betac
Y^\ast_{2m'}(\vgh)Y^\ast_{1m''}(\vbetach)
\left(\!\!
\begin{array}{ccc}
3  & 2 & 1\\
m & m' & m''
\end{array}
\!\!\right)
\left[4+\xg\partial_{\xg}\right]
n_{3m},
\nonumber
\end{align}
\esub
where in the last two terms sums over $m$, $m'$, and $m''$ have to be taken. Here also the 
direction of the velocity vector is still left independent of the coordinate system, so that it is 
valid in any system.

\subsubsection{Derivation using explicit Lorentz transformations}
It is also possible to directly derive the terms $\propto\betac$ using explicit Lorentz transformations of the collision term, Eq.~\eqref{eq:BoltzEq_lim_0}.
First we have to transform the photon distribution from the Newtonian frame into the rest frame of the moving volume element. 
This can be done using the relation, Eq.~\eqref{eq:n_Lorentz_lm}. After applying the collision operator we transform back into the Newtonian frame. 
After all these transformations the term $\propto n$ is back to the starting point, however, the collision integral has to be multiplied by\footnote{We have $\id t=\gamma_{\rm p} (1+\betac\mu') \id t' \equiv [\gamma_{\rm p} (1-\betac\mu)]^{-1} \id t'$ and a factor of $1/\gamma_{\rm p}$ from the transformation of $\Ne$.} $\gamma^{-2}_{\rm p}\,(1+\betac\mu')^{-1}\equiv 1-\betac\mu$ to account for the transformation of $\partial_{\tau'}\rightarrow \partial_\tau$.
This leads to a term $\propto -\betac \muc n(\xe, \muc)$.
For the same reason one has the additional two terms $-\betac\muc (n_0 + n_2/10)$.
In addition one has those terms $\propto \betac$ 
caused by the Lorentz transformation of $n'_0$ and $n'_2$ back into the Newtonian frame.
In total this leads to 
\beal
\label{eq:BoltzEqlm_beta_1_Lorentz}
\left.\pAb{n(\xg, \vgh)}{\tau}\right|_{\betac}
&\approx
-\betac\muc \left( n_0(\xg)+\frac{1}{10} n_{2}(\xg, \vgh)-n(\xg, \vgh)\right) 
\\
&\!\!\!\!\!\!\!\!\!\!\!\!
+\betac c_{10}\left[ 2 + \xg \partial_{\xg} \right] n_{10} Y_{00}
-\frac{\betac}{10}\,c_{2m}\left[ 1 - \xg \partial_{\xg} \right] n_{1m} Y_{2m}
\nonumber\\
&\!\!\!\!\!\!\!\!\!\!
+\frac{\betac}{10}\,c_{3m}\left[ 4 + \xg \partial_{\xg} \right] n_{3m} Y_{2m}
-\betac c_{10} \xg \partial_{\xg} n_{00} Y_{10}
\nonumber\\
&\!\!\!\!\!\!\!\!
-\frac{\betac}{10} c_{2m}\left[ 3 + \xg \partial_{\xg} \right] n_{2m} Y_{1m}
+\frac{\betac}{10} c_{3m}\left[ 2 - \xg \partial_{\xg} \right] n_{2m} Y_{3m}.
\nonumber
\end{align}
The first term is from the transformation of $\partial_{\tau'}$. The next three terms arise from the Lorentz transformation of $n$ from the Newtonian frame into the moving frame. 
The remaining terms are from the back transformation of $n_0 + n_2/10$ into the Newtonian frame.
If we perform the projection of the first term onto the different spherical harmonics, it is straightforward to confirm the equivalence of this expression with Eq.~\eqref{eq:BoltzEqlm_beta_1}.
No additional integrals had to be taken and everything followed from the collision term in the Thomson limit using Lorentz transformations of the photon field.

\subsubsection{Perturbed contributions in $\mathcal{O}(\betac)$}
We now write down the results for different perturbation orders.
In lowest order perturbation theory the contribution $\propto \betac$ vanishes. 
In first order we only have
\beal
\label{eq:BoltzEqlm_beta_1_pert}
\left.\pAb{n^{(1)}(\xg, \vgh)}{\tau}\right|_{\betac}
&\!\approx\!
- \frac{\betac^{(1)}}{\sqrt{3}}  \xg \partial_{\xg} n^{(0)}_{00} Y_{10}(\vgh)
\equiv - \betac^{(1)}  \muc\xg \partial_{\xg} n^{(0)}_{0} 
\end{align}
with $\muc=\vbetach\cdot \vgh$. This is the well known expression which allows one to account for the Doppler effect.
If in zeroth order a distortion is created by global energy release, then this term also modulates the distortions. Otherwise, only temperature perturbation are produced, since $-\xg \partial_{\xg} n^{(0)}_{0}\equiv \mathcal{G}(\xg)$.
Also, one can pair this term with the dipole term in Eq.~\eqref{eq:BoltzEq_lim_0} by introducing $n^{\rm g}_1=n_1+\betac \muc\xg \partial_{\xg} n_{0}$.
The next few sections show that {\it every} dipole term actually appears in this way when accounting for the correct gauge-dependence of the collision term.

To give the second order perturbation result we need to insert $n^{(1)}_{lm}$ into the expressions Eq.~\eqref{eq:BoltzEqlm_beta_1}, with $\beta^{(1)}$.
In addition we have to add  $N^{(1)}_{\rm e}\partial_\tau n^{(1)}(\xg, \vgh) |_{\betac} / N^{(0)}_{\rm e}$ to account for the effect of electron density perturbations, as well as $\betac^{(2)}\partial_\tau n^{(1)}(\xg, \vgh) |_{\betac} / \betac ^{(1)}$.
However, the latter two contributions only source temperature perturbations and therefore can be omitted in the end.

Assuming that up to first order perturbation no spectral distortions are present, the second order terms $\propto \betac$ only depend on $\mathcal{G}\,\Theta_{lm}$ and its first derivatives with respect to $\xg$, which source $y$-type distortions for $l<4$ and $m=0$.
Using $-\xg \partial_{\xg} \mathcal{G}\equiv \mathcal{Y}_{\rm SZ}+3\mathcal{G}$ and $\betac^{(1)} \Theta^{(1)}_{lm}\approx \delta_{m0}\,\betac^{(1)} \Theta^{(1)}_{l0}$ from Eq.~\eqref{eq:BoltzEqlm_beta_1} 
we find
\bsub
\label{eq:BoltzEqlm_beta_1_spect}
\beal
\label{eq:BoltzEqlm_beta_pert_2_ang_av}
\left.\pAb{n^{(2)}_{00}(\xg)}{\tau}\right|_{\betac}
&\approx
-\frac{\betac^{(1)}}{\sqrt{3}}\Theta^{(1)}_{10} \,\mathcal{Y}_{\rm SZ}
\\[0mm]
\left.\pAb{n^{(2)}_{10}(\xg)}{\tau}\right|_{\betac}
&\approx
\frac{3\betac^{(1)}}{\sqrt{3}}\,\Theta^{(1)}_{00}\,\mathcal{G}
+\!\frac{9\betac^{(1)}}{5\sqrt{15}} \,\Theta^{(1)}_{20}\,\mathcal{G}
+\left[
\frac{N^{(1)}_{\rm e}}{N^{(0)}_{\rm e}}\frac{\betac^{(1)}}{\sqrt{3}}
+\frac{\betac^{(2)}}{\sqrt{3}}
\right]
\mathcal{G}
\nonumber\\
&\qquad
+\frac{\betac^{(1)}}{\sqrt{3}} \Theta^{(1)}_{00}\,\mathcal{Y}_{\rm SZ}
+\frac{\betac^{(1)}}{5\sqrt{15}} \,\Theta^{(1)}_{20}\,\mathcal{Y}_{\rm SZ}
\\[0mm]
\left.\pAb{n^{(2)}_{20}(\xg)}{\tau}\right|_{\betac}
&\approx
\frac{6\betac^{(1)}}{5\sqrt{15}} \Theta^{(1)}_{10}\,\mathcal{G}
+\frac{33\betac^{(1)}}{10\sqrt{35}}\Theta^{(1)}_{30} \,\mathcal{G}
\nonumber\\
&\qquad
-\frac{\betac^{(1)}}{5\sqrt{15}} \Theta^{(1)}_{10}\,\mathcal{Y}_{\rm SZ}
-\frac{3\betac^{(1)}}{10\sqrt{35}} \Theta^{(1)}_{30}\,\mathcal{Y}_{\rm SZ}
\\[0mm]
\left.\pAb{n^{(2)}_{30}(\xg)}{\tau}\right|_{\betac}
&\approx
\frac{21\betac^{(1)}}{5\sqrt{35}}\Theta^{(1)}_{20} \,\mathcal{G}
+\frac{4\betac^{(1)}}{3\sqrt{7}}\Theta^{(1)}_{40} \,\mathcal{G}
\nonumber\\
&\qquad
+\frac{3\betac^{(1)}}{10\sqrt{35}} \Theta^{(1)}_{20}\,\mathcal{Y}_{\rm SZ}
\\[0mm]
\left.\pAb{n^{(2)}_{l0}(\xg)}{\tau}\right|_{\betac}
&\approx
\betac^{(1)} \left[ c_{l,0}\Theta^{(1)}_{l-1, 0} +c_{l+1,0}\Theta^{(1)}_{l+1, 0} \right]\mathcal{G}.
\end{align}
\esub
For the problem of spectral distortions only the terms $\propto \mathcal{Y}_{\rm SZ}$ matter, while the other terms are only important for second order corrections to the temperature perturbations.

\subsection{Second order in $\betac$}
\label{sec:betac_2}
Similar to the case of first order in $\betac$, at second order
  in $\betac$ too we have some dependence on the azimuthal angles. 
However, the expressions are rather lengthy and for our problem we need not consider terms 
$\propto \betacsq \Theta$ or higher, so that only the spatially average monopole term matters.
Therefore, in terms of CMB spectral distortions in $\mathcal{O}(\betac^2)$ the only non-negligible moments are 
\bsub
\label{eq:BoltzEqlm_beta2_moments}
\beal
I^{10}_{00}&=\frac{4}{3}\betacsq Y_{00}+\frac{2}{3\sqrt{5}}\betacsq Y_{20}
\\
I^{20}_{00}&=\frac{1}{3}\betacsq Y_{00}+\frac{11}{30\sqrt{5}}\betacsq Y_{20}.
\end{align}
\esub
With this we find
\bsub
\label{eq:BoltzEqlm_beta_2}
\beal
\label{eq:BoltzEqlm_beta2_pert_2_ang_av}
\left.\pAb{n^{(2)}_{00}(\xg)}{\tau}\right|_{\betacsq}
& \approx
\frac{\betacsq}{3} \mathcal{D}_{\xg} n^{(0)}_{00}
\\[2mm]
\left.\pAb{n^{(2)}_{20}(\xg)}{\tau}\right|_{\betacsq}
&\approx
-\frac{4\betacsq}{5\sqrt{5}}  \xg \partial_{\xg} n^{(0)}_{00}
+ \frac{11\betacsq}{30\sqrt{5}}  \mathcal{D}_{\xg} n^{(0)}_{00}.
\end{align}
\esub
%
All other terms vanish up to second order perturbation.
Note that here $\betacsq$ is evaluated in first order perturbation theory.

An alternative way of writing this result is
\bsub
\beal
\left.\pAb{n^{(2)}(\xg, \vgh)}{\tau}\right|_{\betacsq}
& \approx
\betacsq\left\{ 
\left[1+\mu_{\rm p}^2\right] \xg \partial_{\xg} n^{(0)}_{0} 
+ \left[\frac{3}{20}+ \frac{11}{20} \mu_{\rm p}^2\right] x_\gamma^2 \partial^2_{\xg} n^{(0)}_{0} 
\right\}.
\nonumber
\end{align}
\esub
This expression was also given by \citet[][compare their Eq.~(5.35)]{Bartolo2007} and earlier by \citet[][see their Eq.~(14)]{Hu1994pert}.
Again assuming that no distortions are appearing up to first order perturbation theory one has 
$\mathcal{D}_{\xg} n^{(0)}_{0} \approx \mathcal{Y}_{\rm SZ}$ and $- \xg \partial_{\xg} n^{(0)}_{0} \approx \mathcal{G}$.
For our purpose again only the terms $\propto\mathcal{Y}_{\rm SZ}$ matter.

\subsubsection{Derivation using explicit Lorentz transformations}
Like for the terms $\propto \betac$ one can obtain those $\propto \betacsq$ by simple Lorentz transformation of Eq.~\eqref{eq:BoltzEq_lim_0}.
We only care about the monopole in the Newtonian frame and neglect higher multipoles.
As the first step we transform the monopole (assuming that there are no other anisotropies) into the moving frame and apply the scattering operator. 
This yields
\beal
\pAb{n'(\xg', \vgh')}{\tau'}
& \approx
-\frac{9\betacsq}{20} \,d_{20}\,\left[\mathcal{D}_{\xg'} - 4\xg' \partial_{\xg'}\right]n_{00}(\xe') Y_{20}(\vgh')
\nonumber\\
&\qquad\qquad-\betac c_{10} \xg' \partial_{\xg'} n_{00}(\xg')Y_{10}(\vgh').
\end{align}
%
The first term is already second order in $\betac$, so it directly carries over to the Newtonian frame without further modification.
From the second we have the contribution because of the change $\partial_{\tau'} \rightarrow \partial_{\tau}$ plus the Lorentz transformation of this term back into the Newtonian frame.
The derivative $\xg' \partial_{\xg'}$ is Lorentz invariant and just carries over.
If we therefore simply set $n'_{10}=-\betac c_{10} \xg' \partial_{\xg'} n_{00}(\xg')$ we directly have the back transform using relations Eq.~\eqref{eq:n_Lorentz_examples}, with $\betac\rightarrow-\betac$.
Collecting terms $\propto \betacsq$ we find
\beal
\label{eq:step_one}
\left.\pAb{n^{(2)}(\xg, \vgh)}{\tau}\right|_{\betacsq}
& \approx
-\frac{9\betacsq}{20} \,d_{20}\,\left[\mathcal{D}_{\xg} - 4\xg \partial_{\xg}\right]n_{00}(\xg) Y_{20}(\vgh)
\\
&\quad+\betacsq c_{10} \,\muc \xg \partial_{\xg} n_{00}(\xg)Y_{10}(\vgh)
\nonumber\\
&\qquad+\betacsq c^2_{10} [2+\xg \partial_{\xg}] [\xg \partial_{\xg}n_{00}(\xg) ] Y_{00}
\nonumber\\
&\qquad\quad-\betacsq d_{20} [1-\xg \partial_{\xg}] [\xg \partial_{\xg} n_{00}(\xg) ] Y_{20}(\vgh).
\nonumber
\end{align}
Since $\xg \partial_{\xg} \xg \partial_{\xg}  = \xg \partial_{\xg} + \xg^2 \partial^2_{\xg} = \mathcal{D}_{\xg} - 3\xg \partial_{\xg}$, we directly have $[2+\xg \partial_{\xg}] \xg \partial_{\xg} = \mathcal{D}_{\xg} - \xg \partial_{\xg}$ and $-[1-\xg \partial_{\xg}] \xg \partial_{\xg} = \mathcal{D}_{\xg} - 4 \xg \partial_{\xg}$.

From the second term in Eq.~\eqref{eq:step_one} we obtain the contributions $\betacsq [c^2_{10} Y_{00}+d_{20} Y_{00} ] \xg \partial_{\xg} n_{00}(\xg)$ after projection. Here $\betacsq c^2_{10} Y_{00}\xg \partial_{\xg} n_{00}(\xg)$ cancels the drift part from the third term.
Therefore the remaining term for the monopole is $\frac{1}{3}\betacsq \mathcal{D}_{\xg}  n_{00}$.
Finally, if we add the first and last terms of Eq.~\eqref{eq:step_one} we find $\frac{11}{30\sqrt{5} \betacsq} \,\left[\mathcal{D}_{\xg} - 4\xg \partial_{\xg}\right]n_{00}(\xg) Y_{20}(\vgh)$. 
Together with $\betacsq d_{20} Y_{00} \xg \partial_{\xg} n_{00}(\xg)$ from the second term this eventually gives $\betacsq\left[\frac{11}{30\sqrt{5}}\mathcal{D}_{\xg} - \frac{4}{5 \sqrt{5}}\xg \partial_{\xg}\right]n_{00}(\xg) Y_{20}(\vgh)$, which confirms Eq.~\eqref{eq:BoltzEqlm_beta_2}.
We emphasize again, that no additional integrals had to be taken and everything followed from the collision term in the Thomson limit using Lorentz transformations of the photon field.

\subsection{Contributions to the collision term $\propto \betac\The$ and $\propto \betac\nut$}
\label{sec:betac_gauge_first}
So far we have only considered the velocity-dependent terms that arise from the Thomson scattering part of the collision integral, i.e. those $\propto \betac$ and $\propto \betacsq$.
However, also from the Kompaneets part additional gauge-dependent terms arise.
Usually these can be ignored, however, for the thermalization problem at high redshifts it is important to understand the net source of distortions in addition to the evolution of the distortions under Compton scattering.
If we are interested in the higher order terms $\propto \betac\The$ and $\propto \betac\nut$ it is clear that in first order perturbation theory we only need the moments $I^{st}_{00}$ and $K^{s}_{00}$, since terms $\propto \betac \Theta$ are second order.
We find
\bsub
\label{eq:I_lm_betac_gauge}
\beal
{I}^{00}_{00}&= 4\betac \nut \,c_{10} Y_{10},
&
{I}^{10}_{00}&=-\betac \left[10\The -\frac{31}{5}\,\nut \right] c_{10}Y_{10},
\nonumber\\[1mm]
I^{20}_{00} &=-\betac\left[\frac{47}{5}\,\The -\frac{7}{5}\nut\right] c_{10}Y_{10},
&
I^{30}_{00} &=-\frac{7}{5} \betac\The \,c_{10}Y_{10},
\nonumber\\[2mm]
{K}^{0}_{00}&= -8 \betac \nut\, c_{10} Y_{10},
&
{K}^{1}_{00}&= -\frac{62}{5}\,\betac\nut \, c_{10}Y_{10},
\nonumber\\[1mm]
{K}^{2}_{00}&= -\frac{14}{5}\,\betac\nut \, c_{10}Y_{10},
&
I^{40}_{00}&= {K}^{3}_{00} = 0.
\end{align}
\esub
At this point it is easier to go the other way around and first derive the expression using Lorentz transformations, as the grouping of terms and their origin is easier to follow.

If in the Newtonian frame we only have a monopole, then in the rest frame of the moving volume element the photon occupation number has a motion-induced dipole anisotropy of first order in $\betac$, i.e. $n'(\xe') \approx n_0(\xe') + \xe' \partial_{\xe'} n_0(\xe') \,\betac \muc$. If we use Eq.~\eqref{eq:BoltzEqlm_The}, we know that in the moving frame the collision term reads
\beal
\label{eq:BoltzEqlm_The_moving}
\left.\pAb{n'(\xe',\vgh')}{\tau'}\right|_{\The}&\approx
\frac{\The}{{x'}_{\rm e}^{2}} \,\partial_{\xe'} {x'}_{\rm e}^{4} \left[\partial_{\xe'} n_0(\xe')+n_0(\xe')[1+n_0(\xe')]\right]
\nonumber\\
&\!\!\!\!\!\!\!\!\!\!\!\!\!\!\!\!\!\!\!\!\!\!\!\!\!\!
-\frac{2}{5}\The \delta n_1(\xe')
-\frac{2}{5} 
\frac{\The}{{x'}_{\rm e}^{2}} \,\partial_{\xe'} {x'}_{\rm e}^{4} \left[\partial_{\xe'} \delta n_1(\xe')+ \delta n_1(\xe')[1+2n_0(\xe')]\right]
\nonumber\\
&\;\;\;
+\frac{14}{5}\The \delta n_1(\xe')\,\left[\xe'+\partial_{\xe'} {\xe'}^2 n_0(\xe') \right],
\end{align}
where $\delta n_1(\xe')=\betac\muc\,\xe' \partial_{\xe'} n_0(\xe')$.
Those terms $\propto \delta n_1(\xe')$ are already first order in $\betac$ so that they directly carry over to the Newtonian frame.
Then we only have to transform the first term of Eq.~\eqref{eq:BoltzEqlm_The_moving} back into the Newtonian frame.
One contribution is again from the transition $\partial_{\tau' } \rightarrow \partial_\tau$ and a second arises from the back transformation $n_0(\xe') \rightarrow n_0(\xe) - \betac\muc\,\xe \partial_{\xe} n_0(\xe)$.
The operator $\mathcal{D}_{\xg}$ is Lorentz invariant, however the term $\propto n_0(\xe')[1+n_0(\xe')]$ gives another contribution $-\betac \muc\, n_0(\xe)[1+n_0(\xe)]$, because the operator ${x'}_{\rm e}^{-2} \,\partial_{\xe'} {x'}_{\rm e}^{4}$ transforms like $\xe'=\gamma_{\rm p}\xe(1-\betac\muc)$.
This yields 
\beal
%
\left.\pAb{n(\xe,\vgh)}{\tau}\right|_{\betac \The}&\approx
- \betac \muc 
\frac{\The}{x_{\rm e}^{2}} \,\partial_{\xe} x_{\rm e}^{4} \left[\partial_{\xe} n_0 +n_0 (1+n_0 )\right]
\nonumber\\
&\!\!\!\!\!\!\!\!\!\!\!\!\!\!\!\!\!\!\!\!\!\!\!\!\!\!
-
\frac{\The}{x_{\rm e}^{2}} \,\partial_{\xe} x_{\rm e}^{4} \left[\partial_{\xe} \delta n_1 + \delta n_1 (1+ 2n_0 )\right]
%
-\betac \muc
\frac{\The}{x_{\rm e}^{2}} \,\partial_{\xe} x_{\rm e}^{4} n_0 (1+ n_0 )
\nonumber\\
&\!\!\!\!\!\!\!\!\!\!\!\!\!\!\!\!\!\!\!\!
-\frac{2}{5}\The \delta n_1
-\frac{2}{5} 
\frac{\The}{{x}_{\rm e}^{2}} \,\partial_{\xe} {x}_{\rm e}^{4} \left[\partial_{\xe} \delta n_1+ \delta n_1(1+2n_0)\right]
\nonumber\\
&\;\;\;
+\frac{14}{5}\The \delta n_1\,\left[\xe+\partial_{\xe} \xe^2 n_0 \right].
\nonumber
\end{align}
Grouping terms and transforming to $\xg$ we finally have
\beal
\label{eq:BoltzEqlm_Thebeta_fin}
\left.\pAb{n^{(1)}(\xg,\vgh)}{\tau}\right|_{\betac \The}&\approx
- \betac^{(1)} \muc 
\frac{\The^{(0)} }{\xg^{2}} \,\partial_{\xg} \xg^{4} 
\left[\partial_{\xg} n^{(0)}_0 +\frac{\Tz}{\Te^{(0)}} n^{(0)}_0 (1+n^{(0)}_0 )\right]
\nonumber\\
&\!\!\!\!\!\!\!\!\!\!\!\!\!\!\!\!\!\!\!\!\!\!\!\!\!\!\!\!\!
-\betac^{(1)} \muc
\frac{\Thz}{\xg^{2}} \,\partial_{\xg} \xg^{4} n^{(0)}_0 (1+ n^{(0)}_0 )
-\betac^{(1)} \muc \frac{2}{5}\The^{(0)} \xg \partial_{\xg} n^{(0)}_0
\nonumber\\
&\!\!\!\!\!\!\!\!\!\!\!\!\!\!\!\!\!\!\!\!\!\!\!
-\frac{7}{5} \betac^{(1)}\muc
\frac{\The^{(0)}}{\xg^{2}} \,\partial_{\xg} \xg^{4} 
\left[\partial_{\xg} \xg \partial_{\xg} n^{(0)}_0+\frac{\Tz}{\Te^{(0)}} (1+2n^{(0)}_0) \xg \partial_{\xg} n^{(0)}_0\right]
\nonumber\\
&\;\;\;\!\!\!\!\!\!\!\!\!\!\!\!\!\!\!\!\!
+\frac{14}{5}\Thz\, \betac^{(1)} \muc  \,\left[\xg+\partial_{\xg} \xg^2 n^{(0)}_0 \right] \xg \partial_{\xg} n^{(0)}_0.
\end{align}
If we explicitly carry out all the derivatives, turning this expression into a sum over different orders of $\xg^k \partial^k_{\xg} n_0$, we can verify each of the  terms and moments given by Eq.~\eqref{eq:BoltzEqlm_lm} and \eqref{eq:I_lm_betac_gauge}.
As this result shows the first order velocity correction to the generalized Kompaneets equation only affect the dipolar part of the spectrum.
Again, by redefining the dipolar part of the spectrum in the Newtonian frame as 
$n^{\rm g}_1=n_1+\betac\muc \xg \partial_{\xg} n_0$ we can absorb most of these terms when pairing with the dipole terms of Eq.~\eqref{eq:BoltzEqlm_The_pert_b}.
However, additional contributions arising from the Lorentz transformation of the zeroth order collision term, Eq.~\eqref{eq:BoltzEqlm_The_pert_a}, back into the Newtonian frame remain.
These are basically the first and second term of Eq.~\eqref{eq:BoltzEqlm_Thebeta_fin}, as well as $5/7$ of the fourth term. 
%

\subsubsection{Partial confirmation using result for kinematic SZ effect}
We can also check part of the expression~\eqref{eq:BoltzEqlm_Thebeta_fin} with Eq.~(12) of \citet{Sazonov1998}. Assuming $\The\gg \Thz$ and $n_0=\nPl$, the conditions valid for the scattering of CMB photons by the hot electrons inside clusters of galaxies, we find
\beal
%
\left.\pAb{n(\xg,\vgh)}{\tau}\right|^{\rm SZ}_{\betac \The}&\approx
- \betac \muc \The \mathcal{Y}_{\rm SZ}
+\betac \muc \frac{2}{5}\The \mathcal{G}
+\frac{7}{5} \betac \muc \The \mathcal{D}_{\xg} \mathcal{G}
\nonumber\\
&=-\betac \muc \The 
\left\{
\frac{47}{5}\mathcal{Y}_{\rm SZ}
+\frac{138}{5}\mathcal{G}
-\frac{7}{5}\xg\mathcal{G}\left(6\mathcal{G}+\xg\right)  
\right\}.
\nonumber
\end{align}
In the second line we used the relations Eq.~\eqref{eq:diffusion_op_nPl}.
This result is precisely the term $\propto\betac \The$ that was given in Eq.~(12) of \citet{Sazonov1998}, after converting to their notation\footnote{The first two terms in our expression correspond to $10-\frac{47}{5}\,F$, while the last one is equivalent to $\frac{7}{10}(2F^2 + G^2)$ in their notation.}. This term was also obtained by \citet{Nozawa1998SZ}.
%
%

\subsubsection{Terms in second order perturbation theory}
In second order perturbation theory we also have to account for the anisotropies in the photon distribution. We are however only interested in the terms that appear for the average spectrum at each point.
For all terms of Eq.~\eqref{eq:BoltzEqlm_The} which are linear in $n$ it is clear that to first order in $\betac$ only those from the dipolar part of the spectrum can mix into the monopole.
On the other hand for the non-linear terms, $\propto n^2$, also the monopole matters when appearing together with dipole terms, since the final average of expressions $\simeq n_0 \muc n_1$ over photon directions 
gives rise to a monopole term.

With this argument, terms $\propto N^{(1)}_{\rm e}$ and $\The^{(1)}$ can be directly neglected, leading only to a dipolar term in second order perturbation theory.
From Eq.~\eqref{eq:n_Lorentz_examples} we know that Lorentz transformation of the photon occupation number into the moving frame gives $\delta n^{(2)}_{00}(\xe')=\betac^{(1)} c_{10} [2+\xe' \partial_{\xe'}] n^{(1)}_{10}(\xe')$ and $\delta n^{(1)}_{10}(\xe')=\betac^{(1)} c_{10} \xe' \partial_{\xe'} n^{(0)}_{00}(\xe')$. 
Assuming $n^{(0)}\approx \nPl$ and $n^{(1)}\approx\mathcal{G} \Theta^{(1)}$ we therefore have
\beal
\delta n^{(2)}_{0}
&=-\frac{\betac^{(1)}\Theta^{(1)}_{10}}{\sqrt{12\pi}}\left[\mathcal{G}+\mathcal{Y}_{\rm SZ} \right],
&
\delta n^{(1)}_{1}
&=- \betac^{(1)} \muc\, \mathcal{G}.
\nonumber
\end{align}
Inserting this into Eq.~\eqref{eq:BoltzEqlm_The}, and assuming $\The^{(0)}\approx \Thz$, 
we find the following terms that contribute to the evolution of the monopole
\beal
\label{eq:second_order_The_beta_part_I}
\left.\pAb{n^{(2)}_0(\xg)}{\tau}\right|^{\rm in}_{\betac\The}&\approx
\frac{\Thz}{x_{\gamma}^{2}} \,\partial_{\xg} x_{\gamma}^4 
\left\{ \partial_{\xg} \delta n^{(2)}_{0} 
\!+\!  \delta n^{(2)}_{0} \mathcal{A} \right\}
\nonumber\\
&\!\!\!\!\!\!\!\!\!\!\!\!\!\!\!\!\!\!\!\!\!\!\!\!\!
+\frac{8}{25}\Thz \mathcal{D}^\ast_{\xg} \left[ n^{(1)}_{1} \delta n^{(1)}_{1} \right]_0
-\frac{28}{25} \Thz x_\gamma \left[ n^{(1)}_1 \left( 2 \delta n^{(1)}_{1} +\xg \partial_{\xg} \delta n^{(1)}_{1} \right)\right]_0
\nonumber\\
&
-\frac{28}{25} \Thz x_\gamma \left[ \delta n^{(1)}_1 \left( 2 n^{(1)}_{1} +\xg \partial_{\xg} n^{(1)}_{1} \right)\right]_0
\nonumber
\\[2mm]
&=\frac{\betac^{(1)}\Theta^{(1)}_{10}}{\sqrt{12\pi}} \Thz\left[\mathcal{Y}_{\rm SZ}+2\mathcal{H}\right]
-\frac{8}{25}\,\frac{\betac^{(1)}\Theta^{(1)}_{10}}{\sqrt{12\pi}} 
\Thz\mathcal{D}^\ast_{\xg}\mathcal{G}^2
\nonumber\\
&\qquad\quad
-\frac{28}{25} \frac{\betac^{(1)}\Theta^{(1)}_{10}}{\sqrt{12\pi}} \Thz \mathcal{E},
\end{align}
where $\mathcal{H}$ and $\mathcal{E}$ are defined in Eq.~\eqref{eq:def_HE_func} and $\mathcal{D}^\ast_{\xg}=x_{\gamma}^{-2} \,\partial_{\xg} x_{\gamma}^4$.
This is not the final answer, as we still have to account for those contributions appearing from the transformation back into the Newtonian frame. The transformation\footnote{We just have to multiply Eq.~\eqref{eq:BoltzEqlm_The} by $-\betac \muc$ and average over all photon directions.} 
$\partial_{\tau'}\rightarrow \partial_ \tau$ results in
%
\beal
\left.\pAb{n^{(2)}_0(\xg)}{\tau}\right|^{\tau}_{\betac\The}&
\approx
\frac{2}{5} \frac{\betac^{(1)}\Theta^{(1)}_{10}}{\sqrt{12\pi}} \Thz \left[ \mathcal{G} - \mathcal{Y}_{\rm SZ} \right]
-\frac{14}{5} \frac{\betac^{(1)}\Theta^{(1)}_{10}}{\sqrt{12\pi}} \Thz \mathcal{G} \mathcal{B},
\nonumber
\end{align}
%
with $\mathcal{B}=\xg[\mathcal{A}-\mathcal{G}]$.
For the transformation of the occupation number the first term of Eq.~\eqref{eq:BoltzEqlm_The} leads to $\frac{2}{5}\Thz \delta n^{(2)}_{0}$.
The back transformation of the second term adds $2/5$ times the first and $5/2$ times the second term in Eq.~\eqref{eq:second_order_The_beta_part_I}, totalling the coefficients to $7/5$ and $-28/25$, respectively.
However, because the operator $\mathcal{D}^\ast_{\xg}$ transforms as $\xg$ one has to add another term 
$+\frac{2}{5\sqrt{12\pi}} \betac^{(1)}\Theta^{(1)}_{10} \Thz\mathcal{D}^\ast_{\xg}\mathcal{G}\mathcal{A}$.
Note also that $\mathcal{D}^\ast_{\xg}\mathcal{G}^2=-\mathcal{E}$ (cf. Eq.~\eqref{eq:def_HE_func}).
Finally, the last term of Eq.~\eqref{eq:BoltzEqlm_The} gives rise to $-\frac{14}{5\sqrt{12\pi}} \betac^{(1)}\Theta^{(1)}_{10} \Thz \mathcal{G} \mathcal{B}$ and $-\frac{7}{5\sqrt{12\pi}} \betac^{(1)}\Theta^{(1)}_{10} \Thz \mathcal{E}+\frac{14}{5\sqrt{12\pi}} \betac^{(1)}\Theta^{(1)}_{10} \Thz [\mathcal{G}+\mathcal{Y}_{\rm SZ}] \mathcal{B}$.

Adding all these contributions we eventually find
\beal
\label{eq:second_order_The_beta_result_fin}
\left.\pAb{n^{(2)}_0(\xg)}{\tau}\right|_{\betac\The}&
\approx
-\frac{2}{5}\frac{\betac^{(1)}\Theta^{(1)}_{10}}{\sqrt{12\pi}} \Thz
\left[\mathcal{D}_{\xg}\mathcal{G}+3\mathcal{Y}_{\rm SZ}\right]
-\frac{7}{5}\frac{\betac^{(1)}\Theta^{(1)}_{10}}{\sqrt{12\pi}} \Thz\,\mathcal{E}
\nonumber\\
&\!\!\!\!\!\!\!\!\!
+
\frac{7}{5}\frac{\betac^{(1)}\Theta^{(1)}_{10}}{\sqrt{12\pi}} \Thz
\left[\mathcal{Y}_{\rm SZ}+2\mathcal{Y}_{\rm SZ}\mathcal{B}+2\mathcal{H}\right]
-\frac{14}{5}\frac{\betac^{(1)}\Theta^{(1)}_{10}}{\sqrt{12\pi}} \Thz\mathcal{G}\mathcal{B}
\nonumber\\[2mm]
&=
\frac{\betac^{(1)}\Theta^{(1)}_{10}}{\sqrt{12\pi}} \Thz \left[\frac{27}{5}\mathcal{Y}_{\rm SZ} -\frac{2}{5}\mathcal{E} +\frac{26}{5}\mathcal{H}
\right].
\end{align}
To explicitly check this result with the Boltzmann collision term, Eq.~\eqref{eq:BoltzEqlm_lm}, one needs the monopole parts of the moments
\bsub
\label{eq:I_lm_betac1_gauge}
\beal
{I}^{00}_{10}
&=\frac{6}{5}\betac\,c_{10} \left[\The -6\,\nut \right] Y_{00}
+\frac{3}{5}\betac\,c_{20} \left[\The -\nut \right] Y_{20},
\nonumber\\[1mm]
{I}^{10}_{00}
&=2\betac\,c_{10} \left[\frac{47}{5}\The -\frac{39}{10}\,\nut \right] Y_{00}
+\betac\,c_{20} \left[\frac{7}{5}\The -\frac{9}{5}\nut \right] Y_{20},
\nonumber\\[1mm]
{I}^{20}_{00}
&=\betac\,c_{10} \left[\frac{58}{5}\The -\frac{7}{5}\,\nut \right] Y_{00}
+\frac{1}{2}\betac\,c_{20} \left[\frac{29}{5}\The -\nut \right] Y_{20},
\nonumber\\[1mm]
{I}^{30}_{00}
&=\frac{7}{5}c_{10}\, \betac\The Y_{00}
+\frac{1}{2} c_{20}\,\betac\The Y_{20},
\\[2mm]
{K}^{0}_{00}&= \frac{72}{5} c_{10} \betac \nut\, Y_{00}+\frac{6}{5} c_{20} \betac \nut\, Y_{20},
\nonumber\\[1mm]
{K}^{1}_{00}&= \frac{78}{5} c_{10} \betac \nut\, Y_{00}+\frac{18}{5} c_{20} \betac \nut\, Y_{20},
\nonumber\\[1mm]
{K}^{2}_{00}&= \frac{14}{5} c_{10} \betac \nut\, Y_{00}+ c_{20} \betac \nut\, Y_{20}.
\end{align}
\esub
Inserting these together with the expressions Eq.~\eqref{eq:I_lm_betac_gauge} into the Boltzmann equation, Eq.~\eqref{eq:BoltzEqlm_lm}, we can confirm the correctness of Eq.~\eqref{eq:second_order_The_beta_result_fin} by directly computing the final collision term and averaging over all photon directions.

\subsection{Contributions to the collision term $\propto \betac^2\The$ and $\propto \betac^2\nut$}
\label{sec:betac_gauge_second}
Also to second order in $\betac$ additional gauge-dependent terms arise from the Kompaneets part.
Since $\betacsq$ is already second order, we only need the moments
 $I^{st}_{00}$ and $K^{s}_{00}$:
\bsub
\label{eq:I_lm_betac2_gauge}
\beal
{I}^{00}_{00}&= -\frac{5}{3} \nut \betacsq Y_{00} - \frac{4}{3} \nut \betacsq \frac{Y_{20}}{\sqrt{5}},
\nonumber\\[1mm]
{I}^{10}_{00}&=\betacsq \left[8\The -\frac{47}{6}\,\nut \right] Y_{00}
+4\betacsq \left[\The -\frac{47}{30}\,\nut \right] \frac{Y_{20}}{\sqrt{5}},
\nonumber\\[1mm]
{I}^{20}_{00}&=\betacsq \left[16\The -\frac{21}{5}\,\nut \right] Y_{00}
+12 \betacsq \left[\The -\frac{11}{30}\,\nut \right]  \frac{Y_{20}}{\sqrt{5}},
\nonumber\\[1mm]
{I}^{30}_{00}&=\betacsq \left[\frac{28}{5}\The -\frac{7}{15}\,\nut \right] Y_{00}
+ \betacsq \left[6\The -\frac{19}{30}\,\nut \right]  \frac{Y_{20}}{\sqrt{5}},
\nonumber\\[1mm]
{I}^{40}_{00}&=\frac{7}{15}\betacsq \The Y_{00}
+ \frac{19}{30} \betacsq \The  \frac{Y_{20}}{\sqrt{5}},
\\[2mm]
{K}^{0}_{00}&= \frac{10}{3} \betacsq \nut\, Y_{00}+\frac{8}{3} \betacsq \nut\, \frac{Y_{20}}{\sqrt{5}},
\nonumber\\[1mm]
{K}^{1}_{00}&= \frac{47}{3} \betacsq \nut\, Y_{00}+\frac{188}{15} \betacsq \nut\, \frac{Y_{20}}{\sqrt{5}},
\nonumber\\[1mm]
{K}^{2}_{00}&= \frac{42}{5} \betacsq \nut\, Y_{00}+\frac{44}{5} \betacsq \nut\, \frac{Y_{20}}{\sqrt{5}},
\nonumber\\[1mm]
{K}^{3}_{00}&= \frac{14}{15} \betacsq \nut\, Y_{00}+\frac{19}{15} \betacsq \nut\, \frac{Y_{20}}{\sqrt{5}}.
\end{align}
\esub
Let us again go the other way around and first derive the expression using explicit Lorentz transformations.
Neglecting anisotropies with $l>0$, the transformation into the moving frame 
gives 
\bsub
\beal
\delta n'_{0}(\xg')
&\approx \frac{1}{6}\betacsq
\left[\mathcal{D}_{\xg'}-\xg'\partial_{\xg'} \right] n'_{0} \approx 
\frac{1}{6}\betacsq
\left[\mathcal{Y}_{\rm SZ}+\mathcal{G} \right]
\\
\delta n'_{1}(\xg')&\approx
\betac c_{10}\,\xg'\partial_{\xg'}n'_{00} Y_{10}(\muc')
= -\betac \muc' \mathcal{G}'.
\end{align}
\esub
We directly omitted the motion induced quadrupole, since we are only interested in the average spectrum.
For the full Boltzmann collision term in the moving frame we find
%
\beal
\label{eq:step_one_betasq}
\pAb{n'(\xg', \vgh')}{\tau'}
& \approx \The\mathcal{D}_{\xg'}n_0+ \Thz\mathcal{D}^\ast_{\xg'}n_0(1+n_0)
+\frac{4}{25}\Thz \mathcal{D}^\ast_{\xg'}\delta {n'_1}^2
\nonumber\\
&\!\!\!\!\!\!\!\!\!\!\!\!\!\!\!\!\!\!\!\!\!
+\The\mathcal{D}_{\xg'}[\delta n'_0-\frac{2}{5}\delta n'_1] 
+ \Thz \mathcal{D}^\ast_{\xg'}[\delta n'_0-\frac{2}{5}\delta n'_1]\mathcal{A}'
-\frac{2}{5} \The\delta n'_1
\nonumber\\
&+\frac{14}{5}\Thz\, \delta n'_1\mathcal{B}' 
-\frac{28}{25}\Thz\, \delta n'_1\xg'[2+\xg'\partial_{\xg'}]\delta n'_1.
\end{align}
These are all terms that either are already second order in $\betac$ or lead to additional terms upon transforming back to the Newtonian frame.
Those $\propto\delta n'_0$ and $\delta {n'_1}^2$ are already $\propto \betacsq$, so they directly carry over without leading to anything in addition. 
%
%
To obtain the final result in the Newtonian frame we furthermore have to transform $\partial_{\tau'} \rightarrow \partial_\tau$, which implies multiplication of the above expression by $1-\betac\muc$.
In addition one had to take care of $\mathcal{D}^\ast_{\xg'}\rightarrow \gamma_{\rm p}[1-\muc\betac]\mathcal{D}^\ast_{\xg}$, and the transforms of $n'_0$, $\mathcal{B}'$, $\mathcal{A}'$ and $\delta n'_1$ into the Newtonian frame.
After carrying out all these transformations we find
\bsub
\label{eq:betasq_gauge_fin}
\beal
\left.\pAb{n^{(2)}_0(\xg)}{\tau}\right|_{\betacsq\The}&
\!\!\!\!
\approx
\frac{29}{75}\betacsq\,\Thz \mathcal{D}^\ast_{\xg}\mathcal{G}^2
-\frac{2}{3} \betacsq \Thz\mathcal{H}-\frac{1}{3}\betacsq\Thz\mathcal{Y}_{\rm SZ}
+\frac{14}{75}\betacsq\,\Thz \mathcal{E}
\nonumber\\
\label{eq:betasq_gauge_fin_a}
&\!\!\!\!\!\!\!\!\!\!\!\!\!\!\!\!\!\!\!\!\!\!\!\!\!
+\frac{1}{6}\betacsq \Thz\left[\mathcal{Y}_{\rm SZ}+2\mathcal{H}+2\mathcal{E}\right]
+\frac{2}{15}\betacsq \Thz\left[3\mathcal{Y}_{\rm SZ}-\mathcal{H}-\mathcal{E}\right]
\\
&\!\!\!\!\!\!
+\frac{7}{15}\betacsq \Thz
\left[\mathcal{E} +2\left(\mathcal{G}-\mathcal{Y}_{\rm SZ} \right)\mathcal{B}
\right]
\nonumber\\
\label{eq:betasq_gauge_fin_b}
&= -\frac{49}{30}\betacsq \Thz\mathcal{Y}_{\rm SZ} -\frac{7}{5}\betacsq \Thz\mathcal{H}.
\end{align}
\esub
The first five terms in Eq.~\eqref{eq:betasq_gauge_fin_a} are those that directly carried over, plus those  coming from the first two terms in Eq.~\eqref{eq:step_one_betasq}. 
The rest arises because of the terms $\propto \delta n'_1$ in Eq.~\eqref{eq:step_one_betasq}.
We confirmed this result by inserting the moments Eq.~\eqref{eq:I_lm_betac2_gauge} into the Boltzmann equation, Eq.~\eqref{eq:BoltzEqlm_lm}, and averaging over all photon directions.

\subsubsection{Partial confirmation using result for kinematic SZ effect}
Again we can check part of our results using the expression given in connection with the 
SZ effect of clusters of galaxies.
Assuming $\The\gg \Thz$ and $n_0=\nPl$, in the moments Eq.~\eqref{eq:I_lm_betac2_gauge} we can drop all terms $\simeq \nut$.
Computing the terms explicitly and then comparing with the $\betacsq \The$ term of \citet{Nozawa1998SZ}, Eq.~(20) and Eq.~(24)--(25), we obtain their result plus one term $-\frac{1}{2}\betacsq \mathcal{Y}_{\rm SZ}$.
The reason for this residual is that here we explicitly transformed $\Ne$ into the CMB rest frame, while in \citet{Nozawa1998SZ} $\Ne$ is evaluated in the cluster frame.
After redefining $\Ne$ in this way our results again agree.

\subsection{Final expressions for the collision term}
\label{app:fin_exp_coll}

\subsubsection{Final expression for the first order thermalization term}
We now write down the final expression for the temperature-dependent parts of the collision term.
If we introduce
\beal
\label{eq:n_gauge_def}
n^{\rm g}_l=n^{(1)}_l + \delta_{l1}\,\betac^{(1)} \muc \xg \partial_{\xg} n^{(0)}_0 
= n^{(1)}_l - \delta_{l1}\,\betac^{(1)} \muc \left[ \mathcal{G} + \mathcal{T}\right],
\end{align}
with $\mathcal{T}=- \xg \partial_{\xg} \Delta n^{(0)}_0$, then one can absorb a large part of the velocity-dependent terms in Eq.~\eqref{eq:BoltzEqlm_Thebeta_fin} by adding Eq.~\eqref{eq:BoltzEqlm_The_pert_b}.
We find
\beal
\label{eq:BoltzEqlm_The_pert_first_final}
\left.\pAb{n^{(1)}(\xg,\vgh)}{\tau}\right|_{\The}&\approx
\frac{\The^{(0)}}{x_{\gamma}^{2}} \,\partial_{\xg} x_{\gamma}^4 
\left\{ \partial_{\xg} \bar{n}^{\rm g} \!+\! \frac{\Tz}{\Te^{(0)}}  \bar{n}^{\rm g} [ 1 + 2 n^{(0)}_{0} ]\right\}
\nonumber
\\
&\!\!\!\!\!\!\!\!\!\!\!\!\!\!\!\!\!\!\!\!\!\!\!\!
+\left[\frac{N^{(1)}_{\rm e}}{N^{(0)}_{\rm e}}- \betac^{(1)} \muc \right]
 \left.\pAb{n^{(0)}(\xg,\vgh)}{\tau}\right|_{\The}
+\The^{(0)}\sum^3_{l=1}\beta_{l}\,n^{\rm g}_{l}(\xe)
\nonumber\\
&\!\!\!\!\!\!\!\!\!\!\!\!\!\!\!\!\!\!\!\!\!
+ \The^{(1)}\mathcal{D}_{\xg} n^{(0)}_{0}
-\betac^{(1)} \muc \Thz \,\mathcal{D}^\ast_{\xg} n^{(0)}_0 (1+ n^{(0)}_0 )
\nonumber\\
&\!\!\!\!\!\!\!\!\!\!\!\!\!\!\!\!\!\!\!
-\betac^{(1)}\muc
\frac{\The^{(0)}}{\xg^{2}} \,\partial_{\xg} \xg^{4} 
\left\{
\partial_{\xg} \xg \partial_{\xg} n^{(0)}_0+\frac{\Tz}{\Te^{(0)}} (1+2n^{(0)}_0) \xg \partial_{\xg} n^{(0)}_0
\right\}
\nonumber\\
&\!\!\!\!\!\!\!\!\!\!\!\!\!\!\!
+2 \Thz x_\gamma\left[n^{\rm g} - \bar{n}^{\rm g}   \right]  
\left[1+ 2 n^{(0)}_{0} +\xg \partial_{\xg} n^{(0)}_{0}\right],
\end{align}
where $\mathcal{D}^\ast_{\xg}=\xg^{-2} \,\partial_{\xg} \xg^{4}$ and $\bar{n}^{\rm g}=n^{(1)}_0-\frac{2}{5} n^{\rm g}_1+\frac{1}{10}n^{(1)}_2-\frac{3}{70}n^{(1)}_3$.
Since all velocity-dependent terms appear as part of the dipole anisotropy, they do not affect the total energy exchange between electrons and photons. Upon averaging over all photon directions, $\vgh$, we are only left with 
\beal
\label{eq:BoltzEqlm_The_pert_first_final}
\left.\pAb{n^{(1)}_0(\xg,\vgh)}{\tau}\right|_{\The}&\approx
\frac{\The^{(0)}}{x_{\gamma}^{2}} \,\partial_{\xg} x_{\gamma}^4 
\left\{ \partial_{\xg} n^{(1)}_0 \!+\! \frac{\Tz}{\Te^{(0)}}  n^{(1)}_0 [ 1 + 2 n^{(0)}_{0} ]\right\}
\nonumber
\\
&\qquad
+\frac{N^{(1)}_{\rm e}}{N^{(0)}_{\rm e}} \left.\pAb{n^{(0)}_0(\xg,\vgh)}{\tau}\right|_{\The}
+ \The^{(1)}\mathcal{D}_{\xg} n^{(0)}_{0},
\end{align}
which clearly shows that the local equilibrium between photons and electrons is unaffected by the peculiar motion.


\subsubsection{Final expression for the angle-averaged collision term in second order perturbation theory}
\label{app:fin_exp_first}
We have now obtained all terms to consistently write down the final expression for the angle-averaged collision term in second order perturbation theory.
We first introduce the new variables
\bsub
\label{eq:y_Theta_10_gauge}
\beal
\label{eq:y_Theta_10_gauge_a}
3(\hat{\Theta}^{\rm g}_{1})^2&=3\hat{\Theta}_{1}^2-2\betac \hat{\Theta}_{1}+\frac{\betacsq}{3}
=\frac{(3\hat{\Theta}_{1}-\betac)^2}{3}
\\
\label{eq:y_Theta_10_gauge_b}
\ygl{l}&=\sum_{l'=l}^\infty  (2l'+1)\hat{\Theta}^2_{l'},
\end{align}
\esub
where in the definition of $\ygl{l}$ the term $l'=1$ is $3(\hat{\Theta}^{\rm g}_{1})^2$.
Adding Eq.~\eqref{eq:BoltzEqlm_The_pert_2_ang_av_exp}, \eqref{eq:BoltzEqlm_beta_pert_2_ang_av}, \eqref{eq:BoltzEqlm_beta2_pert_2_ang_av}, \eqref{eq:second_order_The_beta_result_fin},
and Eq.~\eqref{eq:betasq_gauge_fin_b}, we find
\beal
\label{eq:Final_EQ}
\left.\pAb{n^{(2)}_0(\xg)}{\tau}\right|_{\rm scatt}&
\approx
-\betac\hat{\Theta}^{\rm g}_{1}\mathcal{Y}_{\rm SZ}
+\frac{\Thz}{x_{\gamma}^{2}} \,\partial_{\xg} x_{\gamma}^4 
\left\{ \partial_{\xg} \Delta n^{(2)}_0 + \Delta  n^{(2)}_0 \mathcal{A}\right\}
\nonumber\\
&
+
\left[\The^{(2)}-\Thz[\Theta^{(2)}_0+\ygl{1} ] \right] \mathcal{Y}_{\rm SZ} 
\\[1mm]
&\quad
- \Thz\left[ \frac{\betacsq}{6} -\frac{17}{5} \hat{\Theta}^{\rm g}_{1}\betac \right] \mathcal{Y}_{\rm SZ}
%
- \Thz\left[ \Theta^{(2),\rm g}_{y,\geq 1} - \frac{16}{5} \hat{\Theta}^{\rm g}_{1}\betac \right] \mathcal{H} 
\nonumber\\[1mm]
&\qquad
- \Thz\left[ - \frac{6}{5}(\hat{\Theta}^{\rm g}_{1})^2 + \frac{1}{2}\hat{\Theta}^2_{2} - \frac{3}{10}\hat{\Theta}^2_{3}
- \frac{2}{5}\hat{\Theta}^{\rm g}_{1}\betac 
\right] \mathcal{E}
\nonumber
\end{align}
for the Boltzmann collision term of the average photon distribution in second order perturbation theory.

\subsubsection{Expression inside the local rest frame}
\label{app:fin_exp_first_rest}
To compute the corrections to the local electron temperature we need the result for the collision term inside the rest frame of the moving gas element.
To obtain the required expression, we start by integrating Eq.~\eqref{eq:BoltzEqlm_The} over all photon directions, which yields
\beal
%
\left.\pAb{n_0}{\tau}\right|_{\The}
&\approx
\The \mathcal{D}_{\xg} n_0 
+ \Thz \left( \mathcal{D}^\ast_{\xg} n_0
+ \mathcal{D}^\ast_{\xg} \left[ \bar{n}^2 \right]_0
+2 \left[ (n-\bar{n})\partial_{\xg} \xg^2 \bar{n} \right]_0 \right).
\nonumber
\end{align}
In second order perturbation theory we find
\beal
\pAb{n^{(2)}_0}{\tau}
&\approx
\The^{(0)} \mathcal{D}_{\xg} n_0^{(2)} + \Thz \mathcal{D}^\ast_{\xg} n_0^{(2)} \mathcal{A}
+\The^{(1)} \mathcal{D}_{\xg} n_0^{(1)} +\The^{(2)} \mathcal{D}_{\xg} n_0^{(0)} 
\nonumber\\
&\quad + \Thz\mathcal{D}^\ast_{\xg} \left[ (\bar{n}^{(1)})^2 \right]_0
+2 \Thz\left[ (n^{(1)}-\bar{n}^{(1)})\partial_{\xg} \xg^2 \bar{n}^{(1)} \right]_0
\nonumber\\[1mm]
&\approx
\Thz \mathcal{D}_{\xg} n_0^{(2)} + \Thz \mathcal{D}^\ast_{\xg} n_0^{(2)} \mathcal{A}
+\Thz (\Theta^{(1)}_0)^2 \mathcal{D}_{\xg} \mathcal{G}
+\The^{(2)} \mathcal{Y}_{\rm SZ} 
\nonumber\\
&
\quad + \Thz \left[ (\bar{\Theta}^{(1)})^2 \right]_0 \mathcal{D}^\ast_{\xg} \mathcal{G}^2
+2 \Thz\left[ (\Theta^{(1)}-\bar{\Theta}^{(1)})\bar{\Theta}^{(1)})\right]_0 
\mathcal{G}\,\partial_{\xg} \xg^2 \mathcal{G}.
\nonumber
\end{align}
In the second line we have used the fact that in first order perturbation theory no distortion is created unless there is significant energy release in zeroth order perturbation theory.

If we want to use this expression to compute $\The^{(2)}$ in the local rest frame we have to transform the CMB spectrum into this frame.
The anisotropies with $l>0$ only appear in first order perturbation theory, so that we only need to substitute $\hat{\Theta}^{(1)}_{1}\rightarrow \hat{\Theta}^{(1)}_{1} - \betac/3\equiv \hat{\Theta}^{\rm g}_{1}$.
Furthermore, $\mathcal{D}^\ast_{\xg} \mathcal{G}^2=2\mathcal{G}\,\partial_{\xg} \xg^2 \mathcal{G} = - \mathcal{E}$, so that the last two terms only give $-\Thz[ \Theta^{(1)}\bar{\Theta}^{(1)}]_0 \mathcal{E}$.

For the second order monopole spectrum one has to insert
\beal
%
n_0^{(2)}
&\approx \Delta n^{(2)}_0+\mathcal{G} \left(\Theta^{(2)}_0+[(\Theta^{(1)})^2]_0\right)+\frac{1}{2} \mathcal{Y}_{\rm SZ}[(\Theta^{(1)})^2]_0
\nonumber\\
&\qquad 
- \betac \left[\hat{\Theta}^{(1)}_1 - \frac{\betac}{6} \right] \left(\mathcal{G}+\mathcal{Y}_{\rm SZ}\right)
\nonumber\\
&=\Delta n^{(2)}_0+\mathcal{G} \left(\Theta^{(2)}_0+\ygl{0}\right)+\frac{1}{2} \mathcal{Y}_{\rm SZ}\,\ygl{0}
+ \betac \left[\hat{\Theta}^{(1)}_1 - \frac{\betac}{6} \right] \mathcal{G},
\nonumber
\end{align}
where $\ygl{l}$ is defined by Eq.~\eqref{eq:y_Theta_10_gauge_b}. 
The terms $\propto \betac$ and $\betacsq$ arise from the transformation of the CMB spectrum into the moving frame, Eq.~\eqref{eq:n_Lorentz_examples}.
Putting all this together and using $\mathcal{D}_{\xg} \mathcal{G}=\mathcal{Y}_{\rm SZ}+\mathcal{H}+\mathcal{E}$ we eventually find
\beal
\label{eq:integrated_Kompaneets}
\pAb{n^{(2)}_0}{\tau/\gamma_{\rm p}}
&\approx
\Thz \mathcal{D}_{\xg} \Delta n_0^{(2)} + \Thz \mathcal{D}^\ast_{\xg} \Delta n_0^{(2)} \mathcal{A}
+\left[\The^{(2)} -\Thz \left(\Theta^{(2)}_0 + \ygl{1}\right) \right] \mathcal{Y}_{\rm SZ} 
\nonumber\\
&
\qquad - \Thz\,\betac \left[\hat{\Theta}^{\rm g}_1 + \frac{\betac}{6} \right]\mathcal{Y}_{\rm SZ}
- \Thz\,\ygl{1}\mathcal{H}
\nonumber\\
& \qquad \qquad
-\Thz\left[ 
- \frac{6}{5}(\hat{\Theta}^{\rm g}_{1})^2 + \frac{1}{2}\hat{\Theta}^2_{2} - \frac{3}{10}\hat{\Theta}^2_{3}
\right]\mathcal{E}.
\end{align}
Comparing with Eq.~\eqref{eq:Final_EQ} we see that the terms $-\betac\hat{\Theta}^{\rm g}_{1}\mathcal{Y}_{\rm SZ} +\Thz \hat{\Theta}^{\rm g}_{1}\betac \left[ \frac{17}{5} \mathcal{Y}_{\rm SZ} + \frac{16}{5}  \mathcal{H}  + \frac{2}{5} \mathcal{E}\right]$ all arise from the transformation back into the Newtonian frame.

\section{Anisotropic Bremsstrahlung and double Compton emission} 
\label{sec:AnisoBRDC}
To treat the thermalization problem in the expanding Universe one also needs to include the effect of Bremsstrahlung and double Compton emission.
These processes are important at high redshifts and allow for adjusting the number of photons, while scattering only leads to redistribution of photons over energy.
In the standard thermalization problem perturbations are ignored and the collision term for BR and DC has the form \citep[see e.g.,][]{Illarionov1975b, Rybicki1979, Lightman1981, Burigana1991}
\beal
\label{eq:BR_DC}
\left.\pAb{n}{\tau}\right|_{\rm e/a}
&\approx\frac{K(\xg, \The)}{x^3_{\gamma}} \left\{ 1- n_0 [e^{\xg \varphi}-1]\right\},
\end{align}
with $\varphi =\Tz/\Te$. In this paper we do not attempt to derive a consistent second order collision term for BR and DC. However, assuming that the above expression describes the evolution of the local spectrum inside the rest frame of a moving gas element, we can account for the effects of Doppler boosting and aberration using Lorentz transformations.
It is expected that like in the case of Compton scattering also the dipole through octupole anisotropies will enter the full BR and DC collision term, but we neglect these correction terms here.

In zeroth order perturbation theory Eq.~\eqref{eq:BR_DC} just carries over. In first order perturbation theory the transformation of the CMB spectrum into the moving frame does not change anything, i.e., ${n'_0}\approx n_0$.
Using $\nu'=\gammac \nu (1-\betac\muc)$ we find
\beal
\label{eq:BR_DC_first}
\left.\pAb{n^{(1)}}{\tau}\right|_{\rm e/a}
&\approx
\left[\left(2-\xg\partial_{\xg} \ln K \right) \betac \muc + \The^{(1)}\partial_{\The} \ln K\right]
\left.\pAb{n^{(0)}}{\tau}\right|_{\rm e/a}
\nonumber\\
&\qquad
-\frac{K}{\xg^3} (e^{\xg}-1) \left[ n^{(1)}_0 -\betac\muc \xg\partial_{\xg} n^{(0)}_0 \right]
\nonumber\\
&\qquad\quad
+\frac{K}{\xg^2} n^{(0)}_0 \varphi ^{(0)}\left(\frac{\The^{(1)}}{\The^{(0)}} +\betac\muc\right) e^{\xg},
\end{align}
where $\left.\partial_\tau n^{(0)}\right|_{\rm e/a}$ is given by Eq.~\eqref{eq:BR_DC} with $n_0=n^{(0)}_0$, $\The=\The^{(0)}$ and $\varphi = \varphi ^{(0)}=\Tz/\Te^{(0)}$.
As this results shows, in first order perturbation theory the photon emission and absorption term acquires a dipolar contribution from the Lorentz transformation in and out of the moving frame.
Note however, that this is not the fully consistent expression, as we mentioned above.

In second order perturbation theory we directly integrate over all directions of the outgoing photon to obtain the averaged emission and absorption term.
We also assume that in zeroth order perturbation theory $\left.\partial_\tau n^{(0)}\right|_{\rm e/a}=0$.
With this we find
\beal
\label{eq:BR_DC_second}
\left.\pAb{n^{(2)}_0}{\tau}\right|_{\rm e/a}
%
&\approx
\frac{K}{\xg^3} (e^{\xg}-1) \mathcal{G} 
\left[ 
\frac{\The^{(2)}}{\Thz} -\left( \Theta^{(2)}_0 + \ygl{1}+\betac\hat{\Theta}^{\rm g}_1 
+\frac{\betacsq}{6}\right)
\right]
\nonumber\\
&\qquad
-\frac{K}{\xg^3} (e^{\xg}-1) \left[ \Delta n^{(2)}_0 + \frac{1}{2}\mathcal{Y}_{\rm SZ}\,\ygl{1}  \right],
\end{align}
where we have used Eq.~\eqref{eq:def_nlm_perturbed_c} and $\Delta n^{(2)}_0=\Delta n^{(2)}_{\rm av}+\Delta n^{(2)}_{\rm s, 0}$.

\end{appendix}

\small

\bibliographystyle{mn2e}
\bibliography{Lit}

\end{document}